%% file: thesis.tex
\documentclass[a4paper,11pt]{report}

\newcommand{\linespacing}{1.5}
\renewcommand{\baselinestretch}{\linespacing}

\usepackage{amssymb}
\usepackage{amsmath}
\usepackage{subfigure}
\usepackage{feynmf}
\usepackage{sparticles}
\usepackage{multirow}
\usepackage{txfonts}
\usepackage{color}

\usepackage{graphicx,color}
\usepackage{epstopdf}
\usepackage[british]{babel}
\usepackage[a4paper,top=2.5cm,bottom=2.5cm,left=4cm,right=2cm,headsep=10pt]{geometry}
\usepackage{fancyhdr}
\definecolor{red}{rgb}{0,0,0}
\definecolor{black}{rgb}{0,0,0}

\usepackage[pdfusetitle,bookmarksnumbered,plainpages=false]{hyperref}
\usepackage{backref}

\begin{document}

\pagenumbering{roman}

\thispagestyle{empty}
\begin{flushright}
\includegraphics[width=6cm]{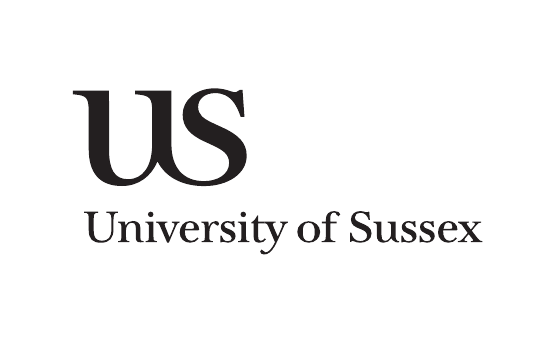}
\end{flushright}	
\vskip40mm
\begin{center}
\huge\textbf{Constraints on Models with Warped Extra Dimensions.}
\vskip7mm
\Large\textbf{Paul Archer}
\normalsize
\end{center}
\vfill
\begin{flushleft}
\large
Submitted for the degree of Doctor of Philosophy \\
University of Sussex	\\
June 2011
\end{flushleft}		

\chapter*{Declaration}
I hereby declare that this thesis has not been and will not be submitted in whole or in part to another University for the award of any other degree.
	

\vskip5mm
Signature:
\vskip20mm
Paul Archer

\thispagestyle{empty}
\newpage
\null\vskip10mm
\begin{center}
\large
\underline{UNIVERSITY OF SUSSEX}
\vskip20mm
\textsc{Paul Archer, Doctor of Philosophy}
\vskip20mm
\underline{\textsc{Constraints on Models with Warped Extra Dimensions.}}
\vskip0mm
\vskip20mm
\underline{\textsc{Summary}}
\vskip2mm
\end{center}
\renewcommand{\baselinestretch}{1.0}
\small\normalsize
It has been known for some time that warped extra dimensions offer a potential explanation of the large hierarchy that exists between the electroweak scale and the Planck scale. The majority of this work has focused on a five dimensional slice of AdS space. This thesis attempts to address the question, what possible spaces offer phenomenologically viable resolutions to this gauge hierarchy problem. In order for a space to offer a potential resolution to the hierarchy problem two conditions must be met: Firstly one should be able to demonstrate that the space can be stabilised such that a small effective electroweak scale (or large effective Planck scale) can be obtained. Secondly one must demonstrate that the space allows for a Kaluza Klein (KK) scale that is small enough such that one does not reintroduce a hierarchy in the effective theory. Here we focus on the second condition and examine the constraints, on the KK scale, coming from corrections to electroweak observables and flavour physics which arise when gauge fields propagate in an additional dimension. We study a large class of possible spaces of different geometries and dimensionalities. In five dimensions it is found that such constraints are generically large. In more than five dimensions it is found that a significant proportion of such spaces suffer from either a high density of KK modes or alternatively strongly coupled KK fields. The latter would not offer viable resolutions to the hierarchy problem. Models in which the Higgs propagates in the bulk are also studied, in the context of models with a `soft wall' and it is found these have significantly reduced constraints from flavour physics as well as a notion of a minimum fermion mass. 


\chapter*{Acknowledgements}
\renewcommand{\baselinestretch}{\linespacing}
\small\normalsize
First and foremost I would like to thank my supervisor, Stephan Huber, for tolerating many an ill timed knock on the door, for being able to continually offer a fresh perspective on a problem and above all for offering three and a half years of continued support which has made studying for this PhD a thoroughly enjoyable experience. My thanks should also go to my collaborators Sebastian J\"{a}ger and Michael Atkins for many an enjoyable and productive conversation. In addition I would like to thank the other members of the Sussex theory group, Mark Hindmarsh, David Bailin, Daniel Litim, Xavier Calmet, Oliver Roston, Christoph Rahmede, Anders Basboll and Jorge Camalich. As well as my fellow students Edouard Marchais, Kevin Falls, Dimitri Skliros, Kostas Nikolakopoulos, Miguel Sopena, Nina Gausmann, Rob Old, Ting-Cheng Yang, Dionysios Fragkakis and all of the member of the Sussex physics department. 

Outside of physics I am greatly indebted to my family. For without their support this simply would never have happened. Also my thanks should go to friends, housemates, climbers including Andy, Alex, Chris, Rachel, Miggy, Catherine, Tom, Jon and those who I haven't mentioned but should of. All have contributed to a very enjoyable few years in Brighton. 

This work has been supported by the Science and Technology Facilities Council.
       

\newpage
\pdfbookmark[0]{Contents}{contents_bookmark}
\tableofcontents
\listoffigures
\phantomsection
\addcontentsline{toc}{chapter}{List of Figures}
\newpage
\section*{Notation and Conventions}
Throughout this thesis we shall use the $(+-\dots-)$ metric convention. We shall also work with natural units with $\hbar=c=1$. We shall denote the full $5+\delta$ dimensional space time using $x^M$ with $M,N$ running over $0,\dots,3,5, \dots, 5+\delta$. The four large dimensions are denoted by $x^\mu$ with $\mu,\nu=0,\dots,3$. Flat tangent space is denoted using the upper case latin indices $A,B$, while colour indices are denoted using the lower case greek indices $\alpha,\beta$. The lower case latin indices, $i,j,k$ are used to simultaneously denote the flavour indices and run over the $\delta$ internal space time indices. We shall denote KK number by $m,n$ and note that in spaces of more than five dimensions $m,n$ will be vectors running from $n=(n_1,\dots,n_{1+\delta})$. The Einstein summation convention is also used throughout.

\phantomsection
\addcontentsline{toc}{chapter}{Notation and Conventions}
\newpage
\pagenumbering{arabic}

 \input{thesis_introduction}
 
\input{KK_ReductionChap}

\input{FlavourChap}

\input{EWChap}
\input{AlernativeSpaces}
\input{SWchap}

\input{Conclusion}

\clearpage
\phantomsection
\addcontentsline{toc}{chapter}{Bibliography}
\bibliography{bib}

\end{document}

%% file: thesis_introduction.tex
\chapter{Introduction}
\label{chap:intro}
Despite its experimental success the standard model (SM) of particle physics simply cannot be a fundamental theory. We will explain this statement in the following section but for now we will simply accept that the SM is a low energy effective theory.  This leads to two further questions, what is it a low energy effective theory of and at what energy scale does it break down? There are a number of compelling extensions to the SM that include extending the space-time symmetry or extending the gauge symmetry of the SM. However here we will consider what is arguably one of the simplest extensions, that of changing the dimensionality of space-time. It is important to remember that in quantum field theory the dimensionality of space-time is a free parameter where as in string theory (one of the few candidates for a fundamental theory) the dimensionality is fixed to either 10 or 26 dimensions. As we shall demonstrate extra dimensions, in one form or another, offer possible explanations of many of the existing problems of the SM and hence are worthy of investigation as a plausible extension of the SM.

The second question relating to the scale at which the SM breaks down is linked to the question of what has broken electroweak (EW) symmetry. If EW symmetry is broken by a SM Higgs then either nature has an inexplicable fine tuning or one requires new physics close to the EW scale in order to explain the associated gauge hierarchy problem. If EW symmetry is not broken by a SM Higgs then one would anticipate new physics at a scale close to the EW scale. Hence it is the gauge hierarchy problem that is really the reason for assuming new physics will be come apparent at the LHC in the next few years.

In considering more than four dimensions one is immediately confronted with two more questions. Firstly there appears to be considerable freedom in choosing the form of the extra dimension. In other words what should the geometry, dimensionality and topology of the extra dimensions be? Secondly since there is considerable empirical evidence to suggest that the universe, in the low energy regime, is effectively four dimensional, then clearly there must be some mass gap (or Kaluza Klein (KK) scale) below which the model is effectively four dimensional. The second question is then what is this scale? Here we will argue that if one seeks to resolve the gauge hierarchy problem using extra dimensions then this KK scale must be close to the EW scale.

This then bring us to the question that this thesis wishes to address.       
\begin{quotation}
If one assumes that the gauge hierarchy problem is resolved, via the gravitational red shifting (or warping) of an extra dimension, then what are the phenomenologically viable dimensionalities, geometries and topologies of the extra dimensions?
\end{quotation}
Note we have already restricted this question to extra dimensional models that resolve the hierarchy problem via warping \cite{Randall:1999ee} rather than those that resolve the problem via volume effects \cite{ArkaniHamed:1998rs, Antoniadis:1998ig}. We will inevitably add further amendments to this question later but first we should clarify the opening statement by listing some of the problems of the SM and looking at how extra dimensions offer potential resolutions. 

\section{The Problems of the Standard Model}
Many may consider the extension of space-time to more than four dimensions to be an extravagant extension of the SM. In order to justify such an addition here we will briefly outline some of the problems of the SM and their potential solutions in extra dimensional models. 
\begin{itemize}
  \item \textbf{The Gauge Hierarchy Problem} In the SM electroweak symmetry is broken at the EW scale ($v\approx 174$ GeV) by a scalar Higgs boson. If the Higgs potential is $V(\Phi)=-m_h^2|\Phi |^2+\lambda |\Phi |^4$ then the EW scale is dependent on the Higgs mass, $v^2=m_h^2/2\lambda$. The problem arises when one considers the one loop correction to the Higgs mass coming from its interaction with a fermion, $\mathcal{L}\supset-\lambda_\psi\Phi\bar{\psi}\psi$,
  \begin{displaymath}
\Delta m_h^2\sim-\frac{\lambda_\psi^2}{8\pi^2}\Lambda^2,
\end{displaymath}
  or its interaction with a gauge field $\mathcal{L}\supset g^2|\Phi |^2A_\mu A^\mu$ or its self interaction,
  \begin{displaymath}
\Delta m_h^2\sim\frac{g^2}{16\pi^2}\Lambda^2\quad\mbox{ or }\quad\Delta m_h^2\sim\frac{\lambda^2}{16\pi^2}\Lambda^2,
\end{displaymath} 
where $\Lambda$ is the UV cut off of the SM. In other words the Higgs mass (and hence the EW scale) is sensitive to both the masses of particles in the theory and the cut off of the theory. Bearing in mind that these are additive corrections, this is already slightly problematic when one takes $\Lambda\sim\mathcal{O}(\mbox{TeV})$. However if one assumes that the SM is valid all the way to the Planck scale (at which gravitational effects become important) then one can see that the Higgs mass would be quadratically sensitive to $M_{\rm{Pl}}\sim 10^{18}$ GeV. Naively this looks as though there is an immense fine tuning at the heart of the SM. We shall look at the extra dimensional resolution of this problem in the next section.  
  \item \textbf{The Fermion Mass Hierarchy Problem} The SM fermions are observed to have masses ranging from the top mass ($m_t\approx 172$ GeV \cite{Nakamura:2010zzi}) to the neutrino masses with a combined mass of $\color{red}\sum_\nu m_\nu< 1.19$ eV \cite{Hannestad:2010yi}. In the SM the neutrino masses cannot be accounted for, while the large range in the other fermion masses can only be explained with an ad hoc hierarchy in the Yukawa couplings. Again we will go through the extra dimensional explanation of this in more detail in section \ref{FlavourChap}. 
  \item \textbf{Baryogenesis} In order for a model to explain how the asymmetry between matter and antimatter developed in the early universe it is necessary that it satisfies three conditions \cite {Sakharov:1967dj}. It must have interactions in which baryon number is not conserved. It must have CP violating processes and it must have interactions occurring out of thermal equilibrium. It is thought that the SM does not provide sufficient CP violation. As we shall discuss in section \ref{FlavourChap} when considering fermions propagating in extra dimensions there are a number of additional sources of CP violation.
  \item \textbf{Dark Energy} The SM cannot explain the late time acceleration of the universe. Once again there is considerable work being done on the 4D effective cosmological constant that can arise when one considers a brane in a higher dimensional spacetime. See \cite{Langlois:2002bb} for a review.
  \item \textbf{Dark Matter} The SM has no dark matter candidate. Where \color{red} as \color{black} extra dimensional scenarios, in which the extra dimensions are not cut off by a brane, can often give rise to massive stable particles that can act as dark matter candidates. See section \ref{sect:orthog}.
  \item \textbf{Gravity} The SM does not include gravity. One of the few candidates for a quantum theory of gravity is string theory which, as already mentioned, necessarily requires extra dimensions.   
\end{itemize}
This is by no means a complete list. One could, for example, include questions related to why is there are three generations of fermions or include the strong CP problem in QCD or the Landau pole in QED. Here we merely wish to point out that extra dimensional scenarios offer plausible explanations to many of the problems of the SM.  

\section{Warped Extra Dimensions and the Gauge Hierarchy Problem}\label{sect:hierarchyResol}
Many of the problems in the previous section can be resolved for a large range of KK scales. However a resolution of the gauge hierarchy problem necessarily requires that the KK scale be close to that of the EW scale. Hence the resolution of this problem is sensitive to many of the constraints which we shall consider in this thesis. For the same reason any resolution of the gauge hierarchy problem should be testable, in the next few years, at the LHC.

Central to the gauge hierarchy problem is the fact that the Higgs mass is quadratically sensitive to the scale of new physics. Hence the problem can potentially be removed if the scale of new physics is the same as the EW scale. Or in other words if all the dimensionful parameters of a theory exist at the same scale then the EW scale will need no fine tuning. At first sight this does not look possible since we observe the EW scale to be $\sim 200$ GeV while the other known dimensionful parameter is the Planck mass $M_{\rm{Pl}}\sim 10^{18}$ GeV. However from an extra dimensional perspective these are four dimensional effective parameters which have been scaled down from some fundamental bare scale.    

To see how these parameters scale we need to consider a specific space. When the Randall and Sundrum (RS) model was originally proposed in \cite{Randall:1999ee} they considered a slice of AdS${}_5$, bounded by two 3 branes at $r=r_{\rm{ir}}$ and $r=r_{\rm{uv}}$ and described by    
\begin{equation}
\label{RSMetric}
ds^2=e^{-2kr}\eta_{\mu\nu}dx^\mu dx^\nu-dr^2,
\end{equation}
where $\eta_{\mu\nu}=\mbox{diag}(+---)$. However in order to address the question posed at the beginning of this section here we wish to consider as generic a space as possible. Having said that considering completely generic spaces not only rapidly leads to equations with no analytical solutions, it also makes it difficult to keep track of the source of the scaling present in the model. We are also partly motivated by the spaces of relevance to the AdS/CFT correspondence that are discussed in the following section. Hence here we shall restrict our study to spaces in which the four large dimensions are warped with respect to (w.r.t) a single `preferred' direction $r$. Hence we shall considers $5+\delta$ dimensional spaces described by    
\begin{equation}
\label{GenMetric}
ds^2=a^2(r)\eta_{\mu\nu}dx^\mu dx^\nu-b^2(r)dr^2-c^2(r)d\Omega_\delta^2,
\end{equation}
 with an internal manifold described by $d\Omega_\delta^2=\gamma_{ij}(\phi)\;d\phi^id\phi^j$, with $i,j$ running from $1\dots \delta$. We will also define the boundaries of this space to be $\phi^i_{\alpha}$ and $\phi^i_{\beta}$. Although for the moment we will not specify whether this boundary has arisen via compactification, or corresponds to an orbifold fixed point or a brane cutting the space off. Note one can always set $b=1$ with a coordinate transformation
 \begin{displaymath}
r\rightarrow\tilde{r}=\int_c^r b(\hat{r})d\hat{r},
\end{displaymath}     
although by not doing so one can analytically represent a greater range of spaces. 

If we now consider the two dimensionful parameters, mentioned above, which in the $5+\delta$ dimensions, with the Higgs propagating in the bulk, will be given by
\begin{displaymath}
S=\int \;d^{1+\delta}x\;\sqrt{G} \left (\frac{1}{2}M_{\rm{fund}}^{3+\delta}\mathcal{R}+|D_MH|^2-m_0^2H^2\right ).
\end{displaymath}
Firstly after taking account of the two derivatives in the Ricci scalar then the four dimensional effective Planck mass will be given by the effective coupling of the flat graviton zero mode, see section \ref{sect:GravKK},  
\begin{equation}
\label{MPlanck}
M_{\rm{Pl}}^2=M_{\rm{fund}}^{3+\delta}\int d^{1+\delta}x\; a^2bc^\delta\sqrt{\gamma}.
\end{equation}
Secondly while the four dimensional kinetic term term will scale as $a^2bc^\delta\sqrt{\gamma}|D_\mu H|^2$ the mass term will scale as $a^4bc^\delta\sqrt{\gamma}m_0^2H^2$. Hence after a suitable field redefinition the four dimensional effective Higgs mass \color{red} term  \color{black} will be given by
\begin{equation}
\label{HiggsMassGen}
m_h^2\int d^{1+\delta}x\; H^2=m_0^2\int d^{1+\delta}x\; a^2H^2.
\end{equation}
Note, after a KK decomposition, this expression would typically not be diagonal w.r.t KK number. It should also be pointed out that radiative corrections to $m_0$ will also scale in the same way. Although one could ask the question what should the cut off of the higher dimensional theory be? A cut off that is significantly larger than $m_0$ can potentially lead to the reintroduction of a `little hierarchy'. But if $m_0\sim M_{\rm{Pl}}$ then one is in the regime of quantum gravity and such a question cannot be answered at this time.  However if one assumes the cut off is not too large then the gauge hierarchy problem can potentially be resolved when $m_0\sim M_{\rm{fund}}$.

Clearly there are a number of ways this can be achieved. One possibility, considered in \cite{ArkaniHamed:1998rs, Antoniadis:1998ig}, is to use the volume of an extra dimension with $a=1$ to suppress $M_{\rm{fund}}$ down to the EW scale. However this typically requires that the model has parameters that are not at the scale of the rest of the theory. In particular the inverse of the radius of the extra dimension is often not at the EW scale. Hence we shall not consider this option here.   

An arguably more compelling idea is to consider spaces in which all the dimensionful parameters (including both the curvature and radii of the space\footnote{In the RS model it is found that the proper distance between between the branes, $l=\int_{r_{\rm{ir}}}^{r_{\rm{uv}}}\; b dr$, differs from the curvature, $k$, by a factor of $\ln ( \Omega)$. If the KK mass scale is given by $M_{\rm{KK}}=\frac{k}{\Omega}$ then $l=\frac{\ln ( \Omega)}{M_{\rm{KK}}\Omega}$ and hence if the KK scale differs significanly from the EW scale then \color{red} $l \nsim M_{\rm{fund}}^{-1}$\color{black}. A similar situation arises in all the backgrounds considered in this thesis.}) are at the same order of magnitude, which in turn implies
\begin{displaymath}
M_{\rm{fund}}^{1+\delta}\int d^{1+\delta}x a^2bc^\delta\sqrt{\gamma}\sim \mathcal{O}(1).
\end{displaymath}
This then implies that only via the warp factor $a$ can the hierarchy problem be resolved. The idea, introduced in \cite{Randall:1999ee}, is to have the Higgs localised at the tip of a warped throat i.e. $H\delta(r-r_{\rm{ir}})$ and so $m_h^2=a(r_{\rm{ir}})^2m_0^2$. Hence the hierarchy problem can be resolved when the warp factor is
\begin{equation}
\label{Warpfactor}
\Omega\equiv a(r_{\rm{ir}})^{-1}\sim 10^{15}.
\end{equation}
We shall repeatedly use this definition of the warp factor throughout this thesis. This is just one aspect of resolving the gauge hierarchy problem. The second aspect is demonstrating that the space can be stabilised, such that a large warp factor can be obtained, with out fine tuning. This was demonstrated for the RS model in \cite{Goldberger:1999uk}. Here we shall not consider aspects of stabilisation. While, it is arguably reasonable to assume that spaces with no large radii and no large contributions to the curvature can always be stabilised with out fine tuning, it is none the less a significant assumption.

The third aspect, which will be the central topic of this thesis, is related to the size of the KK scale. As we shall discuss in the next chapter, in any phenomenologically viable extra dimensional scenario there is a KK scale related to the curvature and radii of the extra dimension. The four dimensional effective Higgs vacuum expectation value (VEV) will be quadratically sensitive to this scale and hence if this scale is too large then one risks reintroducing the hierarchy problem. On the other hand one relies on this scale to suppress extra dimensional contributions to existing observables. In other words, in order for warped extra dimensions to offer a viable resolution of the gauge hierarchy problem then one must explain why we have seen no sign of them with out resorting to a very large KK scale.  
  
\section{The AdS/CFT Conjecture}
Before we address this problem in greater detail we should first offer further explanation as to why we are considering warped spaces and in particular spaces which are related to AdS${}_5$. In 1997 there was a surge of interest in AdS${}_5$ following the proposal of the AdS/conformal field theory (CFT) conjecture \cite{Maldacena:1997re}. The original conjecture states that solutions of type IIb string theory on AdS${}_5\times S^5$ are dual to a four dimensional $\mathcal{N}=4$ super Yang-Mills theory $\color{red}\mathrm{SU}(N)$. Although this dualism has still not been proven, a considerable number of quantities have been computed on each side of the dualism and found to be in agreement \cite{Aharony:1999ti, Polchinski:2010hw}. In particular one can match the symmetries, as well as the spectra of possible states. Also bulk fields, $\Phi$, on the gravity side will be dual to operators, $\mathcal{O}$ of the conformal field theory. The value of the bulk field at the AdS boundary\footnote{An interesting feature of AdS space is that it has a boundary. In particular if the space is described by $ds^2=\frac{R^2}{z^2}\left (\eta^{\mu\nu}dx_\mu dx_\nu-dz^2\right )$ with $0<z<\infty$ then light can reach the boundary at $z=0$ in a finite period of time, although massive particles cannot. Hence boundary condition must be properly defined here.}   will act as a source for the CFT operator, i.e. with $\Phi |_{\rm{AdS\; boundary}}=\phi_0$ then
\begin{displaymath}
\int \mathcal{D}\phi_{\rm{CFT}}e^{-S_{\rm{CFT}}[\phi_{\rm{CFT}}] -\int d^4x\phi_0\mathcal{O}}=\int_{\phi_0}\mathcal{D}\Phi e^{-S_{\rm{Bulk}}[\Phi]}.
\end{displaymath}
Hence one can also match the amplitudes of correlation functions across the duality. Further still the KK modes of these bulk fields are then dual to bound states of the CFT.

Undoubtably this duality influenced the proposal of the RS model, however there are a number of significant differences between the RS model and the scenario described above. Firstly in the RS model the space is cut off by two branes in the IR and UV. This is supposed to be dual to cut offs imposed on the CFT which in turn break the conformal symmetry \cite{ArkaniHamed:2000ds, Rattazzi:2000hs}. A feature of cutting off the space with two branes is that the fifth component of the metric gives rise to a scalar field called the radion. This radion typically gains a mass once the inter-brane separation has been stabilised \cite{Goldberger:1999uk} and is interpreted as the pseudo-Nambu-Goldstone boson of the breaking of the conformal symmetry \cite{Gherghetta:2010cj}. It is also interesting to note that the Higgs, being localised on the IR brane, would be dual to a bound state of a near conformal field theory. In other words the RS model is very closely related to walking technicolour.  

Another significant difference with the RS model is that it has no $S^5$ internal manifold. In the original AdS/CFT the volume of the $S^5$ was important in determining the number of colours in the CFT. In particular the ten dimensional effective theory of type IIb string theory includes a five form Ramond-Ramond field strength tensor $F_5$, while in the dual theory the number of colours ($N$) is given by the flux, of this field, through the internal space \cite{Aharony:1999ti},
\begin{displaymath}
\int _{S^5} F_5=N.
\end{displaymath}              
Hence it is not really clear what field theory the RS model is dual to. 

One can then ask what happens when one changes the number of colours? In \cite{Klebanov:2000hb, Klebanov:2000nc}, $\color{red}\mathrm{SU}(2M)\times \color{red}\mathrm{SU}(M)$ \color{red}is considered \color{black} and the numbers of colours \color{red}is \color{black} reduced by repeatedly using the Seiberg duality in a `duality cascade' \cite{Strassler:2005qs}. The result on the gravity side \color{red}is \color{black}  a `deformed conifold' in which the radius of the internal space is shrinking towards the IR. We shall consider this space in more detail in section \ref{sect:KlebStras}. Work has been done in extending these `duality cascades' to RS-like scenarios \cite{Abel:2010vb} although the full phenomenological implications have yet to be properly explored.

The point is that, while on one hand warped extra dimensions potentially offer a neat resolution to the gauge hierarchy problem and equivalently a new way of exploring technicolour. On the other hand, just as in technicolour models there are unknowns relating to the number of colours and number of flavours\color{red}, In \color{black} warped extra dimensions there are unknowns relating to the geometry of the space. For the bulk of this thesis we shall not be too concerned with what field theory is dual to the particular space we are studying but rather consider the spaces purely from the extra dimensional perspective. Hence this thesis will focus on `bottom up' deformations of the RS model in an attempt \color{red} to \color{black} investigate what models can be used to resolve the gauge hierarchy problem. 
        
\section{Outline}
As mentioned here we shall focus on the constraints on the KK scale coming from existing observables. In particular constraints coming from EW observables and flavour physics. The approach will be to look at corrections to such observables arising from models with extra dimensions, with the emphasis on trying to consider as generic an extra dimension as possible. In section \ref{chap:KKRed.} we shall lay much of the groundwork by carrying out the KK decomposition for the scalar field, gauge field, fermion field and graviton. We shall then move on, in section \ref{FlavourChap}, to consider the description of flavour that exists in models with warped extra dimensions and the associated constraints from flavour changing neutral currents (FCNC's). Likewise in section \ref{EWChap} we shall consider the constraints coming from EW observables. In section \ref{AlernativeSpaces} we shall apply these generic results to different possible spaces. Starting with five dimensions we shall demonstrate why large constraints from EW observables are a generic feature of models that resolve the hierarchy problem. We will then consider three examples of possible spaces, firstly a class of spaces where we have simply assumed that the internal space is scaling as a power law, secondly a particular string theory solution, which was mentioned above \cite{Klebanov:2000nc} and thirdly a class of spaces that arise as the classical solutions of the Einstein equations with an anisotropic bulk cosmological constant. In section \ref{SWmodel} we shall consider deforming the RS model by changing the cut off of the space. This critically allows for the Higgs to propagate in the extra dimension and still resolve the hierarchy problem. We conclude in section \ref{chap:conc}.     

%% file: KK_ReductionChap.tex
\chapter{The Kaluza-Klein Reduction}
\label{chap:KKRed.}

The principle focus of this thesis is to assess whether or not models with warped extra dimensions offer a viable resolution to many of the existing problems of the SM, in particular the gauge hierarchy problem. The approach will simply be to match the low energy effective theory to existing experimental results and hence arrive at the allowed range for the free parameters in the higher dimension theory. One of the central tools required for this is the KK decomposition, that allows the field to be split into that which is dependent on the four large dimensions and that which is dependent on the extra dimensions. Hence the extra dimensions can be integrated out to a four dimensional effective theory. In this chapter, in order to lay the groundwork for later studies, we will go through the KK decomposition for the particles considered in this thesis.  

\section{The Scalar Field}
We will start by considering the simplest case of a scalar field propagating in the extra dimension described by Eq. \ref{GenMetric},
\begin{equation}
\label{ }
S=\int d^{5+\delta}x\; \sqrt{G} \left (|\partial_M\Phi|^2\color{red} -\color{black}V(\Phi)\right ).
\end{equation}
The equations of motion are then given by
\begin{equation}
\label{ }
a^{-2}\partial_\mu\partial^\mu\Phi-a^{-4}bc^{-\delta}\partial_5 (a^4b^{-1}c^\delta\partial_5\Phi)-\frac{c^{-2}}{\sqrt{\gamma}}\partial_i(\sqrt{\gamma}\gamma^{ij}\partial_j\Phi)\color{red}+\color{black}\partial_\Phi V(\Phi)=0.
\end{equation}
If we now use the Klein Gordon equation to define the, on shell, four dimensional effective mass, $\partial_\mu\partial^\mu \Phi=-m^2\Phi$, then one can see that this effective mass is nothing but the eigenvalues of the Laplacian of the internal space,
\begin{equation}
\label{ScalarEOM}
\Phi^{\prime\prime}+\frac{(a^4b^{-1}c^\delta)^{\prime}}{a^4b^{-1}c^\delta}\Phi^{\prime}+\frac{b^2}{c^2}\frac{1}{\sqrt{\gamma}}\partial_i(\sqrt{\gamma}\gamma^{ij}\partial_j\Phi)\color{red}-\color{black}b^2\partial_\Phi V(\Phi)+\frac{b^2}{a^2}m^2\Phi=0.
\end{equation}
Where we have now denoted derivatives w.r.t $r$ as $^{\prime}$. When writing down the equations of motion it is important not to forget the accompanying boundary terms which, in this case, require that
\begin{equation}
\label{ScalarBCS}
\left [\Phi a^4b^{-1}c^\delta\sqrt{\gamma}\Phi^{\prime}\right ]_{r=r_{\rm{ir}}}^{r=r_{\rm{uv}}}+\sum_{i=1}^{\delta}\left [\Phi a^4bc^{\delta-2}\sqrt{\gamma}\gamma^{ij}\partial_j\Phi\right ]_{\phi_i=\phi^i_{\alpha }}^{\phi_i=\phi^i_{\beta}}=0.
\end{equation}
Hence one is forced to impose either Neumann (NBC's) or Dirichlet boundary conditions (DBC's). In order to now solve Eq. \ref{ScalarEOM}, one would carry out a KK decomposition or separation of variables,
\begin{equation}
\label{ }
\Phi=\sum_n\;f^{(\Phi)}_n(r)\;\Theta^{(\Phi)}_n(\phi_1,\dots,\phi_\delta)\;\Phi^{(n)}(x_\mu).
\end{equation} 
\color{red} Here the KK number subscript $n$ is shorthand notation for $1+\delta$ subscripts, $n=(n_1,n_2,\dots n_{1+\delta})$.  \color{black} It is important to realise that this is a completely arbitrary choice and one is free to decompose the field however one likes. However some decompositions are clearly more sensible than others. In particular if a decomposition is chosen such that
\begin{equation}
\label{ScalarOrthog}
\int d^{1+\delta}x\; a^2bc^{\delta}\sqrt{\gamma}f^{(\Phi)}_nf^{(\Phi)}_m\Theta^{(\Phi)}_n\Theta^{(\Phi)}_m=\delta_{nm},
\end{equation}
then the kinetic term and mass term would both be orthogonal w.r.t the KK number, $n$. Hence, if one neglects interaction terms (i.e. taking $V(\Phi)=\color{red} M^2\Phi^2$), the four dimensional effective action is then given by
\begin{equation}
\label{ }
S=\int d^4x \sum_n\left (|\partial_\mu\Phi^{(n)}|^2-m_n^2\Phi^{(n)\;2}\right )
\end{equation} 
where
\begin{equation}
\label{ScalarEqn}
f^{(\Phi)\;\prime\prime}_n+\frac{(a^4b^{-1}c^\delta)^{\prime}}{a^4b^{-1}c^\delta}f^{(\Phi)\;\prime}_n-b^2M^2f^{(\Phi)}_n-\frac{b^2}{c^2}\alpha_nf^{(\Phi)}_n+m_n^2\frac{b^2}{a^2}f^{(\Phi)}_n=0
\end{equation}
and
\begin{equation*}
\label{ }
-\frac{1}{\sqrt{\gamma}}\partial_i(\sqrt{\gamma}\gamma^{ij}\partial_j\Theta^{(\Phi)}_n)=\alpha_n\Theta^{(\Phi)}_n.
\end{equation*}
In other words, when one reduces the higher dimensional theory to a four dimensional effective theory one obtains an infinite `KK tower' of particles of masses $m_n$. This tower will include all possible eigenfunctions of Eq. \ref{ScalarEqn}  and hence the multiplicity of KK particles will increase with dimensionality. Typically, with a KK decomposition defined by Eq. \ref{ScalarOrthog}, interaction terms in the higher dimensional theory will not be orthogonal w.r.t KK number and hence will mix particles of different KK number. 

If we now take the specific example of the five dimensional RS model (Eq. \ref{RSMetric}), then with $a(r)=e^{-kr}$ and $b(r)=1$, Eq. \ref{ScalarEqn} reduces to
\begin{equation*}
\label{ }
f^{(\Phi)\;\prime\prime}_n-4kf^{(\Phi)\;\prime}_n-M^2f^{(\Phi)}_n+e^{2kr}m_n^2f^{(\Phi)}_n=0.
\end{equation*}  
 This can be solved to give
 \begin{equation*}
\label{ }
f^{(\Phi)}_n=Ne^{2kr}\left (\mathbf{J}_\nu\Big(\frac{m_ne^{kr}}{k}\Big )+\beta\mathbf{Y}_\nu\Big(\frac{m_ne^{kr}}{k}\Big )\right ),
\end{equation*}
where $\nu=\frac{\sqrt{4k^2+M^2}}{k}$. Once $N$ and $\beta$ have been fixed by imposing the boundary conditions and the orthogonality condition (Eq. \ref{ScalarOrthog}) then it is found that the KK masses are given by
\begin{equation}
\label{RSscalarmass}
m_n\sim X_n\frac{k}{e^{kR}}=X_n\frac{k}{\Omega}
\end{equation}
where $X_n$ is typically an order one number dependent on the boundary conditions and the roots of the Bessel functions. For example, with NBC's and a large warp factor, $X_n\sim (n+\frac{2\nu-3}{4})$ \cite{Gherghetta:2000qt}. The point is that there is a natural scale, related to both the curvature and the size of the extra dimension, that determines the mass of these KK particles.  Hence when we talk of a low energy theory we are referring to energies lower than this KK scale. In other words, at energies much lower than this scale, the theory will appear four dimensional. If the extra dimension was too large, the KK scale would be reduced to an energy level that is easily probed and the model would be in conflict with everyday observation.

The situation becomes more interesting when one considers more than one additional dimension. Since now there exists the possibility of multiple KK scales all determined by the radii and geometries of the extra spaces. For example, if the $\delta$ additional dimensions were compactified over a $\delta$ dimensional sphere of radius $R_\phi$ then in Eq. \ref{ScalarEqn} \cite{sandberg1978}
\begin{displaymath}
\alpha_n=\frac{l(l+\delta-1)}{R_\phi^2}
\end{displaymath}   
and hence the spacing in $m_n$ will be dependent on two scales, $\frac{k}{\Omega}$ and $\frac{1}{R_\phi}$. We shall return to this point with a more concrete example in section \ref{AnIsoLamb}. Having covered the basic features of KK reduction for the scalar field we will now consider gauge fields propagating in the extra dimensions.

\section{The Gauge Field}
One has to be a little careful when considering bulk gauge fields in that, when writing down the Lagrangian, it is necessary to account for the unphysical degrees of freedom that can be removed by gauge transformations. In other words it is necessary to include a gauge fixing term. Before we consider the $D$ dimensional case it is useful to first look at gauge fields in five dimensions.

\subsection{Gauge Fields in Five Dimensions}\label{sect:5dGauge}
We start by considering a $U(1)$ gauge field in five dimensions described by 
\begin{equation*}
\label{ }
S=\int d^5x\;\sqrt{G}\left ( -\frac{1}{4}F_{MN}F^{MN}\right )
\end{equation*} 
which after expanding and integrating by parts gives
\begin{equation*}
\label{ }
S=\int d^5x \left (-\frac{b}{4}F_{\mu\nu}F^{\mu\nu}-\frac{1}{2}\partial_5(a^2b^{-1}\partial_5A_\mu)A^\mu-\partial_\mu A^\mu\partial_5(a^2b^{-1}A_5)+\frac{a^2}{2b}\partial_\mu A_5\partial^\mu A^5\right ).
\end{equation*}      
At first sight this looks problematic since it appears that the fifth component of the gauge field will appear, in the four dimensional effective theory,  as a massless scalar field mixing with the four dimensional gauge field through a term with a single derivative. However such mixing can be cancelled with the inclusion with a gauge fixing term. In particular, working in the $\rm{R}_\xi$ gauge, with the gauge fixing term 
\begin{equation}
\label{ }
\mathcal{L}_{G.F.}=-\frac{1}{2\xi}b\left ( \partial_\mu A^\mu-\xi b^{-1}\;\partial_5(a^2b^{-1}A_5)\right)^2,
\end{equation} 
 the expanded action can now be given as
 \begin{eqnarray}
S+S_{G.F.}=\int d^5x \Bigg (-\frac{b}{4}F_{\mu\nu}F^{\mu\nu}-\frac{b}{2\xi}(\partial_\mu A^\mu)^2-\frac{1}{2}\partial_5(a^2b^{-1}\partial_5A_\mu)A^\mu\nonumber\\
+\frac{a^2}{2b}\partial_\mu A_5\partial^\mu A^5-\xi\frac{1}{2b}\left (\partial_5(a^2b^{-1}A_5)\right )^2\Bigg ).\label{5DexpanGauge}
\end{eqnarray}
Now the last term in Eq. \ref{5DexpanGauge} would act as a mass term for the scalar field but it is completely gauge dependent. If one was to move into the unitary gauge ($\xi\rightarrow\infty$) then these scalars would become infinitely heavy and hence should be considered unphysical. In other word one can always carry out a gauge transformation such that $A_5$ is constant. Note one can carry out an analogous gauge fixing to remove any boundary term involving $A_5$ \cite{Csaki:2005vy}. 

Further still, when one reduces to a four dimensional theory, the third term in Eq. \ref{5DexpanGauge} will appear as a mass term for the gauge fields and hence break the gauge symmetry. In particular if one makes the KK decomposition
\begin{equation}
\label{5DGaugeKKdecomp}
A_\mu=\sum_{n}f_{n}(r)A_\mu^{(n)}(x^\mu),
\end{equation}  
 such that 
 \begin{equation}
\label{ }
\int dr\; bf_{n}f_{m}=\delta_{nm},
\end{equation}
then the four dimensional KK gauge fields will gain masses, $m_n$, where
\begin{equation}
\label{5DGaugeEqn}
f_{n}^{\prime\prime}+\frac{(a^2b^{-1})^{\prime}}{a^2b^{-1}}f_{n}^{\prime}+m_n^2\frac{b^2}{a^2}f_{n}=0.
\end{equation}
What we are essentially seeing, by going from five dimensions to four dimensions, is the Higgs mechanism as work. I.e. the 5D gauge symmetry is broken, by the boundary conditions, giving rise to a tower of unphysical scalar fields (or goldstone bosons), $A_5$, which are then `eaten' by the gauge KK modes, which in turn gives them masses. 

The possible exception to this is the zero mode, i.e. the KK mode corresponding to $m_0=0$, in which the gauge symmetry is not broken in the four dimensional effective theory. In this case, whether or not the $A_5^{(0)}$ scalar can be gauged away is dependent on the boundary conditions \cite{Kubo:2001zc, Hall:2001tn}.  This has led to a considerable amount of work being done on ascribing the $A_5^{(0)}$ component of a non abelian gauge field to the Higgs, that then breaks electroweak symmetry, see for example \cite{Oda:2004rm, Cacciapaglia:2005da, Hosotani:2006qp, Medina:2007hz}. Although the idea of using the boundary conditions of an extra dimension to dynamically break a gauge symmetry is much older \cite{Manton79, RandjbarDaemi83, Hosotani83a, Hosotani83b}.  

\subsection{Gauge Fields in $D$ Dimensions}
Having looked at the five dimensional case we will now extend this to look at an abelian field in $D$
dimensions
\begin{equation}
\label{AbGaugeLag}
S=\int d^{D}x\;\sqrt{G}\left ( -\frac{1}{4}F_{MN}F^{MN}\right ).
\end{equation}    
For ease of notation, we will briefly work in a more generic metric of the form
\begin{equation}
\label{mostGenMet }
ds^2=a^2(\phi_i)\eta_{\mu\nu}dx^\mu dx^\nu-\gamma_{ij}d\phi^id\phi^j,
\end{equation} 
where now $i,j$ run from $5$ to $D$. Eq. \ref{AbGaugeLag} can then be expanded to
\begin{eqnarray}
S=\int d^Dx\;\bigg ( -\frac{1}{4}\sqrt{\gamma}F_{\mu\nu}F^{\mu\nu}-\frac{1}{2}A^\mu\partial_i(\sqrt{\gamma}a^2\gamma^{ij}\partial_jA_\mu)-\partial_i(\sqrt{\gamma}\gamma^{ij}a^2A_j)\partial_\mu A^\mu\nonumber\\+\frac{1}{2}\sqrt{\gamma}a^2\gamma^{ij}(\partial_\mu A_i)(\partial^\mu A_j)^\dag-\frac{1}{4}a^4\gamma^{ij}\gamma^{kl}F_{ik}F_{jl}\bigg ).
\end{eqnarray}
As in the five dimensional case it seems that the extra components of the gauge field would appear, in the four dimensional theory as $D-4$ scalar field mixing with the gauge fields. Once again the terms linear in $\partial_\mu A^\mu$ can be cancelled with the gauge fixing term,
\begin{equation}
\label{ GFterm}
S_{G.F.}=\int d^Dx\; -\frac{1}{2\xi}\sqrt{\gamma}\left (\partial_\mu A^\mu-\frac{\xi}{\sqrt{\gamma}}\partial_i(\sqrt{\gamma}a^2\gamma^{ij}\partial_j A_\mu)\right )^2,
\end{equation} 
yielding the action
\begin{eqnarray}
S+S_{G.F.}=\int d^Dx\; \bigg [\sqrt{\gamma}\left (-\frac{1}{4}F_{\mu\nu}F^{\mu\nu}-\frac{1}{2\xi}(\partial_\mu A^\mu)^2\right )-\frac{1}{2}A^\mu \partial_i\left (\sqrt{\gamma}a^2\gamma^{ij}\partial_j A_\mu\right ) \nonumber\\
+\frac{1}{2}\sqrt{\gamma}a^2\gamma^{ij}(\partial_\mu A_i)(\partial^\mu A_j)^\dag-\frac{1}{4}a^4\gamma^{ij}\gamma^{kl}F_{ik}F_{jl}-\frac{\xi}{2\sqrt{\gamma}}\partial_i\left (\sqrt{\gamma}\gamma^{ij}a^2A_j\right )\partial_k\left (\sqrt{\gamma}\gamma^{kl}a^2A_l\right )\bigg ].\nonumber
\end{eqnarray}
If we once again define the four dimensional effective mass, of the gauge field, by the equations of motion,
\begin{displaymath}
\partial^\nu F_{\nu\mu}+\frac{1}{\xi}\partial_\mu (\partial_\nu A^\nu)=-m^2 A_\mu,
\end{displaymath}
then the equations of motion for the four dimensional gauge field are then given by
\begin{equation}
\label{gaugeEOM}
\frac{1}{\sqrt{\gamma}}\partial_i\left (\sqrt{\gamma} a^2\gamma^{ij}\partial_j A_\mu\right )+m^2 A_\mu=0.
\end{equation}
As we shall see in the following sections, the low energy phenomenology and corrections to the SM are typically dominated by the exchange of KK gauge fields. The effective couplings and masses of these fields will be largely determined by this equation.  Hence we shall see that this equation is critical in determining which spaces offer a viable resolution to the gauge hierarchy problem. Although before we look at this equation in more detail we should include the other $D-4$ equations of motion corresponding to the additional components of the gauge field or the `gauge scalars'.
\begin{equation}
\label{Eq.gaugescalar}
\partial_\mu\partial^\mu A_i=\frac{a^{-2}\gamma_{ij}}{\sqrt{\gamma}}\partial_l\left (a^4\sqrt{\gamma}\gamma^{hj}\gamma^{kl}F_{hk}\right )+\xi \partial _i\left (\frac{1}{\sqrt{\gamma}}\partial_k\left (a^2\sqrt{\gamma}\gamma^{kl}A_l\right )\right )
\end{equation}   
Here we see a significant difference between the five dimensional gauge field and the higher dimensional gauge field. Notably that, where as in five dimensions the gauge scalar mass term was completely gauge dependent and hence were not physical particles, in more than five dimension there is both a gauge dependent and a gauge independent mass term. In other words one cannot gauge away all the gauge scalars. We shall look in more detail at the gauge scalars in section \ref{sect:Gaugescalars} but now we shall continue considering the 4D gauge fields.  

Working with the space described by Eq. \ref{GenMetric} and again making a KK decomposition that leaves the kinetic term diagonal w.r.t KK number, i.e.
\begin{equation}
\label{GaugeKKdecomp}
A_\mu=\sum_n f_n(r)\Theta_n(\phi_1,\dots,\phi_\delta)A_\mu^{(n)}(x^\mu)
\end{equation}
such that
\begin{equation}
\label{GaugeOrthog}
\int d^{1+\delta}x\; bc^{\delta}\sqrt{\gamma}f_nf_m\Theta_n\Theta_m=\delta_{nm},
\end{equation}
where once again $n$ denotes all $1+\delta$ KK numbers. Eq. \ref{gaugeEOM} is then given by
\begin{equation}
\label{GaugeEOM}
f_n^{\prime\prime}+\frac{(a^2b^{-1}c^\delta)^{\prime}}{(a^2b^{-1}c^\delta)}f_n^{\prime}-\frac{b^2}{c^2}\alpha_nf_n+\frac{b^2}{a^2}m_n^2f_n=0
\end{equation}
and
\begin{equation}
\label{GaugLapl}
-\frac{1}{\sqrt{\gamma}}\partial_{\phi_i}\left (\sqrt{\gamma}\gamma^{ij}\partial_{\phi_i}\Theta_n\right )=\alpha_n\Theta_n.
\end{equation}
Once again the 4D effective theory contains $1+\delta$ towers of KK gauge modes of masses $m_n$. Only for the zero mode will the gauge symmetry be left unbroken. Hence the SM gauge fields are always attributed to the zero modes. In all spaces considered here, the eigenvalues of the Laplacian operator will be positive ($\alpha_n\geqslant 0$) and hence it is important to note that the SM gauge fields will always correspond to KK modes in which $\alpha_n=0$. 

Before moving on to the gauge scalars, we will again give a brief example of a solution using the 5D RS model (Eq. \ref{RSMetric}) in which the eigenfunctions of Eq. \ref{5DGaugeEqn} are given by
\begin{equation}
\label{RSgaugeProfile}
f_n(r)=Ne^{kr}\left (\mathbf{ J}_1\left (\frac{m_ne^{kr}}{k}\right )-\frac{\mathbf{J}_0(\frac{m_n}{k})}{\mathbf{Y}_0(\frac{m_n}{k})}\mathbf{Y}_1\left (\frac{m_ne^{kr}}{k}\right )\right ),
\end{equation}  
where we have imposed NBC's on the UV brane, $r=0$. If one imposes NBC's on the IR brane $r=R$ and assumes a large warp factor ($\Omega=e^{kR}$) then the mass eigenvalues are approximately given by \cite{Gherghetta:2000qt} 
\begin{displaymath}
m_n\approx(n-\frac{1}{4})\pi \frac{k}{e^{kR}}\sim X_n\frac{k}{\Omega}.
\end{displaymath} 
Comparing this result with Eq. \ref{RSscalarmass}, one can see that typically spin 1 and spin 0 KK modes will have different masses but that these masses will be of the same order of magnitude, i.e. $M_{\rm{KK}}=\frac{k}{\Omega}$. One can see that this will typically be the case for all spaces since the only term, in Eq \ref{ScalarEqn} and Eq. \ref{GaugeEOM}, that is different is the coefficient in front of $f_n^\prime$. So the terms that determine the KK mass scale, i.e. $\frac{b^2}{c^2}$, $\frac{b^2}{a^2}$ and $\alpha_n$, are the same for the two equations and hence gauge fields and scalars will have the same KK scales.   

\subsection{The Gauge Scalars}\label{sect:Gaugescalars}
Turning now to the gauge scalars, in order to determine what can and cannot be `gauged away' we would like to split apart the gauge dependent part of the equations of motion. By noting that $F_{hk}=\partial_hA_k-\partial_kA_h$ then one can expand Eq. \ref{Eq.gaugescalar} to
\begin{eqnarray*}
\partial_\mu\partial^\mu A_i=\frac{a^{-2}\gamma_{ij}}{\sqrt{\gamma}}\partial_l\left (a^4\sqrt{\gamma}\gamma^{hj}\gamma^{kl}\partial_hA_k\right )-\frac{a^{-2}\gamma_{ij}}{\sqrt{\gamma}}\partial_l\left (a^4\sqrt{\gamma}\gamma^{hj}\gamma^{kl}\partial_kA_h\right )\\+\xi \partial _i\left (\frac{1}{\sqrt{\gamma}}\partial_k\left (a^2\sqrt{\gamma}\gamma^{kl}A_l\right )\right ).
\end{eqnarray*} 
Hence the unphysical gauge dependent fields will be given by 
\begin{equation}
\label{ }
\partial_\mu\partial^\mu\left (\sum_iA_i\right )=\xi\sum_i \partial _i\left (\frac{1}{\sqrt{\gamma}}\partial_k\left (a^2\sqrt{\gamma}\gamma^{kl}A_l\right )\right ).
\end{equation}
Likewise one can find the physical gauge independent part by taking the derivative of Eq. \ref{Eq.gaugescalar} and using that the derivatives commute. This then gives
\begin{equation}
\label{ }
\partial_\mu\partial^\mu\left (\partial_iA_j-\partial_jA_i\right )=\partial_i\left (\frac{a^{-2}\gamma_{ji}}{\sqrt{\gamma}}\partial_l\left (a^4\sqrt{\gamma}\gamma^{hi}\gamma^{kl}F_{hk}\right )\right )-\partial_j\left (\frac{a^{-2}\gamma_{ij}}{\sqrt{\gamma}}\partial_l\left (a^4\sqrt{\gamma}\gamma^{hj}\gamma^{kl}F_{hk}\right )\right ).
\end{equation}  
Hence one can define the unphysical Goldstone boson by $\varphi=\sum_i A_i$ while the physical gauge scalars will be given by $\Phi_{ij}=A_i-A_j$. At first sight this looks like one has gone from $1+\delta$ gauge field components to $\left(\begin{array}{c}1+\delta \\2\end{array}\right)$ combinations of physical scalars. Although after taking into account that they will not all be independent, one finds that there is in fact $\delta$ gauge scalars or, after a KK decomposition, $\delta$ towers of scalars.

Having said this the KK decomposition is surprisingly involved. Firstly in more than six dimensions one has mixing between the different gauge scalars and the Goldstone bosons. However there is also a difficulty in finding an orthogonal decomposition that leaves the kinetic term canonically normalised, i.e,
\begin{displaymath}
\int d^{1+\delta}x a^2\sqrt{\gamma}\gamma^{ij}\partial_\mu A_i\partial^\mu A_j=\partial_\mu\varphi\partial^\mu\phi +\sum_{i=1}^\delta \partial_\mu \Phi_i\partial^\mu \Phi_i.
\end{displaymath} 
Of course this is not impossible but rather algebraically awkward. Hence as far as the author is aware the KK decomposition has only been completed for six dimensional models in flat space \cite{Burdman:2005sr, Cacciapaglia:2009pa} and AdS${}_D$ in \cite{McDonald:2009hf}. Of course one does not have to work in the unitary gauge. In \cite{Cacciapaglia:2011hx} the Feynman gauge ($\xi=1$) was used and it was found one does not obtain a mixing between $A_5$ and $A_6$. Although one must then include the corresponding ghost terms. Alternatively one can work with the phase of Wilson loops as in models with gauge-Higgs unification. Since it is difficult to eliminate the gauge dependent part, arguably it would be better to leave it in and check that it drops \color{red} out \color{black} of any physical result. The point we really wish to make is, no matter what method one uses, these gauge scalars will probably make a contribution to existing observables and this contribution is not straight forward to calculate for spaces such as AdS${}_5\times S^\delta$. 

Further still the non Abelian gauge scalars will gain an effective potential at tree level and it is feasible that such fields will gain a VEV upon compactification. The scale of this potential VEV will of course be dependent on the geometry and so could prove significant in excluding possible spaces as resolutions of the hierarchy problem. Unfortunately, a full study of such gauge scalars has proved beyond the scope of this thesis and hence here we will leave the computation of  \color{red} their \color{black} phenomenological implications to future work.   
  
\subsection{The Position / Momentum Space Propagator} \label{PosMomProp}
Up to now we have defined the KK decomposition by matching the higher dimensional theory to the 4D effective theory and defined our KK masses by the 4D equations of motion. One may be concerned at this purely `on-shell' description. To be a little clearer about the physics behind the KK decomposition in this section we will compute the propagator for $A_\mu$ in the higher dimensional theory. Here we will work in just five dimensions although it is straight forward to work in more. Following \cite{Randall:2001gb}, Eq. \ref{5DexpanGauge} can be expanded, in the unitary gauge ($\xi\rightarrow\infty$), to
\begin{equation*}
S=\int d^5x\; \frac{b}{2}A_\mu\left (\eta^{\mu\nu}\partial^2-\partial^\mu \partial^\nu-b^{-1}\eta^{\mu\nu}\partial_5(a^2b^{-1}\partial_5)\right )A_\nu.
\end{equation*} 
Here we are interested in the extra dimensional contribution to the gauge propagator and so we will not go through the full Faddeev-Popov quantisation. Rather we simply note that, after Fourier transforming just w.r.t the four large dimensions, the full derivation is identical to the four dimensional case, which can be found in many textbooks. Hence one can read off the propagator as 
\begin{equation}
\label{ }
<A^{\mu}(r)A^{\nu}(\tilde{r})>=-iG_p(r,\tilde{r})\left (\eta^{\mu\nu}-\frac{p^\mu p^\nu}{p^2}\right )
\end{equation}  
where $p^\mu$ is the four momentum and 
\begin{equation}
\label{5DGaugeProp}
\left (\frac{1}{b}\partial_5(a^2b^{-1}\partial_5) +p^2\right )G_p(r,\tilde{r})=\frac{1}{b}\delta(r-\tilde{r}).
\end{equation}
As with the KK decomposition, in order to proceed further, it is necessary to specify a particular space. Once again we shall use the 5D RS model (Eq. \ref{RSMetric}) as a specific example. This gives the most general solution of Eq. \ref{5DGaugeProp} as
  \begin{equation}
\label{ }
G_p(r,\tilde{r}) =
\begin{cases}
 e^{kr}\left (A\mathbf{J}_1(\frac{pe^{kr}}{k})+B\mathbf{Y}_1(\frac{pe^{kr}}{k})\right ) & \text{for } r<\tilde{r} \\
 e^{kr}\left (C\mathbf{J}_1(\frac{pe^{kr}}{k})+D\mathbf{Y}_1(\frac{pe^{kr}}{k})\right ) & \text{for } r>\tilde{r}
\end{cases}.
\end{equation}
Although clearly these solutions must match up, i.e. $G^{>}_p(r,r)=G^{<}_p(r,r)$ and so after defining $u\equiv\min(r,\tilde{r})$ and $v\equiv\max(r,\tilde{r})$ then
\begin{displaymath}
G_p(u,v)=Ne^{k(u+v)}\left (A\mathbf{J}_1\left(\frac{pe^{ku}}{k}\right )+B\mathbf{Y}_1\left (\frac{pe^{ku}}{k}\right )\right )\left (C\mathbf{J}_1\left (\frac{pe^{kv}}{k}\right )+D\mathbf{Y}_1\left (\frac{pe^{kv}}{k}\right )\right ).
\end{displaymath}
One also has a constraint coming from integrating over Eq. \ref{5DGaugeProp}
\begin{displaymath}
\lim_{\epsilon\rightarrow 0}\int^{\tilde{r}+\epsilon}_{\tilde{r}-\epsilon} dr\left(\partial_r^2-2k\partial_r+e^{2kr}p^2\right )G_p (r,\tilde{r}) =[\partial_vG_p(u,v)-\partial_uG_p(u,v)]_{u=v}=e^{2kv}.
\end{displaymath}
This then fixes the normalisation constant to be
\begin{displaymath}
N=-\frac{1}{pe^{kv}(AD-BC)\left (\mathbf{J}_0\left (\frac{pe^{kv}}{k}\right )\mathbf{Y}_1\left (\frac{pe^{kv}}{k}\right )-\mathbf{Y}_0\left (\frac{pe^{kv}}{k}\right )\mathbf{J}_1\left (\frac{pe^{kv}}{k}\right )\right )}=\frac{ \pi}{2k(BC-AD)},
\end{displaymath}
where in the second step we have used that $x(\mathbf{Y}_0(x)\mathbf{J}_1(x)-\mathbf{Y}_1(x)\mathbf{J}_0(x))=\frac{2}{\pi}$. Finally we must impose boundary conditions. If one imposes NBC's at both the UV and IR brane then this fixes the remaining coefficients to be
\begin{eqnarray*}
A=\mathbf{Y}_0\left (\frac{p}{k}\right )\hspace{0.5cm}B=-\mathbf{J}_0\left (\frac{p}{k}\right )\hspace{0.5cm}C=\mathbf{Y}_0\left (\frac{pe^{kR}}{k}\right )\hspace{0.5cm}D=-\mathbf{J}_0\left (\frac{pe^{kR}}{k}\right )
\end{eqnarray*}
and hence
\begin{equation}
\label{ }
G_p(u,v)=\frac{\pi e^{k(u+v)}\left (A\mathbf{J}_1\left(\frac{pe^{ku}}{k}\right )+B\mathbf{Y}_1\left (\frac{pe^{ku}}{k}\right )\right )\left (C\mathbf{J}_1\left (\frac{pe^{kv}}{k}\right )+D\mathbf{Y}_1\left (\frac{pe^{kv}}{k}\right )\right )}{2k(BC-AD)}.
\end{equation}
So when one imposes boundary conditions, the denominator of the propagator becomes periodic w.r.t $p$ and hence the propagator gains an infinite number of poles. It is straightforward to check that these poles correspond to the KK masses found by imposing the same boundary conditions on Eq. \ref{RSgaugeProfile}. In other words the act of compactification, or what ever is responsible for giving rise to a discrete KK spectrum, will equivalently give rise to multiple poles in the propagator.    

\begin{figure}[ht!]
    \begin{center}
        \subfigure[$p=0.1\;M_{\rm{KK}}$]{%
           \label{fig:second}
           \includegraphics[width=0.45\textwidth]{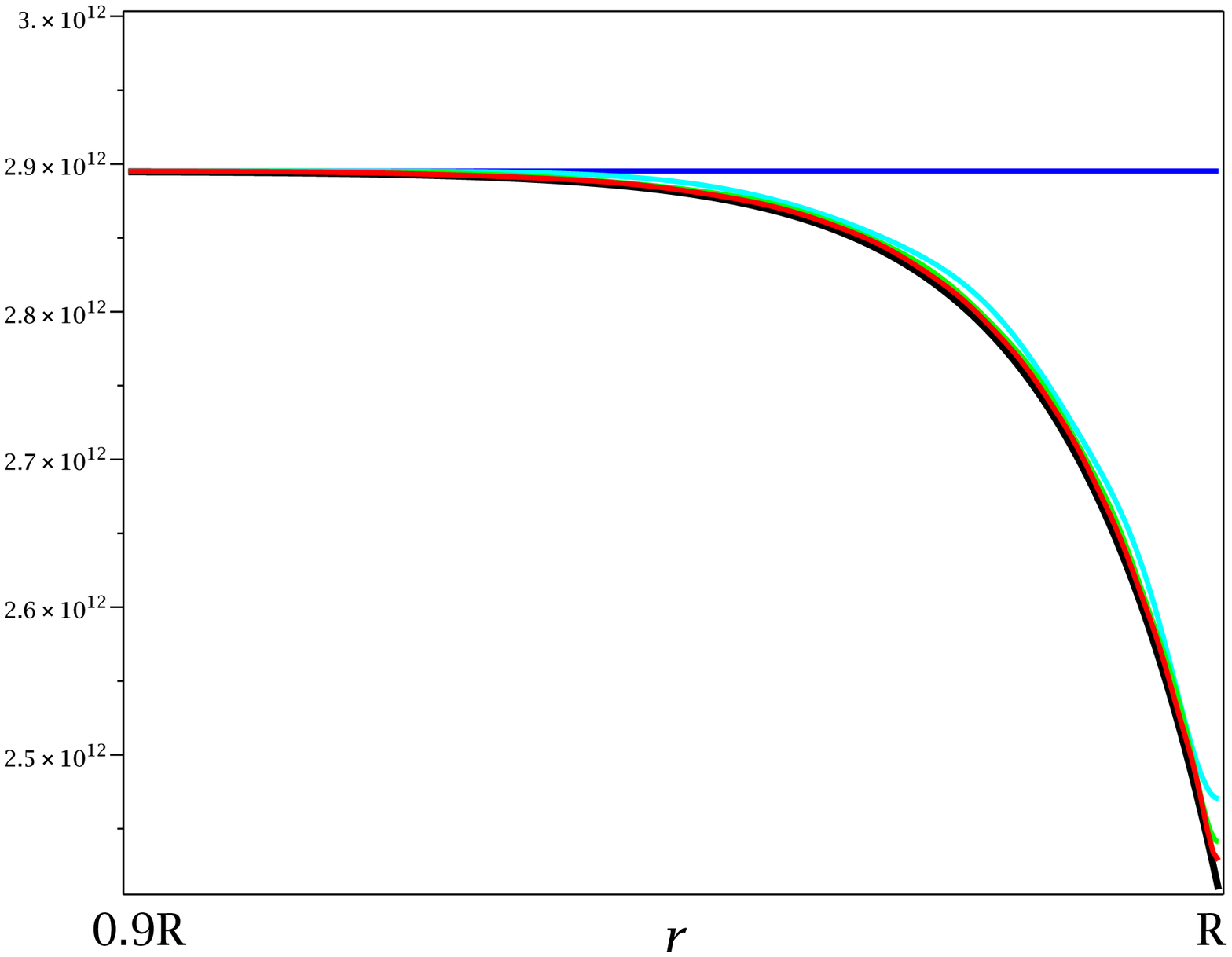}
        }
        \subfigure[$p=10\; M_{\rm{KK}}$]{%
           \label{fig:second}
           \includegraphics[width=0.45\textwidth]{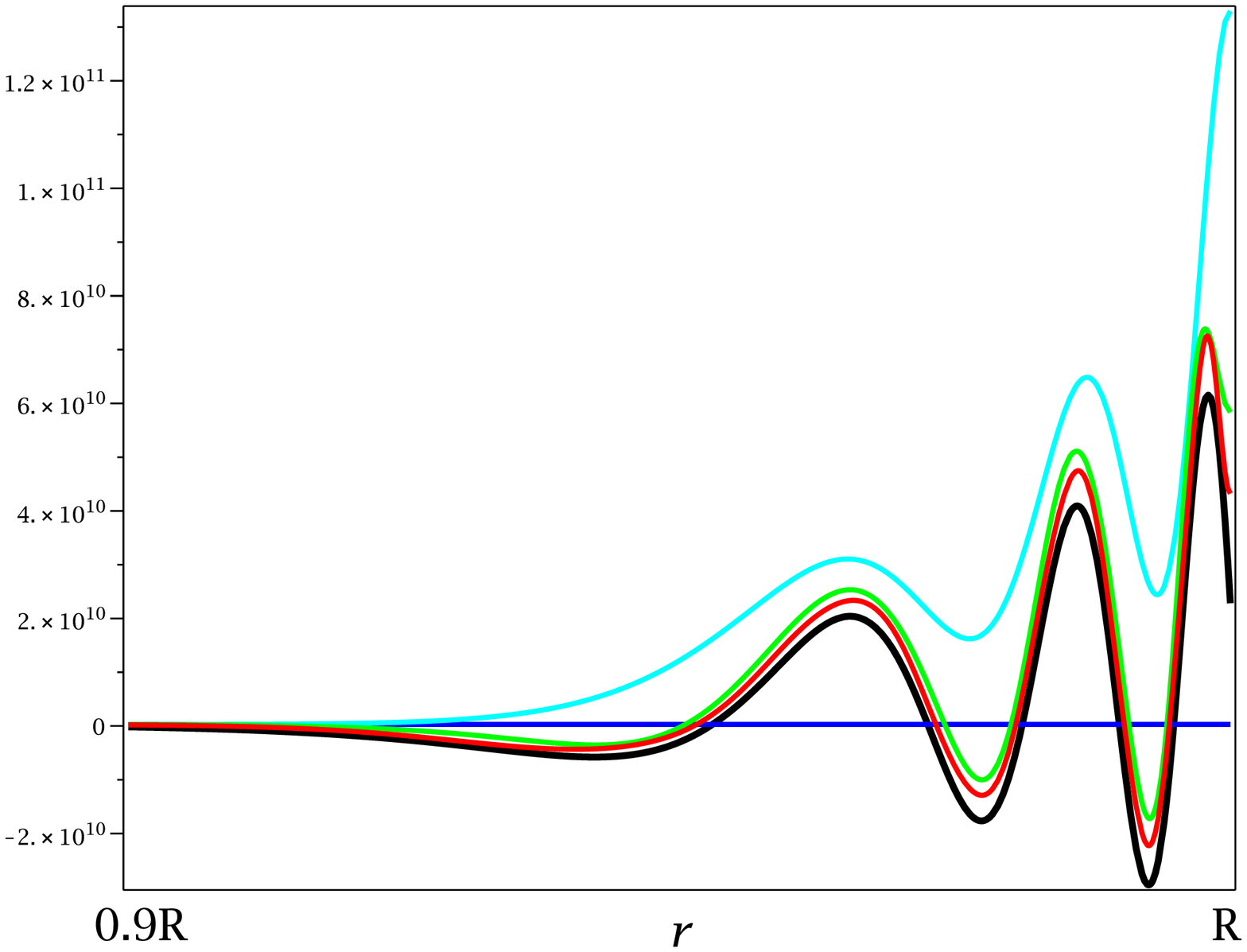}
        }
   \label{fig:PropSum}
    \end{center}
    \caption[Comparison Between KK Sum and Propagator]{The comparison between the propagator $G_p(r,r)$ (in black) and the KK sum $\sum_n\frac{f_n(r)f_n(r)}{p^2-m_n^2}$. The sum is taken over the zero mode (blue), zero mode and three KK modes (cyan), zero mode and six KK modes (green) and zero mode and ten KK modes (red). Here only the IR tip of the space is plotted. $M_{\rm{KK}}\equiv \frac{k}{\Omega} =1$ TeV and $\Omega=e^{kR}=10^{15}$. } \label{fig:PropSum}
\end{figure}

One can equate the two equivalent methods by noting that
\begin{equation}
\label{GaugePropKKequiv}
\int dr \int d\tilde{r}\; G_p(u,v)=\sum_n \int dr \int d\tilde{r} \frac{f_n(r)f_n(\tilde{r})}{p^2-m_n^2}.
\end{equation}
This equivalence is plotted in figure \ref{fig:PropSum}. When one is working with a KK expansion, it is typically both impractical and unnecessary to work with the full infinite tower. Even if the $10^{100}$th KK mode was still weakly coupled one would expect the four dimensional effective theory to have broken down along time before reaching it. So one should ask how many KK modes should be included? In figure \ref{fig:PropSum} it can be seen that, at momenta much lower than the KK scale, the propagator will be completely dominated by the zero mode with the exception of a small deviation in the IR tip of the space. This deviation can be included by just including the first few KK modes. However as one works at higher and higher momenta, the higher KK modes will increasingly contribute to the propagator and hence the 4D effective theory should include more KK modes.     

\section{The Fermions}\label{sect:fermions}
Having considered gauge fields we will now move on to look at fermions. Fermions, in $D$ dimensions, are representations of the $\color{red}\mathrm{SO}(1,D-1)$ Lie algebra. It can be shown, in odd dimensions, that there is only one irreducible representation and that is a $2^{\frac{D-1}{2}}$ component Dirac representation \cite{Ortin:2004ms}. Although in even dimensions there is also a $2^{\frac{D}{2}-1}$ component Weyl representation that arises due to the algebra being isomorphic to two sub-algebras, e.g. in 4D $\color{red}\mathrm{SO}(1,3)\cong \color{red}\mathrm{SU} (2)\times \color{red}\mathrm{SU} (2)$. As a slight side note it is these enlarged Lorentz symmetries and representations that can be used, in six dimensions, to elegantly explain both proton stability \cite{Appelquist:2001mj} and the number of fermion generations \cite{Dobrescu:2001ae}. They also give rise to a potential problem in reproducing the SM. Notably that the low energy four dimensional effective theory should be comprised of 2 component Weyl spinors. The now standard solution to this is to compactify the space over an orbifold chosen such that the zero modes of the additional 2 components Weyl spinors are forbidden by the resulting boundary conditions. However defining such an orbifold, in more than five dimension, is far from trivial. Before looking at the higher dimensional case we shall once again first look at the five dimensional case. 

\subsection{Fermions in Five Dimensions}\label{sect:5dFerm}
If we begin by considering the action for a massive Dirac spinor in a five dimensional version of Eq. \ref{GenMetric}.   
\begin{equation}
\label{5dFermact}
S=\int dx^5\;\sqrt{G}\left [i\bar{\Psi}\Gamma^M\nabla_M\Psi-M\bar{\Psi}\Psi\right ],
\end{equation}
where $\Gamma^M=E^M_A\gamma^A$ are the Dirac matrices in curved space. \color{red}The \color{black}  F\"{u}nfbein are defined by $G_{MN}=E_M^A\eta_{AB}E_N^B$ and are given as $E^M_A=\rm{diag}(a^{-1},a^{-1},a^{-1},a^{-1},b^{-1})$. The covariant derivative must now also include a spin connection term, $\nabla_M=D_M+\omega_M$, which is computed using \cite{ZinnJustins02}
\begin{displaymath}
\omega_M=\omega_M^{AB}\gamma_{AB}=\frac{1}{2}\omega_M^{AB}[\gamma_A,\gamma_B]
\end{displaymath}
 where 
 \begin{displaymath}
\omega_M^{AB}=\frac{1}{2}E^{AN}\left(\partial_ME_N^B-\partial_NE_M^B\right )+\frac{1}{4}E^{AN}E^{BP}\left (\partial_P E_N^C-\partial_NE_P^C\right )E_M^C-(A\leftrightarrow B).
\end{displaymath}
For this metric (Eq. \ref{GenMetric}) the spin connection is computed to be 
\begin{displaymath}
\omega_\mu=\frac{1}{2}b^{-1}a^{\prime}\gamma_5\gamma_\mu\quad\mbox{and}\quad\omega_5=0.
\end{displaymath}
Here we will use the chiral representation of the Dirac matrices, \cite{WessBagger}, in which $\gamma^\mu=\left (\begin{array}{cc} 0  & \sigma^\mu   \\\bar{\sigma}^\mu &   0\end{array}\right )$ where $\sigma^\mu=\left \{\mathbb{I},\sigma^i\right \}$ and $\bar{\sigma}^\mu=\left \{\mathbb{I},-\sigma^i\right \}$ are the Pauli sigma matrices, while $\gamma_5$ is given by $\gamma_5=\gamma_0\gamma_1\gamma_2\gamma_3=\left (\begin{array}{cc} i  & 0   \\ 0 & -i \end{array}\right )$. It is the same $\gamma_5$ which is used to define chirality $i\gamma_5\Psi_{L,R}=\mp\Psi_{L,R}$ and allows for the five dimensional vector like Dirac fermion to be expressed in terms of two component Weyl representations,
\begin{displaymath}
\Psi=\left (\begin{array}{c}\psi_L \\\psi_R   \end{array}\right ).
\end{displaymath}
Eq. \ref{5dFermact} can now be expanded in terms of two component Weyl spinors,
\begin{eqnarray}
S=\int d^5x \bigg ( ia^3b\bar{\psi}_L\bar{\sigma}^\mu D_\mu\psi_L+ia^3b\bar{\psi}_R\sigma^\mu D_\mu\psi_R+2a^3a^{\prime}\bar{\psi}_R\psi_L-2a^3a^{\prime}\bar{\psi}_L\psi_R\nonumber\\
+a^4\bar{\psi}_RD_5\psi_L-a^4\bar{\psi}_LD_5\psi_R-M(\bar{\psi}_R\psi_L+\bar{\psi}_L\psi_R)\bigg )\label{5DWeylAction}
\end{eqnarray}
Before going further it is worth taking a brief moment to review orbifolds. \newline

Up to now, when we have considered the 5D RS model, we have essentially considered an interval $r\in[0, R]$ without being too specific about the precise compactification. For reasons we shall see in a moment, the standard compactification is that of a $S^1/Z_2$ orbifold in which the action is invariant under $r\rightarrow r+2R$ and $r\rightarrow -r$. Such transformations also imply $r\rightarrow r-2R$ and $r\rightarrow 2R-r$. If we now consider how an arbitrary field transforms under such transformations, in particular $\varphi(r+2R)=T\varphi(r)$ and $\varphi(-r)=Z\varphi(r)$. Combining the above transformations then implies $TZ=ZT^{-1}$ which then gives rise to a second $Z_2$ symmetry $(\tilde{Z})^2=(TZ)(ZT^{-1})=1$ \cite{Csaki:2005vy}. It is these two $Z_2$ symmetries that allow us to then consider fields defined over the interval $r\in[0, R]$ and not $r\in[0, 2R]$. Naively one could suspect that one could write down non trivial boundary conditions that satisfy Eq. \ref{ScalarBCS}, however we now see that this is not the case in an orbifolded space. Since the parity of the field under these $Z_2$ symmetries will determine the boundary conditions of the field. So  a field with a positive parity will be even over the interval and have NBC's, $\partial_r \varphi |_{0,R}=0$, while a field with negative parity will be odd with DBC's, $\varphi |_{0,R}=0$. An important consequence of this is that fields with DBC's will not have zero modes and hence the parity under the $Z_2$ orbifold will determine the particle content of the low energy theory.\newline

If we now return to the case of fermions. There is a subtlety, in that, when writing down a set of consistent boundary conditions, one is doing so for the fundamental four component Dirac field and not the apparent two component Weyl representation. Hence for a given fermion field there is one set of boundary conditions which can be expressed in terms of two sets of boundary conditions on the two Weyl spinors. If we now look at how the Weyl spinors (in Eq. \ref{5DWeylAction}) transform under the $S_1/Z_2$ transformations then clearly the terms $\bar{\psi}_{L,R}\partial_5\psi_{R,L}$ is only invariant if, under the $Z_2$ transformation, $\psi_{L}$ has positive parity and $\psi_R$ has negative parity or vice versa. Bearing in mind the absence of a zero mode for fields with DBC's one arrives at the point of this discussion. Notably that, quite generically, one finds that the low energy theory of a five dimensional fermion compactified over a $S_1/Z_2$ orbifold is inherently chiral. Also it should be noted that in order for the bulk mass term, $M\bar{\Psi}\Psi$ in Eq. \ref{5dFermact}, to be invariant under the $S_1/Z_2$ transformations it is necessary for the mass term to undergo discrete jumps at the orbifold fixed points \cite{Bagger:2001qi}.           

Finally we can now continue with the KK expansion of Eq. \ref{5DWeylAction} by defining the KK decomposition
\begin{equation}
\label{FermKKdecomp}
\psi_{L,R}=\sum_{n=0}a^{-2}\,f_{L,R}^{(n)}(r)\,\psi_{L,R}^{(n)}(x^\mu)
\end{equation}
such that
\begin{equation}
\label{FermOrthog}
\int dr \,\frac{b}{a}\,f_{L,R}^{(n)}\,f_{L,R}^{(m)}=\delta_{nm}
\end{equation}
to give the 4D effective action as
\begin{equation}
\label{4dfermact}
S=\int dx^4\,\sum_n\left [i\bar{\psi}_L^{(n)}\bar{\sigma}^\mu D_\mu\,\psi_L^{(n)}+i\bar{\psi}_R^{(n)}\sigma^\mu D_\mu\,\psi_R^{(n)}-m_n\left (\bar{\psi}_L^{(n)}\psi_L^{(n)}+\bar{\psi}_R^{(n)}\psi_R^{(n)}\right )\right ],
\end{equation}
where $m_n$ is obtained from the coupled equations of motion
\begin{eqnarray}
f_R^{(n)\,\prime}+bMf_R^{(n)}=\frac{b}{a}f_L^{(n)}m_n\label{fermequat1} \\
-f_L^{(n)\,\prime}+bMf_L^{(n)}=\frac{b}{a}f_R^{(n)}m_n\label{fermequat2}.
\end{eqnarray}
These two equations can of course be combined to give a second order differential equation in terms of just $f_{L,R}$. If we now consider the parity under the $Z_2$ symmetries of the orbifold and hence choose either $f_L$ or $f_R$ to be an even field (positive parity), then the other Weyl spinor's resulting boundary condition is already fixed. For example if $f_R$ has positive parity then $f^{(n)}_R|_{0,R}=0$ and
\begin{eqnarray}
\left [f_R^{(n)\,\prime}\right ]_{0,R}=\left [m_n\frac{b}{a}f_L^{(n)}\right ]_{0,R} \\
\left [f_L^{(n)\,\prime}\right ]_{0,R}=\left [bMf_L^{(n)}\right ]_{0,R}
\end{eqnarray}
With such boundary conditions the zero mode of $\psi_L$ has the general solution
\begin{equation}
\label{5DfermionZeromode}
f_L^{(0)}(r)=\frac{\exp \left [\int_c^r b(\tilde{r})Md\tilde{r}\right ]}{\sqrt{\int_{0}^{R}\frac{b}{a}\exp \left [\int_c^r 2b(\tilde{r})Md\tilde{r}\right ]}},
\end{equation}
while $\psi_R$ has no zero mode. One can also see that the fermion profile is exponentially sensitive to the bulk mass parameter, $M$. It is this profile which will essentially determine the 4D effective coupling of the fermions and so, as we shall see in the next chapter, this generic dependence will prove critical to the warped extra dimensional description of flavour. 

Although here we have considered the boundary conditions, necessary for a low energy chiral theory, arising from an orbifold there is nothing intrinsically wrong in simply considering an interval with the appropriate boundary conditions imposed by hand. However such an approach is a little ad hoc. When we consider fermions in more than five dimensions the number of components that need to be removed from the low energy theory increases significantly and hence one is forced to look for increasingly complicated orbifolds. Also if one is to construct a description of flavour, analogous to that of the 5D RS model, it is necessary to include a bulk mass term. It has proved beyond the scope of this thesis to generalise this description of flavour to $D$ dimensional generic spaces. None the less we will now consider fermions in more than five dimensions in order to demonstrate some of the problems and their potential solutions.

\subsection{Fermions in Flat Six Dimensional Space}\label{sect:6dFerm}
If we start by looking at the requirement that an extra dimensional description of flavour would require a bulk mass term. For simplicity we shall work with a flat six dimensional space and include the three possible mass terms
\begin{equation}
\label{6DFermionAct}
S=\int d^6x \left (i\bar{\chi} \Gamma^M\partial_M \chi-M\bar{\chi}\chi-M_5\bar{\chi}\Gamma^5\chi-M_6\bar{\chi}\Gamma^6\chi\right ) 
\end{equation}               
One could envisage the mass terms $M^5$ and $M^6$ appearing from, for example, some gauge Higgs unification scenario in which the additional components of a gauge field gain a VEV. Although here we shall just include them as generic mass terms.  Here $\chi$ is an eight component Dirac spinor and the representations of the Dirac matrices used here are \cite{Cacciapaglia:2009pa}
\begin{displaymath}
\Gamma^\mu=\left (\begin{array}{cc} \gamma^\mu  & 0   \\ 0 &   \gamma^\mu \end{array}\right )\quad \Gamma^5=\left (\begin{array}{cc} 0  & -\gamma^5   \\ -\gamma^5 & 0 \end{array}\right )\quad \Gamma^6=\left (\begin{array}{cc} 0  & i\gamma^5   \\ -i\gamma^5 & 0 \end{array}\right ) 
\end{displaymath}
which of course satistfy $\{\Gamma_A,\Gamma_B\}=2\eta_{AB}$. One can also define the the chirality matrix
\begin{equation}
\label{7Dchirality}
\Gamma^7=\Gamma^0\Gamma^1\Gamma^2\Gamma^3\Gamma^5\Gamma^6=\left (\begin{array}{cc} -i\gamma^5 & 0   \\ 0& i\gamma^5 \end{array}\right )
\end{equation}
such that $\Gamma^7\chi_{\pm}=\pm\chi_\pm$. Hence one can split the eight component Dirac spinor such that
\begin{displaymath}
\chi_+=\left (\begin{array}{c}\Psi_{L+} \\\Psi_{R+}   \end{array}\right )\quad \chi_-=\left (\begin{array}{c}\Psi_{R-} \\\Psi_{L-}   \end{array}\right )\quad\mbox{and}\quad \Psi_L=\left (\begin{array}{c}\psi_{L} \\0   \end{array}\right )\quad \Psi_R=\left (\begin{array}{c}0\\\psi_{R}    \end{array}\right )
\end{displaymath} 
where again $\psi_{L,R}$ are 2 component Weyl spinors and $i\gamma_5\Psi_{L,R}=\mp\Psi_{L,R}$. Eq. \ref{6DFermionAct} can now be expanded to
\begin{eqnarray*}
S=\int d^6x \bigg [\;i\bar{\psi}_{L\pm}\bar{\sigma}^\mu\partial_\mu\psi_{L\pm} +i\bar{\psi}_{R\pm}\sigma^\mu\partial_\mu\psi_{R\pm} +\bar{\psi}_{L\pm}\partial_5\psi_{R\pm}-\bar{\psi}_{R\pm}\partial_5\psi_{L\pm}\mp i\bar{\psi}_{L\pm}\partial_6\psi_{R\pm}\\
\mp i\bar{\psi}_{R\pm}\partial_6\psi_{L\pm} -M\left (\bar{\psi}_{L\pm}\psi_{R\mp}+\bar{\psi}_{R\pm}\psi_{L\mp}\right )
-iM_5\left (\bar{\psi}_{L\pm}\psi_{R\pm}-\bar{\psi}_{R\pm}\psi_{L\pm}\right )\\-M_6\left (\pm\bar{\psi}_{L\pm}\psi_{R\pm}\pm\bar{\psi}_{R\pm}\psi_{L\pm}\right )\bigg ].
\end{eqnarray*}
Where a $\pm$ has been included both terms are present. This give rise to the first problem. The mass term $M$ mixes both $\chi_{\pm}$ and $\Psi_{L,R}$ chiralities and hence such a mass term would require both to be present in the higher dimensional theory. In order to reproduce a low energy SM it would be necessary to ensure that the only Weyl spinors to gain a zero mode would be a doublet under $\color{red}\mathrm{SU} (2)$, i.e. $\psi_L$, and a singlet under $\color{red}\mathrm{SU}(2)$, $\psi_R$. Hence the inclusion of a mass term $M$ would require boundary conditions that removed three out of four possible zero modes. Clearly this problem becomes increasingly severe as one increases the dimensionality.   

If we now define an effective 4D mass by $i\gamma^\mu\partial_\mu \Psi=m \Psi$ then one can see that the higher dimensional fermion, $\chi$, is now described by four coupled equations of motion.
\begin{eqnarray}
\left (\partial_5\mp i\partial_6-iM_5\mp M_6\right )\psi_{R\pm}-M\psi_{R\mp}+m\psi_{L\pm}=0\nonumber\\
\left (\partial_5\pm i\partial_6-iM_5\pm M_6\right )\psi_{L\pm}+M\psi_{L\mp}-m\psi_{R\pm}=0\label{6DfermionsEOMs}
\end{eqnarray}  
If we now make the KK decomposition
\begin{displaymath}
\psi_{L,R\pm}=\sum_n \mathcal{F}_{L,R\pm}^{(n)}(r,\phi)\psi^{(n)}_{L,R\pm}
\end{displaymath}
and consider the zero mode $m_n=0$ then Eq. \ref{6DfermionsEOMs} can be combined to
\begin{equation}
\label{ }
\left (\partial_5^2+\partial_6^2-2iM_5\partial_5-2iM_6\partial_6-M_5^2-M_6^2-M^2 \right )\mathcal{F}_{L,R\pm}^{(0)}=0.
\end{equation}
The most general solution can then be expressed as the linear combination 
\begin{displaymath}
\mathcal{F}_{L,R\pm}^{(0)}=\sum_le^{iM_5r}e^{iM_6\phi}\left (A_l e^{\frac{1}{2}\sqrt{2M^2+4l^2}r}+B_l e^{-\frac{1}{2}\sqrt{2M^2+4l^2}r}\right )\left (C_l e^{\frac{1}{2}\sqrt{2M^2-4l^2}\phi}+D_l e^{-\frac{1}{2}\sqrt{2M^2-4l^2}\phi}\right )
\end{displaymath}
There is of course still only one zero mode and once the bulk mass terms have been uniquely defined and the solution is reapplied to Eq \ref{6DfermionsEOMs} then the solution will be uniquely defined. For example if one assumes $M$ and $\psi$ are real then the solution reduces to  
\begin{displaymath}
\mathcal{F}_{R\pm}^{(0)}=Ne^{iM_6\phi} e^{(M+iM_5)r}\quad\mbox{     and     }\quad\mathcal{F}_{L\pm}^{(0)}=Ne^{iM_6\phi} e^{(iM_5-M)r}.
\end{displaymath}
Hence the bulk mass terms do not just determine in which direction the fermions are localised they also determine the complex phase of the fermions. It is important to realise that by fixing the basis of the Dirac algebra and by assuming $M$ is real one is essentially selecting a `prefered' direction in the extra dimensions with no physical motivation for doing so. In other words in six dimensions, if one neglects the $M_5$ and $M_6$ and assumes real mass terms, then the zero modes will be flat in one direction but not the other. This was also found in \cite{Biggio:2003kp}. Naively this looks like one would arrive at an absurd situation in which the physical results are dependent on which Dirac matrices are being used. It is possible that this is related to the fact that the $\color{red}\mathrm{SO}(1,5)$ Lie algebra contains an $\color{red}\mathrm{SO}(1,3)\times \color{red}\mathrm{U}(1)$ subgroup, where the $\color{red}\mathrm{U}(1)$ symmetry is associated with rotations in the $x_5$ and $x_6$ coordinates \cite{Appelquist:2001mj}. Here it is suspected that any space that yields different results for different representations of the Lie algebra must break this $\color{red}\mathrm{U}(1)$ symmetry.  
   
The point is that, in extending the 5D RS description of flavour to more than five dimensions, one must not only deal with a significantly enlarged parameter space but one must also deal with a number of other subtleties including anomaly cancelation \cite{Dobrescu:2001ae} as well as what representations of the Lie algebra are permitted in different topologies. While these are potentially interesting areas of future study they are also beyond the scope thesis.      \newline 

Turning now to the question of finding a suitable orbifold in which to reproduce a four dimensional chiral theory. In six dimensions work has been done in this direction by considering a $T^2/Z_4$ orbfold \cite{Dobrescu:2004zi} and a $T^2/Z_2$ orbifold \cite{Antoniadis:2001cv, Ponton:2001hq, Appelquist:2000nn}. However this work does not include a bulk mass term and hence could not be used to construct a model of flavour. Alternatively one could localise the fermions on a six dimensional hypersurface in a seven dimensional space \cite{Appelquist:2002ft}. The situation becomes more promising in odd dimensions where, as explained in \cite{McDonald:2009hf}, one can compactify over an orbifold of the form 
\begin{displaymath}
\left (T^2/Z_2\times\dots\times T^2/Z_2\right )\times S^1/Z_2.
\end{displaymath}  
The advantage is there is now multiple parities associated with each fermion. For example in seven dimensions ( with an orbifold $T^2/\tilde{Z}_2\times S^1/Z_2$) the fermions would carry two parities corresponding to $(\tilde{Z}_2,Z_2)=(\pm,\pm)$. Hence if one decomposes the eight component vectorial fermion into two component Weyl fermions w.r.t the chirality matrix (Eq. \ref{7Dchirality}) then one finds that the parities are distributed as
\begin{displaymath}
\chi=\left (\begin{array}{c}\psi_{R-}(+,-) \\\psi_{L-}(-,-)\\\psi_{L+}(+,+)\\\psi_{R+}(-,+)\end{array}\right )
\end{displaymath}      
and hence only one two component Weyl spinor would gain a zero mode. This method can be used in any odd dimension greater than five. However before one extends this method to completely generic odd spaces, described by Eq. \ref{GenMetric}, one has yet another problem to solve. In particular it is necessary to decompose the higher dimensional fermion in to Weyl spinors with chirality defined w.r.t $\color{red}\mathrm{SO}(1,3)$. This happens naturally when the space has a Lorentz symmetry $\color{red}\mathrm{SO}(1,D-1)$ but when one has two fibered spaces, e.g. $AdS_5\times \mathcal{M}^\delta$ the Lorentz algebra is now $\color{red}\mathrm{SO}(1,4)\times \color{red}\mathrm{SO}(\delta)$. In such a case it is far from trivial to split up the higher dimensional fermions into two component Weyl spinors that are representations under $\color{red}\mathrm{SO}(1,3)$.  \newline

In theory there is no fundamental reason that one cannot consider fermions in more than five dimensions but it is rather that, as far as the author is aware, many of the details of the KK decomposition are yet to be explicitly worked out. None the less one can see a number of generic features of fermions in higher dimensional theories, notably the feature of a bulk mass term determining in which direction a fermion profile will sit. However for the remainder of this thesis we will primarily consider fermions in five dimensions. 

\section{The Gravitational Sector}\label{sect:GravKK}
For completeness we shall briefly consider the gravitational sector. This thesis is primarily concerned with the constraints on the low energy theory arising from existing observables. Naively one could suspect that, since the gravitational coupling is so much weaker than the other forces, one need not be concerned with it. However here we will briefly demonstrate firstly that the coupling of the graviton KK modes can be of sufficient size to be of relevance to low energy physics \cite{Davoudiasl:1999jd, Randall:1999vf}. Although we will also demonstrate that, when one has a discrete KK spectrum, the constraints from existing gravitational tests (i.e. tests on the inverse square law) are not significant.

\subsection{KK Modes of the Graviton}
In order to arrive at the KK profiles of the graviton the standard approach is to consider small pertubations from a Minkowski background in the four large dimensions
\begin{equation}
\label{GravitonPertubations}
ds^2=a^2(r)\left (\eta_{\mu\nu}+h_{\mu\nu}\right )dx^\mu dx^\nu-b^2(r)dr^2-c^2(r)d\Omega_\delta^2,
\end{equation}
Considering the Einstein-Hilbert action and fixing the gauge such that $\partial_Mh^{MN}=0$ and $h_M^N=0$ then it can be shown \cite{Csaki:2000fc} that variation of the action yields the equations of motion
\begin{equation}
\label{ }
\frac{1}{\sqrt{G}}\partial_M\left (\sqrt{-G}G^{MN}\partial_Nh_{\mu\nu}\right )=0.
\end{equation}
In other words the graviton profile is essentially the same as that of a massless scalar field.  So if one made the KK decomposition
\begin{displaymath}
h_{\mu\nu}(x^\mu,r,\phi)=\sum_n\tilde{h}_{\mu\nu}^{(n)}\mathcal{F}_{\mu\nu}^{(n)}(r,\phi)
\end{displaymath} 
then the equations of motion split to the analogue of Eq. \ref{ScalarEqn}
\begin{equation}
\label{ }
\mathcal{F}_{\mu\nu}^{(n)\;\prime\prime}+\frac{(a^4b^{-1}c^\delta)^{\prime}}{a^4b^{-1}c^\delta}\mathcal{F}_{\mu\nu}^{(n)\;\prime}+\frac{b^2}{c^2\sqrt{\gamma}}\partial_{\phi_i}\left (\sqrt{\gamma}\gamma^{ij}\partial_{\phi_i}\mathcal{F}_{\mu\nu}^{(n)}\right )+m_n^2\frac{b^2}{a^2}\mathcal{F}_{\mu\nu}^{(n)}=0
\end{equation}
and
\begin{equation}
\label{GravitonEOM}
\Box\tilde{h}^{(n)}_{\mu\nu}+m_n^2\tilde{h}^{(n)}_{\mu\nu}=0
\end{equation}
where $\Box=\eta^{\mu\nu}\partial_\mu\partial_\nu$. If once again we take the example of the 5D RS model (Eq. \ref{RSMetric}) then the graviton zero mode, which will be responsible for low energy gravity, will have the profile
\begin{equation}
\label{ }
\mathcal{F}_{\mu\nu}^{(0)}=\sqrt{\frac{2k}{1-e^{-2kR}}}\approx \sqrt{k}
\end{equation}
whereas the graviton KK modes will be described by
\begin{equation}
\label{ }
\mathcal{F}_{\mu\nu}^{(n)}\approx\frac{\sqrt{k}e^{2kr}\mathbf{J}_2(\frac{m_ne^{kr}}{k})}{e^{kR}\sqrt{\mathbf{J}^2_2(\frac{m_ne^{kR}}{k})-\mathbf{J}_1(\frac{m_ne^{kR}}{k})\mathbf{J}_3(\frac{m_ne^{kR}}{k})}}.
\end{equation}
These profiles will then determine the scale of the effective graviton coupling. For instance, if we consider the coupling to a particle localised on the IR brane ($r=R$), then one can see that the KK graviton's coupling will be enhanced by a factor of $\Omega$ from that of the zero mode, $\frac{ \mathcal{F}_{\mu\nu}^{(n)}(R)}{\mathcal{F}_{\mu\nu}^{(0)}}\sim e^{kR}$. In other words if warped extra dimensions are the sole resolution to the gauge hierarchy problem then one would anticipate gravitational effects becoming apparent at the LHC. Although if $\Omega$ is less than $10^{15}$, as in the little RS model \cite{Davoudiasl:2008hx, PhysRevD.79.076001, PhysRevD.77.124046}, then one can safely neglect gravity. It should also be noted that the coupling between fields localised towards the UV brane and KK gravitons would be suppressed by a factor of $\Omega$. Hence constraints from gravity mediated FCNC's would be heavily suppressed as would many of the contribution to EW observables.

\subsection{The Newtonian Limit}    
To get an idea of how Eq. \ref{GravitonEOM} reduces to \color{red} Newton's \color{black} inverse square law we introduce a source mass term on the IR brane, $m_s\delta^4(x)\delta(r-r_{\rm{ir}})\delta^\delta(\phi-\phi_{\rm{ir}})$ \cite{ArkaniHamed:1999hk}. We also consider a very large speed of light. Hence we let $\Box\rightarrow -\nabla^2$ and also only consider the scalar quantity $\tilde{h}^{(n)}_{00}$ which we denote $U_n$ \cite{Carroll:1997ar}. Eq. \ref{GravitonEOM} now reduces to a Poisson's equation
\begin{equation}
\label{ }
-\nabla^2U_n+m_n^2U_n=-\frac{1}{M_{\rm{fund}}^{3+\delta}}m_s\delta^4(x)\delta(r-r_{\rm{ir}})\delta^\delta(\phi-\phi_{\rm{ir}})
\end{equation}    
This can then be solved to give the total potential at a point $x=\tilde{r}$
\begin{eqnarray*}
U(\tilde{r})=\sum_{n=0}\frac{m_se^{-m_n\tilde{r}}}{M_{\rm{fund}}^{3+\delta}\tilde{r}}|\mathcal{F}_{00}^{(n)}(r_{\rm{ir}},\phi_{\rm{ir}})|^2\hspace{2.2cm}\\
=\frac{1}{M_{\rm{Pl}}^2}\frac{m_s}{\tilde{r}}\left (1+\sum_{n=1}e^{-m_n\tilde{r}}\left |\frac{\mathcal{F}_{00}^{(n)}(r_{\rm{ir}},\phi_{\rm{ir}})}{\mathcal{F}_{00}^{(0)}(r_{\rm{ir}},\phi_{\rm{ir}})}\right |^2\right )
\end{eqnarray*}
where in the second line we have used both the normalisation of the graviton zero mode and Eq. \ref{MPlanck}. Hence where one has a discrete KK spectrum the corrections to the Newtonian potential are suppressed by the $e^{m_n \tilde{r}}$ term. In scenarios in which the graviton KK modes tend towards a continuous spectrum, such as the RS model with a single brane, then the sum becomes an integral and the contribution can become significant \cite{Randall:1999vf, Kirtis:2002ca}. 

For these reasons, in the remainder of this thesis, we will not consider the gravitational sector. It turns out that in the scenarios that we are considering here the dominant constraints come from the tree level exchange of KK gauge modes. We will now move on to look at these constraints over the next two chapters.   

%% file: FlavourChap.tex
\chapter{Flavour in Warped Extra Dimensions}
\label{FlavourChap}

Although originally proposed as a resolution of the gauge hierarchy problem, it was quickly realised that the 5D RS model also provides a quite neat description of flavour \cite{Grossman:1999ra, Pomarol:1999ad, Huber:2000ie}. In this chapter, this description will be outlined along with some of the problems which can arise. Although before we get into the details of the model it is worth explaining what is meant by a `description' of flavour.

The fermion masses are observed to range from the top mass of $\approx 172$ GeV \cite{Nakamura:2010zzi} to the neutrino masses with a combined mass of $\color{red}\sum_\nu m_\nu< 1.19$ eV  (or  $\color{red} \sum_\nu m_\nu< 0.44$ eV if one assumes neutrinos are hot dark matter candidates) \cite{Hannestad:2010yi}. Also the mixing between the different generations in the quark sector \cite{Nakamura:2010zzi,  Bona:2007vi}, 
 \begin{eqnarray}
V_{us}=0.2253\pm0.0007\quad\quad V_{cb}=0.0410^{+0.0011}_{-0.0007}\quad\quad V_{ub}=0.00347^{+0.00016}_{-0.00012}\nonumber\\
V_{cd}=0.2252\pm0.0007\quad\quad V_{ts}=0.0403^{+0.0011}_{-0.0007}\quad\quad V_{td}=0.00862^{+0.00026}_{-0.00020}\label{expMixangle}
\end{eqnarray} 
is observed to be slightly smaller than in the lepton sector \cite{Valle:2006vb}
\begin{equation} 
\label{PMNSangle}
\sin^2\theta_{12}=0.24-0.40\quad\quad \sin^2\theta_{23}=0.34-0.68\quad\quad\sin^2\theta_{13}\le0.04\quad\quad\mbox{(at $3\sigma$)}.
\end{equation}  
\color{red} Although recent results, from T2K and MINOS, disfavour a $\theta_{13}=0$ hypothesis at $2.5\sigma$ and $1.5\sigma$ respectively \cite{Abe:2011sj, Adamson:2011qu}. \color{black} In addition the amount of CP violation, that has been observed in the decays of kaons and B mesons, is not sufficient to give rise to baryogenesis \cite{Farrar:1993sp}. So a `neat description' of flavour should be able to offer a natural explanation of why the fermion masses range over so many orders of magnitude, as well as why the mixing angles take the values that they do. It should also include additional sources of CP violation while at the same time \color{red} ensuring \color{black} all additional contributions to existing sources of CP violation are small enough to be in agreement with experiment. Likewise, to avoid conflict with experiment, FCNC's must be suppressed in the low energy theory. Ideally a description of flavour would also explain why there exists three generations although, as already mentioned, one must go to at least six dimensions to achieve this and here we will focus on five dimensions \cite{Dobrescu:2001ae}. 

Of course the next question is what is meant by natural? There is no clear cut answer to this question. Here we will demonstrate that warped extra dimensions offer a plausible explanation to many of these problems but, as we have already seen, considering fermions in extra dimensions leads to a significantly enhanced parameter space. However here we will argue that even with an enlarged parameter space these models are still reasonably predictive and critically all the new free parameters are taken to be at the same order of magnitude. 

\section{The RS Description of Flavour}\label{sect:RSflavour}
Our starting point is the observation, made in section \ref{sect:fermions}, that in five dimensions, fermions are vector like objects and one can include a bulk mass term. Quite generically this mass term causes the fermion zero mode to sit towards either the IR or UV end of the space (Eq. \ref{5DfermionZeromode}). Turning now to the RS model  the space is specified by two dimensionful parameters, the curvature $k$ and the size $R^{-1}$, which are both taken to be of roughly the same order of magnitude. Hence one would anticipate that the bulk mass terms are also at approximately the same order of magnitude and so we parameterise the bulk mass terms (in Eq. \ref{5DfermionZeromode}) by
\begin{displaymath}
M=ck
\end{displaymath}      
where $c$ is assumed to be $\mathcal{O}(1)$. The fermion zero modes are then given by Eq. \ref{5DfermionZeromode}
\begin{equation}
\label{RSfermZero}
\color{red} f_{L,R}^{(0)\;i}(r)=\sqrt{\frac{k(1\mp2c_i)}{\exp (kR\mp 2c_ikR)-1}}\exp(\mp c_ikr),
\end{equation}
where $i$ is now a flavour index running from $1$ to $3$ and the $\pm$ refer to $f_L$ and $f_R$ respectively. By looking at how the function scales relative to the normalisation constant it is straight forward to see that $f_L^{(0)}$ is localised towards the UV (IR) brane when $c>\frac{1}{2}$ ($c<\frac{1}{2}$) and like wise $f_R^{(0)}$ is sitting towards the UV (IR) brane when $c>-\frac{1}{2}$ ($c<-\frac{1}{2}$). If we now consider the Yukawa coupling to a Higgs localised on the IR brane. 
\begin{equation}
\label{RSFermionMassMat}
S=\int d^5x \sqrt{G}\;Y_{ij} \Phi\cdot\bar{\psi}^i_L\psi^j_R\delta(r-R)\supset\sum_{n,m}\int d^4x \frac{\lambda_{ij}\Omega v}{k}f_L^{(n)\;i}(R)f_R^{(m)\;j}(R)\bar{\psi}^{(n)}_L\psi^{(m)}_R
\end{equation}
where in the second step we have parameterised the 5D Yukawa couplings in terms of dimensionless $\mathcal{O}(1)$ couplings $\lambda_{ij}=Y_{ij}k$.\newline In breaking electroweak symmetry, $\Phi\rightarrow\left (\begin{array}{c}0 \\ v_0+H \end{array}\right )$. The Higgs VEV is then rescaled by warping, $v_0=\Omega v$, such that $v\approx 174$ GeV. In the gauge eigenstate the masses of the fermion zero modes will then be dominated by the mass term
\begin{equation}
\label{RSMassMat}
M_{ij}=\lambda_{ij} v \sqrt{\frac{(1-2c_L^i)(1+2c_R^j)}{(\Omega^{1-2c_L^i}-1)(\Omega^{1+2c_L^i}-1)}}\;\Omega^{1-c_L^i+c_R^j}
\end{equation} 
Up to now we have assumed that $\lambda$ is $\mathcal{O}(1)$, however one can get an idea of the upper bound on $\lambda$ by considering some naive dimensional analysis. Following the arguments of \cite{Csaki:2008zd}, since $Y$ has mass dimension $-1$, one can infer that the loop corrections to the Yukawa coupling should be $\sim Y(Y E\Omega)^2/(16\pi^2)$. One would like to retain perturbative control until energies involving $N$ KK modes, i.e $E\sim N (2\;\mbox{to}\;3) M_{\rm{KK}}\sim N (2\;\mbox{to}\;3) k/\Omega$. So we require that $ (Y E \Omega)^2/(16\pi^2)\sim (\lambda N (2\;\mbox{to}\;3))^2)/ (16\pi^2)\lesssim 1$ and hence if $\lambda \gtrsim (4\pi)/((2\;\mbox{to}\;3)N)$ one loses perturbative control of the low energy theory and these constraints become meaningless. With the conservative bound of $N =2$ then one requires $\lambda\lesssim 3$.  

\begin{figure}[t!]
\begin{center}
\includegraphics[width=0.9\textwidth]{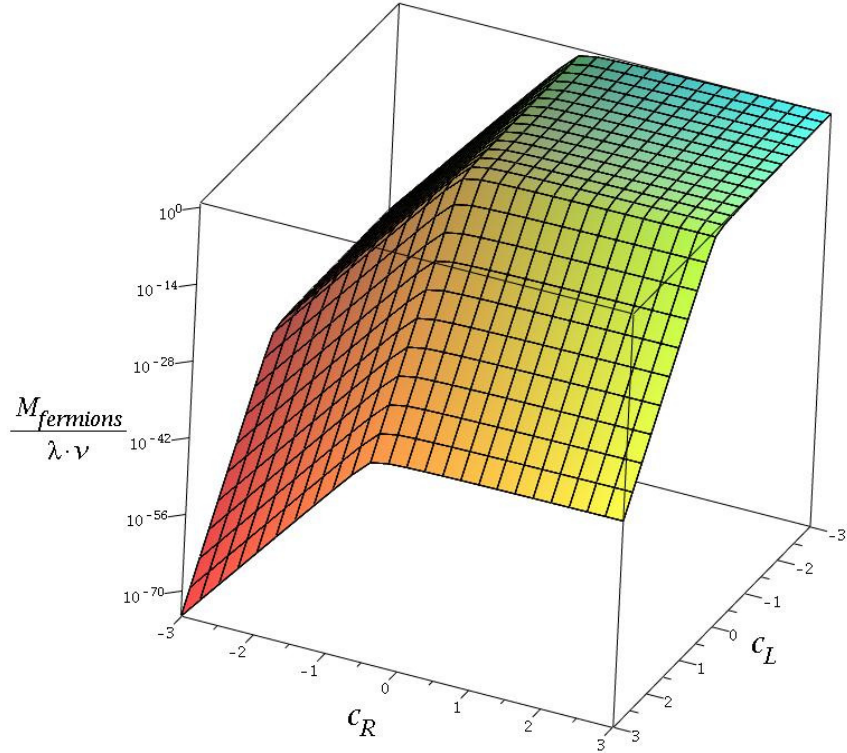}
\caption[Masses of Fermion Zero Modes in the RS Model]{The range of possible masses of the fermion zero modes in the RS model with $\Omega=10^{15}$.}
\label{RSMM}
\end{center}
\end{figure}

The range of possible zero mode fermion masses have then been plotted in figure \ref{RSMM}. Firstly one can see that with small changes in the bulk mass parameters, $c$, one can easily generate fermions with masses many orders of magnitude lower than $v$.  However it can be seen that, if a non-perturbative low energy theory is to be avoided, then one cannot have fermions with zero modes much larger than $\sim 500-600$ GeV. This is not in conflict with models with a fourth generation of fermions, for which EW fits seem to favour $M_{t^{\prime},\; b^{\prime}}\lesssim 500$ GeV \cite{Hashimoto:2010at}. Although, since the measurement of the invisible Z decay width has pretty much excluded a light fourth generation ($M_{\nu^{\prime}}\leqslant 45.6$ GeV), this does raise the question why are there no very light fermions. From figure \ref{RSMM}, one can see that even with $|c|<1$, one can easily generate fermions of masses 20 orders of magnitude lower than the higgs VEV. Yet even if the lightest neutrino is massless, the majority of the SM fermions appear to have bulk mass terms close to $c_L\sim0.5-0.7$ with no obvious reason as to why.

Moving on to consider the unitary matrices that will diagonalise the mass matrix in Eq. \ref{RSFermionMassMat}. If the fermion masses, in the eigenstate in which gauge couplings are flavour diagonal, are given by
\begin{displaymath}
M_{ij}^{(n,m)}=\frac{\lambda_{ij}\Omega v}{k}f_{L\;i}^{(n)}(R)f_{R\;j}^{(m)}(R)+m_n\delta^{nm}\delta_{ij}f_{L\;i}^{(n)}(R)f_{R\;i}^{(n)}(R)
\end{displaymath} 
where $i,j$ run over the three flavour indices and $m,n$ run over the KK number. The physical masses are then found by acting with the unitary transformations $\tilde{\psi}_{L,R}= U_{L,R}\psi_{L,R}$ such that $U_L^\dag MU_R$ is diagonal. As explained in \cite{Cheng&Li}, $U_{L,R}$ is found by diagonalising $MM^\dag$.
Hence at tree level there is a mixing between the KK fermions and the fermion zero modes. One would expect that this would lead to a violation of unitarity in the observed CKM matrix caused by a combination of the truncation to zero modes as well as a modification in the fermion couplings. However in practice this potential deviation from unitarity is found to be small \cite{Huber:2003tu} and dominated by the deviation in the fermion couplings. Bearing in mind that corrections from the KK fermions to the fermion zero mode masses will be of order $\mathcal{O}(v^2/M_{\rm{KK}}^2)$, then one can see that, if $M_{\rm{KK}}\gg v$, it is reasonable to take a zero mode approximation and neglect the fermion KK modes.    

If one considered just the zero modes then the mass matrix has a clear product structure, i.e. $M_{ij}\sim a_ib_j$, which can be diagonalised by $(U_L)_{ij}\sim\frac{a_i}{a_j}$ and $(U_R)_{ij}\sim\frac{b_i}{b_j}$ \cite{Huber:2003tu}. Hence one would anticipate that Eq. \ref{RSFermionMassMat} would be diagonalised by
\cite{Huber:2003tu, Blanke:2008zb, Casagrande:2008hr}
\begin{equation}
\label{RSUnitary}
(U_{L,R})_{ij}\sim\omega_{ij}\frac{f_{L,R}^i(R)}{f_{L,R}^j(R)}
\end{equation} 
where $\omega_{ij}$ is related to the Yukawa couplings, $\lambda_{ij}$. Hence, if one assumes purely anarchic couplings (i.e. there is no hierarchical structure to $\lambda_{ij}$), the RS model predicts that the \color{red} off diagonal terms \color{black} in the CKM matrix ($V_{\rm{CKM}}=U_L^{u\;\dag}U_R^{d}$) will obey the relationship
\begin{equation}
\label{CKMpred}
V_{ub}\sim V_{us}V_{cb} \quad\quad\mbox{and}\quad\quad V_{td}\sim V_{ts}V_{cd}.
\end{equation}      
This is in partial agreement with the observed values in Eq. \ref{expMixangle}. Although one would anticipate such a relation arising in any model of flavour in which the SM effective Yukawa's arose from some product structure between $\psi_L$ and $\psi_R$. It also does not hold completely, since one would then anticipate $V_{ub,\;td}\sim  0.009$. \color{red} Nonetheless \color{black} it is a promising zeroth order result.

\section{Flavour Changing Neutral Currents} \label{sect:FCNC}
\begin{figure}[ht!]
    \begin{center}
        \subfigure[SM contribution to FCNC's.]{%
           \label{fig:SMbox}
           
    \begin{fmffile}{figures/feynSMBOX} 	
  \fmfframe(1,4)(1,4){ 	
   \begin{fmfgraph*}(180,80) 
    \fmfleft{i1,i2}	
    \fmfright{o1,o2}    
    \fmflabel{$\psi$}{i1} 
    \fmflabel{$\psi$}{i2} 
    \fmflabel{$\psi$}{o1} 
    \fmflabel{$\psi$}{o2} 
    \fmf{fermion}{i1,v1,v2,i2} 
    \fmf{fermion}{o2,v4,v3,o1} 
    \fmf{photon,label=$W_\mu$}{v1,v3}
    \fmf{photon,label=$W_\mu$}{v2,v4} 
   \end{fmfgraph*}
  }
\end{fmffile}

        }
        \subfigure[FCNC's arising at tree level from the exchange of KK gauge bosons.]{%
           \label{fig:KKgaugeexchange}
      \begin{fmffile}{figures/feynKKexchange} 	
  \fmfframe(1,4)(1,4){ 	
   \begin{fmfgraph*}(180,80) 
    \fmfleft{i1,i2}	
    \fmfright{o1,o2}    
    \fmflabel{$\psi$}{i1} 
    \fmflabel{$\psi$}{i2} 
    \fmflabel{$\psi$}{o1} 
    \fmflabel{$\psi$}{o2} 
    \fmf{fermion}{i1,v1,i2} 
    \fmf{fermion}{o1,v2,o2} 
    \fmf{photon,label=$A_\mu^{(n)}$}{v1,v2} 
   \end{fmfgraph*}
  }
\end{fmffile}

        }
   \label{fig:FCNC}
    \end{center}
    \caption[Feynman Diagrams Contributing to FCNC's]{Feynman diagrams contributing to FCNC's.}
\end{figure}
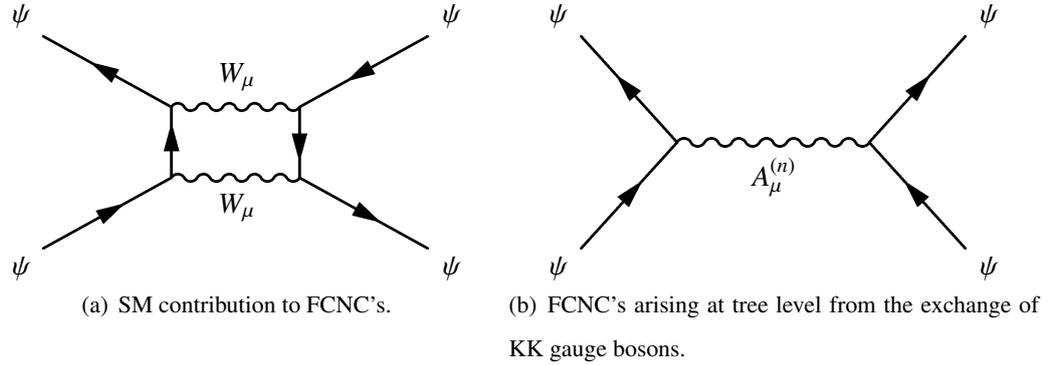

Inherent in the above description of the hierarchy of fermion masses is the requirement that different generations of fermions are located at different points throughout the space, i.e. have different bulk masses. This would then give rise to non universal gauge fermion couplings and FCNC's at tree level. Such processes occur in the SM at loop level (see figure \ref{fig:SMbox}) but are heavily suppressed by the GIM mechanism. They have also typically been measured to a reasonable level of precision and found to be in agreement with the SM, particularly with regards to $K^0-\bar{K}^0$ mixing. Hence FCNC's often give rise to very stringent constraints on new physics.    

Here we will primarily study kaon physics however it is straight forward to extend this analysis to, for example, $B$ physics \cite{Bauer:2009cf, Goertz:2011nx}. As already mentioned, in extra dimensional scenarios FCNC's arise, at tree level, through the exchange of gauge fields with non universal gauge fermion couplings (see figure \ref{fig:KKgaugeexchange}). Hence it is convenient to integrate out the full tower \color{red} of \color{black} gauge fields to arrive at the 4D effective Hamiltonian written in terms of dimension six, four fermion operators, see e.g. \cite{Bagger:1997gg}  
\begin{equation}
\label{ }
\mathcal{H}_{\mbox{eff}}^{\triangle S=2}=\sum_{i=1}^5C_iQ_i^{sd}+\sum_{i=1}^3\tilde{C}_i\tilde{Q}_i^{sd}
\end{equation}
where
\begin{equation*}
\color{red}\begin{array}{ll}
Q_1^{sd}=(\bar{d}_L\gamma^\mu s_L)(\bar{d}_L\gamma_\mu s_L)  & \tilde{Q}_1^{sd}=(\bar{d}_R\gamma^\mu s_R)(\bar{d}_R\gamma_\mu s_R)   \\
 Q_2^{sd}=(\bar{d}_R s_L)(\bar{d}_R s_L)  &    \tilde{Q}_2^{sd}=(\bar{d}_L s_R)(\bar{d}_L s_R) \\
 Q_3^{sd}=(\bar{d}^\alpha_R s^\beta_L)(\bar{d}^\beta_R s^\alpha_L) & \tilde{Q}_3^{sd}=(\bar{d}^\alpha_L s^\beta_R)(\bar{d}^\beta_L s^\alpha_R)\\
Q_4^{sd}=(\bar{d}_Rs_L)(\bar{d}_Ls_R) & \\ 
Q_5^{sd}=(\bar{d}^\alpha_Rs^\beta_L)(\bar{d}^\beta_Ls^\alpha_R) &
\end{array}
\end{equation*}
and $\alpha$ and $\beta$ are colour indices. The size of most of the relevant observables are then determined from the Wilson coefficients of these operators. The current, model independent,  bounds on these coefficients are given by \cite{Bona:2007vi} as;
\begin{table}
\begin{center}
\begin{tabular}{|cc|cc|}
\hline
  Parameter & $95\%$ allowed range (GeV${}^{-2}$) &Parameter & $95\%$ allowed range (GeV${}^{-2}$)  \\
  \hline
  \hline
 Re $C_1$  & $[-9.6,\;9.6]\;\times 10^{-13} $& Im $C_1$  & $[-4.4,\;2.8]\;\times 10^{-15} $  \\
 Re $C_2$  & $[-1.8,\;1.9]\;\times 10^{-14} $& Im $C_2$  & $[-5.1,\;9.3]\;\times 10^{-17} $  \\
 Re $C_3$  & $[-6.0,\;5.6]\;\times 10^{-14} $& Im $C_3$  & $[-3.1,\;1.7]\;\times 10^{-16} $  \\
 Re $C_4$  & $[-3.6,\;3.6]\;\times 10^{-15} $& Im $C_4$  & $[-1.8,\;0.9]\;\times 10^{-17} $  \\
 Re $C_5$  & $[-1.0,\;1.0]\;\times 10^{-14} $& Im $C_5$  & $[-5.2,\;2.8]\;\times 10^{-17} $  \\
\hline
\end{tabular}
\end{center}
  \centering 
  \caption[]{\color{red}Model independent bounds on coefficients of four fermion operators \cite{Bona:2007vi} \color{black}.}\label{ }
\end{table}
We now consider the gauge fermion interactions in the five dimensional SM, after gauging away the fifth component of the gauge field and working in the gauge eigenstates, 
\begin{equation}
\label{ }
S=\int d^5x\; a^3b\left (g(W_\mu^+J_{W+}^\mu+W_\mu^-J_{W-}^\mu+Z_\mu J_Z^\mu)+eA_\mu J_{\rm{EM}}^\mu+g_sG_\mu^aJ_{\rm{QCD}}^{\mu \; a}\right ).
\end{equation}   
Here we are working post spontaneous symmetry breaking with the gauge fields and couplings defined in the next chapter. The fermion currents of relevance to kaon physics are then given by
\begin{eqnarray*}
J_\mu^{W+}\supset \frac{1}{\sqrt{2}}\bar{u}_{Li}\gamma_\mu d_{Li}\hspace{1cm}J_\mu^{W-}\supset \frac{1}{\sqrt{2}}\bar{d}_{Li}\gamma_\mu u_{Li}  \hspace{4cm}\\
J_\mu^Z\supset\frac{1}{c_w}\left (\bar{u}_{Li}(\frac{1}{2}-\frac{2}{3}s_w^2)\gamma_\mu u_{Li}+\bar{d}_{Li}(-\frac{1}{2}+\frac{1}{3}s_w^2)\gamma_\mu d_{Li}+\bar{u}_{Ri}(-\frac{2}{3}s_w^2)\gamma_\mu u_{Ri}+\bar{d}_{Ri}(\frac{1}{3}s_w^2)\gamma_\mu d_{Ri}\right )\\
J_\mu^{\rm{EM}}\supset +\frac{2}{3}\bar{u}_{Li}\gamma_\mu u_{Li}+\frac{2}{3}\bar{u}_{Ri}\gamma_\mu u_{Ri}-\frac{1}{3}\bar{d}_{Li}\gamma_\mu d_{Li}-\frac{1}{3}\bar{d}_{Ri}\gamma_\mu d_{Ri}\hspace{3cm}\\
J_\mu^{\rm{QCD}\;a}\supset \bar{u}_{Li}\tau^a\gamma_\mu u_{Li}+\bar{u}_{Ri}\tau^a\gamma_\mu u_{Ri}+\bar{d}_{Li}\tau^a\gamma_\mu d_{Li}+\bar{u}_{di}\tau^a\gamma_\mu u_{di}\hspace{3cm},
\end{eqnarray*}
where $s_w$ is the weak mixing angle. Rather than integrating out each KK gauge field individually, here it is easier to integrate out the extra dimensional component of the gauge propagator obtained in section \ref{PosMomProp}. Hence the dimension six operators that appear in the low energy effective Hamiltonian, from diagrams of the form of figure \ref{fig:KKgaugeexchange}, will be given by
 \begin{eqnarray*}
\mathcal{H}_{\mbox{eff}}\supset \frac{1}{2}\int dr\int d\tilde{r} \;a^3(r)b(r)\Bigg [ g^2J_Z^\mu(r) G_p^{(Z)}(u,v)J_\mu^{Z}(\tilde{r})+e^2J_{\rm{EM}}^\mu(r)G_p^{(A)}J^{\rm{EM}}_\mu(\tilde{r})\\+g_s^2J_{\rm{QCD}}^\mu(r)G_p^{(G)}J^{\rm{QCD}}_\mu(\tilde{r})\Bigg ] a^3(\tilde{r})b(\tilde{r}).
\end{eqnarray*} 
We now rotate into the mass eigenstate with the unitary matrices $U_{L,R}$ and define the integral
\begin{eqnarray}
\mathbb{I}_{\psi_k\chi_l\xi_m\sigma_n}^{(A)}=\sum_{i,j=1}^3(U_\psi^\dag)^{ki}(U_\chi)^{il}\Bigg[\int_R^\infty dr\int_R^\infty d\tilde{r}\frac{b(r)}{a(r)}f_\psi^i(r) f_\chi^i(r)G_p^{(A)}(u,v)\nonumber\\
\times f_\xi^j(\tilde{r})f_\sigma^j(\tilde{r})\frac{b(\tilde{r})}{a(\tilde{r})}\Bigg ](U_\xi^\dag)^{mj}(U_\sigma)^{jn}\label{fourfermInt}
\end{eqnarray}  
where $\psi,\chi,\xi,\sigma=L,R$ and $i,j,k,l,m,n$ are flavour indices. If we now use the Fierz identities
\begin{equation*}
\begin{array}{l}
\left(\bar{\psi}_{L,R}\gamma^\mu\psi_{L,R}\right )\left(\bar{\chi}_{L,R}\gamma_\mu\chi_{L,R}\right )=\left(\bar{\psi}_{L,R}\gamma^\mu\chi_{L,R}\right )\left(\bar{\chi}_{L,R}\gamma_\mu\psi_{L,R}\right )\\
\left(\bar{\psi}_{R}\gamma^\mu\psi_{R}\right )\left(\bar{\chi}_{L}\gamma_\mu\chi_{L}\right )=-2\left (\bar{\psi}_R\chi_L\right )\left (\bar{\chi}_L\psi_R\right )\\
\left(\bar{\psi}^\alpha\gamma^\mu\tau^a_{\alpha\beta}\psi^\beta\right )\left(\bar{\chi}^\gamma\gamma_\mu\tau^a_{\gamma\delta}\chi^\delta\right )=\frac{1}{2}\left (\left (\bar{\psi}^\alpha\gamma^\mu\psi^\beta\right )\left (\bar{\psi}^\beta\gamma_\mu\chi^\alpha\right )-\frac{1}{N_C}\left (\bar{\psi}^\alpha\gamma^\mu\psi^\alpha\right )\left (\bar{\psi}^\beta\gamma_\mu\chi^\beta\right )\right )
\end{array}
\end{equation*}
then the Wilson coefficients for the tree level exchange of KK gauge fields are given by
\begin{equation*}
\begin{array}{l}
C_1=\frac{g_s^2}{6}\mathbb{I}_{L_dL_sL_dL_s}^{(G)}+\frac{g^2}{2c_w^2}\bigg(\frac{1}{2}-\frac{1}{3}s_w^2\bigg)^2\mathbb{I}_{L_dL_sL_dL_s}^{(Z)}+\frac{e^2}{18}\mathbb{I}_{L_dL_sL_dL_s}^{(A)}\\
\tilde{C}_1=\frac{g_s^2}{6}\mathbb{I}_{R_dR_sR_dR_s}^{(G)}+\frac{g^2 s_w^4}{18c_w^2}\mathbb{I}_{R_dR_sR_dR_s}^{(Z)}+\frac{e^2}{18}\mathbb{I}_{R_dR_sR_dR_s}^{(A)}\\
C_4=-g_s^2\;\mathbb{I}_{L_dL_sR_dR_s}^{(G)}\\
C_5=\frac{g_s^2}{3}\mathbb{I}_{L_dL_sR_dR_s}^{(G)}+\frac{2g^2s_w^2}{3c_w^2}\bigg (\frac{1}{2}-\frac{1}{3}s_w^2\bigg )\mathbb{I}_{L_dL_sR_dR_s}^{(Z)}-\frac{2e^2}{9}\mathbb{I}_{L_dL_sR_dR_s}^{(A)}
\end{array}
\end{equation*}
with $C_2=C_3=0$. Having obtained these coefficients the observables of interest can be computed. For example two important observables in the $K^0-\bar{K}^0$ system is the difference in masses of the two eigenstates $K_L$ and $K_S$, $\Delta m_K=3.483\pm 0.006\times 10^{-12}$ MeV and the measure of CP violation in the system $|\epsilon_K|=2.228\pm0.011\times 10^{-3}$ \cite{Nakamura:2010zzi}. These are given by \cite{Bauer:2009cf, Blanke:2008zb}
\begin{equation}
\label{ }
\Delta m_K=2 \rm{Re}<K^0|\mathcal{H}_{\mbox{eff}}^{\triangle S=2}|\bar{K}^0>
\end{equation}
and
\begin{equation}
\label{ }
\epsilon_K=\frac{\kappa_\epsilon e^{i\varphi_\epsilon}}{\sqrt{2}(\triangle m_K)_{\rm{exp}}}\rm{Im}<K^0|\mathcal{H}_{\mbox{eff}}^{\triangle S=2}|\bar{K}^0>
\end{equation} 
where $\varphi_\epsilon=43.51^\circ$ and $\kappa_\epsilon=0.92$, while
\begin{eqnarray*}
<K^0|Q_1^{sd}|\bar{K}^0>=\frac{m_Kf_K^2}{3}B_1\hspace{2.3cm}\\
<K^0|Q_4^{sd}|\bar{K}^0>=\left (\frac{m_K}{m_d+m_s}\right )^2\frac{m_Kf_K^2}{4}B_4\\
<K^0|Q_5^{sd}|\bar{K}^0>=\left (\frac{m_K}{m_d+m_s}\right )^2\frac{m_Kf_K^2}{12}B_5
\end{eqnarray*}
and $m_K=497.6$ MeV, $f_K=156.1$ MeV. The hadronic matrix elements and quark masses are run to a scale of the mass of the first gauge KK mode ($\sim (2-3)\times M_{\rm{KK}}$) \cite{Buras:1998raa}.
\begin{table}
\begin{center}
\begin{tabular}{|c|cccc|}
\hline
   & 1 TeV & 3 TeV & 10 TeV & 30 TeV \\
 \hline  
  $B_1$ & 0.4074 & 0.3953 & 0.3837 & 0.3744 \\
  $B_2$ & 0.5235 & 0.4993 & 0.4765 & 0.4585 \\
  $B_3$ & 0.00736 & -0.01963 & -0.04018 & -0.05349 \\
 $B_4$ & 0.938 & 0.938 & 0.938 & 0.938 \\
   $B_5$& -0.3358 & -0.3725  & -0.4041 & 0.4273 \\
\hline
\end{tabular}
\end{center}
  \caption[]{\color{red}Hadronic matrix elements for kaons.\color{black}}\label{ }
\end{table}
 While there is considerable uncertainty in the SM prediction for these observables, for example $(\epsilon_K)_{\rm{SM}}=2\pm0.5 \times 10^{-3}$ \cite{Charles:2004jd}, the point is these observables are suppressed by the GIM mechanism and hence must also be suppressed in any beyond the standard model scenario. 
 
For fear of \color{red} losing \color{black} sight of the physics behind these slightly cumbersome expressions it is worth rewriting Eq. \ref{fourfermInt} in terms of the gauge field KK expansion Eq. \ref{GaugePropKKequiv},
\begin{equation}
\label{ILLLL}
\mathbb{I}_{L_kL_lL_mL_h}^{(A)}=f_0^2\sum_n\sum_{i,j=1}^3(U_L^\dag)^{ki}(U_L)^{il}\frac{F_{L\;\psi}^{(n)\;i}F_{L\;\psi}^{(n)\;j}}{p^2-m_n^2}(U_L^\dag)^{mj}(U_L)^{jh}
\end{equation}
where we have introduced the relative coupling of the KK gauge field to the fermion zero mode defined to be
\begin{equation}
\label{Fpsi}
F_{L,R\;\psi}^{(n)\;i}=\frac{\int dr\; \frac{b}{a}f_{L,R}^{(0)\;i}f_{n}f_{L,R}^{(0)\;i}}{f_0}
\end{equation}
where $f_n$ is the KK gauge profile (Eq. \ref{GaugeEOM}) and $f_0$ is the flat profile of the zero mode of a massless gauge field
\begin{equation}
\label{fzero}
f_0=\frac{1}{\sqrt{\int b dr}}.
\end{equation}
If one neglects EW corrections then the dimensionful five coupling can be equated to $g_s^2f_0^2\approx 4\pi \alpha_s$, $g^2f_0^2\approx 4\pi \alpha/s_w^2$ and $e^2 f_0^2=4\pi \alpha$. Hence $F_\psi$ gives the gauge fermion coupling relative to the SM coupling. These couplings have been plotted in figure \ref{RelFermCoupl} for the RS model.

\begin{figure}[ht!]
    \begin{center}
        \subfigure[]{%
           \label{RSFpsi}
           \includegraphics[width=0.45\textwidth]{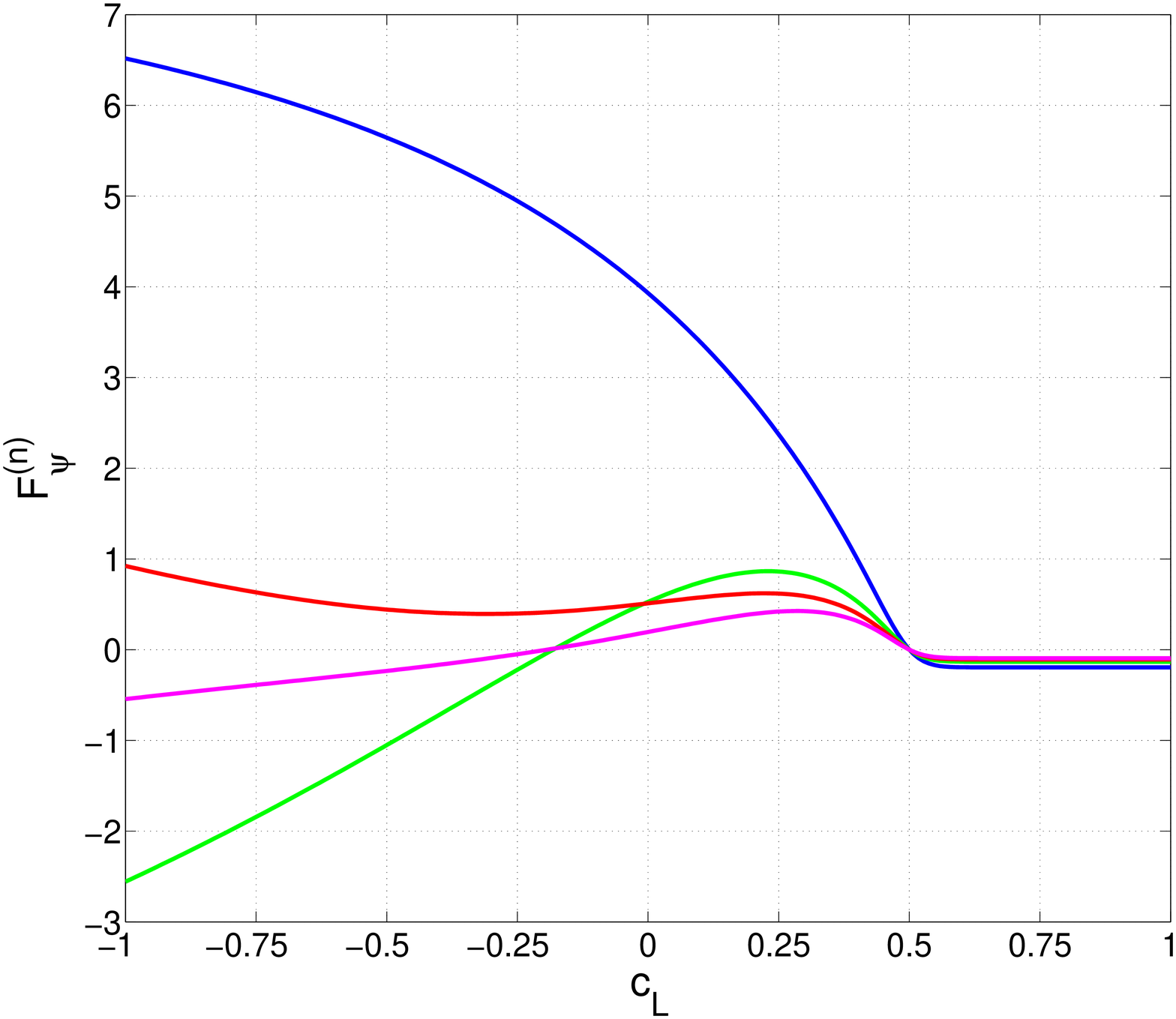}
        }
        \subfigure[]{%
           \label{fig:ZfermCoup}
           \includegraphics[width=0.45\textwidth]{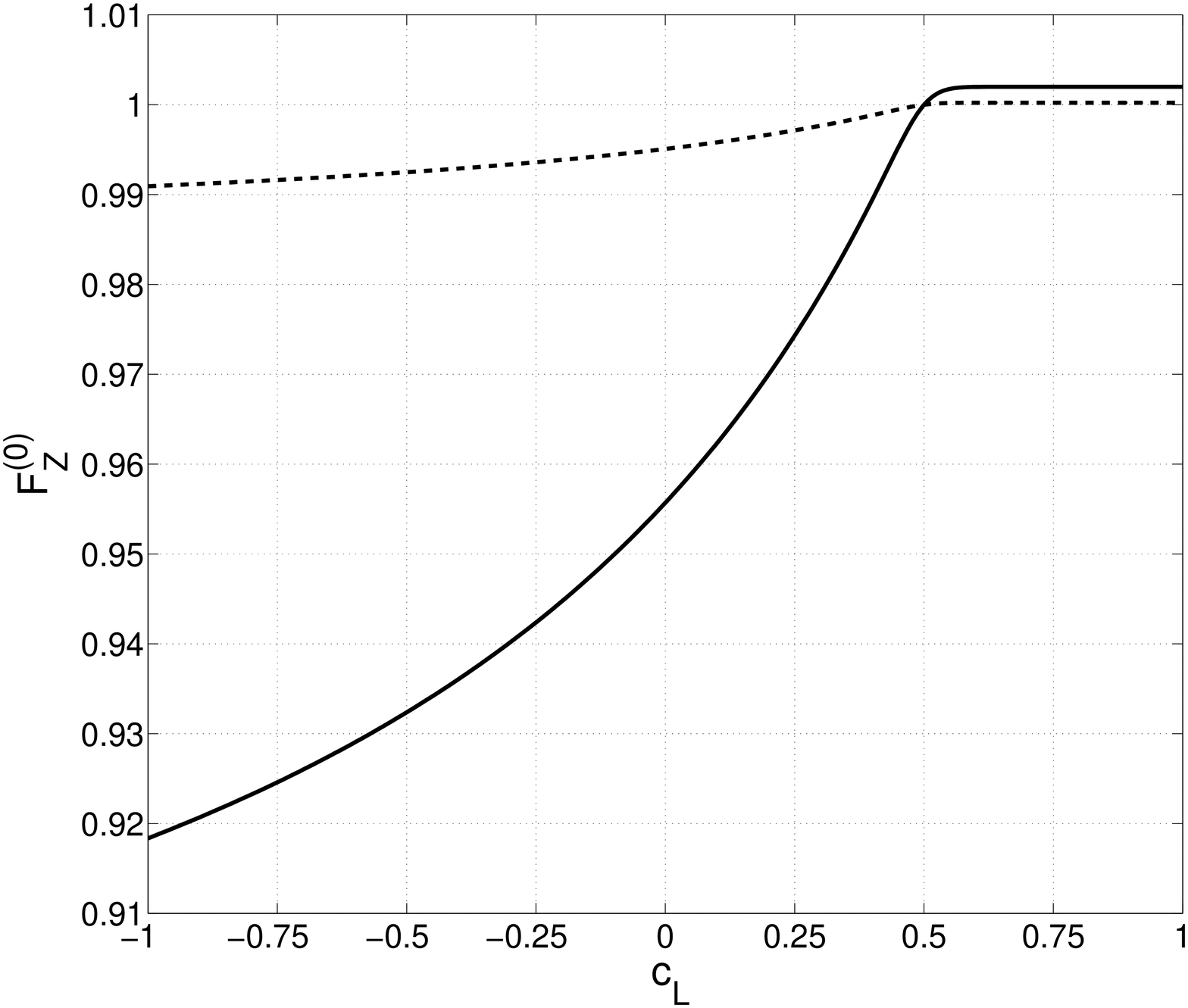}
        }
   \label{RelFermCoupl}
    \end{center}
    \caption[Gauge Fermion Couplings in the RS Model ]{The relative gauge fermion coupling for the RS model with $M_{\rm{KK}}=1$ TeV (solid line) and $M_{\rm{KK}}=3$ TeV (dashed line). Plotted on the left is the first (blue), second (green), third (red) and fourth (magenta) gauge KK modes. On the right is the coupling of the zero mode of the Z boson. Note on the left hand side the relative coupling for $3$ TeV is indistinguishable from coupling for $1$ TeV. $\Omega=10^{15}$. } \label{RelFermCoupl}
\end{figure}    

Turning back to Eq. \ref{ILLLL} one can see that if one had universal couplings (i.e. $F_\psi$ was constant for all generations) then the unitary matrices would act on each $F_\psi^i$ and the off diagonal terms in $\mathbb{I}_{L_kL_lL_mL_h}^{(A)}$ would be zero. Hence the extent to which FCNC's are suppressed is determined by the size of the non universalities in the gauge fermion couplings. We can now see the so called `RS-GIM' mechanism. As mentioned in the previous section the light fermions will sit towards the UV tip of the space ($c_L>0.5$), while the KK gauge fields will sit towards IR tip of the space and hence the gauge fermion couplings will be quite universal (see figure \ref{RSFpsi} with $c_L>0.5$). In other words the same mechanism that is used in order to generate fermion masses many orders of magnitude lower than the Higgs VEV would also suppress the FCNC's. In fact it can be shown that all four fermion operators are suppressed, to some extent, by this mechanism \cite{Huber:2000ie}. The only operators for which the suppression is not sufficient are the operators responsible for proton decay, $qqql$. However these operators can be suppressed further by using the enlarged Lorentz symmetry that one obtains in spaces of more than five dimensions \cite{Appelquist:2001mj}.       

\section{A Numerical Analysis of the RS Model}
In order to see the extent to which FCNC's are suppressed here we shall once again restrict our analysis to the RS model. The price to pay for using extra dimensions to describe flavour is one has a significantly enhanced parameter space. If one ignores, for the moment, the possibility of different geometries then in addition to the Yukawa couplings one also has the bulk mass parameters.  Assuming flavour diagonal bulk mass terms then the quark sector is then described by two $3\times 3$ complex Yukawas and 9 bulk mass parameters giving a 45 dimensional parameter space. While in the 5D theory this is only 9 more than the SM. However one can no longer absorb many of these parameters with field redefinitions and hence these parameters become physical.

At the very least one should fit to the six quark masses, the three CKM mixing angles and the CP violating phase of the CKM or equivalently the Jarlskog invariant. Ideally one should also fit to all observables taken from $D^0-\bar{D}^0$ mixing, $K^0-\bar{K}^0$ mixing and $B_{d,s}-\bar{B}_{d,s}$ mixing as well as their many possible decays. Hence while such models of flavours are constrained and hence predictive, it is in practice computationally challenging to carry out a comprehensive study. Also, since many of these processes have relatively large experimental uncertainties, such a study would be dominated by the mixing angles, quark masses and Jarlskog invariant.

The approach taken here is to find points in parameter space that give the correct masses, mixing angles and Jarlskog invariant and then compute relevant observables, in particular $\epsilon_K$. The standard approach to this sort of analysis is to find a `natural' set of bulk mass parameters ($c$ values) and then to randomly generate a large number of \color{red} 5D \color{black} Yukawa couplings. One can then get an idea of the typical size of a given observable \cite{Huber:2003tu, Agashe:2004cp}. To be more specific, here we define a natural set of bulk mass parameters to be those that give the correct masses and mixing angles if there is no hierarchy in the Yukawa couplings. In other words if one generated a large number of anarchic Yukawa couplings the average masses and mixing angles should fit the observed values.

\color{red} With this definition, there is still not a unique set of natural $c$ values. \color{black} One can see, from Eq. \ref{RSUnitary}, that the mixing angles are determined from the spacing between the $c$ values, in particular the $c_L$ values. Likewise given a set of $c_L$ values one can always find a set $c_R$ values that give the right masses. Hence in theory one can slide the correctly spaced $c$ values from the UV to the IR branes. To investigate this, here we will consider four possible $c_L$ configurations which all have roughly the right spacing 
\begin{equation}
\label{ }
\begin{array}{cc}
  (A)\quad c_L=[0.69,\; 0.63,\; 0.49] & (B)\quad c_L=[0.66,\; 0.60,\; 0.42]  \\
  (C)\quad c_L=[0.63,\; 0.57,\; 0.34] & (D)\quad c_L=[ 0.60,\; 0.52,\; 0.25]
\end{array}
\end{equation}       
Our method is as follows;

\begin{figure}[ht!]
    \begin{center}
        \subfigure[]{%
            \label{fig:RScuR3}
            \includegraphics[width=0.44\textwidth]{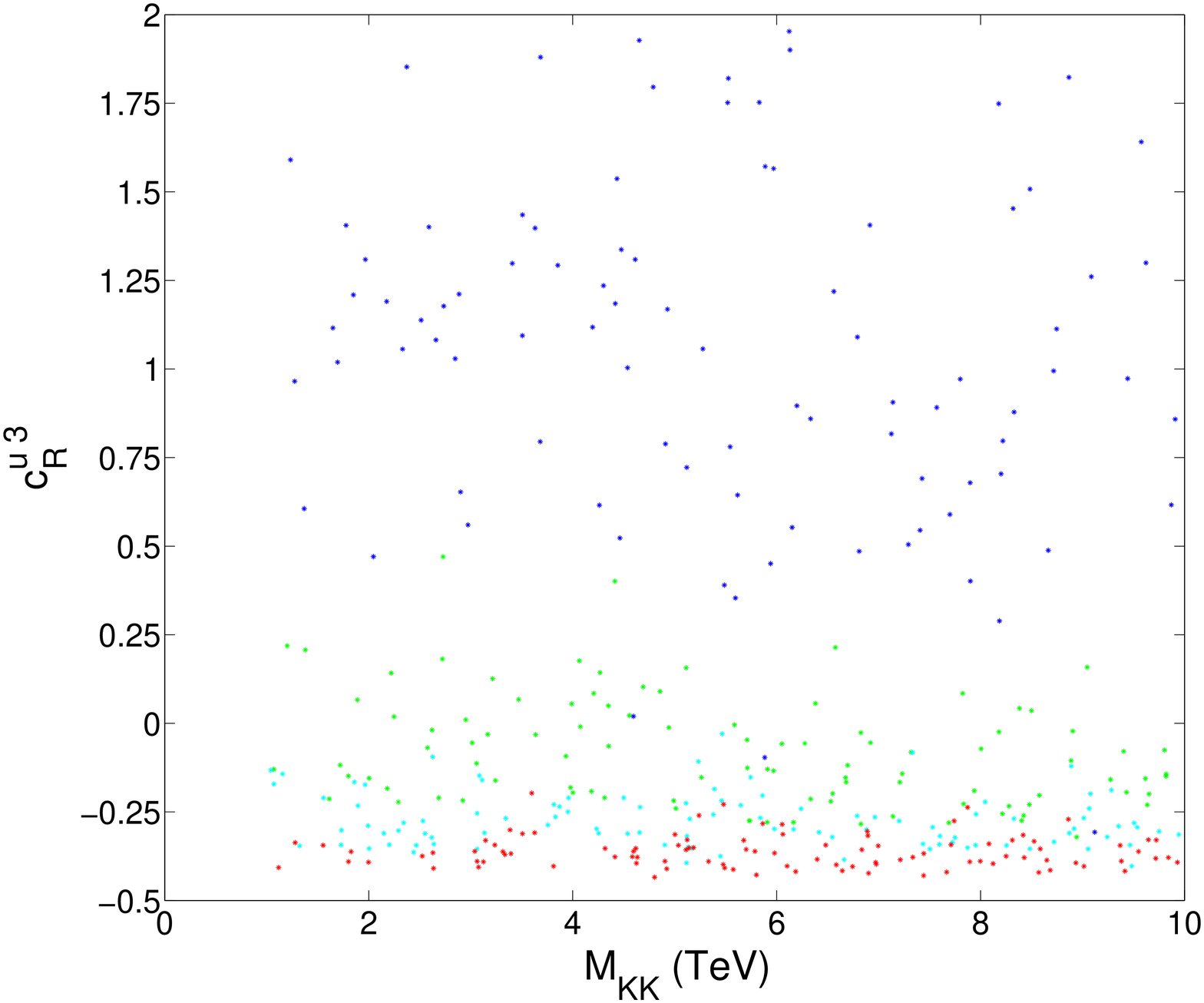}
        }
         \subfigure[]{%
            \label{fig:RScuR3}
            \includegraphics[width=0.44\textwidth]{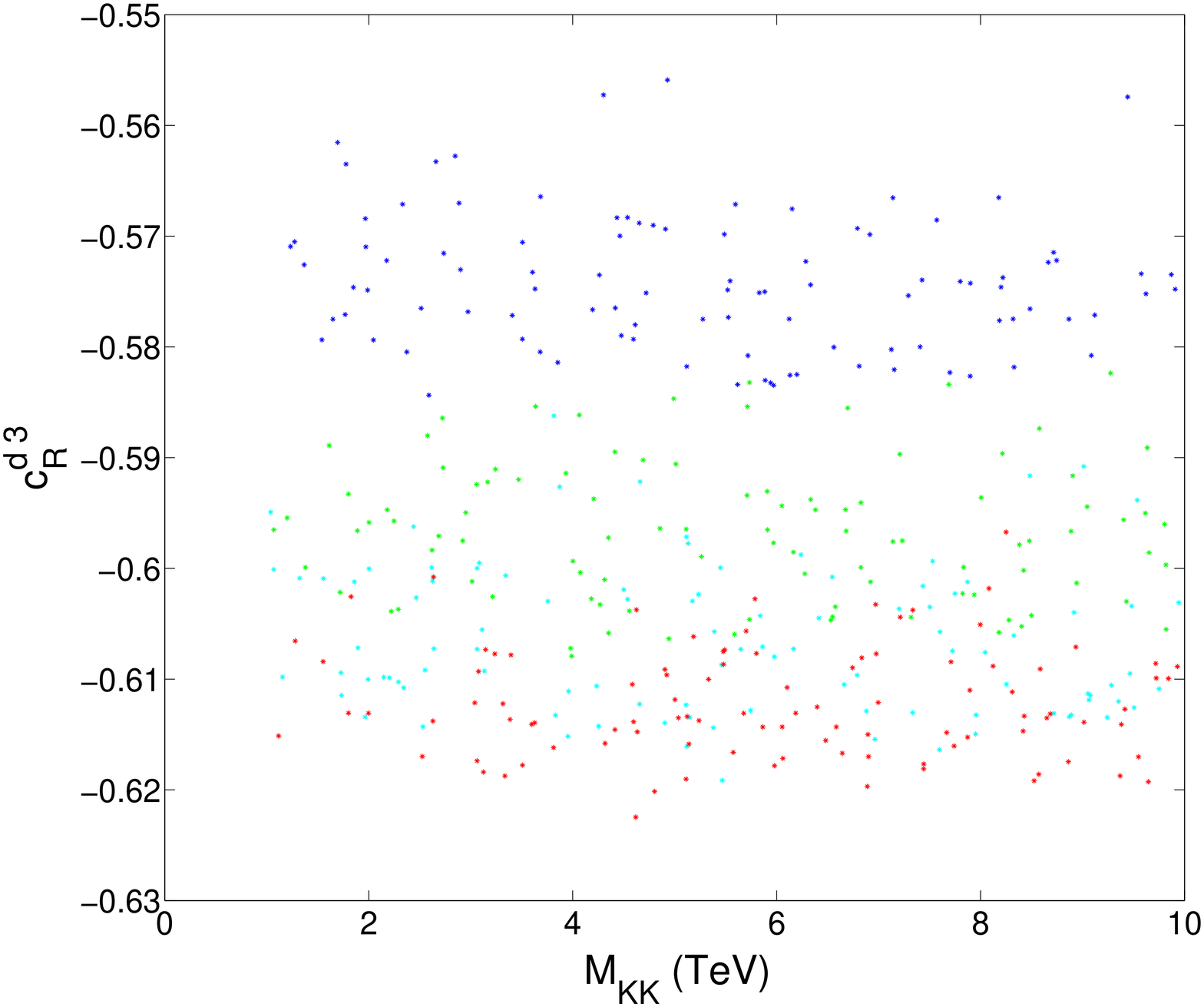}
        }
        \\ 
         \subfigure[]{%
            \label{fig:RScuR3}
            \includegraphics[width=0.44\textwidth]{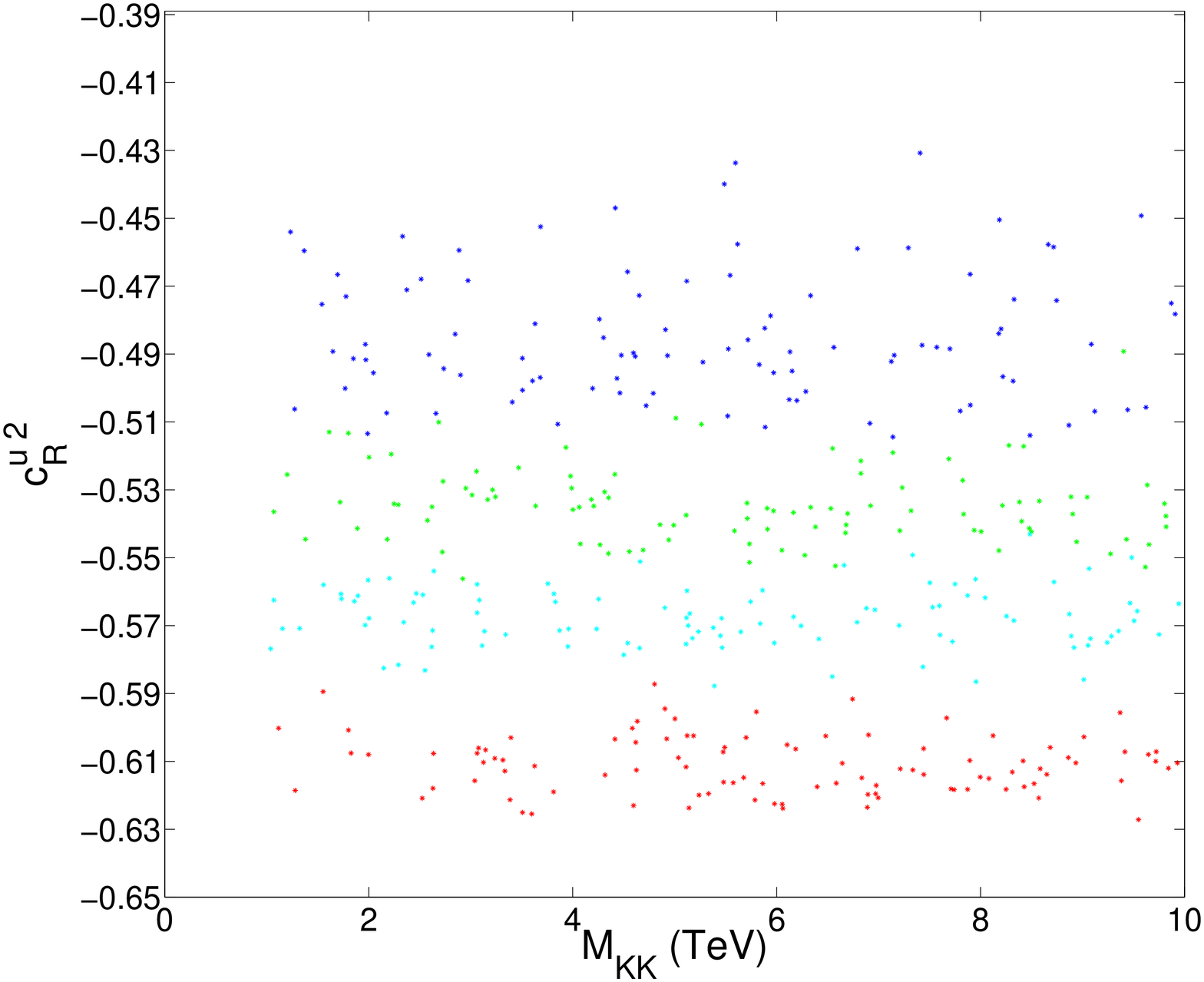}
        }
        \subfigure[]{%
            \label{fig:RScuR3}
            \includegraphics[width=0.44\textwidth]{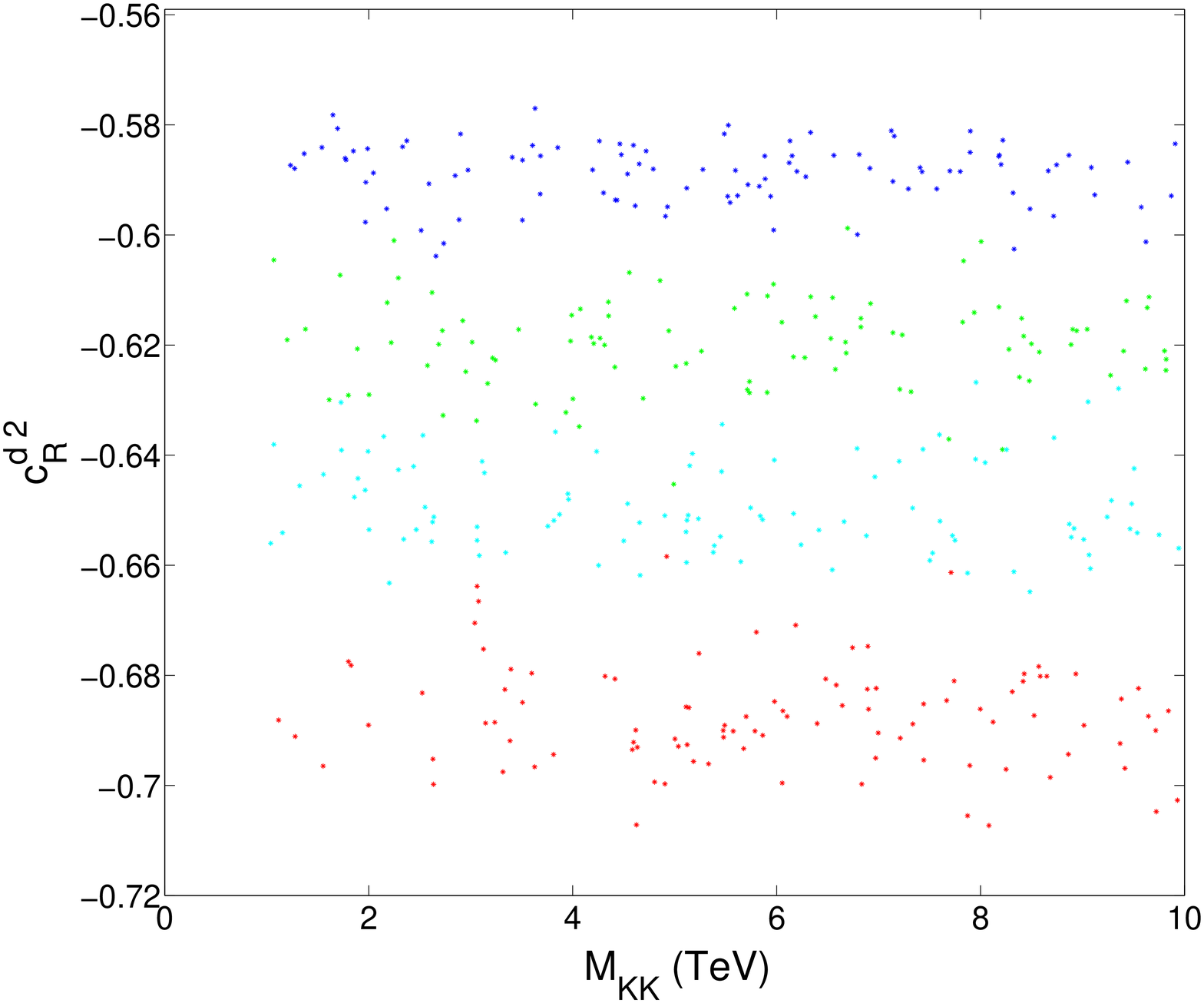}
        }
   \label{fig:RScvalues}
    \end{center}
    \caption[Natural Bulk Mass Parameter in the RS Model]{Natural bulk mass parameter in RS model for configuration (A) in blue, (B) in green, (C) in cyan and (D) in red. The $c_R^1$ values are equivalently spaced to $c_R^2$ but with $c_R^{u\;1}\in[-0.64,-0.76]$ and $c_R^{d\;1}\in[-0.62,-0.72]$.} \label{fig:RScvalues}
\end{figure}

\begin{itemize}
  \item We first need to solve for the natural $c_R$ values. To do this we randomly generate 10 sets of 2 complex Yukawas matrices such that $1\leqslant |\lambda_{ij} |\leqslant 3$ and for each solve for the corresponding $c_R$ values by fitting to the quark masses, run to a scale of the mass of the first KK gauge field.\footnote{ The light quark masses are run from their values at $2$ GeV using $m_q(\mu)=m_q(2\mbox{ GeV})\eta_4^{\frac{12}{25}}\eta_5^{\frac{12}{23}}\eta_6^{\frac{4}{7}}$, where $\eta_4=\frac{\alpha_s(m_b)}{\alpha_s(2 \mbox{ GeV})}$, $\eta_5=\frac{\alpha_s(m_t)}{\alpha_s(m_b)}$ and $\eta_5=\frac{\alpha_s(\mu)}{\alpha_s(m_t)}$. The running of $\alpha_s$ is taken from \cite{Buras:1998raa}.} To avoid any hierarchical Yukawas the $c_R$ value used are then the median average. These values are plotted in figure \ref{fig:RScvalues}.
  \item Having obtained a set of $c$ values we proceed to find 50 random Yukawa matrices that all give the correct masses, mixing angles and Jarlskog invariant. For these 50 points the four fermion coefficients are computed and relevant observables are calculated. As already mentioned some of the tightest constraint often comes from $\epsilon_K$ which is plotted in figure \ref{RSepkFig }.
  
  \begin{figure}
\begin{center}
\includegraphics[width=0.7\textwidth]{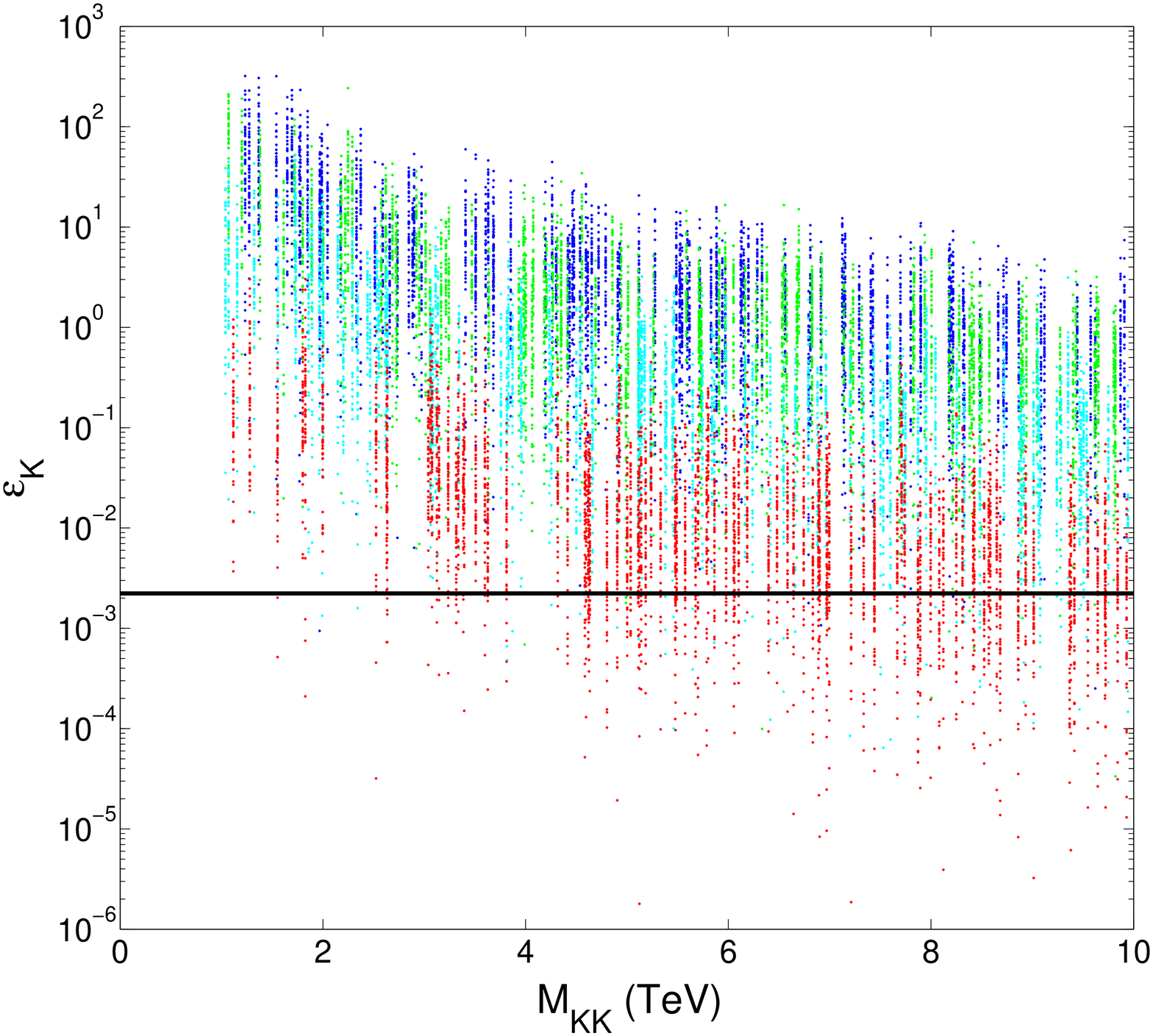}
\caption[$\epsilon_K$ in the RS Model ]{$\epsilon_K$ in the RS model. The colours are the same as those used in figure \ref{fig:RScvalues}. The mass of the first KK gauge field will be $\approx\;2.45\;M_{\rm{KK}}$.  }
\label{RSepkFig }
\end{center}
\end{figure}

 \begin{figure}
\begin{center}
\includegraphics[width=0.7\textwidth]{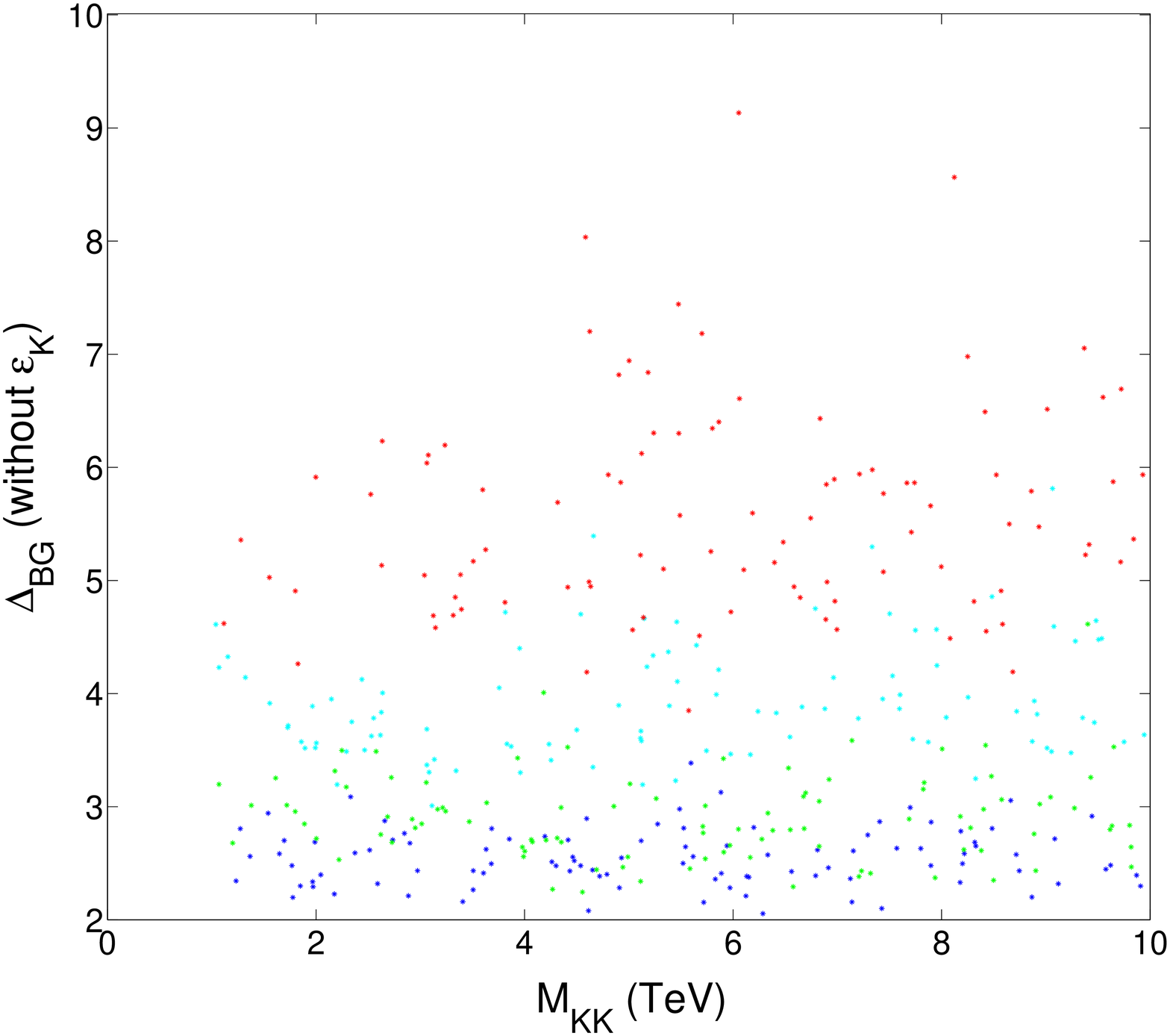}
\caption[The Fine Tuning Parameter for the RS Model ]{The Fine Tuning Parameter for the RS model. The colours are the same as those used in figure \ref{fig:RScvalues}.  }
\label{fig:FTRS }
\end{center}
\end{figure}

  \item We also compute the fine tuning parameter \cite{Barbieri88}  
\begin{equation}
\label{ }
\Delta_{BG}(O_i,p_j)=\left |\frac{p_j}{O_i}\frac{\Delta O_i}{\Delta p_j}\right |,
\end{equation}
 in which the observables $O_i$ run over the masses, mixing angles and Jarlskog invariant, while the input parameters run over the Yukawa couplings. While $p_j$ runs over the input parameters, in particular the Yukawa couplings. The average over the fifty points is plotted in figure \ref{fig:FTRS }. This will, as the name suggests,  give a measure of the sensitivity of the output parameters to changes in the input parameters. Conventionally one requires that $\Delta_{BG}<10$.  
\item Finally this is repeated for a 100 random KK scales in the hope that one will make a relatively unbiased scan over parameter space.  
\end{itemize}

Bearing in mind that the black constraint in figure \ref{RSepkFig } represents the observed value, which would include the SM contribution $2\pm0.5\times 10^{-3}$ \cite{Charles:2004jd}, then $\epsilon_K$ looks to be providing a significant constraint on the RS description of flavour. Analogous studies computed that this constraint forces the mass of the first KK gauge field to be $\gtrsim 10$ TeV $ \cite {Bauer:2009cf}, \gtrsim 22$ TeV \cite{Csaki:2008zd} or $\gtrsim 30$ TeV \cite{Blanke:2008zb} depending on what one considers an acceptable level \color{red} of \color{black} tuning.  Essentially the problem lies in the fact that in order to generate the heavy quark masses, in particular the top, some of the zero mode profiles have to sit towards the IR brane and in the region where the fermion couplings are not universal. Hence the constraints can be reduced by sending $c_R^{u\;3}$ as close to $-0.5$ as possible (i.e. configuration (D)), but this typically results in requiring slightly more tuning to achieve a valid point.

It is important to realise that all this assumes that there is no hierarchy in the Yukawa couplings. As already mentioned it is necessary to include a slight hierarchy in order to generate the small value of $V_{ub}$. Although clearly if one introduces too large a hierarchy then one has not explained the fermion mass hierarchy using the \color{red}fermion's \color{black}  location. Likewise one can reduce these constraints by either increasing the overall size of the Yukawa couplings which can lead to a loss of perturbative control of the model or alternatively introducing some flavour symmetry into the Yukawa couplings \cite{Santiago:2008vq}. Although it is not clear if this flavour symmetry is compatible with the assumption of diagonal bulk mass terms. 

We should also comment on the fact that here we have focused on $\Delta S=2$ observables in the kaon sector. More comprehensive studies have found that including $\Delta F=1$ observables narrows the allowed parameter space. In particular observables which are \color{red} more \color{black} sensitive to operators such as $(\bar{q}^i_L\gamma^\mu q^j_L)(\bar{q}^i\gamma_\mu q^i)$ and $(\bar{q}^i_R\gamma^\mu q^j_R)(\bar{q}^i\gamma_\mu q^i)$ will be more sensitive to where one slides the $c$ values. Hence one would anticipate that including any potential future measurements of, for example, $\Delta B=1$ decays will significantly reduce the viable parameter space.    

While these constraints look serious it is worth bearing in mind that, due to the enlarged parameter space, these constraint are very dependent on what one considers an acceptable level of tuning. One should also compare these constraints with those of flat universal extra dimensions. There analogous constraints force the KK scale to be $M_{\rm{KK}}\gtrsim 5000$ TeV \cite{Delgado:1999sv}. We shall now move on to consider the constraints coming from the EW sector. Typically one finds that these constraints can not be significantly reduced by finding special points in parameter space, although they can be reduced a little \cite{Delaunay:2010dw}.

%% file: EWChap.tex
\chapter{Constraints from EW Observables}
\label{EWChap}
Here we will consider models in which electroweak symmetry is broken by a Higgs localised towards the IR tip of the space. It is worth mentioning that extra dimensional models offer a number of alternatives to including the Higgs as a fundamental scalar. One possibility which was mentioned in section \ref{sect:5dGauge} is to ascribe the Higgs to the fifth component of a gauge field \cite {Oda:2004rm, Cacciapaglia:2005da, Hosotani:2006qp, Medina:2007hz}. Another possibility, referred to as Higgsless models \cite{Csaki:2003dt, Csaki:2003zu}, is to choose boundary conditions such that only the photon gains a zero mode. The W and Z bosons will then be combinations of the first KK modes and hence massive. A third possibility, which has not attracted as much attention, is to consider geometries in which $\int dr b$ is divergent and $f_0=N$ is not normalisable. This forces the gauge fields to gain a mass, however it leads to a non renormalisable theory and as far as the author is aware a realistic model is yet to be formulated \cite{Shaposhnikov:2001nz}. While these are interesting possibilities they are simply not studied \color{red} here, although \color{black} the Higgsless model in generic backgrounds has been considered in \cite{Delgado:2007ne}. 

\section{The Location of the Higgs}\label{sect:HiggsLocat}
If one accepts that EW symmetry is broken by a Higgs one must then determine the location of that Higgs. In the original RS model the Higgs is strictly localised to a 3-brane in the IR tip of the five dimensional space. From the holographic perspective, this would be dual to a four dimensional conformal field theory in which the conformal symmetry is broken by an IR cut off. The Higgs must then emerge, in the low energy theory, as a bound state out of the strong dynamics of the field theory \cite{ArkaniHamed:2000ds}. Hence this picture is very closely related to the walking technicolor model of EW symmetry breaking. However if the Higgs is free to propagate in the bulk then this scenario would be dual to a field theory with additional operators and hence the Higgs must be viewed as a \color{red} mixture of bound states and fundamental fields. \color{black} Also, with the Higgs propagating in the bulk, it can become more difficult to resolve the gauge hierarchy problem.    

If we now consider a scenario with more than five dimensions in which, in order to resolve the gauge hierarchy problem, the Higgs is at least partly localised in \color{red} the IR, this then presents two possibilities. Firstly, \color{black} that the Higgs could be localised to a 3-brane and hence not free to propagate in the internal manifold. So the EW sector would be described by
\begin{equation}
\label{EWact3brane}
S=\int d^Dx \sqrt{-G}\Big [-\frac{1}{4}A_{MN}^aA^{MN\;a}-\frac{1}{4}B_{MN}B^{MN}\Big ]
+\int d^4x\sqrt{-g_{\rm{ir}}}\left [|D_\mu\Phi|^2-V(\Phi)\right ]
\end{equation}
where $g_{\rm ir}^{\mu\nu}$ is the induced metric $G_{MN}\delta_\mu^M \delta_\nu^N\delta(r-r_{\rm{ir}})\delta^{(\delta)}(\phi-\phi_{\rm{ir}})$. $A_{MN}$ and $B_{MN}$ are the field strength tensors for the $\color{red}\mathrm{SU} (2)$ and $\color{red}\mathrm{U}(1)$ gauge fields. 
 
The second option is that the Higgs is localised to a codimension one brane, i.e. localised only w.r.t $r$, and is free to propagate in the internal manifold. Hence the EW sector would be described by
\begin{equation}
\label{EWactcodim1}
S=\int d^Dx \sqrt{-G}\left (-\frac{1}{4}A_{MN}^aA^{MN\;a}-\frac{1}{4}B_{MN}B^{MN}+\frac{\delta(r-r_{\rm{ir}})}{b}\Big [|D_\mu\Phi|^2 -V(\Phi)\Big ]\right ).
\end{equation}

In computing the low energy effective action we are faced with two equivalent options, either to carry out the KK decomposition pre spontaneous symmetry breaking (SSB) or post SSB. Post SSB the boundary mass term will induce a non trivial BC but the resulting gauge field mass matrix will be diagonal w.r.t KK number. On the other hand pre SSB one can use the original Neumann or Dirichlet BCs but the Higgs terms will mix the KK modes, and hence the mass matrix will gain off diagonal terms. It is the latter option that we use here. Let us return to our two possible Higgs localisations. The first terms of  the actions can be expanded as described in Eq. \ref{GaugeKKdecomp}, so the covariant derivative is now given as
\begin{equation}
\label{ }
D_\mu=\partial_\mu+\sum_n\left (-igf_n\Theta_nA_\mu^{a(n)}\tau^a-iYg^{\prime}f_n\Theta_nB_\mu^{(n)}\right ).
\end{equation}
After SSB the Higgs gains a VEV, $\Phi\rightarrow\frac{a(r_{\rm{ir}})^{-1}}{\sqrt{2}}\left (\begin{array}{ c}0\\v+H\end{array}\right )$. We now perform the usual field redefinitions
\begin{eqnarray}
A_\mu^{1(n)}=\frac{1}{\sqrt{2}}\left (W_\mu^{+(n)}+W_\mu^{-(n)}\right )\qquad A_\mu^{2(n)}=\frac{i}{\sqrt{2}}\left (W_\mu^{+(n)}-W_\mu^{-(n)}\right )\nonumber\\
A_\mu^{3(n)}=cZ_\mu^{(n)}+sA_\mu^{(n)}\qquad B_\mu^{(n)}=cA_\mu^{(n)}-sZ_\mu^{(n)}\nonumber
\end{eqnarray}
where
\begin{equation}
c\equiv\frac{g}{\sqrt{g^2+g^{\prime 2}}}\qquad s\equiv\frac{g^{\prime}}{\sqrt{g^2+g^{\prime 2}}} .
\end{equation} 
The gauge field mass matrices can then be computed from the Higgs kinetic term. We can now see the difference between the two scenarios. When the Higgs is localised to a codimension one brane (Eq. \ref{EWactcodim1}) and if one assumes the Higgs VEV has a flat profile, then the KK modes with $\alpha_n\neq 0$ will not mix with the gauge zero modes (with $\alpha_n=0$) due to the orthogonality relation Eq. \ref{GaugeOrthog}. Hence only the KK modes with $\alpha_n=0$ will contribute to the mass of the W and Z boson zero mode obtained from
  \begin{equation}
\label{MassTermscodim1}
|D_\mu\Phi|^2\supset\sum_{n,m}\frac{g^2v^2}{4}f_nf_mW_\mu^{+(n)}W^{-(m)\mu}+\frac{(g^2+g^{\prime 2})v^2}{8}f_nf_mZ_\mu^{(n)}Z^{\mu (m)}
\end{equation}
where $f_n=f_n(r_{\rm{ir}})$. On the other hand, where the Higgs is localised to a 3-brane, Eq. \ref{EWact3brane}, the orthogonality relations cannot generally be applied and hence one gets a mixing between all of the KK modes. The zero mode masses of the W and Z bosons will hence receive corrections from all of these modes  
 \begin{equation}
\label{MassTerms3brane}
|D_\mu\Phi|^2\supset\sum_{n,m}\frac{g^2v^2}{4}f_n\Theta_nf_m\Theta_mW_\mu^{+(n)}W^{-(m)\mu}+\frac{(g^2+g^{\prime 2})v^2}{8}f_n\Theta_nf_m\Theta_mZ_\mu^{(n)}Z^{\mu (m)}
\end{equation}
where $\Theta_n=\Theta_n(\phi_{\rm{ir}})$. This scenario gives rise to two problems. Firstly one must know the geometry of the internal manifold before estimating the corrections to the EW observables. This can introduce more free parameters and more model dependence into any analysis making the model less predictive. The second, more serious, problem is that in order to compute any observable one must sum over all the KK modes that are contributing. Where such a sum involves `more than one KK tower', this sum will involve a sum over multiple KK numbers and will typically be divergent. For example in two flat dimensions
\begin{displaymath}
\sum_n\frac{1}{m_n^2}\sim\sum_{n,m}\frac{1}{n^2+m^2}
\end{displaymath}  
which is divergent. Hence one is forced to impose an arbitrary cut off in the KK number. Any results will typically be dependent on this cut off \cite{Appelquist:2000nn}. One would hope that such divergences would be resolved by some high energy physics. However such high energy physics would also offer corrections to the KK profiles and equivalently the low energy phenomenology. The size of such corrections would of course be model dependent. For example the corrections to the graviton propagator, when this divergence is resolved by a UV fixed point, was found to give sizeable corrections to the low energy phenomenology \cite{Gerwick:2010kq, Gerwick:2011jw}. Hence here we would argue that when the Higgs is not localised to a codimension one brane the EW constraints, computed using this perturbative method, is susceptible to unknown inaccuracies. 

There is also a third problem with considering particles localised to branes of codimension two or higher. Notably matching solutions of the Einstein equation across these branes typically give rise to singular solutions. While there are known solution to this problem such as in six dimensions the inclusion of Gauss Bonnet terms \cite{Bostock:2003cv}. None the less the construction of realistic models with codimension two branes is not trivial.

\section{A Perturbative Approach to EW Corrections}
In carrying out the KK decomposition pre SSB one finds that the coupling to the Higgs mixes the KK modes and the W and Z mass matrices gain off diagonal terms. In particular
 \begin{equation}
\label{WZmassmatrixcodim1}
(M_W^2)_{mn}=m_n^2\delta_{mn}+\frac{g^2v^2}{4}f_nf_m\qquad(M_Z^2)_{mn}=m_n^2\delta_{mn}+\frac{(g^2+g^{\prime 2})v^2}{4}f_nf_m.
\end{equation} 
when the Higgs is localised to a codimension one brane or  when the Higgs is localised to a brane of codimension two or higher then
 \begin{equation}
\label{massmatrix}
(M_W^2)_{mn}=m_n^2\delta_{mn}+\frac{g^2v^2}{4}f_n\Theta_nf_m\Theta_m\quad(M_Z^2)_{mn}=m_n^2\delta_{mn}+\frac{(g^2+g^{\prime 2})v^2}{4}f_n\Theta_nf_m\Theta_m.
\end{equation} 
Bearing in mind the multiple solutions of Eq. \ref{GaugLapl}, $f_n\Theta_nf_m\Theta_m$ will now be block matrices. In other words, for notational simplicity, we have again denoted multiple KK number indices as $n$. At tree level the W and Z zero mode masses will then be the lowest eigenvalues of these matrices. When the KK scale is larger than the EW scale one can use a perturbative approximation to diagonalise these mass matrices \cite{Goertz:2008vr, Archer:2010hh}. In particular given an $N\times N$ symmetric matrix
 \begin{displaymath}
M=\left(\begin{array}{ccccc}A_{1} & B_{12} & B_{13} & \cdots & B_{1N} \\B_{12} & A_2 & B_{23} & \cdots &  \\B_{13} & B_{23} & A_3 &  &  \\\vdots &\vdots  &  & \ddots &  \\B_{1N} &  &  &  & A_N\end{array}\right)
\end{displaymath}
where $A_n\gg B_{nm}$. Then the nth eigenvalue is approximately
\begin{equation}
\label{eigenvalue}
\lambda_n\,\approx A_n-\sum_{i\neq n}^N\frac{B_{ni}^2}{A_i-A_n}+\sum_{i\neq n}^N\sum_{j\neq i}^N\frac{B_{ni}B_{nj}B_{ij}}{(A_i-A_n)(A_j-A_n)}+\mathcal{O}(A^{-3})
\end{equation}
If $M$ is diagonalised by the unitary matrix $U$ i.e.
 \begin{displaymath}
U^{\dag}MU=D=\delta_{nm}\left (A_n-\sum_{i\neq n}^N\frac{B_{ni}^2}{A_i-A_n}+\sum_{i\neq n}^N\sum_{j\neq i}^N\frac{B_{ni}B_{nj}B_{ij}}{(A_i-A_n)(A_j-A_n)}+\dots\right ).
\end{displaymath}
Then by acting on each side with $U$, one finds that
\begin{displaymath}
U_{nn}=1\quad\mbox{and}\quad U_{in}\approx\frac{-B_{in}}{A_i-A_n}+\sum_{j\neq i}^N\frac {B_{nj}B_{ij}}{(A_i-A_n)(A_j-A_n)}+\mathcal{O}(A^{-3})
\end{displaymath}
or\begin{equation}
\label{Uinv}
U^{-1}\approx\left(\begin{array}{cccc}1 & -\frac{B_{12}}{A_2-A_1} & -\frac{B_{13}}{A_3-A_1} & \cdots \\\frac{B_{12}}{A_2-A_1} & 1 & -\frac{B_{23}}{A_3-A_2} &  \\\frac{B_{13}}{A_3-A_1} & \frac{B_{23}}{A_3-A_2} & 1 &  \\\vdots & \vdots &  & \ddots\end{array}\right).
\end{equation}
We define the quantities
\begin{equation}
\label{bareWZmass}
m_z^2\equiv\frac{(g^2+g^{\prime 2})v^2}{4}f_0^2\Theta_0^2\qquad \mbox{and}\qquad m_w^2\equiv\frac{g^2v^2}{4}f_0^2\Theta_0^2
\end{equation}
as well as 
\begin{equation}
\label{Fn}
F_n \equiv \frac{f_n(r_{\rm{ir}})\Theta_n(\phi_{\rm{ir}})}{f_0(r_{\rm{ir}})\Theta_0(\phi_{\rm{ir}})}.
\end{equation}   
As mentioned in the previous section, if one is working in five dimensions or if the Higgs is localised to a codimension one brane then clearly $F_n=f_n(r_{\rm{ir}})/f_0(r_{\rm{ir}})$. The mass eigenvalues will then be given by
\begin{equation}
\left (M_{W/Z}^2\right )_{nn}\approx m_n^2+m_{w/z}^2F_n^2-\sum_{m\neq n}\frac{\left (m_{w/z}^2F_nF_m\right )^2}{m_m^2-m_n^2+m_{w/z}^2\left (F_m^2-F_n^2\right )}+\mathcal{O}(m_n^{-4})\label{massgaugestate}
\end{equation}
where the mass matrices have been diagonalised by
\begin{equation}
\left (U_{W/Z}\right )_{nn}\approx 1\quad\mbox{and}\quad \left (U_{W/Z}\right )_{mn}\approx-\frac{m_{w/z}^2F_nF_m}{m_m^2-m_n^2+m_{w/z}^2(F_m^2-F_n^2)}+\mathcal{O}(m_n^{-4}). \label{UnitaryMatrix}
\end{equation} 

\section{Estimating Corrections to EW Observables}
\subsection{Fixing the Input Parameters} \label{sect:FixEWInput}
The expressions in Eq. \ref {massgaugestate} and Eq. \ref{UnitaryMatrix} are of course dependent on the unknown higher dimensional couplings $g$, $g^{\prime}$ and the Higgs VEV $v$. So before one can compute the relevant corrections to the EW observables it is necessary to fix the input parameters. In an ideal world one would carry out a full $\chi^2$ test including all \color{red} known \color{black} observables. However three observables stand out as being known to a significantly higher precision and therefore one would anticipate that they would dominate any $\chi^2$ test. These observables are \cite{Amsler:2008zzb},
\begin{equation}
\label{ }
\hat{\alpha}(M_Z)^{-1}=127.925\pm0.016\quad G_F=1.166367(5)\times10^{-5} \;\mbox{GeV}^{-2}\quad \hat{M}_Z=91.1876\pm0.0021\; \mbox{GeV}.
\end{equation}
By matching to observables at the Z pole, we are assuming that the running of the fine structure \color{red} constant, below the Z mass, is dominated by SM physics. Arguably if this were not the case \color{black} the size of the EW corrections would mean the model was not phenomenologically viable. The Z mass has been primarily  taken from the location of the Z pole in LEP one data and hence can be matched directly to the lowest eigenvalue of Eq.  \ref{massgaugestate}.
\begin{equation}
\label{ZMass}
\hat{M}_Z^2=\left (U_Z(M_Z^2)U_Z^{-1}\right )_{00}
\end{equation}
However the fine structure constant and the Fermi constant have been most precisely measured using the anomalous magnetic moment of the electron and the rate of muon decay $\mu^-\rightarrow \nu_\mu e^-\bar{\nu}_e$. In both cases it is important to consider the location of the fermions. 

If one allowed the fermions to propagate in the bulk and made the KK decomposition $\psi_{L,R}=\sum_n f^{(n)}_{L,R}\Theta_{L,R}^{(n)} \psi_{L,R}^{(n)}$ then one could quite generically define the gauge fermion coupling to be
\begin{equation}
\label{eq:GaugeFerm }
f_\psi^{(n,l,m)}=\int d^{1+\delta}a^3bc^\delta\sqrt{\gamma} f^{(n)}_{L,R}\Theta_{L,R}^{(n)}f_l\Theta_lf^{(m)}_{L,R}\Theta_{L,R}^{(m)}.
\end{equation}
 Of course if the fermions were localised to branes of codimension one or higher then one would simply include delta functions in the above integral and reduce the number of KK numbers included in $n,m$. The extreme case would be if the fermions were localised to 3-branes, in which case 
 \begin{equation}
\label{ }
f_\psi^{(n,l,m)}=f_\psi^{(l)}=f_l(r_{\rm{ir/uv}})\Theta_l(\phi_{\rm{ir/uv}}).
\end{equation} 
It was demonstrated in the previous two chapters that when considering bulk fermions one can quite generically include a bulk mass term which determines where the zero mode profile would be peaked. It is also a generic feature of warped models, that resolve the gauge hierarchy problem, that the KK modes would be peaked towards the IR tip of the space. Therefore if one anticipates that the light fermions are sitting towards UV tip of the space then they will have approximately universal couplings to potential KK modes, be that gauge modes, gauge scalar modes or KK Higgs modes (see figure \ref{RelFermCoupl}). Of course one could envisage scenarios in which this is not the case, but arguably such a model would not be phenomenologically viable due to potentially large FCNC's. Hence here we will make the assumption that
\begin{displaymath}
f_\psi^{(0,n,0)}=f_{e^-}^{(0,n,0)}=f_{\mu^-}^{(0,n,0)}=f_{\nu}^{(0,n,0)}=\dots\mbox{  all light fermions}
\end{displaymath}
With this assumption it is possible to fit the remaining two observables which, at tree level, would be given by
\begin{eqnarray}
\sqrt{4\pi\alpha(M_Z)}=\frac{gg^{\prime}}{\sqrt{g^2+g^{\prime2}}}f_\psi^{(0,0,0)}\label{alpha}\equiv e f_\psi^{(0,0,0)}\\
4\sqrt{2}G_f=g^2(f_\psi^{(0,m,0)})^\dag\,(M_W^2)_{mn}^{-1}f_\psi^{(0,n,0)}\label{Gf}.
 \end{eqnarray} 
Here there has been a slight abuse of notation, in that the $M_W^2$ in Eq. \ref{Gf} would be the mass matrix of all the negatively charged bosons that couple to the light leptons, where as Eq. \ref{massgaugestate} only the charged gauge fields that mix with the W zero mode. Where these two expressions are not equivalent one should be careful to include the additional contributions to the EW observables.   

\subsection{Computing Corrections to EW Observables}
Having fixed the input parameters one can proceed to compute the constraints arising from EW observables. The first approach, we shall consider, to EW precision tests is to simply compare observables with SM predictions. In particular here we shall focus on Z pole observables measured at SLC and LEP one. The advantage with focusing on the Z pole is here one would anticipate that any observables would be dominated by the Z zero mode. Hence such observables will be less sensitive to unknown aspects of a given model such as, for example, the contribution from the gauge scalars. Corrections to such observables occur at tree level for three reasons. Firstly one gets a corrections to the Z mass (Eq. \ref{massgaugestate}), secondly one gets corrections to the Z coupling (plotted for the RS model in figure \ref{fig:ZfermCoup})    
\begin{equation}
\label{fZzero}
f_Z^{(n)}=U_Z^{-1}f_\psi^{(0,n,0)}
\end{equation}   
and thirdly one gets corrections to the ratio of the couplings, essentially due to $f_W^{(0)}\neq f_Z^{(0)}\neq f_\psi^{(0)}$. The relevant Z pole observables include the Z mass and partial decay width \cite{Nakamura:2010zzi, Han:2008es}
\begin{equation}
\label{ }
\Gamma_{Z\rightarrow\bar{\psi}\psi}=\frac{(g^2+g^{\prime\;2})f_Z^{(0)\;2}M_Z}{48\pi}\left (g_V^{\psi\;2}+g_A^{\psi\; 2}\right )
\end{equation}  
as well as the polarised asymmetry measured at SLC
\begin{equation}
\label{ }
A_\psi=\frac{2g_V^\psi g_A^\psi}{g_V^{\psi\;2}+g_A^{\psi\; 2}}
\end{equation}
and also the ratios 
\begin{displaymath}
R_l=\frac{\Gamma_{Z\rightarrow had}}{\Gamma_{Z\rightarrow \bar{l}l}}.
\end{displaymath}
The Z couplings are given by table \ref{SMZcoupl}\newline
\begin{table}[h!]
\begin{center}
\begin{tabular}{c|cc}
   $\psi$&$g_V^\psi$ & $g_A^\psi$ \\
   \hline
   $\nu_e$, $\nu_\mu$, $\nu_\tau$ & $+\frac{1}{2}$ &$ +\frac{1}{2}$ \\
   $e$, $\mu$, $\tau$ & $-\frac{1}{2}+2s_Z^2$  & $-\frac{1}{2}$\\
    $u$, $c$, $t$ & $+\frac{1}{2}-\frac{4}{3}s_Z^2$  & $+\frac{1}{2}$\\
     $d$, $s$, $b$ & $-\frac{1}{2}+\frac{2}{3}s_Z^2$  & $-\frac{1}{2}$
\end{tabular}
\end{center}
\centering 
  \caption[]{\color{red}The couplings of the Z boson, to SM fields. \color{black}}\label{SMZcoupl}
\end{table}

with 
\begin{equation}
\label{ }
s_Z^2=\frac{g^{\prime\;2}}{g^2+g^{\prime\;2}}.
\end{equation}
$\hat{s}_Z^2$ can also be used as a `pseudo' observable obtained by taking a best fit over a range of Z pole data $\hat{g}^{\prime2}(M_Z)/[\hat{g}^2(M_Z)+\hat{g}^{\prime2}(M_Z)]$. Although the weak mixing angle can also be defined using
\begin{displaymath}
s_{p}^2(1-s_{p}^2)\equiv s_p^2c_p^2\equiv\frac{\pi \alpha(M_Z)}{\sqrt{2} G_F M_Z^2}\quad\mbox{ or }\quad s_{\rm{eff}}^{\rm{lept}}\equiv\frac{1}{4}\left (1-\frac{g_V^e}{g_A^e}\right ).
\end{displaymath} 
The approach taken in this analysis is to then compute the tree level contribution to a particular observable and insist that the deviation from the tree level SM result is within $2\sigma$ of the experimental uncertainty in the observable. The observables found to give the tightest constraints are listed in table \ref{EWobservables}.  

\begin{table}
  \centering 
  \begin{tabular}{|c|c|c|c|}
\hline
  EW Observable & Observed Value  &  SM prediction & Tree level SM prediction\\
\hline
  $M_W$(GeV) & $80.428\pm0.039$ & $80.375\pm0.015$ & $79.82\pm0.0048 $\\
   $\Gamma_Z$ (GeV)&$2.4952\pm0.0023$  & $2.4968\pm0.0010$ &$2.41625\pm0.00031$ \\
   $\Gamma (had)$ (GeV)&$1.7444\pm0.0020$ &$1.7434\pm0.0010$ &$1.6687\pm0.00025$\\
   $\Gamma (inv)$ (MeV)&$499.0\pm1.5$ &$501.59\pm0.08$ &$497.65\pm0.0346$\\ 
   $\Gamma (l^+l^-)$ (MeV)&$83.984\pm0.086$ &$83.988\pm0.016$ &$83.29\pm0.0083$\\ 
   $R_e$ &$20.804\pm0.05$ &$20.758\pm0.011$ &$20.03\pm 0.0010$\\ 
   $A_e$ &$0.15138\pm0.00216$ &$0.1473\pm0.0011$ &$0.1304\pm 0.00045$\\
   $\hat{s}_Z^2$ &$0.23119\pm0.00014$ &- &$0.2336\pm 0.00006$\\  
\hline
\end{tabular}
  \caption[Electroweak observables]{ The SM predictions and Observed values of eight EW observables. The tree level SM predictions have been computed directly from the input parameters $\alpha(M_Z)$, $G_f$ and $M_Z$.}\label{EWobservables}
\end{table}   

\begin{figure}[t!]
\begin{center}
\includegraphics[width=6in] {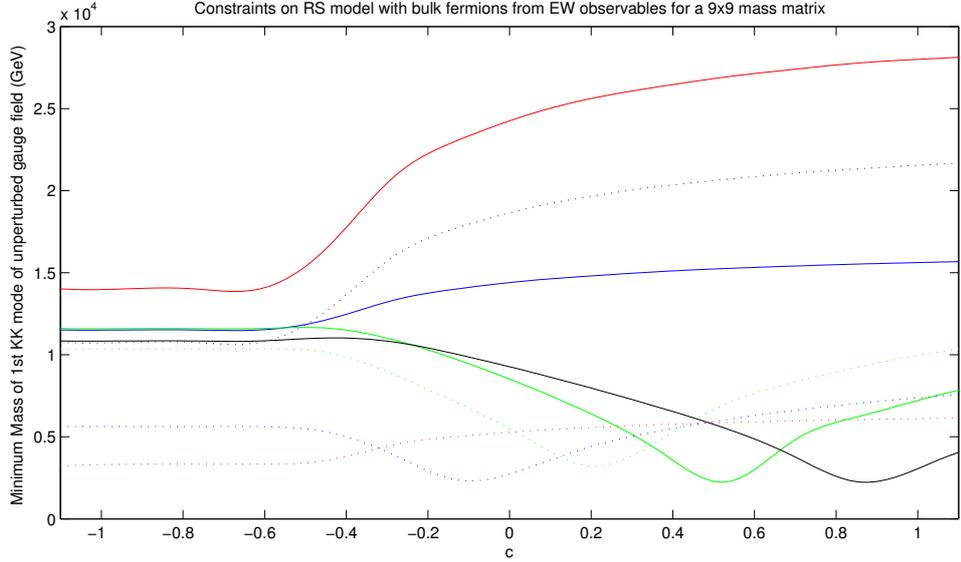}
\caption[Constraints on the RS Model from EW Observables] {\footnotesize The constraints on the 5D RS model with the SM propagating in the bulk. Plotted is the lower bound on the first KK gauge mass ($\sim2.45M_{\rm{KK}}$) arising from comparison with experimental error on the EWO's; $s_Z^2$ (Red line), $M_W$ (blue line), $\Gamma_Z$ (green line), $\Gamma_{had}$ (black line),  $R_e$ (blue dots), $\Gamma_{inv}$ (red dots), $\Gamma_{l^+l^-}$ (green dots) and $A_e$ (black dots).  On the horizontal axis is plotted the 5D Dirac mass, in a slight change of notation, $M=-ck$ and hence $c<-0.5$($>-0.5$) results in the fermions being localised towards the UV (IR) brane. $\Omega=10^{15}$ \cite{Archer:2010hh}.}
\label{RSconstraintsfig}
\end{center}
\end{figure}
As way of an example the constraints from each observable is plotted in figure \ref{RSconstraintsfig} for the 5D RS model. In all cases considered it is found that the tightest constraint comes from the weak mixing angle. It would be useful to try and find an analytical expression for such corrections. 

If we begin by combining Eq. \ref{alpha} and Eq. \ref{Gf} to get
\begin{displaymath}
s_Z^2=\frac{\pi\alpha}{\sqrt{2}G_f}\frac{f_\psi^{(m)}(M_W^2)^{-1}_{nm}f_\psi^{(n)}}{(f_\psi^{(0)})^2}
\end{displaymath}
But the inverse of the W mass matrix has the relatively simple form
\begin{displaymath}
(M_W^2)^{-1}=\left(\begin{array}{ccccc}(M_W^2)^{-1}_{00} & -\frac{F_1}{m_1^2} & -\frac{F_2}{m_2^2} & \cdots & -\frac{F_n}{m_n^2} \\-\frac{F_1}{m_1^2} & \frac{1}{m_1^2} & 0 & \vdots &  \\-\frac{F_2}{m_2^2} & 0 & \frac{1}{m_2^2} &  &  \\\vdots & \vdots &  & \ddots &  \\-\frac{F_n}{m_n^2}&  &  &  & \frac{1}{m_n^2}\end{array}\right)
\end{displaymath}
where
\begin{displaymath}
(M_W^2)^{-1}_{00}=\frac{1}{m_w^2}+\sum_{n=1}\frac{F_n^2}{m_n^2}.
\end{displaymath}
Here we have expressed the inverse of the mass matrix using only KK number although clearly the result generalises to when `more than one KK tower' is contributing. We have also assumed that only W bosons are contributing to muon decay. Although again it is straight forward to include additional particles. Hence if we define
\begin{equation}
\label{Fpsi}
F_\psi^{(n)}\equiv\frac{f_\psi^{(0,n,0)}}{f_\psi^{(0,0,0)}}
\end{equation}
then the weak mixing angle is given by
\begin{equation}
\label{szapprox1}
s_Z^2=\frac{\pi\alpha}{\sqrt{2}G_f}\left (\frac{1}{m_w^2}+\sum_{n=1}\frac{\left(F_n-F_\psi^{(n)}\right )^2}{m_n^2}\right ).
\end{equation}
We can solve for $m_w^2$ by noting that $m_w^2=m_z^2(1-s_Z^2)$ and using that 
\begin{displaymath}
\hat{M}^2_Z\approx m_z^2\left (1-\sum_{n=1}\frac{m_z^2F_n^2}{m_n^2}+\mathcal{O}(m_n^{-4})\right ).
\end{displaymath}
The weak mixing angle is then given by
\begin{equation}
\label{WeakMixingAngle}
s_Z^2\approx s_{p}^2\left (1-\frac{c_p^2}{c_p^2-s_p^2}\sum_{n=1}\left [\frac{m_z^2F_n^2}{m_n^2}-\frac{m_w^2\left(F_n-F_\psi^{(n)}\right )^2}{m_n^2}\right ]+\mathcal{O}(m_n^{-4})\right ).
\end{equation}
It is possible to express all of the above observables using these expansions. For example the W mass is given from Eq. \ref{massgaugestate}) to be 
\begin{displaymath}
M_W^2\approx m_w^2\left (1-\sum_{n=1}\frac{m_w^2F_n^2}{m_n^2}+\mathcal{O}(m_n^{-4})\right )\equiv m_w^2\left (1-\triangle M_W^2\right ).
\end{displaymath}
Again one can express
\begin{displaymath}
m_W^2=c_p^2\hat{M}_Z^2\left (1+\triangle M_Z^2\right )\left (1+\triangle c_z^2\right )
\end{displaymath}
and therefore  
\begin{equation*}
\label{ }
M_W^2\approx c_p^2\hat{M}_Z^2\left (1+\sum_{n=1}\left [\frac{(m_z^2-m_w^2)F_n^2}{m_n^2}+\frac{s_p^2}{c_p^2-s_p^2}\left (\frac{m_z^2F_n^2}{m_n^2}-\frac{m_w^2\left(F_n-F_\psi^{(n)}\right )^2}{m_n^2}\right )\right ]+\mathcal{O}(m_n^{-4})\right ).
\end{equation*}
Note the W mass is always shifted up, relative to the SM prediction, which is in agreement with current best fits on the W mass. 

This approach of comparing to a selection of Z pole observables offers a deliberately conservative method. Arguably one has more control over the assumptions made. In particular the neglecting of the beyond the standard model loop effects and the assumption of universal gauge fermion coupling. As discussed in the previous chapter, assuming the muon and electron couplings are universal is far more reasonable than assuming that the top coupling is universal. None the less it is possible that this approach is too pessimistic. It does not include anomalous results such as the anomalous magnetic moment of the muon and it is possible that if a full $\chi^2$ test was completed a more favourable point in parameter space could be found. 

\subsection{The S and T Parameters}
The more conventional approach to computing the size of constraints from the EW observables is to compute the Peskin Takeuchi $S$ and $T$ parameters \cite{Peskin:1991sw}. The idea is that when the scale of new physics is much higher than the EW scale then one can integrate out the additional particles, typically using the on shell equations of motions. This new physics can then be described by a set of dimension six operators which can then be fitted to a selection of observables in a model independent fashion. It was this approach which was used in section \ref{sect:FCNC}. Of the 80 Lorentz invariant operators listed in \cite{Buchmuller:1985jz}, that conserve lepton and baryon number, many offer no physical contribution to EW observables since their effects can be absorbed by suitable field redefinitions. Those that do contribute fall into five classes 
\begin{itemize}
  \item The four fermion operators, for a full list and their contributions see for example \cite{Han:2004az} 
  \begin{displaymath}
   \mathcal{O}_{\psi\chi}=(\bar{\psi}\gamma^\mu\psi)(\bar{\chi}\gamma^\mu\chi) 
  \end{displaymath}
  \item Operators that modify the gauge fermion couplings, such as
  \begin{displaymath}
   \mathcal{O}_{\Phi\psi}=i(\Phi^\dag D_\mu \Phi)(\bar{\psi}\gamma^\mu\psi)
  \end{displaymath}
  \item Operators that modify the anomalous magnetic moment of the lepton
  \begin{displaymath}
   \mathcal{O}_{eA}=i(\bar{e}\sigma^a\gamma^\mu D^\nu e)A_{\mu\nu}^a \hspace{1cm} \mathcal{O}_{eB}=i(\bar{e}\gamma^\mu D^\nu e)B_{\mu\nu}
  \end{displaymath}
   \item Operators that modify the triple gauge boson coupling
   \begin{displaymath}
   \mathcal{O}_{A}=\epsilon^{abc}A_\mu^{a\nu}A_\nu^{b\lambda}A_\lambda^{c\mu}
  \end{displaymath}
  \item And operators that modify the gauge boson propagator
  \begin{displaymath}
   \mathcal{O}_{\Phi}=(\Phi^\dag D_\mu \Phi) (\Phi^\dag D^\mu \Phi)\hspace{1cm} \mathcal{O}_{AB}=(\Phi^\dag \sigma^a \Phi )A_{\mu\nu}^a B^{\mu\nu}.
  \end{displaymath}
\end{itemize} 
These operators would of course be present in the SM but one would anticipate additional contributions coming from many beyond the SM scenarios which would in turn give rise to constraints on a given model. Of these operators it is typically found that the tightest constraints come from the last two `oblique' operators, $ \mathcal{O}_{\Phi}$ and $\mathcal{O}_{AB}$. The rescaled coefficients of such operators are then the $S$ and $T$ parameters
\begin{equation}
\label{STdim6}
S=\frac{\alpha}{4scv^2}C_{AB}, \hspace{1.5cm} T=-\frac{2\alpha}{v^2} C_\Phi.
\end{equation}
The complete dominance of these two operators has led to many fits, to EW observables, being parameterised entirely in terms of these oblique operators. If one is working purely with oblique corrections, it is often convenient to express the EW corrections in terms of effective polarisation amplitudes such that the 4D effective Lagrangian is given by 
\begin{eqnarray}
\mathcal{L}=-\frac{1}{2}(1-\Pi_{WW}^{\prime})W_{\mu\nu}^+W^{\mu\nu}_--\frac{1}{4}(1-\Pi_{ZZ}^{\prime})Z_{\mu\nu}Z^{\mu\nu}-\frac{1}{4}(1-\Pi_{\gamma\gamma}^{\prime})A_{\mu\nu}A^{\mu\nu}-\Pi_{\gamma Z}^{\prime}A^{\mu\nu}Z_{\mu\nu}\nonumber\\
+\left (\frac{g_4^2v^2}{4}+\Pi_{WW}(0)\right )W_\mu^+W^\mu_-+\frac{1}{2}\left (\frac{(g_4^2+g_4^{\prime\;2})v^2}{4}+\Pi_{ZZ}(0)\right )Z_\mu Z^\mu+\sum_\Psi i\bar{\Psi}\gamma^\mu D_\mu\Psi.\label{OblAction}\hspace{0.4cm}
\end{eqnarray}
Such a parameterisation has arisen by expanding the corrections to the gauge propagator in powers of momentum
\begin{displaymath}
\Pi(p^2)=\Pi(0) +p^2\Pi^{\prime}(0)+\frac{(p^2)^2}{2}\Pi^{\prime\prime}(0)+\dots
\end{displaymath} 
Of course changing the parameterisation of the effective Lagrangian will not change the scale of the corrections and so one can completely equivalently define the $S$ and $T$ parameters by
\begin{eqnarray*}
S=16\pi \left (\Pi_{33}^{\prime}-\Pi_{3Q}^{\prime}\right ) \\
T=\frac{4\pi}{s^2c^2M_Z^2}\left (\Pi_{11}(0)-\Pi_{33}(0)\right ). 
\end{eqnarray*} 
What is important for this parameterisation is to note that it assumes that the corrections to the gauge fermion vertex is negligible. \color{red} One \color{black} can consider the next term, in the expansion of the polarisation amplitude, in order to parameterise any additional contribution to the gauge propagator from such gauge fermion couplings. This has given rise to additional `oblique' coefficients being defined \cite{Barbieri:2004qk,Cacciapaglia:2006pk, Grojean:2006nn}
\begin{eqnarray*}
U=16\pi\left (\Pi^{\prime}_{11}-\Pi^{\prime}_{33}\right )\hspace{1.0cm}V=\frac{M_W^2}{2}\left (\Pi_{33}^{\prime\prime}-\Pi_{WW}^{\prime\prime}\right )\\
W=\frac{M_W^2}{2}\Pi_{33}^{\prime\prime}\hspace{1.0cm}X=\frac{M_W^2}{2}\Pi_{30}^{\prime\prime}\hspace{1cm}Y=\frac{M_W^2}{2}\Pi_{00}^{\prime\prime}.
\end{eqnarray*} 

Returning to the extra dimensional scenario, then as already mentioned the W and Z gauge fermion coupling will receive corrections, at tree level, due to the deformation of the zero mode profile (Eq. \ref{fZzero}). Therefore before one can compare to fits based on oblique corrections it is necessary to remove such corrections. If one again assumes universal gauge fermion coupling, or equivalently that all the fermions are sitting at the same point, then one can make a field redefinition to absorb these non oblique corrections \cite{Agashe:2003zs}
\begin{eqnarray*}
W_\mu^{(0)}\rightarrow \left (1+\sum_n\frac{m_w^2F_nF_\Psi^{(n)}}{m_n^2+m_w^2(F_n^2-1)}+\mathcal{O}(m_n^{-4})\right )W_\mu^{(0)}\\
Z_\mu^{(0)}\rightarrow \left (1+\sum_n\frac{m_z^2F_nF_\Psi^{(n)}}{m_n^2+m_z^2(F_n^2-1)}+\mathcal{O}(m_n^{-4})\right )Z_\mu^{(0)}. 
\end{eqnarray*}
One can then simply match the resulting Lagrangian for the gauge zero modes to Eq. \ref{OblAction} to get
\begin{eqnarray*}
\Pi_{11}(0)=\frac{\Pi_{WW}(0)}{g_4^2}\approx\frac{v^2}{4}\sum_n\frac{m_w^2(2F_nF_\psi^{(n)}-F_n^2)}{m_n^2+m_w^2(F_n^2-1)}\\
\Pi_{33}(0)=\frac{\Pi_{ZZ}(0)}{g_4^2+g_4^{\prime\;2}}\approx\frac{v^2}{4}\sum_n\frac{m_z^2(2F_nF_\psi^{(n)}-F_n^2)}{m_n^2+m_z^2(F_n^2-1)}\\
\Pi_{3Q}^{\prime}=\Pi_{\gamma Z}^{\prime}=0\\
\Pi_{11}^\prime=\frac{\Pi_{WW}^{\prime}}{g_4^2}\approx-\frac{v^2}{4}\sum_n\frac{2F_nF_\psi^{(n)}}{m_n^2+m_w^2(F_n^2-1)}\\
\Pi_{33}^\prime=\frac{\Pi_{ZZ}^{\prime}}{g_4^2+g_4^{\prime\;2}}\approx-\frac{v^2}{4}\sum_n\frac{2F_nF_\psi^{(n)}}{m_n^2+m_z^2(F_n^2-1)}
\end{eqnarray*}
and hence the tree level contribution to the S and T parameters are
\begin{eqnarray}
S\approx -\frac{4M_Z^2c^2s^2}{\alpha}\sum_n\frac{2F_nF_\psi^{(n)}}{m_n^2+m_z^2(F_n^2-1)}+\mathcal{O}(m_n^{-4})\nonumber\\
T\approx\frac{1}{\alpha}\sum_n \frac{m_w^2(2F_nF_\psi^{(n)}-F_n^2)}{m_n^2+m_w^2(F_n^2-1)}-\frac{m_z^2(2F_nF_\psi^{(n)}-F_n^2)}{m_n^2+m_z^2(F_n^2-1)}+\mathcal{O}(m_n^{-4}).\label{eq:ST}
\end{eqnarray}
The higher order terms that contribute to $U$, $V$, $W$ etc. are zero at tree level. The above expressions come with a number of warnings. The sums in the above expression would be over all the KK gauge modes that mix with the zero modes via the Higgs interaction. As already mentioned, one could envisage many scenarios in which one has heavy particles, such as gauge scalars or orthogonal KK towers, that contribute to a given EW observable at tree level but would not be include in the above sum. Such particles would probably contribute to the $S$ and $T$ parameters at loop level, through for example the triple gauge boson coupling, but \color{red} nonetheless \color{black} one would anticipate that the above expressions would under estimate the EW corrections. In addition to this the assumption of universal gauge fermion coupling is probably only valid for the light fermions.   

Having said that the $S$ and $T$ parameters do offer a simple and convenient measure of the scale of EW corrections. \color{red} The current bounds on $S$ and $T$, quoted in \cite{Nakamura:2010zzi}, are $\rm{S}=0.01\pm0.1$ and $\rm{T}=0.03\pm0.11$ ($M_H=117$ GeV) although a recent fit has put the bounds at $\rm{S}=0.02\pm0.11$ and $\rm{T}=0.05\pm0.12$ ($M_H=120$ GeV) \cite{Haller:2010zb}. Let us consider, for example, the 5D RS model with $F_n\approx 8.3$, $m_1\approx 2.45\; M_{\rm{KK}}$, and $m_2\approx 5.56\; M_{\rm{KK}}$. \color{black} If the fermions are localised on the IR brane, i.e. $F_\psi^{(n)}=F_n$,  then it is straight forward to check that the tightest bound comes from the $S$ parameter for which the $2\sigma$ bound is saturated when $M_{\rm{KK}}\gtrsim 11$ TeV in agreement with \cite{Csaki:2002gy, Huber:2000fh}. However if the fermions are localised towards the UV brane then $F_\psi^{(n)}\approx -0.2$ (see figure \ref{RSFpsi}) and the tightest bound comes from the $T$ parameter for which the $2\sigma$ bound is saturated when $M_{\rm{KK}}\gtrsim 4$ TeV, \cite{Huber:2001gw, Burdman:2002gr, carena-2003-68}.

\section{A Bulk Custodial Symmetry}\label{sect:Cust}
The observation that the mass of the first KK gauge field in the RS model must be greater than about 10 TeV gave rise to what has been labeled a little hierarchy problem. Since the Higgs mass would be quadratically sensitive to this scale, it starts becoming questionable whether the RS model has in fact fully resolved the gauge hierarchy problem. Unlike the constraints from flavour physics, it is difficult to see how one can reduce the constraints by tuning the free parameters of the theory. One possible resolution to this problem is to note that in addition to the bulk Lagrangian one should also include a brane localised Lagrangian at the fixed points of the orbifold. As was noted in \cite{Carena:2002dz, Carena:2004zn} if such a Lagrangian includes large gauge field kinetic terms then the constraints are significantly reduced. While this is a plausible explanation we will now consider what is arguably a more `natural' explanation that is not dependent on the size of the free parameters of the model.\newline

If we start by considering the four dimensional SM in which the Higgs is a complex doublet under $\color{red}\mathrm{SU} (2)$
\begin{displaymath}
\Phi=\begin{pmatrix}
     \pi_2+i\pi_1  \\
      \sigma-i\pi_3
\end{pmatrix}.
\end{displaymath}
The Higgs Lagrangian 
\begin{displaymath}
\color{red}\mathcal{L}=|\partial_\mu\Phi |^2+\mu^2|\Phi |^2-\lambda|\Phi |^4
\end{displaymath}
is then invariant under a global $\color{red}\mathrm{SO}(4)\cong \color{red}\mathrm{\color{red}\mathrm{SU} } (2)\times \color{red}\mathrm{SU} (2)$ which is broken by the Higgs gaining a VEV ($\langle\sigma\rangle^2=v^2$) to $\color{red}\mathrm{SU} (2)$. In other words before one gauges the SM it is invariant under an `accidental' global $\color{red}\mathrm{SU} (2)\times \color{red}\mathrm{SU} (2)$ symmetry which is referred to as a custodial symmetry \cite{WeinbergVol2}. 

Under the holographic description of the RS model, global symmetries of the four dimensional theory should appear as gauge symmetries in the bulk \cite{ArkaniHamed:2000ds}. So arguably in the bulk one should not describe the EW sector by $\color{red}\mathrm{SU} (2)_L\times \color{red}\mathrm{U}(1)$ but rather $\color{red}\mathrm{SU} (2)_R\times \color{red}\mathrm{SU} (2)_L\times \color{red}\mathrm{U}(1)$.

\subsection{The Model}
The basic idea, first presented in \cite{Agashe:2003zs},  is to extend the bulk EW gauge symmetry to $\color{red}\mathrm{SU} (2)_R\times \color{red}\mathrm{SU} (2)_L\times \color{red}\mathrm{U}(1)_X\times P_{LR}$. The $P_{LR}$ is a discrete symmetry that interchanges the $\color{red}\mathrm{SU} (2)_L$ and $\color{red}\mathrm{SU} (2)_R$ gauge groups and fixes the couplings to be the same $g_L=g_R\equiv g$. The coupling of the $\color{red}\mathrm{U}(1)_X$ field is denoted $g^{\prime}$. We will denote the $\color{red}\mathrm{SU} (2)_R$, $\color{red}\mathrm{SU} (2)_L$ and $\color{red}\mathrm{U}(1)_X$ fields as $\tilde{A}_M^a$, $A_M^a$ and $X_M$. One then essentially has two Higgs mechanisms at work. Firstly on the UV brane one has a `Planckian Higgs' that breaks $\color{red}\mathrm{SU} (2)_R\times U(1)_X\rightarrow \color{red}\mathrm{U}(1)_Y$ and hence mixing   
\begin{displaymath}
\tilde{Z}_M=c^{\prime}\tilde{A}_M^3-s^{\prime}X_M\quad\mbox{and}\quad B_M=s^{\prime}\tilde{A}_M^3+c^{\prime} X_M,
\end{displaymath}
where
\begin{equation}
\label{CustMixang}
s^{\prime}=\frac{g^{\prime}}{\sqrt{g^2+g^{\prime\,2}}}\mbox{   and   }c^{\prime}=\frac{g}{\sqrt{g^2+g^{\prime\,2}}}.
\end{equation}
Although it is more intuitive to think in terms of a second Higgs gaining a VEV at the Planck scale, one can equally achieve this mixing using just boundary conditions \cite{Albrecht:2009xr, Casagrande:2010si}. In the limit that this VEV goes to infinity the BC's become
\begin{displaymath}
\tilde{Z}_M:\;(-,+)\quad \tilde{A}^{1,2}_M:\;(-,+)\quad B_M:\;(+,+)\quad A^a_M:\;(+,+).
\end{displaymath}
Although we have denoted these BC's in terms of the parity under an orbifold, they refer simply to BC's w.r.t $r$ on the UV and IR brane. For ease of notation we shall also denote $f_n(r)\Theta_n(\phi)$ as $f_n$. Hence the covariant derivative is now given by 
\begin{displaymath}
D_\mu \Phi=\partial_\mu\Phi+\sum_n\left (-ig\tau^aA_\mu^{a(n)}f_n-ig\tau^{1,2}\tilde{A}_\mu^{1,2(n)}\tilde{f}_n-igc^{\prime}\tau^3\tilde{Z}_\mu^{(n)}\tilde{f}_n-igs^{\prime}\tau^3B_\mu^{(n)}f_n\right )\Phi.
\end{displaymath}
The remaining $\color{red}\mathrm{SU} (2)_L\times \color{red}\mathrm{U}(1)_Y$ symmetry is then broken by a Higgs (now a bi-doublet under $\color{red}\mathrm{SU} (2)_L\times \color{red}\mathrm{SU} (2)_R$) on the IR brane which gains a VEV at the EW scale, 
\begin{displaymath}
\Phi\rightarrow\left(\begin{array}{cc}0 & -\frac{v+H}{2} \\\frac{v+H}{2} & 0\end{array}\right).
\end{displaymath}
This then leads to the usual EW symmetry breaking and the field redefinitions
\begin{eqnarray}
Z_\mu^{(n)}=cA_\mu^{3(n)}-sB_\mu^{(n)},\hspace{2.5cm} A_\mu^{(n)}=sA_\mu^{3(n)}+cB_\mu^{(n)},\nonumber\\
W_\mu^{\pm\,(n)}=\frac{1}{\sqrt{2}}\left (A_\mu^{1(n)}\mp iA_\mu^{2(n)}\right ),\qquad \tilde{W}_\mu^{\pm\,(n)}=\frac{1}{\sqrt{2}}\left (\tilde{A}_\mu^{1(n)}\mp i\tilde{A}_\mu^{2(n)}\right ),\nonumber
\end{eqnarray}     
where
\begin{equation}
\label{ }
 s=\frac{s^{\prime}}{\sqrt{1+s^{\prime\,2}}}\mbox{   and   }c=\frac{1}{\sqrt{1+s^{\prime\,2}}}.
\end{equation}
If we now carry out the KK decomposition, then the mass terms of the gauge fields will be given by 
\begin{equation*}
\left (\begin{array}{cccc}W_\mu^{+\,(0)}& W_\mu^{+\,(1)}&\tilde {W}_\mu^{+\,(1)}&\dots\end{array} \right )\mathcal{M}_{\rm{charged}}^2\left (\begin{array}{c}W_\mu^{-\,(0)}\\W_\mu^{-\,(1)}\\\tilde{W}_\mu^{-\,(1)}\\\vdots\end{array}\right )
\end{equation*}
\begin{equation*}
\frac{1}{2}\left (\begin{array}{cccc}Z_\mu^{(0)}& Z_\mu^{(1)}&\tilde {Z}_\mu^{(1)}&\dots\end{array} \right )\mathcal{M}_{\rm{neutral}}^2\left (\begin{array}{c}Z_\mu^{(0)}\\Z_\mu^{(1)}\\\tilde{Z}_\mu^{(1)}\\\vdots\end{array}\right )
\end{equation*}
where
\begin{equation}\label{Mcharge}
\mathcal{M}_{\rm{charged}}^2=\left(\scriptsize\begin{array}{cccccc}\frac{g^2v^2}{4}f_0f_0 & \frac{g^2v^2}{4}f_0f_1 & -\frac{g^2v^2}{4}f_0\tilde{f}_1 & \frac{g^2v^2}{4}f_0f_2 & -\frac{g^2v^2}{4}f_0\tilde{f}_2 & \cdots \\\frac{g^2v^2}{4}f_0f_1 & m_1^2+\frac{g^2v^2}{4}f_1f_1 & -\frac{g^2v^2}{4}f_1\tilde{f}_1 & \frac{g^2v^2}{4}f_1f_2 & -\frac{g^2v^2}{4}f_1\tilde{f}_2 & \cdots \\-\frac{g^2v^2}{4}f_0\tilde{f}_1 & -\frac{g^2v^2}{4}f_1\tilde{f}_1 & \tilde{m}_1^2+\frac{g^2v^2}{4}\tilde{f}_1\tilde{f}_1 & -\frac{g^2v^2}{4}\tilde{f}_1f_2 & \frac{g^2v^2}{4}\tilde{f}_1\tilde{f}_2 &  \\\frac{g^2v^2}{4}f_0f_2 & \frac{g^2v^2}{4}f_1f_2 & -\frac{g^2v^2}{4}\tilde{f}_1f_2 & m_2^2+\frac{g^2v^2}{4}f_2f_2 & -\frac{g^2v^2}{4}f_2\tilde{f}_2 &  \\-\frac{g^2v^2}{4}f_0\tilde{f}_2 & -\frac{g^2v^2}{4}f_1\tilde{f}_2 & \frac{g^2v^2}{4}\tilde{f}_1\tilde{f}_2 & -\frac{g^2v^2}{4}f_2\tilde{f}_2 & \tilde{m}_2^2+\frac{g^2v^2}{4}\tilde{f}_2\tilde{f}_2 &  \\\vdots & \vdots &  &  &  & \ddots\end{array}\right)
\end{equation}
\begin{equation}\label{Mneutral}
\mathcal{M}_{\rm{neutral}}^2=\left(\scriptsize\begin{array}{cccccc}\frac{g^2v^2}{4c^2}f_0f_0 & \frac{g^2v^2}{4c^2}f_0f_1 & -\frac{g^2v^2c^\prime}{4c}f_0\tilde{f}_1 & \frac{g^2v^2}{4c^2}f_0f_2 & -\frac{g^2v^2c^\prime}{4c}f_0\tilde{f}_2 & \cdots \\\frac{g^2v^2}{4c^2}f_0f_1 & m_1^2+\frac{g^2v^2}{4c^2}f_1f_1 & -\frac{g^2v^2c^\prime}{4c}f_1\tilde{f}_1 & \frac{g^2v^2}{4c^2}f_1f_2 & -\frac{g^2v^2c^\prime}{4c}f_1\tilde{f}_2 & \cdots \\-\frac{g^2v^2c^\prime}{4c}f_0\tilde{f}_1 & -\frac{g^2v^2c^\prime}{4c}f_1\tilde{f}_1 & \tilde{m}_1^2+\frac{g^2v^2c^{\prime\,2}}{4}\tilde{f}_1\tilde{f}_1 & -\frac{g^2v^2c^\prime}{4c}\tilde{f}_1f_2 & \frac{g^2v^2c^{\prime\,2}}{4}\tilde{f}_1\tilde{f}_2 &  \\\frac{g^2v^2}{4c^2}f_0f_2 & \frac{g^2v^2}{4c^2}f_1f_2 & -\frac{g^2v^2c^\prime}{4c}\tilde{f}_1f_2 & m_2^2+\frac{g^2v^2}{4c^2}f_2f_2 & -\frac{g^2v^2c^\prime}{4c}f_2\tilde{f}_2 &  \\-\frac{g^2v^2c^\prime}{4c}f_0\tilde{f}_2 & -\frac{g^2v^2c^\prime}{4c}f_1\tilde{f}_2 & \frac{g^2v^2c^{\prime\,2}}{4}\tilde{f}_1\tilde{f}_2 & -\frac{g^2v^2c^\prime}{4c}f_2\tilde{f}_2 & \tilde{m}_1^2+\frac{g^2v^2c^{\prime\,2}}{4}\tilde{f}_1\tilde{f}_1 &  \\\vdots & \vdots &  &  &  & \ddots\end{array}\right).
\end{equation}   
Here $\tilde{f}_n$ has $(-+)$ BC's and mass eigenvalues $\tilde{m}_n^2$ and $f_n$ has $(++)$ BC's and mass eigenvalues $m_n^2$. Again $f_n$ now denotes the gauge Higgs coupling $f_n(r_{\rm{ir}})\Theta_n(\phi_{\rm{ir}})$. All the discussion of section \ref{sect:HiggsLocat} is equally applicable here and hence once again $f_nf_m$ can potentially correspond to a block matrix.

We can proceed exactly as before by perturbatively diagonalising the mass matrices to get the corrections to the zero mode masses 
\begin{eqnarray}
\hat{M}^2_W\approx m_w^2\left (1-\sum_n\left [\frac{m_w^2F_n^2}{m_n^2-m_w^2(F_n^2-1)}+\frac{m_w^2\tilde{F}_n^2}{\tilde{m}_n^2+m_w^2(\tilde{F}_n^2-1)}\right ]+\mathcal{O}(m_n^{-4})\right )\nonumber\\
\hat{M}^2_Z\approx \frac{m_w^2}{c^2}\left (1-\sum_n\left [\frac{\frac{m_w^2}{c^2}F_n^2}{m_n^2-\frac{m_w^2}{c^2}(F_n^2-1)}+\frac{c^{\prime\;2}m_w^2\tilde{F}_n^2}{\tilde{m}_n^2+m_w^2(c^{\prime\;2}\tilde{F}_n^2-c^{-2})}\right ]+\mathcal{O}(m_n^{-4})\right )\label{CustWZmasses}
\end{eqnarray}
where as before $m_w=\frac{g^2v^2}{4}f_0^2$.

\subsection{Electroweak Constraints}
We saw in the previous section that of particular relevance to the size of the EW constraints is the difference between the corrections to the W and Z masses. At zeroth order in four momentum this difference gives the $T$ parameter, while at first order this correction gives the $U$ parameter. This difference is also prevalent in expressions such as Eq. \ref{WeakMixingAngle}.

Bearing in mind the relations $1-\frac{1}{c^2}=-s^{\prime\;2}$ and $1-c^{\prime\;2}=s^{\prime\;2}$ then one can see that if $F_n=\tilde{F}_n$ and $m_n=\tilde{m}_n$ this difference, in Eq. \ref{CustWZmasses}, is exactly zero at first order in $m_n^{-2}$. In fact this cancellation continues to the next order of the expansion as well,
\begin{eqnarray*}
\triangle M_Z^2-\triangle M_W^2\approx\sum_{n=1}\Bigg [\frac{\frac{m_w^2}{c^2}F_n^2}{m_n^2+\frac{m_w^2}{c^2}(F_n^2-1)}-\frac{m_w^2F_n^2}{m_n^2+m_w^2(F_n^2-1)}+\frac{c^{\prime\,2}m_w^2\tilde{F}_n^2}{\tilde{m}_n^2+m_w^2(c^{\prime\,2}\tilde{F}_n^2-\frac{1}{c^2})}\\
\qquad\qquad\qquad\qquad-\frac{m_w^2\tilde{F}_n^2}{\tilde{m}_n^2+m_w^2(\tilde{F}_n^2-1)}\Bigg ]+\sum_{n=1}\sum_{m\neq n}\Bigg [\frac{\left (\frac{1}{c^4}-1\right )m_w^4F_n^2F_m^2}{m_n^2m_m^2}+\frac{(c^{\prime\,4}-1)m_w^4\tilde{F}_n^2\tilde{F}_m^2}{\tilde{m}_n^2\tilde{m}_m^2}\\
\qquad\qquad\qquad\qquad\qquad\qquad\qquad\qquad+\frac{\left (\frac{c^{\prime\,2}}{c^2}-1\right )m_w^4F_n^2\tilde{F}_m^2}{m_n^2\tilde{m}_m^2}+\frac{\left (\frac{c^{\prime\,2}}{c^2}-1\right )m_w^4F_m^2\tilde{F}_n^2}{m_m^2\tilde{m}_n^2}\Bigg ]+\mathcal{O}(m_n^{-6}),
\end{eqnarray*}
where now the last four terms cancel using the relations $\frac{1}{c^4}-1=2s^{\prime\,2}+s^{\prime\,4}$, $c^{\prime\,4}-1=-2s^{\prime\,2}+s^{\prime\,4}$ and $\frac{c^{\prime\,2}}{c^2}-1=-s^{\prime\,4}$. So the custodial symmetry is said to be protecting the $T$ parameter, hence the name.

Even if these cancellations occur at every order in the expansion, as one would expect, it is important to note that in any realistic scenario the $T$ parameter is never exactly zero. The reason for this is simply due to the fact that we do not observe the additional $\color{red}\mathrm{SU} (2)$ force and hence it is always necessary to break the custodial symmetry. Here this breaking is done by imposing different boundary conditions on the $\color{red}\mathrm{SU} (2)_R$ fields than on the $\color{red}\mathrm{SU} (2)_L$ fields which in turn causes $\tilde{F}_n$ and $\tilde{m}_n$ to be different from $F_n$ and $m_n$.  Although in the RS model this difference is typically small, e.g. $m_1\approx 2.45\; M_{\rm{KK}}$ while $\tilde{m}_1\approx 2.40\; M_{\rm{KK}}$.

In addition to this the custodial symmetry does not generically protect the $S$ parameter or the corrections  to the gauge fermion couplings. Such couplings arise from the covariant derivative in $i\bar{\psi}\gamma^\mu D_\mu\psi$ with
 \begin{eqnarray*}
D_\mu=\partial_\mu+\sum_{n}\bigg(-igf_\psi^{(n)}(T_L^+W_\mu^{+(n)}+T_L^-W_\mu^{-(n)})-ig\tilde{f}_\psi^{(n)}(T_R^+\tilde{W}_\mu^{+(n)}+T_R^-\tilde{W}_\mu^{-(n)}) \nonumber\\
-igsQf_\psi^{(n)}A_\mu^{(n)}-i\frac{g}{c}(T_L^3-s^2Q)f_\psi^{(n)}Z_\mu^{(n)}-i\tilde{f}_\psi^{(n)}(gc^{\prime}T_R^3-gs^{\prime}Q_X)\tilde{Z}_\mu^{(n)}\bigg ).
\end{eqnarray*}
Where $T_{L,R}^a$, and $Q_X$ are the charges under $\color{red}\mathrm{SU} (2)_{L,R}$ and $\color{red}\mathrm{U}(1)_X$, while $Q=T_L^3+T_R^3+Q_X$ and $T_{L,R}^\pm=(T_{L,R}^1\pm i T_{L,R}^2)$. The charge assignments in models with a custodial symmetry are a little subtle since they must not only reproduce observed SM charges they must also adhere to the $P_{LR}$ symmetry. In particular this symmetry was introduced in order to protect corrections to the $Zb_L\bar{b}_L$ coupling \cite{Agashe:2006at} and hence the left handed quarks must be embedded in a bi-doublet of $\color{red}\mathrm{SU} (2)_L\times \color{red}\mathrm{SU} (2)_R$ with $T_L^3=-T_R^3$. Here we are not concerned with a full phenomenological study but rather an estimate of the size of the EW constraints.  If one is to parameterise these constraints in terms of the oblique parameters then it is necessary to perform a field redefinition such that the corrections to the gauge fermion couplings are absorbed. So here we refer readers to \cite{Albrecht:2009xr, Casagrande:2010si} for a complete list of charge assignments and in what representations the SM particles sit. With such charge assignments it is found that the corrections to the gauge fermion couplings, of the fermions of relevance to EW observables $\hat{\alpha}$ and $G_f$, can again be approximately absorbed by the field redefinitions \cite{Carena:2006bn} 
\begin{eqnarray*}
W_\mu^{(0)}\rightarrow \left (1+\sum_n\frac{m_w^2F_nF_\Psi^{(n)}}{m_n^2+m_w^2(F_n^2-1)}+\mathcal{O}(m_n^{-4})\right )W_\mu^{(0)}\\
Z_\mu^{(0)}\rightarrow \left (1+\sum_n\frac{\frac{m_w^2}{c^2}F_nF_\Psi^{(n)}}{m_n^2+\frac{m_w^2}{c^2}(F_n^2-1)}+\mathcal{O}(m_n^{-4})\right )Z_\mu^{(0)}. 
\end{eqnarray*}
As before this leads to
\begin{eqnarray*}
\Pi_{11}(0)\approx\frac{v^2}{4}\sum_n\;\frac{m_w^2(2F_nF_\psi^{(n)}-F_n^2)}{m_n^2+m_w^2(F_n^2-1)}-\frac{m_w^2\tilde{F}_n^2}{\tilde{m}_n^2+m_w^2(\tilde{F}_n^2-1)}\\
\Pi_{33}(0)\approx\frac{v^2}{4}\sum_n\;\frac{\frac{m_w^2}{c^2}(2F_nF_\psi^{(n)}-F_n^2)}{m_n^2+\frac{m_w^2}{c^2}(F_n^2-1)}-\frac{m_w^2c^{\prime\;2}\tilde{F}_n^2}{\tilde{m}_n^2+m_w^2(c^{\prime\;2}\tilde{F}_n^2-c^{-2})}\\
\Pi_{3Q}^{\prime}=\Pi_{\gamma Z}^{\prime}=0\\
\Pi_{11}^\prime\approx-\frac{v^2}{4}\sum_n\frac{2F_nF_\psi^{(n)}}{m_n^2+m_w^2(F_n^2-1)}\\
\Pi_{33}^\prime\approx-\frac{v^2}{4}\sum_n\frac{2F_nF_\psi^{(n)}}{m_n^2+\frac{m_w^2}{c^2}(F_n^2-1)}.
\end{eqnarray*}
Hence, as found in \cite{ Casagrande:2008hr, Delgado:2007ne, Casagrande:2010si}, the tree level contribution to $S$ parameter is approximately the same for the two models
 \begin{eqnarray}
S\approx -\frac{4M_Z^2c^2s^2}{\alpha}\sum_n\frac{2F_nF_\psi^{(n)}}{m_n^2+\frac{m_w^2}{c^2}(F_n^2-1)}+\mathcal{O}(m_n^{-4})\nonumber\\
T\approx\frac{1}{\alpha}\sum_n\Bigg (\frac{m_w^2(2F_nF_\psi^{(n)}-F_n^2)}{m_n^2+m_w^2(F_n^2-1)}-\frac{m_w^2\tilde{F}_n^2}{\tilde{m}_n^2+m_w^2(\tilde{F}_n^2-1)}-\frac{\frac{m_w^2}{c^2}(2F_nF_\psi^{(n)}-F_n^2)}{m_n^2+\frac{m_w^2}{c^2}(F_n^2-1)}\nonumber\\+\frac{m_w^2c^{\prime\;2}\tilde{F}_n^2}{\tilde{m}_n^2+m_w^2(c^{\prime\;2}\tilde{F}_n^2-c^{-2})}+\mathcal{O}(m_n^{-4})\Bigg ).\label{eqn:STcust}
\end{eqnarray}
So in the RS model with fermions localised towards the UV brane, $F_\psi^{(n)}\approx -0.2$, the $2\sigma$ constraints from the $S$ parameter forces $M_{\rm{KK}}\gtrsim 1.5-2$ TeV \cite{Agashe:2003zs}. It is interesting to note that studies that have included all the dimension six observables and taken into account the non-universality of the heavy quarks gauge coupling have resulted in a slight reduction in this constraint to $M_{\rm{KK}}\gtrsim 1$ TeV \cite{Carena:2007ua}.

Alternatively one can again compare directly with observables, where once again the tightest constraint, \color{red} on the KK scale, \color{black} comes from the weak mixing angle,    
\begin{displaymath}
s_Z^2=\frac{\pi\alpha}{\sqrt{2}G_f}\frac{f_\psi^{(m)}(\mathcal{M}_{\rm{charged}}^2)^{-1}_{nm}f_\psi^{(n)}}{(f_\psi^{(0)})^2}.
\end{displaymath}
Repeating the method of the previous section, but now with the enlarged mass matrix, one finds that
\begin{equation}
\label{SZapp}
s_Z^2\approx s_p^2\left (1+\frac{c_p^2}{c_p^2-s_p^2}\sum_{n=1}\left [\frac{m_w^2F_\psi^{(n)\,2}}{m_n^2}-\frac{2m_w^2F_nF_\psi^{(n)}}{m_n^2}+s^{\prime\, 2}m_w^2\left (\frac{\tilde{F}_n^2}{\tilde{m}_n^2}-\frac{F_n^2}{m_n^2}\right )+\mathcal{O}(m_n^{-4})\right ]\right ).
\end{equation}  
In the RS model it is found that, with the fermions localised toward the UV brane, the constraint from weak mixing angle are slightly higher than from the $S$ parameter, $M_{\rm{KK}}\gtrsim 2.5$ TeV. 

\section{A Midpoint Summary}
The focus of this thesis has been to examine what possible spaces offer phenomenologically viable resolutions to the gauge hierarchy solutions, however up to now we have only focused on the 5D RS model. For such a model we have found that typically the largest constraints, \color{red} on the size of the KK scale, \color{black} come from $K^0-\bar{K}^0$ mixing although such constraints are dependent on the level of tuning one considers acceptable. We have also found that potentially large corrections to the EW observables can be removed by extending the model to include a bulk custodial symmetry. However in extending this analysis to completely generic spaces one immediately runs into questions that are very model dependent. In particular how localised are the fermions, Higgs and gauge fields and what are the boundary conditions of such fields? The answers to such questions will in turn determine the answers to questions such as should one be considering Higgs mediated FCNC's \cite{Azatov:2009na}, gauge-scalar mediated FCNC's or even gravity mediated FCNC's.      
Are some EW observables receiving contributions that are not included in the corrections to the gauge propagator? Or more specific questions such as, in the above expressions, when one sums over $n$ how many KK towers are included in the sum. Such uncertainties are an unfortunate consequence of an attempts to consider completely generic spaces and must be answered for each model individually.

None the less we are now able to make a number of statements which one would anticipate would hold for most spaces that resolve the gauge hierarchy problem. 
\begin{itemize}
 \item Firstly we have broadly demonstrated that the tree level contribution to low energy phenomenology, from the KK modes of gauge fields, is essentially characterised by three quantities. Firstly there is the relative gauge Higgs coupling, which we define to be the VEV of
 \begin{equation}
\label{FnDefinition}
F_nF_m\equiv\frac{\int d^{1+\delta}x bc^\delta\sqrt{\gamma}f_n\Theta_n\Phi^\dag f_m\Theta_m \Phi }{\int d^{1+\delta}x bc^\delta\sqrt{\gamma}f_0\Theta_0\Phi^\dag f_0\Theta_0 \Phi },
\end{equation}
although when the Higgs is localised in the IR tip of the space,
\begin{equation}
\label{FnPrac}
F_n=\frac{f_n(r_{\rm{ir}})}{f_0(r_{\rm{ir}})}\quad\mbox{   or   }\quad F_n=\frac{f_n(r_{\rm{ir}})\Theta_n(\phi_{\rm{ir}})}{f_0(r_{\rm{ir}})\Theta_0(\phi_{\rm{ir}})},
\end{equation}
depending on whether the Higgs is localised to a codimension one brane or a lower dimension brane. \color{red} The second and third quantities are the relative gauge fermion coupling,
 \begin{equation}
\label{FpsiDefinition }
F_\psi^{(n)}\equiv \frac{\int d^{1+\delta}xa^3bc^\delta\sqrt{\gamma} f^{(0)}_{L,R}\Theta_{L,R}^{(0)}f_n\Theta_nf^{(0)}_{L,R}\Theta_{L,R}^{(0)}}{\int d^{1+\delta}xa^3bc^\delta\sqrt{\gamma} f^{(0)}_{L,R}\Theta_{L,R}^{(0)}f_0\Theta_0f^{(0)}_{L,R}\Theta_{L,R}^{(0)}},
\end{equation}
and the mass of the KK gauge fields, $m_n$. \color{black} Of course it is always possible that a given observable will receive additional contributions from other aspects of an extra dimensional model. However here we suspect the above contributions will always be present.
 \item The contribution, from the KK gauge fields, to the scale of FCNC's will be determined by the level of universality of $F_\psi^{(n)}$, w.r.t different flavours. While the scale of the contribution to EW constraints will be determined by the magnitude of $F_\psi^{(n)}$ and $F_n$.
  \item In section \ref{sect:fermions} it was demonstrated that in five and six dimensions the direction that the fermion zero mode is peaked towards is determined by a bulk mass term. It is reasonable to assume that if one can include such bulk mass terms and still obtain a 4D chiral theory then these bulk mass terms will always determine the location of the fermion zero modes. 
  \item It is a generic feature of warped spaces, that resolve the gauge hierarchy problem, that the KK profiles of massless fields will \color{red} typically \color{black} be peaked towards the IR tip of the space. We will elaborate on this statement in more detail in the next chapter.
  \item Bearing in mind the form of Eq. \ref{5DfermionZeromode} then, with the KK gauge modes sitting towards the IR brane, in five dimensions $F_\psi^{(n)}$ will always run from $F_n$ to $f_n(r_{\rm{uv}})/f_0(r_{\rm{uv}})$ (see figure \ref{RSFpsi}). Hence, assuming that there is no limit to the size of the bulk mass parameter, then there will always exist two regions in which $F_\psi^{(n)}$ is universal, one with the fermions sitting towards the UV brane and the other with the fermions sitting towards the IR brane. The RS GIM mechanism is based on the observation that the light fermions would naturally sit in the UV region of universality. 
  \item While it is very possible that this also holds for spaces of more than five dimensions, this is not certain. For example it is possible that complex bulk mass terms could result in sinusoidal $\Theta_{L,R}^{(0)}$ profiles which would lead to a more complicated relation between $F_\psi^{(n)}$ and the bulk mass parameters. An interesting possibility, that has been beyond the scope of this thesis to investigate, is that in more than five dimensions there may exist more than two regions of universality. Of course FCNC's would only be suppressed if all the SM fermions were in the same region.
  \item None the less if one assumes
  \begin{displaymath}
\frac{\int d^\delta x\;\sqrt{\gamma}\Theta_{L,R}^{(0)}\Theta_n\Theta_{L,R}^{(0)}}{\int d^\delta x\;\sqrt{\gamma}\Theta_{L,R}^{(0)}\Theta_0\Theta_{L,R}^{(0)}}\quad\sim \quad\mathcal{O}(1)
\end{displaymath}
then one would expect that $F_\psi^{(n)}\leqslant F_n$. Hence the value of $F_n$ is critical in determining the size of the EW constraints. It would not be correct to say that the size of EW constraints are determined by the value of $F_n$ since there will always be potential for unknown contributions from, for example, the gauge scalars. However one can say that if a given space has a large $F_n$ value then one can expect that large EW constraints will force the KK scale to be quite large and arguably such a space can not be said to resolve the gauge hierarchy problem.  
 \end{itemize}
In the RS model $F_n\approx 8.3$ while in models with a universal extra dimension $F_n=\sqrt{2}$, hence these models have much lower EW constraints. We will now move on to ask the question do all \color{red} spaces \color{black} that resolve the gauge hierarchy problem have large $F_n$ values.

%% file: AlernativeSpaces.tex
\chapter{Alternative Spaces}
\label{AlernativeSpaces} 

\section{$F_n$ in Five Dimensions}\label{sect:5dFn}
We start by considering generic five dimensional warped spaces, described by 
\begin{equation}
\label{ }
ds^2=a(r)^2\eta^{\mu\nu}dx_\mu dx_\nu-b^2(r)dr^2,
\end{equation}
that resolve the gauge hierarchy problem. In order for a space to offer a potential resolution to the gauge hierarchy problem a number of features are required. Firstly we obviously require a large hierarchy between the effective Higgs VEV and the effective Planck scale. So if one assumes that there is no volume scaling between the fundamental Planck scale and the 4D effective Planck scale, that the Higgs is localised at $r=r_{\rm{ir}}$ and we define $a(r_{\rm{uv}})=1$ we require that $a(r_{\rm{ir}})^{-1}=\Omega\sim 10^{15}$. Of course one must be able to stabilise the space such that a large warp factor is achieved without fine tuning \cite{Goldberger:1999uk}. Here we are not interested in finding all the possible spaces that resolve the hierarchy problem but rather studying the phenomenological implications of spaces that do. Hence we will not consider the stabilisation mechanism but assume the space has already been stabilised. Secondly we require that all the dimensionful parameters of the theory have approximately the same order of magnitude. We are referring to the fundamental parameters such as the curvature, proper distance between branes and fundamental Planck mass ($M_{\rm{fund}}$ in Eq. \ref{MPlanck}) rather than the effective scales such as $M_{\rm{KK}}$. Thirdly, since one would anticipate that the Higgs mass would be sensitive to $M_{\rm{KK}}$, we require that $M_{\rm{KK}}\sim\mathcal{O}(\rm{TeV})$. 

We now consider a gauge field, with NBC's, propagating in this space described by Eq. \ref {5DGaugeEqn} which can be rewritten as
\begin{displaymath}
\left (a^2b^{-1}\exp\left [\int_c^r\frac{f^{\prime\prime}(\tilde{r})}{f^{\prime}(\tilde{r})}d\tilde{r}\right ]\right )^{\prime}=-bm_n^2\frac{f}{f^{\prime}}\exp\left [\int_c^r\frac{f^{\prime\prime}(\tilde{r})}{f^{\prime}(\tilde{r})}d\tilde{r}\right ].
\end{displaymath}
We can now solve for $a(r)$ in terms of the gauge field profiles,  
\begin{equation}
\label{a2}
a^2(r)=\frac{-Km_n^2b(r)\int_c^rb(\tilde{r})f_n(\tilde{r})d\tilde{r}}{f_n^{\prime}(r)},
\end{equation}
where $K$ is a constant of integration. However orthogonality with the zero mode implies that 
\begin{equation}
\label{orthogCon}
\int_{r_{\rm{ir}}}^{r_{\rm{uv}}}bf_ndr=0
\end{equation} 
and hence under NBC's Eq. \ref{a2} is ill defined at the boundaries. So at the boundaries 
\begin{displaymath}
a^2(r_{\rm{ir/uv}})=\lim_{\epsilon\rightarrow 0}a^2(r_{\rm{ir/uv}}+\epsilon)=\frac{-Km_n^2b^2(r_{\rm{ir/uv}})f_n(r_{\rm{ir/uv}})}{f_n^{\prime\prime}(r_{\rm{ir/uv}})}.
\end{displaymath}
Using that $a(r_{\rm{uv}})=1$ and $a(r_{\rm{ir}})^{-1}=\Omega$ then one can eliminate $K$ and $m_n$ to get
\begin{equation}
\label{HierCon}
\Omega^2=\frac{b^2(r_{\rm{uv}})f_n(r_{\rm{uv}})f_n^{\prime\prime}(r_{\rm{ir}})}{b^2(r_{\rm{ir}})f_n(r_{\rm{ir}})f_n^{\prime\prime}(r_{\rm{uv}})}.
\end{equation}
Such an expression can be more simply obtained by simply considering Eq. \ref {5DGaugeEqn} at the boundaries. One can always perform a coordinate transformation 
\begin{displaymath}
\color{red} r\rightarrow\tilde{r}=\int_c^r b(\hat{r})d\hat{r} 
\end{displaymath} 
such that $b=1$ and hence one finds that a gauge field, with NBC's, propagating in a space with a warp  factor $\Omega$ must satisfy
\begin{equation}
\label{5DHierarchyCon}
\Omega^2=\frac{f_n(r_{\rm{uv}})f_n^{\prime\prime}(r_{\rm{ir}})}{f_n(r_{\rm{ir}})f_n^{\prime\prime}(r_{\rm{uv}})}.
\end{equation}
Hence if at the boundaries of the space $f_n\gg f_n^{\prime\prime}$ then one would anticipate in order to generate a large warp factor then $f_n(r_{\rm{uv}})\gg f_n(r_{\rm{ir}})$ and the gauge profiles will be peaked towards the UV. Alternatively if at the boundaries $f_n^{\prime\prime}\gg f_n$ then $f^{\prime\prime}_n(r_{\rm{ir}})\gg f^{\prime\prime}_n(r_{\rm{uv}})$ and the profiles will peak towards the IR brane.

Alternatively we can consider the Ricci scalar of the space which (with $b=1$) is given by
\begin{displaymath}
\mathcal{R}=-4\left (3\left(\frac{a^{\prime}}{a}\right )^2+2\frac{a^{\prime\prime}}{a}\right )
\end{displaymath}
which if one plugs in Eq. \ref {5DGaugeEqn} then one finds that the profiles are related to the curvature by
\begin{equation}
\label{ }
\mathcal{R}=-\frac{3}{f_n^{\prime\;2}}\left [(f_n^{\prime\prime})^2+\frac{2m_n^2}{a^2}f_nf_n^{\prime\prime}+\frac{m_n^4}{a^4}f_n^2\right ]-\frac{8a^{\prime\prime}}{a}.
\end{equation}
Clearly if a space has a large hierarchy between the curvature at the IR and UV tips of the space then it has not really resolved the hierarchy problem. Hence we require that $\mathcal{R}|_{r_{\rm{ir}}}\sim\mathcal{R}|_{r_{\rm{uv}}}$ or equivalently
\begin{eqnarray*}
-3\left (\frac{f_n^{\prime\prime}(r_{\rm{ir}})}{f_n^{\prime}(r_{\rm{ir}})}\right )\left [1+2m_n^2\Omega^2\frac{f_n(r_{\rm{ir}})}{f_n^{\prime\prime}(r_{\rm{ir}})}+m_n^4\Omega^4\left (\frac{f_n(r_{\rm{ir}})}{f_n^{\prime\prime}(r_{\rm{ir})}}\right )^2\right ]-8a^{\prime\prime}(r_{\rm{ir}})\Omega \quad\\
\sim\quad-3\left (\frac{f_n^{\prime\prime}(r_{\rm{uv}})}{f_n^{\prime}(r_{\rm{uv}})}\right )\left [1+2m_n^2\frac{f_n(r_{\rm{uv}})}{f_n^{\prime\prime}(r_{\rm{uv}})}+m_n^4\left (\frac{f_n(r_{\rm{uv}})}{f_n^{\prime\prime}(r_{\rm{uv}})}\right )^2\right ]-8a^{\prime\prime}(r_{\rm{uv}}) .
\end{eqnarray*} 
Once again this appears to \color{red} be satisfied \color{black} when either $f_n(r_{\rm{uv/ir}})\gg f^{\prime\prime}_n(r_{\rm{uv/ir}})$ and $f_n(r_{\rm{uv}})\gg f_n(r_{\rm{ir}})$ or $f_n^{\prime\prime}(r_{\rm{uv/ir}})\gg f_n(r_{\rm{uv/ir}})$ and $f^{\prime\prime}_n(r_{\rm{ir}})\gg f^{\prime\prime}_n(r_{\rm{uv}})$. In other words in order to `pull down' the large warp factors, the profiles \color{red} will typically \color{black} be heavily localised towards either the UV or IR brane. To see that the former option is not viable one must simply consider Eq. \ref{5DGaugeEqn} giving
\begin{displaymath}
f_n^{\prime\prime}(r_{\rm{ir}})+m_n^2\Omega^2f_n(r_{\rm{ir}})=0\quad\mbox{ and }\quad f_n^{\prime\prime}(r_{\rm{uv}})+m_n^2f_n(r_{\rm{uv}})=0
\end{displaymath}
and note that $m_n^2\Omega^2$ will be of the order of the Planck mass squared. Clearly it is straightforward to repeat this for other types of fields. \color{red} The point of all this is to emphasize that, in five dimensions, a consequence of a large warp factor is that the KK profiles will generally be very localised towards the IR brane. In theory, one could write down counter examples although such examples would probably be quite contrived. Bearing in mind Eq. \ref{5DfermionZeromode}, then if one can include a fermion bulk mass term and the KK gauge profiles are heavily IR localised, then there will also always be a region in which $F_\psi^{(n)}$ will be universal. \color{black} In other words the RS-GIM mechanism is also a by product of a large warp factor. If we consider $F_n$ with the normalisation factors separated out,
\begin{equation}
 \label{fnsmall}
F_n=\frac{\sqrt{\int b\;dr}\, \tilde{f}_n(r_{\rm{ir}})}{\sqrt{\int b \;\tilde{f}_n^2\,dr}},
\end{equation} 
then one can see that in order to achieve, say, $F_n\approx 1$ one needs the profiles to be close to flat.  Therefore five dimensional models that resolve the hierarchy problem via warping will typically have large $F_n$ values and hence suffer from large corrections to EW observables. This conclusion was also reached using a different method in \cite{Delgado:2007ne}. Of course the inclusion of a bulk custodial symmetry will result in the EW constraints being linearly dependent on $F_n$ as well as the value of $F_\psi^{(n)}$. One would anticipate that the magnitude of $F_\psi^{(n)}$ would be sensitive to the bulk mass parameters and hence significantly more model dependent.

\begin{figure}
\begin{center}
\includegraphics[width=4in]{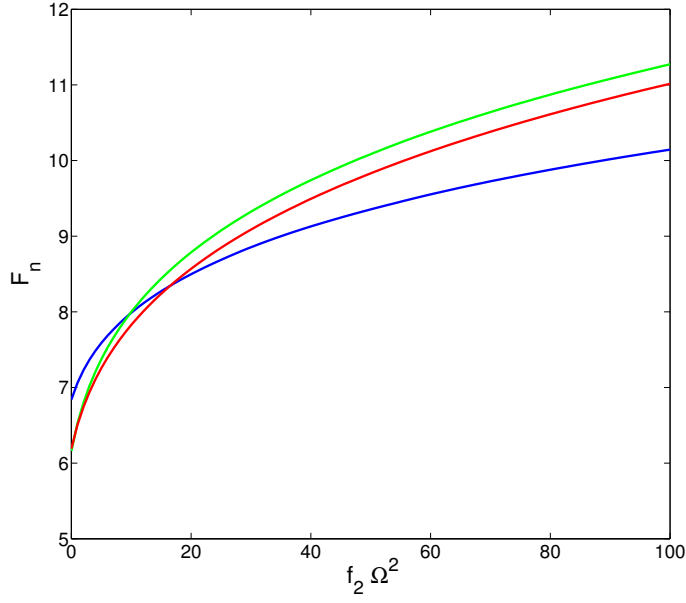}
\caption[Gauge Higgs Coupling in 5D IR Deformed Spaces]{The relative gauge Higgs couplings of the first (blue), second (green) and third (red) gauge KK mode. The equivalent RS value is about 8.3. Here $\Omega=\frac{R}{R^{\prime}}=10^{15}$ and $\frac{R^{\prime}}{R^2}= 1$ TeV.}
\label{MassGapCouplingfig}
\end{center}
\end{figure}

To finish with a specific example one can consider a five dimensional space which is approximately AdS$_5$ but has been deformed in the IR
\begin{equation}
\label{ }
ds^2=h^{-\frac{1}{2}}(r)\eta^{\mu\nu}dx_\mu dx_\nu-h^{\frac{1}{2}}(r)dr^2\quad \mbox{with}\quad h(r)=\frac{R^4}{R^{\prime 4}+\mathrm{ f_2}R^2r^2+r^4}.      
\end{equation} 
This space was introduced in \cite{Shiu:2007tn} as a five dimensional approximation of the Klebanov Strassler solution \cite{Klebanov:2000hb}, which will be discussed in more detail in section \ref{sect:KlebStras}. AdS$_5$ is regained in the limit that $f_2=R^{\prime}=0$. Using Eq. \ref{5DGaugeEqn} it is straight forward to calculate the relative gauge Higgs coupling. This has been plotted in figure \ref{MassGapCouplingfig}. One finds that there is little deviation from the RS values and hence one would anticipate that the size of the EW constraints will not differ significantly from that of the RS model. This was then confirmed in \cite{McGuirk:2007er}.

\section{$F_n$ in More Than Five Dimensions} 
When one considers more than five dimensions, it is far less certain that $F_n$ must be large. Considers again a gauge field with NBC's propagating in a space that resolves the hierarchy problem. Then from Eq. \ref{GaugeEOM} it is straight forward to demonstrate that the higher dimensional analogue of Eq. \ref{HierCon} is
\begin{equation}
\label{ }
\Omega^2=\frac{b^2(r_{\rm{uv}})f_n(r_{\rm{uv}})\left(f_n^{\prime\prime}(r_{\rm{ir}})-\frac{b^2(r_{\rm{ir}})}{c^2(r_{\rm{ir}})}\alpha_n f_n(r_{\rm{ir}})\right )}{b^2(r_{\rm{ir}})f_n(r_{\rm{ir}})\left(f_n^{\prime\prime}(r_{\rm{uv}})-\frac{b^2(r_{\rm{uv}})}{c^2(r_{\rm{uv}})}\alpha_n f_n(r_{\rm{uv}})\right )}.
\end{equation}
Now there are two new unknown scales, notably the warping of the internal manifold $c(r)$ and the eigenvalues of the Laplacian of the internal space $\alpha_n$. Hence it is far more difficult to say what the generic forms of the profiles will be. One could consider the KK modes for which $\alpha_n=0$, i.e. the modes which are flat w.r.t $\phi$, which would have the KK numbers $n=(n_1,0,\dots,0)$. One can use the arguments of the previous section to demonstrate that these profiles must be localised towards the IR tip of the space. However when it comes to computing $F_n$
\begin{equation}
\label{Fn}
F_n=\frac{\sqrt{\int d^{\delta+1}x\; bc^\delta\sqrt{\gamma}}\, \tilde{f}_n(r_{\rm{ir}})\tilde{\Theta}_n(\phi_{\rm{ir}})}{\sqrt{\int d^{\delta+1}x\; bc^\delta\sqrt{\gamma}\tilde{f}_n^2\tilde{\Theta}_n^2}}.
\end{equation}  
one critically finds that one must integrate over the unknown volume factor of the internal space, $c^\delta\sqrt{\gamma}$. Of course finding that it is difficult to make generic statements about $D$ dimensional spaces where one does not know the geometry of the space should not be a surprising result. Hence we will now move on to examine a number of possible warped spaces that all could potentially offer resolutions to the gauge hierarchy problem.

\subsection{A Space With an Internal Manifold Which Varies as a Power Law}\label{sect:powerGrow}
For our first example we will simply assume the form of the space to be
\begin{equation}
\label{ }
ds^2=\frac{r^2}{R^2}\eta_{\mu\nu}dx^\mu dx^\nu-\frac{R^2}{r^2}\left (dr^2+\frac{r^{2+a}}{R^a}d\Omega_\delta^2\right ).
\end{equation}  
where again $d\Omega_\delta^2=\gamma_{ij}d\phi^id\phi^j$. Note that as $a$ goes to zero the space tends towards a space of AdS${}_5\times\mathcal{M}^\delta$, studied in \cite{Davoudiasl:2002wz,Davoudiasl:2008qm}. Here we bound the space with a UV brane at $r_{\rm{uv}}=R$ and a IR brane $r_{\rm{ir}}=R^{\prime}$, such that  $R^{\prime}\leqslant r \leqslant R$ and $\Omega=\frac{R}{R^{\prime}}$. However as described in section \ref{sect:hierarchyResol}, the size of the warp factor required to resolve the hierarchy problem is dependent on the scaling of the Planck mass, Eq. \ref{MPlanck}
\begin{equation}
\label{ }
M_{\rm{P}}^2\sim\left (\int d^\delta x\sqrt{\gamma}\right )\frac{2}{4+a\delta}R^{\delta+1}\left(1-\left (\frac{1}{\Omega}\right )^{2+\frac{a\delta}{2}}\right )M_{\rm{fund}}^{\delta+3}.
\end{equation}
If we assume that $\left (\int d^\delta x\sqrt{\gamma}\right )$ is of order one and $2+\frac{a\delta}{2}>0$ then, when $R\sim M_{\rm{P}}^{-1}$, a warp factor of $\Omega\sim10^{15}$ would resolve the hierarchy problem. The gauge field wavefunctions would then be described by Eq. \ref{GaugeEOM}
\begin{equation}
\label{geqnRr}
f_n^{\prime\prime}+\frac{6+a\delta}{2r}f_n^{\prime}-\frac{R^a}{r^{2+a}}\alpha_nf_n+m_n^2\frac{R^4}{r^4}f_n=0.
\end{equation}
If we now consider KK modes with $\alpha_n=0$ and hence $\Theta_n=$ constant, then Eq. \ref{geqnRr} can be solved to give
\begin{equation}
\label{ }
f_n=\frac{N}{r^{1+\frac{a\delta}{4}}}\left (\mathbf{J}_{-1-\frac{a\delta}{4}}\left (\frac{R^2m_n}{r}\right )+\beta \mathbf{Y}_{-1-\frac{a\delta}{4}}\left (\frac{R^2m_n}{r}\right )\right ).
\end{equation}
Note that the profile will be localised towards the IR provided that $a\delta>-4$. This point exactly coincides with the point at which the four dimensional Planck mass begins to be scaled by the volume of the space and hence the warp factor required to resolve the hierarchy problem changes. If we now define $\hat{m}_n\equiv\frac{R^2m_n}{R^{\prime}}$ then under Neumann BC's 
\begin{displaymath}
\beta=-\frac{\mathbf{J}_{-\frac{a\delta}{4}}(\hat{m}_n\Omega^{-1})}{\mathbf{Y}_{\frac{-a\delta}{4}}(\hat{m}_n\Omega^{-1})}
\end{displaymath}
and
\begin{displaymath}
\mathbf{\mathbf{J}}_{-\frac{a\delta}{4}}(\hat{m}_n)+\beta \mathbf{Y}_{-\frac{a\delta}{4}}(\hat{m}_n)=0.
\end{displaymath}
The normalisation constant is given by
\begin{displaymath}
N^{-2}=\frac{R^{1+\delta-\frac{a\delta}{2}}}{(R^{\prime}\hat{m}_n)^2}\bigg [\frac{x^2}{2}\bigg (\mathbf{J}_v(x)^2-\mathbf{J}_{v-1}(x)\mathbf{J}_{v+1}(x)+\beta^2\bigg(\mathbf{Y}_v(x)^2-\mathbf{Y}_{v-1}(x)\mathbf{Y}_{v+1}(x)\bigg)
\end{displaymath}
\begin{displaymath}
\hspace{3cm}+\beta\bigg(2\mathbf{Y}_v(x)\mathbf{J}_v(x)-\mathbf{Y}_{v-1}(x)\mathbf{J}_{v+1}(x)-\mathbf{J}_{v-1}(x)\mathbf{Y}_{v+1}(x)\bigg)\bigg )\bigg ]_{\hat{m}_n\Omega^{-1}}^{\hat{m}_n}.
\end{displaymath}
Where $v=-1-\frac{a\delta}{4}$. If we assume that $\beta$ is small and $\Omega$ is large then we can make the approximation
\begin{displaymath}
f_n(R^{\prime})\sim\frac{\sqrt{2}}{\sqrt{R^{1+\delta}}}\Omega^{\frac{a\delta}{4}}.
\end{displaymath}
The zero mode, on the other hand, is given by
\begin{displaymath}
f_0=\left\{\begin{array}{cc}  \frac{1}{\sqrt{R^{1+\delta}\ln (\Omega)}} & \mbox{for $a=0$ or $\delta=0$}, \\
\frac{1}{\sqrt{\frac{2R^{1+\delta}}{a\delta}\left [1-\Omega^{-\frac{a\delta}{2}}\right ]}} & \mbox{otherwise}.
\end{array}\right.
\end{displaymath}
Putting this together we find that when $\Omega$ is large and $\beta$ is small the gauge Higgs coupling (for modes with $\alpha_n=0$) is given by
\begin{equation}
\label{ }
F_n\sim\left\{\begin{array}{cc} \sqrt{2\ln \Omega} & \mbox{for $a=0$ or $\delta=0$} \\
\frac{2}{\sqrt{a\delta}}\Omega^{\frac{a\delta}{4}} & \mbox{otherwise}
\end{array}\right.
\end{equation}

\begin{figure}[t!]
\begin{center}
\includegraphics[width=5in]{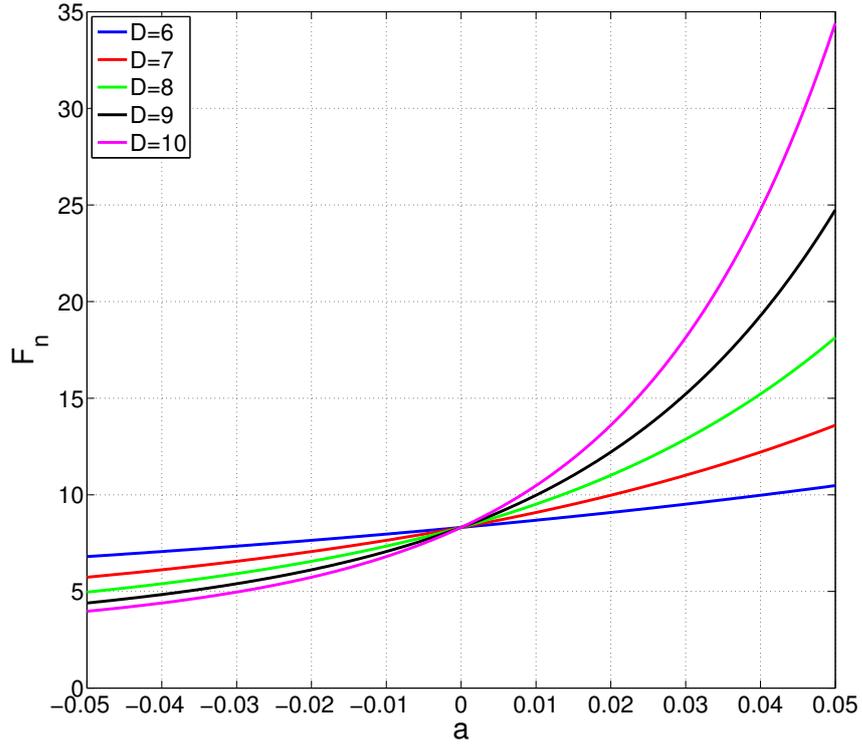}
\caption[$F_n$ for a Space with an Internal Manifold with a Varying Radius] {The relative gauge Higgs coupling for gauge fields propagating in a $D$-dimensional space with a internal manifold growing as $r^a$. Here $\Omega=\frac{R}{R^{\prime}}=10^{15}$ and $\frac{R^{\prime}}{R^2}= 1$ TeV. \cite{Archer:2010hh} }
\label{FnRa}
\end{center}
\end{figure}
This is only an approximate result which tends to break down for $\beta\gtrsim-0.2$, but  the exact couplings are shown in figure {\ref{FnRa}} and it does emphasize an important result. Notably that in spaces where the size of the internal manifold grows (or shrinks), towards the UV, the gauge Higgs coupling scales as a function of the warp factor. This effect is found to be present for all KK modes including KK modes that are not excitations in the internal space ($\alpha_n=0$). Since here we have focused on such modes, the above result is completely independent of the radius of the internal space.    

In the case where the space shrinks towards the UV, the coupling will asymptote to one, while when the internal spaces are growing towards the UV the coupling blows up and the KK gauge modes would become strongly coupled. In this case the tree level perturbative EW analysis done here would not be valid but it is reasonable to assume the EW constraints would be large.

One should not really be surprised by this result since it is just the volume enhancement / suppression of the couplings as used in the ADD model. Although it is not exactly the same since here it is the relative coupling that is of importance and hence a universal scaling of both the zero mode and KK mode would have no effect. Although it is in warped scenarios, where the volume effect in the overlap integral scales the KK modes differently from that of the zero mode, that we get this significant scaling of the couplings. This effect has also been observed in \cite{McDonald:2009hf}. 

Clearly this scaling of the couplings would dramatically alter the low energy phenomenology, of a model with warped extra dimensions, from that of the RS model. Although here we have just simply assumed the form of the space. An alternative possibility is to look at solutions obtained from string theory. In particular, for a second example, we will consider gauge fields propagating in a popular solution of type IIb supegravity \cite{Klebanov:2000hb}.

\subsection{The Klebanov-Strassler Solution}\label{sect:KlebStras}
Both AdS$_D$ and AdS$_5\times\mathcal{M}^\delta$ spaces suffer from a conical singularity in the IR and hence in order to consider QFT's propagating in such a background it is necessary to cut the space off with an IR brane. The Klebanov- Strassler solution \cite{Klebanov:2000hb}, corresponding to the dual of a field theory in which the number of colours is being reduced, has attracted significant attention since it has no such singularity and hence \color{red} requires \color{black} no IR cut off. The IR tip of the space is a rounded off or deformed conifold. Unfortunately it has a quite complex form and hence it can only really be investigated numerically. Here we will again consider gauge fields propagating in this solution while neglecting their backreaction on the background. The metric is then given by 
\begin{equation}
\label{KSmetric }
ds_{10}^2=h^{-\frac{1}{2}}(\tau)\eta_{\mu\nu}dx^\mu dx^\nu - h^{\frac{1}{2}}(\tau)ds_{6}^2
\end{equation} 
where
\begin{eqnarray}
h(\tau)&=&2^{\frac{2}{3}}(g_sM\alpha^{\prime})^2\epsilon^{\frac{-8}{3}}I(\tau) \nonumber\\[.2cm]
I(\tau)&=&\int_{\tau}^{\infty}dx \frac{x\coth x-1}{\sinh^2x}(\sinh(2x)-2x)^{\frac{1}{3}}\nonumber,
\end{eqnarray}
\color{red} where $g_s$ is the string coupling, $M$ is the number fractional D3 branes and $\alpha^\prime$ is a normalisation factor $\sim (g_sM)^2$ . \color{black} While the internal manifold is described by 
\begin{displaymath}
ds_{6}^2=\frac{1}{2}\epsilon^{\frac{4}{3}}K(\tau)\left [ \frac{1}{3K^3(\tau)}(d\tau^2+(g^5)^2) +\cosh^2(\frac{\tau}{2})[(g^3)^2+(g^4)^2]+\sinh^2(\frac{\tau}{2})[(g^1)^2+(g^2)^2]\right ]
\end{displaymath}
where $g^1$ to  $g^5$ are a diagonal combinations of the angular coordinates, $\epsilon$ is a parameter specifying the conic radius at which deforming begins and
\begin{equation}
\label{Ktau }
K(\tau)=\frac{(\sinh(2\tau)-2\tau)^{\frac{1}{3}}}{2^{\frac{1}{3}}\sinh(\tau)}.\nonumber
\end{equation}
In the IR we assume the Higgs is localised to a point $\tau_{ir}$ close to $\tau=0$ while we still cut the space off at $\tau_{uv}$ such that $\color{red}h(\tau_{uv})=1$. For small $\tau$ you can make the approximation $\tau$ $(3K^3(\tau))^{-1}\approx \cosh^2(\tau)$ and write the internal manifold as 
\begin{equation}
\label{ds6approx}
ds_6^2\approx\frac{1}{2}\epsilon^{\frac{4}{3}}K(\tau)\left [\frac{d\tau^2}{3K^3(\tau)} +\cosh^2(\frac{\tau}{2})d\Omega_3^2+\sinh^2(\frac{\tau}{2})d\Omega_2^2\right ], 
\end{equation}
where $d\Omega_3^2$ and $d\Omega_2^2$ are the line elements for $S^3$ and $S^2$. Hence the space is growing towards the UV and so one would anticipate the gauge Higgs couplings to become large. Once again we allow the gauge fields to propagate in the bulk and focusing on the KK modes, with $\alpha_n=0$, such that their profiles are given by Eq. \ref{GaugeEOM},
\begin{equation}
\label{ }
f_n^{\prime\prime}+\left (\frac{h^{\prime}(\tau)}{2h(\tau)}+2\coth(\tau)\right )f_n^{\prime}+\frac{\epsilon h(\tau)}{6K^2(\tau)}m_n^2f_n=0,
\end{equation}
where now the orthogonality relation is given by
\begin{equation}
\label{ }
\int d\tau \frac{h^{\frac{3}{2}}\sinh^2(\tau)\epsilon^4}{24}f_nf_m=\delta_{nm}.
\end{equation}

\begin{figure}[t!]
\begin{center}
\includegraphics[width=5in]{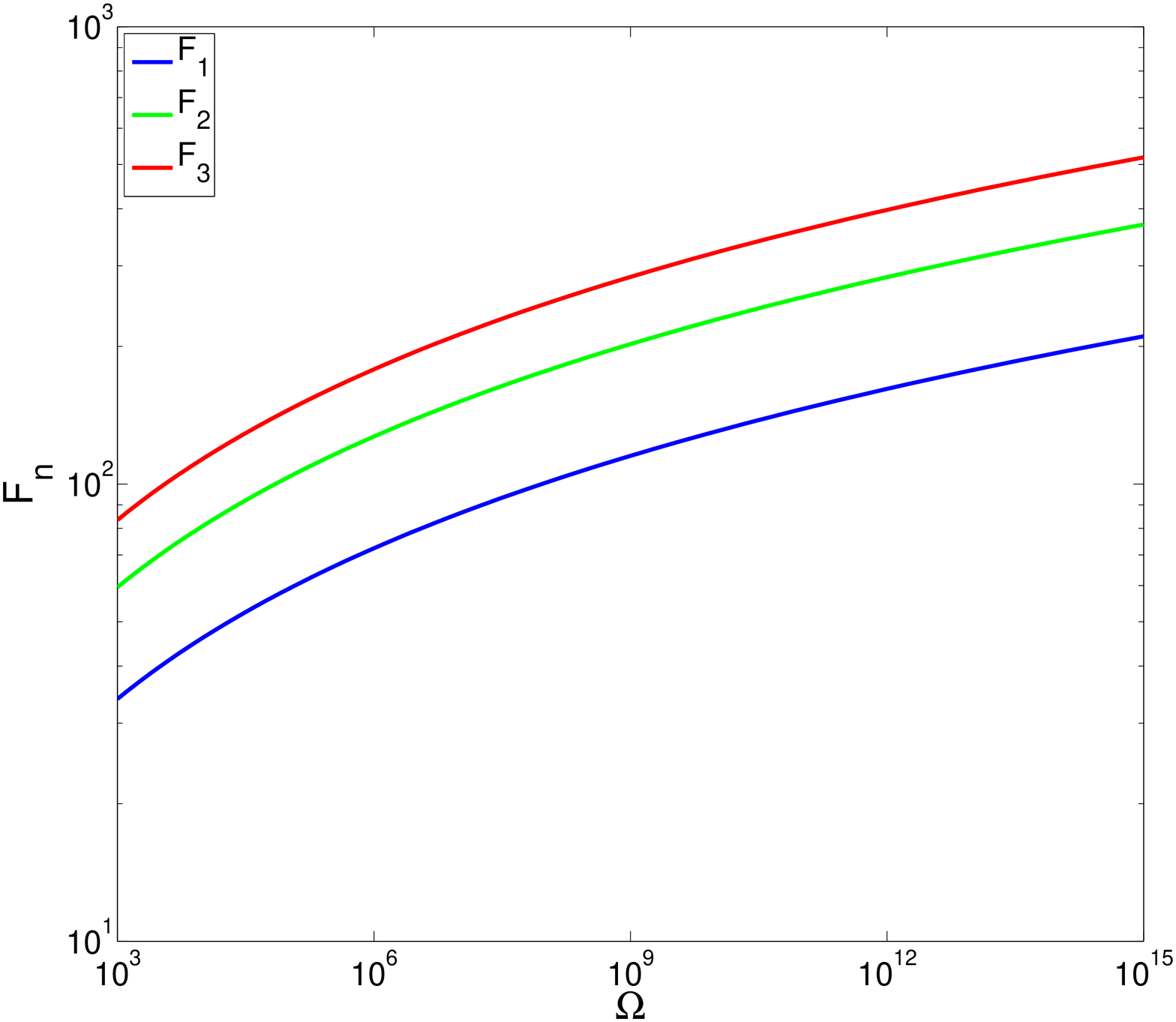}
\caption [$F_n$ in the Klebanov-Strassler Solution]{The relative gauge Higgs coupling as a function of the warp factor. Here $\epsilon$ is taken to be $10^{-35}$ but the results are not significantly dependent on this value. \cite{Archer:2010hh}}
\label{KSlogOm}
\end{center}
\end{figure}

Due to the inaccuracies associated with using splines in differential equations we also approximate 
\begin{displaymath}
I(\tau)\approx\frac{I_0+a\tau^8(\tau-\frac{1}{4})}{1+b\tau^2+c\tau^8e^{\frac{4\tau}{3}}}
\end{displaymath}
where $I_0$, $a$, $b$ and $c$ are choosen so as to fit with the exact function. In practice we take $I_0\approx0.71805$, $a\approx0.03$, $b\approx0.278$ and $c\approx0.0126$. The gauge Higgs couplings are then plotted in figure \ref{KSlogOm}. As expected the relative couplings become large, e.g.~for $\Omega=10^{15}$ we have $F_1\approx210$. Also note that unlike in the RS model the KK modes do not couple with approximately equal strength. Hence the KK modes of bulk gauge fields would be increasingly strongly coupled to fields localised on the IR brane. While the perturbative approach taken in this thesis would not be valid in this scenario, one would anticipate that if such fields were coupling to any SM particles then the resulting constraints, \color{red} on the size of the KK scale, \color{black} would be large. 

So if one wishes the Klebanov Strassler solution to offer a phenomenologically viable resolution to the gauge hierarchy problem then clearly one would require that either there are no additional bulk fields or if there were that they decouple from the SM fields. This is not straight forward to achieve.

\section{A Bulk Anisotropic Cosmological Constant}\label{AnIsoLamb}
What we would like to do is to be able to study a range of possible spaces but with the knowledge that such space could potentially be realised. Hence for our central example we shall consider the classical solutions of the Einstein equations that emerge from the presence of an anisotropic bulk cosmological constant \cite{Archer:2010bm}. 

\subsection{The Model} In particular we shall consider what is arguably the simplest possible extension of the Randall and Sundrum Model. It concerns the classical solutions of a $4+1+\delta$ dimensional space, described by  
\begin{equation}
\label{EHaction}
S=\int d^{5+\delta}x\sqrt{G}\left [\Lambda-\frac{1}{2}M^{3+\delta}\mathcal{R}+\mathcal{L}_{\rm{bulk}}\right].
\end{equation}
The co-ordinates run over $(x_\mu,r,\theta_1\dots\theta_\delta)$, we also collectively write $\theta=(\theta_1\dots\theta_\delta)$. We allow for an anisotropic bulk cosmological constant of the form
\begin{equation}\label{lambda}
\Lambda=\begin{pmatrix}
\Lambda \eta_{\mu\nu } &&&& \\
&\Lambda_5 &&& \\
&&\Lambda_\theta &&\\
&&&\ddots &\\
&&&&\Lambda_\theta\\
\end{pmatrix}.
\end{equation}
It should be stressed that this is very much a toy model. It has been motivated in an attempt to parameterise our ignorance of the bulk energy momentum tensor in a fashion such that one can still find useful solutions. If, for example, we were to consider such a cosmological constant arising from an $r$, $\theta_1$ dependent VEV of some bulk scalar, in 6D,
\begin{displaymath}
S=\int d^6x\sqrt{-G}\left [-\frac{1}{2}M^4\mathcal{R}+\Lambda+\frac{1}{2}|\partial_M\Phi|^2-V(\Phi)\right ],
\end{displaymath}
the corresponding energy-momentum tensor would be
\begin{equation*}
\scriptsize
T_{MN}=\begin{pmatrix}
\left [-\Lambda-\frac{1}{2}\Phi^{\prime\;2}-\frac{1}{2}\dot{\Phi}^2+V(\Phi)\right ]\eta_{\mu\nu } && \\
&-\Lambda+\frac{1}{2}\Phi^{\prime\;2}-\frac{1}{2}\dot{\Phi}^2+V(\Phi)& \\
&&-\Lambda-\frac{1}{2}\Phi^{\prime\;2}+\frac{1}{2}\dot{\Phi}^2+V(\Phi)\\
\end{pmatrix}.
\textstyle
\end{equation*}
Here $^\prime$ and $\dot{}$ denotes derivative w.r.t $r$ and $\theta_1$. Hence Eq. \ref{lambda} could be realised when the derivatives dominate over the potential and the VEV is linearly dependent on $r$ and $\theta_1$. Arguably any realistic approach should allow for a energy-momentum tensor dependent on $r$ and $\theta$. However this would lead to warping in multiple directions, complicated wavefunctions and an enlarged parameter space. Although here we consider very much a bottom up approach, it is of course partially motivated by the AdS throats that arise from the AdS/CFT correspondence. Hence we again use a metric ansatz warped w.r.t a single `preferred' direction $r$,         
\begin{equation}\label{ansatz}
ds^2=e^{-2A(r)}\eta_{\mu\nu}dx^\mu dx^\nu-dr^2-\sum_{i=1}^{\delta}e^{-2C(r)}d\theta_i^2.
\end{equation}
For simplicity we take the internal geometry to be toroidal orbifolds and we take $\theta_i\in [0,R_\theta]$. Neglecting the contribution from any bulk fields the Einstein equations then follow from Eq. \ref{EHaction}.
\begin{eqnarray}
6A^{\prime\;2}-3A^{\prime\prime}+3\delta A^{\prime}C^{\prime}-\delta C^{\prime\prime}+\frac{(\delta+1)!}{2!(\delta-1)!}C^{\prime\;2}=-\frac{\Lambda}{M^{3+\delta}}\\
6A^{\prime\;2}+\frac{\delta !}{2!(\delta-2)!}C^{\prime\; 2}+4\delta A^{\prime}C^{\prime}=-\frac{\Lambda_5}{M^{3+\delta}}\nonumber\\
4(\delta-1)A^{\prime}C^{\prime}-(\delta-1)C^{\prime\prime}+\frac{\delta !}{2!(\delta-2)!}C^{\prime\; 2}+10A^{\prime\;2}-4A^{\prime\prime}=-\frac{\Lambda_\theta}{M^{3+\delta}}.\label{einstein}
\end{eqnarray}

\begin{figure}[!t]
\begin{center}
\includegraphics[width=5.5in]{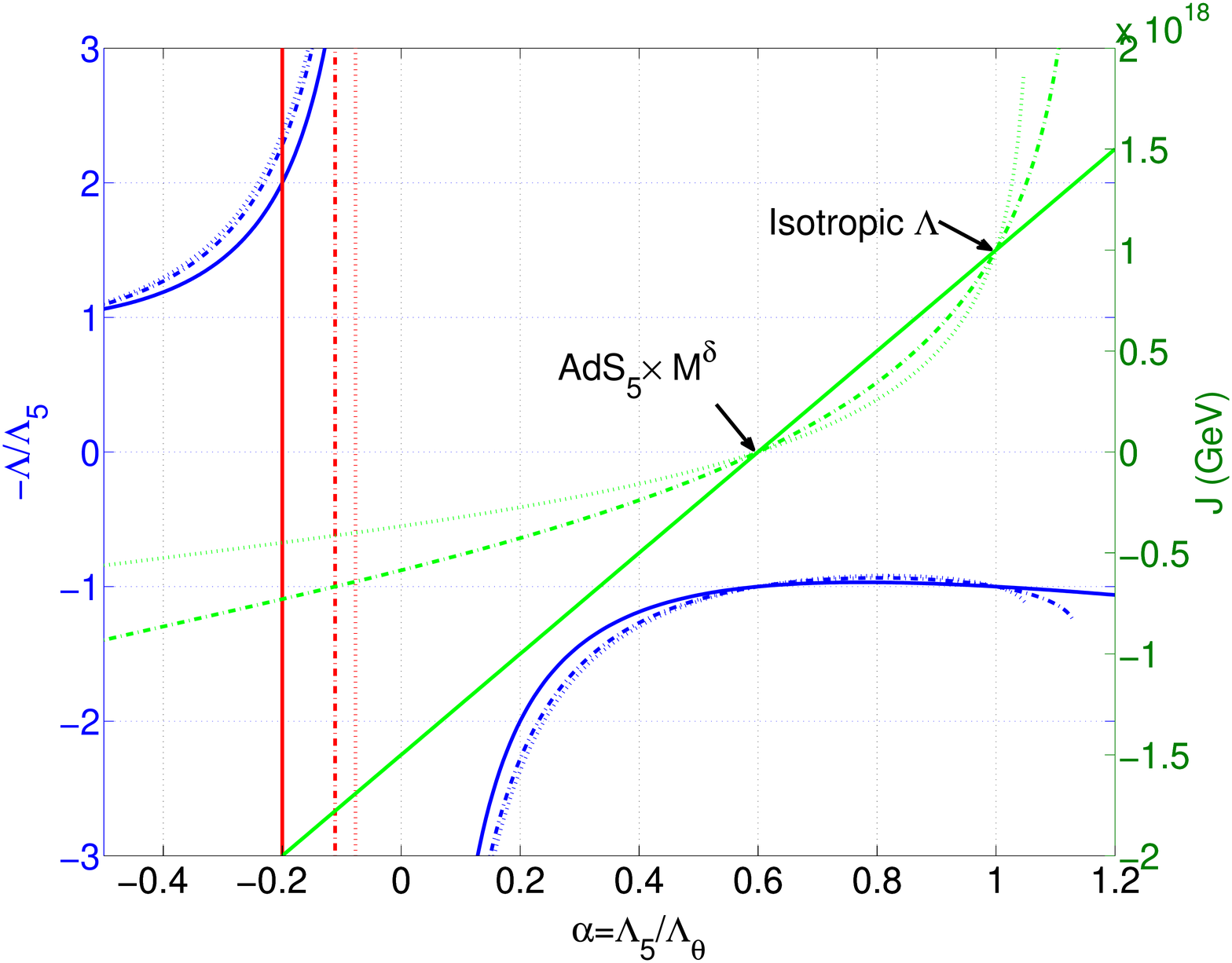}
\caption[Solutions of the Einstein Equations with an Anisotropic Cosmological Constant]{The solutions to the Einstein equations (\ref{einstein}) for 6D (solid lines), 8D(dash-dot lines) and 10D (dot-dot lines). $J$ is plotted in green and $-\frac{\Lambda}{\Lambda_5}$ in blue. The red lines correspond to $2k+\delta J=0$ and hence the spaces to the left of them will have exponentially suppressed fundamental Planck masses. Here we have fixed $\Omega\equiv e^{kR}=10^{15}$ and $M_{\rm{KK}}\equiv \frac{k}{\Omega}=1$ TeV. \cite{Archer:2010bm}}
\label{JLamb}
\end{center}
\end{figure}

\begin{figure}[!t]
\begin{center}
\includegraphics[width=6in]{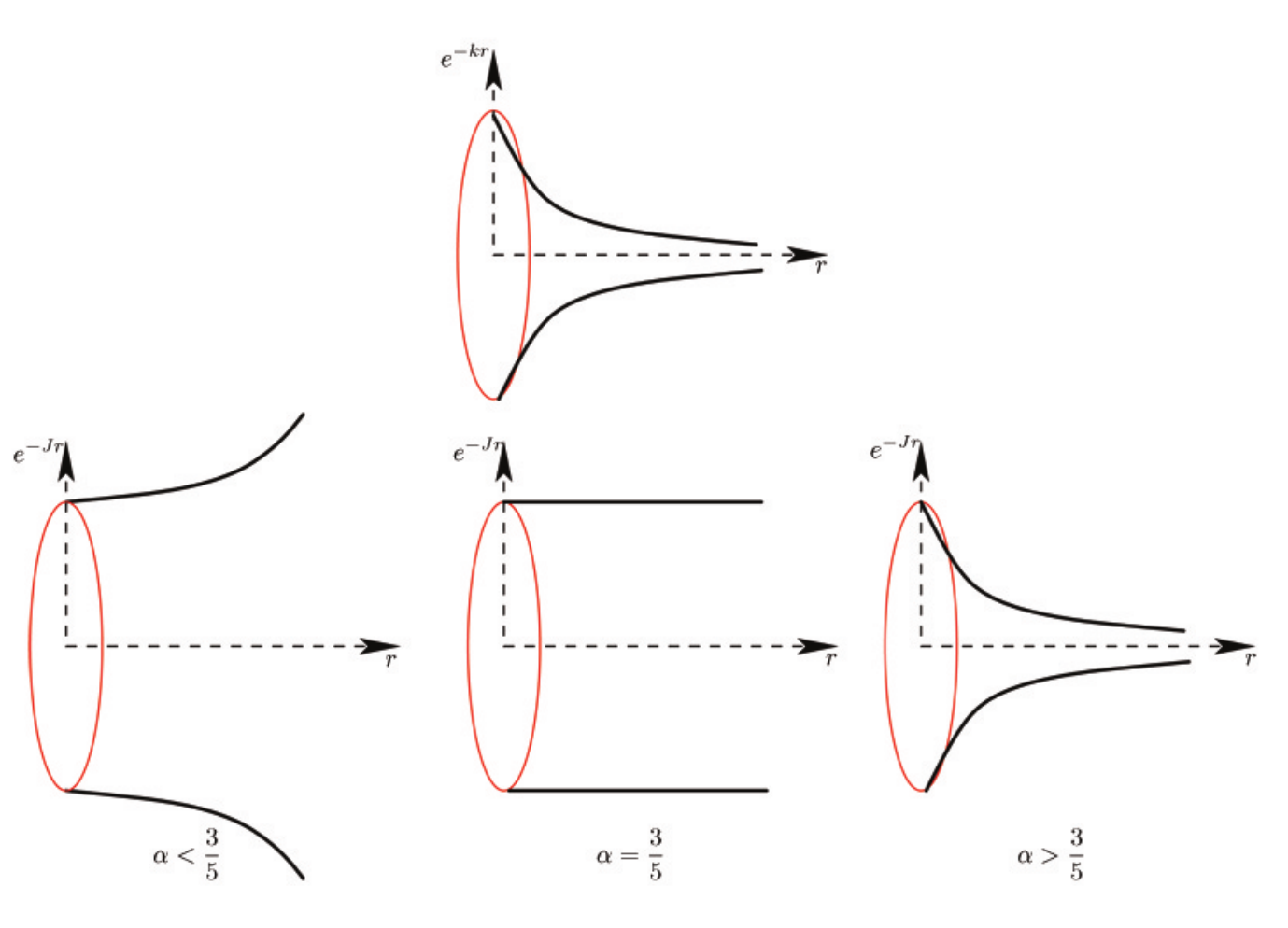}
\caption[The Warping of the Radii of the Additional Dimensions]{The warping of the radii of the additional dimensions.}
\label{throats}
\end{center}
\end{figure}

Only two of these equations are independent and admit the solution (although not unique solution)
\begin{equation*}
A(r)=kr\qquad \mbox{and} \qquad C(r)=Jr,
\end{equation*}
where $k$ and $J$ are constants, depending on $\Lambda$, $\Lambda_5$ and $\Lambda_\theta$. Here we introduce the parameter
\begin{equation*}
\alpha\equiv\frac{\Lambda_5}{\Lambda_\theta}.
\end{equation*}
Then in 6D, as found in \cite{Kogan:2001yr}, Eq. \ref{einstein} leads to
\begin{displaymath}
k=\pm\sqrt{-\frac{\Lambda_5}{10\alpha M^4}},\qquad J=\frac{5}{2}\alpha k-\frac{3}{2}k,\qquad \Lambda=\left (\frac{3}{8\alpha}+\frac{5}{8}\alpha\right )\Lambda_5.
\end{displaymath}
See also \cite{Multamaki:2002cq} for 6D work in this direction. In more than six dimensions the equivalent equations have a more complicated form and are plotted in figure \ref{JLamb}. There are essentially three classes of spaces as illustrated in figure \ref{throats}. When $\alpha<\frac{3}{5}$, the radii of the $\delta$ additional dimensions increase as one moves towards the IR brane and one would anticipate the KK modes of bulk gauge field having reduced couplings to matter on the IR brane. Conversely, where $\alpha>\frac{3}{5}$, the additional dimensions are shrinking towards the IR brane and one would anticipate enhanced couplings. Note that an isotropic cosmological constant ($\Lambda=\mbox{diag}(\Lambda,\Lambda,\dots,\Lambda)$) would belong to this class of solutions and would naively have large EW corrections. The third solution corresponding to $\alpha=\frac{3}{5}$ (or $\Lambda= \mbox{diag}(\Lambda\eta_{\mu\nu},\frac{3}{5}\Lambda,\Lambda,\dots,\Lambda)$) would correspond to the $\rm{AdS}_5\times M^{\delta}$ space studied in \cite{Davoudiasl:2002wz, Davoudiasl:2008qm}. Note also that spaces with $\alpha<0$ are valid but would correspond to a partially deSitter space with $\Lambda_5>0$.  
 
Like the RS model these spaces are all spaces of constant curvature with a Ricci scalar given by
\begin{equation}
\label{ }
\mathcal{R}=-20k^2-8\delta kJ-\delta(\delta+1)J^2.
\end{equation} 
and, like the RS model, these spaces also suffer from a conical singularity as $r\rightarrow\infty$ (and $r\rightarrow-\infty$ when $\alpha<\frac{3}{5}$). Hence such a space must be cut off in both the IR and the UV. How the space is cut off in the IR is critical in determining the EW scale where as the relation between the UV cut off and the IR cut off partially determines the KK scale. In the RS model \cite{Randall:1999ee} the space is cut off with two codimension one \color{red}3-branes. In doing so one finds that the tension on such branes are finely tuned w.r.t each other. Naively one would continue to use 3-branes \color{black} to cut off the space.

However it has been known for some time \cite{PhysRevD.23.852} that matching the solutions across a brane, with a tension, of codimension two or higher typically gives rise to a singularity at the location of the brane. There are known solutions that avoid this singularity particularly in six dimensions \cite{Bayntun:2009im}, for example, including a bulk Gauss-Bonnet term \cite{Bostock:2003cv}. This problem becomes increasingly severe for branes of codimension three and higher \cite{Gherghetta:2000jf, Charmousis:2001hg}. Also, as discussed we would like to localise the Higgs just w.r.t $r$. Hence for the moment we will consider an interval running from $r\in [0,R]$ with a IR codimension one brane at $r=R$ and a UV codimension one brane at $r=0$. 

In order to resolve the hierarchy problem we require that the fundamental Higgs mass and the curvature, $k$ and $J$, are of the same order of magnitude as that of the fundamental Planck mass. If the Higgs is localised on the IR brane then the effective mass is suppressed by gravitational red shifting Eq. \ref{HiggsMassGen},
\begin{equation*}
m_{h}=e^{-kR}m_{0}\equiv\Omega^{-1}m_{0},
\end{equation*}
while the Planck mass is given by Eq. \ref{MPlanck}
\begin{equation*}
M_P^2=M^{3+\delta}\frac{(1-e^{-(2k+\delta J)R})(R_{\theta})^\delta}{(2k+\delta J)}.
\end{equation*}
Hence when $2k+\delta J>0$, as in the 5D RS model, the volume effects are of order one and a warpfactor $\Omega\sim 10^{15}$ is required to resolve the hierarchy problem (assuming $R_{\theta}^{-1}\sim M \sim M_P$). Conversely if $2k+\delta J<0$ then one moves towards a `warped ADD' scenario in which the Planck mass is suppressed by the volume of the additional dimensions, hence introducing an additional hierarchy between the AdS curvature and the Planck mass. Also recent work has indicated that there may be problems with unitarity in lowering the Planck scale to about a TeV \cite{Atkins:2010re} and hence here we focus on spaces with $2k+\delta J>0$. Of course it is always possible that the fundamental scale is neither the Planck scale nor the EW scale but some intermediate scale. In other words one could, for example, increase the size of $R_\theta$ and hence the required warp factor would be reduced. If this was the case the basic results of this section would not change significantly although the numbers of course would. Having said that, throughout this section we fix the warp factor to be $\Omega=10^{15}$. It is this that results in the infinity at $\alpha=0$ in figure \ref{JLamb}.

\subsection{Bulk Gauge Fields}
We now move on to consider gauge fields propagating in the above possible spaces 
\begin{equation}   
ds^2=e^{-2kr}\eta_{\mu\nu}dx^\mu dx^\nu-dr^2-\sum_{i=1}^{\delta}e^{-2Jr}d\theta_i^2.
\end{equation} 
The KK profiles will be given from Eq. \ref{GaugeEOM} by
\begin{equation}
f_n^{\prime\prime}-(2k+\delta J)f_n^{\prime}-\sum_{i=1}^\delta e^{2Jr}\frac{l_i^2}{R_\theta^2}f_n+e^{2kr}m_n^2f_n=0.\label{gaugeeqn}
\end{equation}
Here we have assumed a toroidal internal space, i.e. $\partial_i^2\Theta_n=-\frac{l_i^2}{R_\theta^2}\Theta_n$ where $l_i$ would be an integer KK number appearing as a component of $n=(n_1,l_1,\dots,l_\delta)$. The corresponding orthogonality relation is given by   
\begin{equation}\label{gaugeOrthog}
\int drd^\delta\theta \;e^{-\delta Jr}f_nf_m\Theta_{n}\Theta_{m}=\delta_{nm}.
\end{equation}
Eq. \ref{gaugeeqn} is not analytically solvable although we can look for approximate solutions by making the substitution 
\begin{equation}\label{Xdef}
x^2=\frac{e^{2kr}m_n^2-e^{2Jr}\sum_{i=1}^\delta\frac{l_i^2}{R_\theta^2}}{\gamma^2} 
\end{equation}
such that $x^{\prime}=\gamma x$ and hence
\begin{equation}\label{gammaeqn}
\gamma^\prime=\left(\frac{ke^{2kr}m_n^2-Je^{2Jr}\sum_{i=1}^\delta\frac{l_i^2}{R_\theta^2}}{e^{2kr}m_n^2-e^{2Jr}\sum_{i=1}^\delta\frac{l_i^2}{R_\theta^2}}\right )\gamma-\gamma^2.
\end{equation}
Note that $\gamma$ is typically going to be of the same order of magnitude as $J$ and $k$. In the IR region, where one would expect the gauge fields to be naturally localised, $\gamma \approx$ constant and Eq. \ref{gaugeeqn} can be approximated to 
\begin{equation*}
\ddot{f}+\left (1-\frac{2k+\delta J}{\gamma}\right )\frac{1}{x}\dot{f}+f\approx 0
\end{equation*}
where $\dot{}$ denotes derivative w.r.t. $x$. This can now be solved to give
\begin{equation}
f(x)\approx Nx^{\frac{1}{2}\frac{2k+\delta J}{\gamma}}\left (\mathbf{J}_{-\frac{1}{2}\frac{2k+\delta J}{\gamma}}(x)+\beta \mathbf{Y}_{-\frac{1}{2}\frac{2k+\delta J}{\gamma}}(x) \right ).
\end{equation}
The mass eigenvalues will then be largely determined by the zeros of the Bessel functions and hence, assuming $\frac{1}{2}\frac{2k+\delta J}{\gamma}\sim \mathcal{O}(1)$, the KK masses would roughly follow from Eq. \ref{Xdef} to be
\begin{equation}\label{MNapprox}
m_n\sim X_n\frac{\sqrt{\gamma^2+\frac{e^{2JR}}{R_\theta^2}\sum_i^{\delta}l_i^2}}{e^{kR}}.
\end{equation}
Here $X_n$ will be some $\mathcal{O}(1)$ number dependent on the roots of the Bessel functions, the boundary conditions and any UV effects. Clearly the spacing of KK modes will depend on the warping of the additional dimensions. If for example we take $R_\theta^{-1}\sim M_P$, $M_{\rm{KK}}\equiv\frac{k}{\Omega}\sim 1$ TeV and $\Omega\sim10^{15}$ then:
\begin{itemize} 
\item If $J>0$ (i.e. $\alpha>\frac{3}{5}$) then $\frac{e^{JR}}{R_\theta}\gg\gamma\sim k$ and the $l_i\neq 0$ KK modes would gain masses far larger than $M_{\rm{KK}}$ and essentially decouple from the low energy theory. Hence one would be left with just the KK tower corresponding to the $l_i=0$ modes.
\item If $J\sim 0$ then $\frac{e^{JR}}{R_\theta}\sim\gamma$ and the $l_i\neq 0$ KK modes would have masses of $\mathcal{O}(M_{\rm{KK}})$. Hence the splitting in the KK spectrum would be apparent at the KK scale. For a specific example of the graviton KK spectrum in this case see \cite{Davoudiasl:2002wz, Davoudiasl:2008qm}.
\item If $J<0$ then the masses of the $l_i\neq 0$ modes would not shift much from those of the $l_i=0$ modes. Hence one would introduce a fine splitting in the KK spectrum. 
\end{itemize}

\begin{figure}[h!]
\begin{center}
\subfigure[$R_\theta=R$]{%
           \label{fig:second}
           \includegraphics[width=0.45\textwidth]{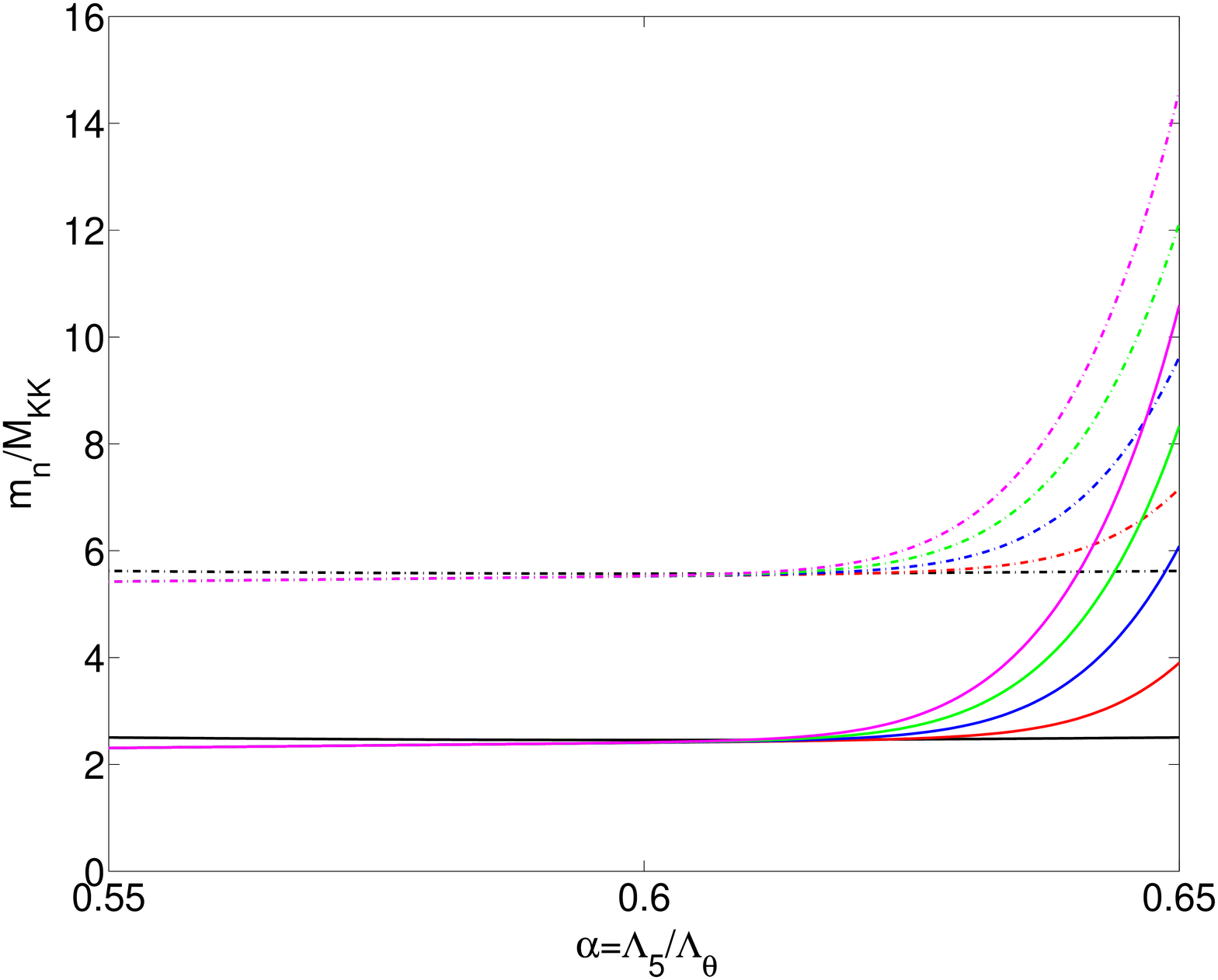}
        }
        \subfigure[$R_\theta=0.1R$]{%
           \label{fig:second}
           \includegraphics[width=0.45\textwidth]{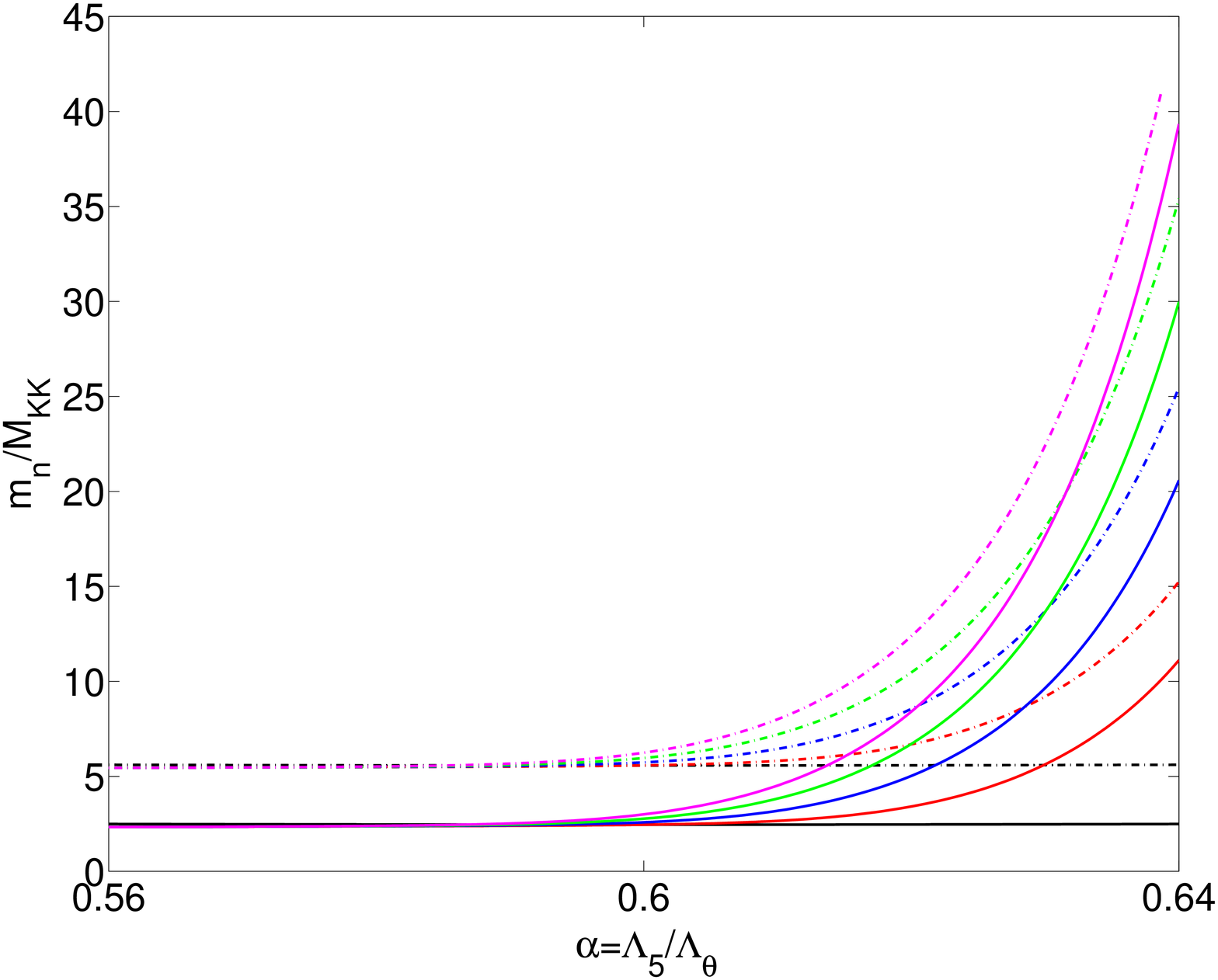}
        }
\caption[KK Masses of $l_1 \neq 0$ KK Modes in 6D]{The first two mass eigenvalues of gauge KK modes in six dimensions, for $l_1=0$ (black), $l_1=1$ (red), $l_1=2$ (blue), $l_1=3$ (green) and $l_1=4$ (purple).  Here $R_\theta=R$ on the left and $R_\theta=0.1R$ on the right. $\Omega\equiv e^{kR}=10^{15}$ and $M_{\rm{KK}}\equiv\frac{k}{\Omega}=1$TeV. \cite{Archer:2010bm}}
\label{gaugedegfig}
\end{center}
\end{figure}  

\begin{figure}[h]
\begin{center}

\subfigure[NBC's]{%
           \label{fig:second}
           \includegraphics[width=0.45\textwidth]{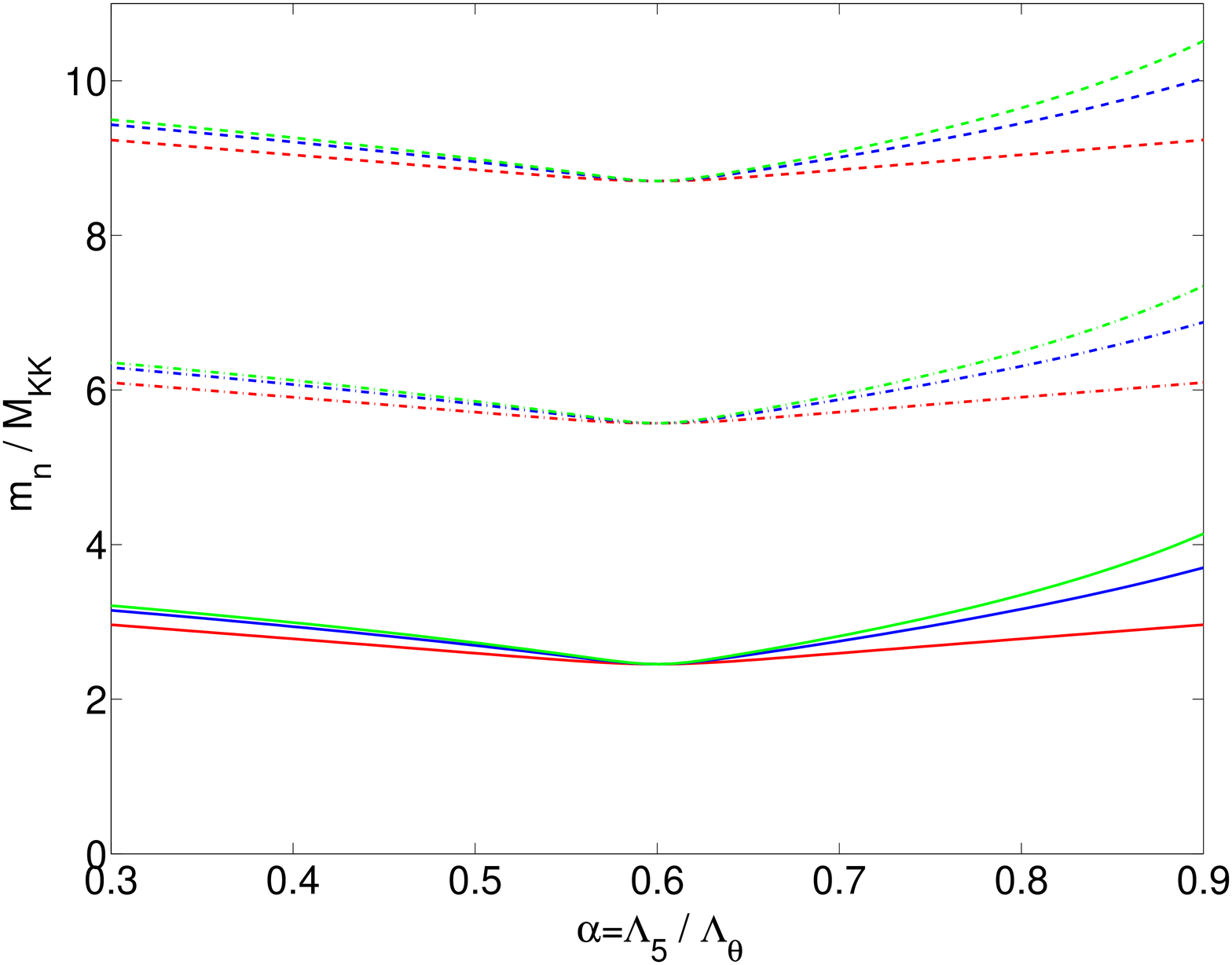}
        }
        \subfigure[DBC's]{%
           \label{fig:second}
           \includegraphics[width=0.45\textwidth]{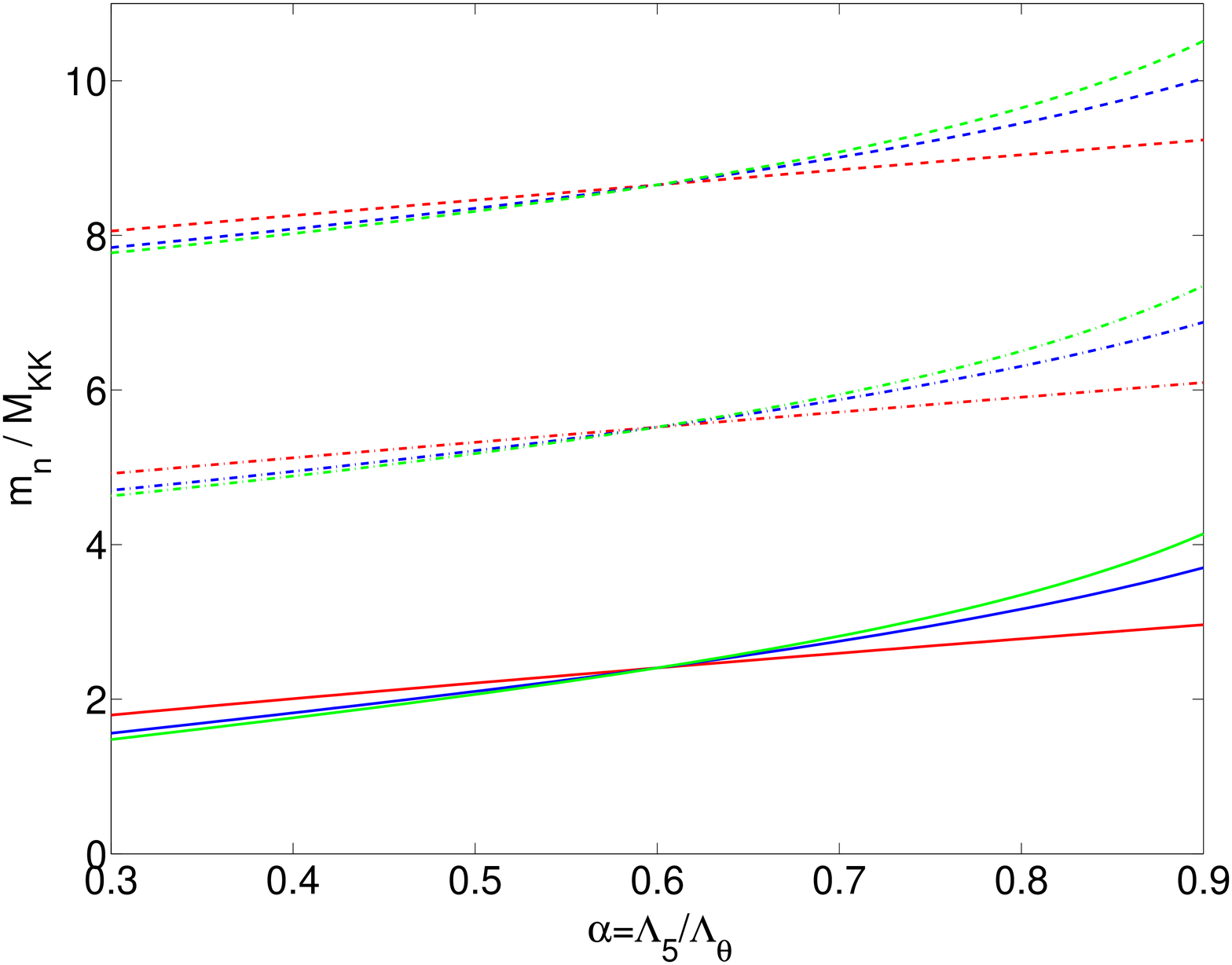}
        }
\caption[KK Masses of $l_i=0$ KK Modes]{First three mass eigenvalues of the $l_i=0$ gauge KK modes with the (IR,UV) boundary conditions $(++)$, on the left and $(-+)$ on the right. The gauge field propagates in 6D (red), 8D (blue) and 10D (green). Here $\Omega\equiv e^{kR}=10^{15}$ and $M_{\rm{KK}}\equiv\frac{k}{\Omega}=1$TeV. \cite{Archer:2010bm} }
\label{gaugemassesfig}
\end{center}
\end{figure}

One should be skeptical about the validity of neglecting the contribution from the UV region of the space to $\gamma$ and hence the KK scale. However Eq. \ref{MNapprox} is found to be in rough agreement with numerical solutions of (\ref{gaugeeqn}), plotted for six dimensions, in figure \ref{gaugedegfig}. It is also possible to analytically solve (\ref{gaugeeqn}) for the $l_i=0$ modes to get,
\begin{equation}\label{gaugewave}
f_n(r)=Ne^{vr}\left [\mathbf{J}_{-\frac{v}{k}}\left(\frac{m_ne^{kr}}{k}\right )+\beta \mathbf{Y}_{-\frac{v}{k}}\left (\frac{m_ne^{kr}}{k}\right )\right ],
\end{equation}
where $v=\frac{1}{2}(2k+\delta J)$ and $N$ is computed from (\ref{gaugeOrthog}). The mass eigenvalues are plotted in figure \ref{gaugemassesfig}. 

\begin{figure}[h!]
\begin{center}
\subfigure[$F_n$]{%
           \label{Fnbig}
           \includegraphics[width=0.45\textwidth]{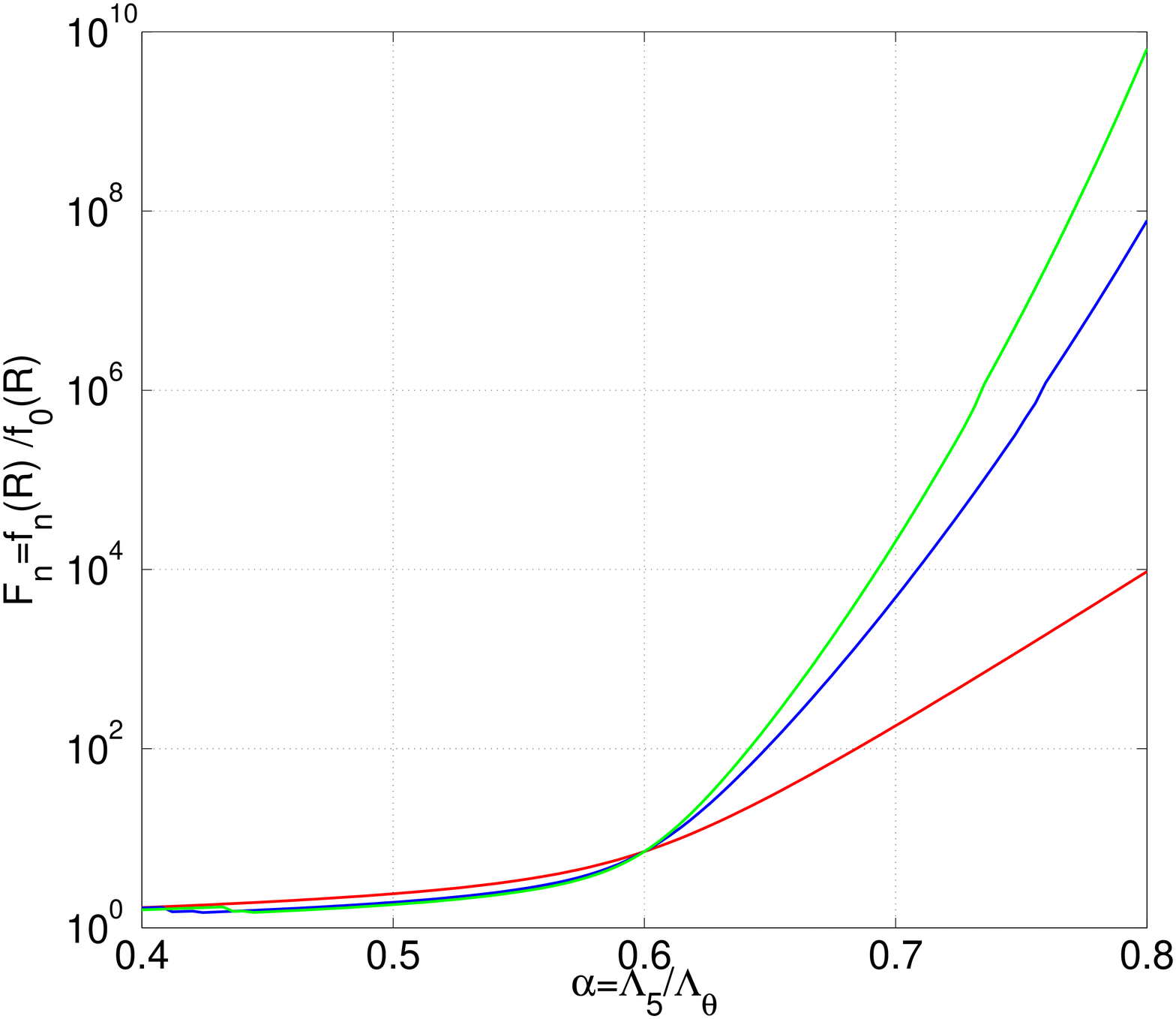}
        }
 \subfigure[Coupling to fields on UV brane]{%
           \label{fig:UVcouplings}
           \includegraphics[width=0.45\textwidth, height=5.6cm]{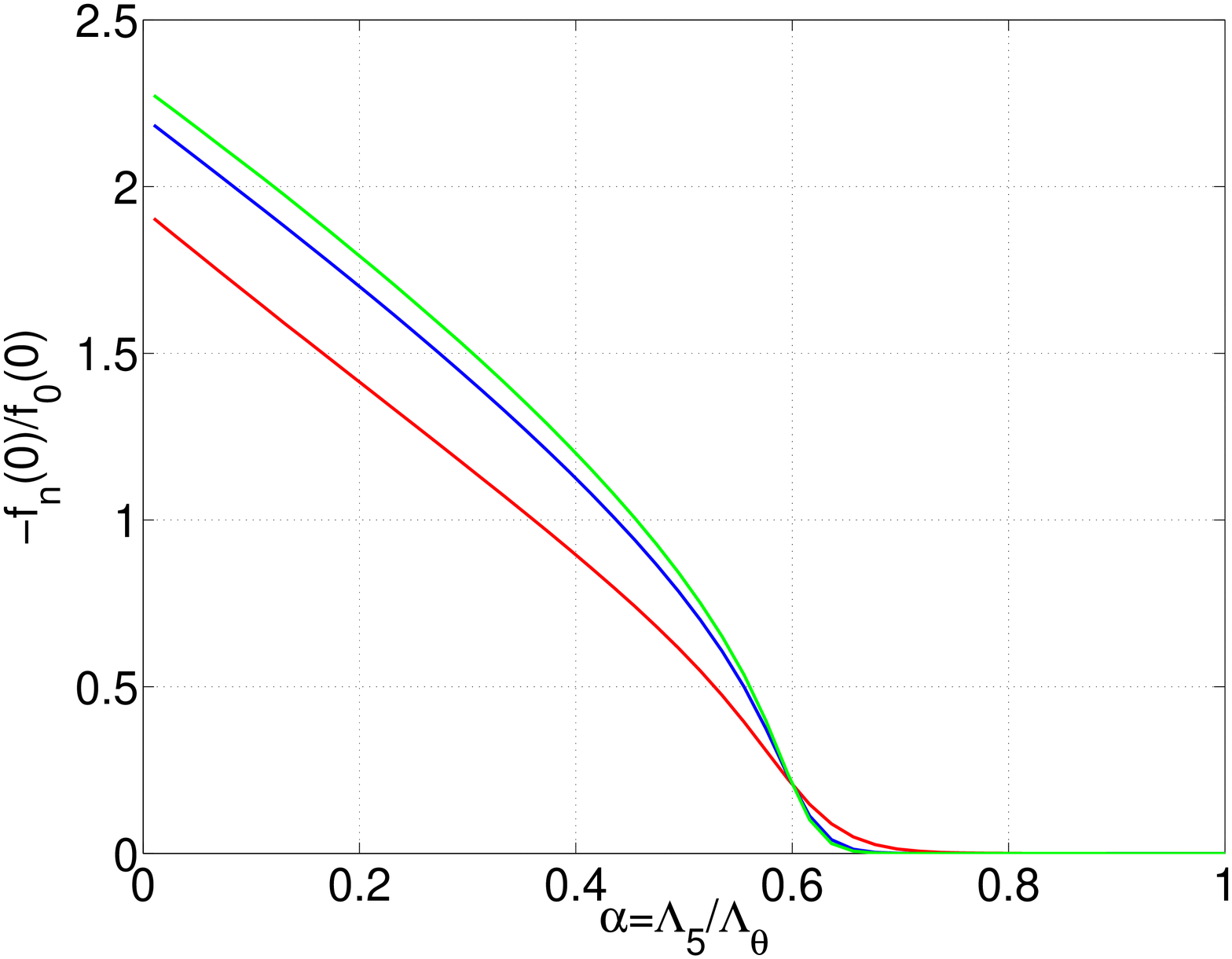}
        }
    \subfigure[$F_n$ for small $\alpha$]{%
           \label{Fnsmall}
           \includegraphics[width=0.45\textwidth]{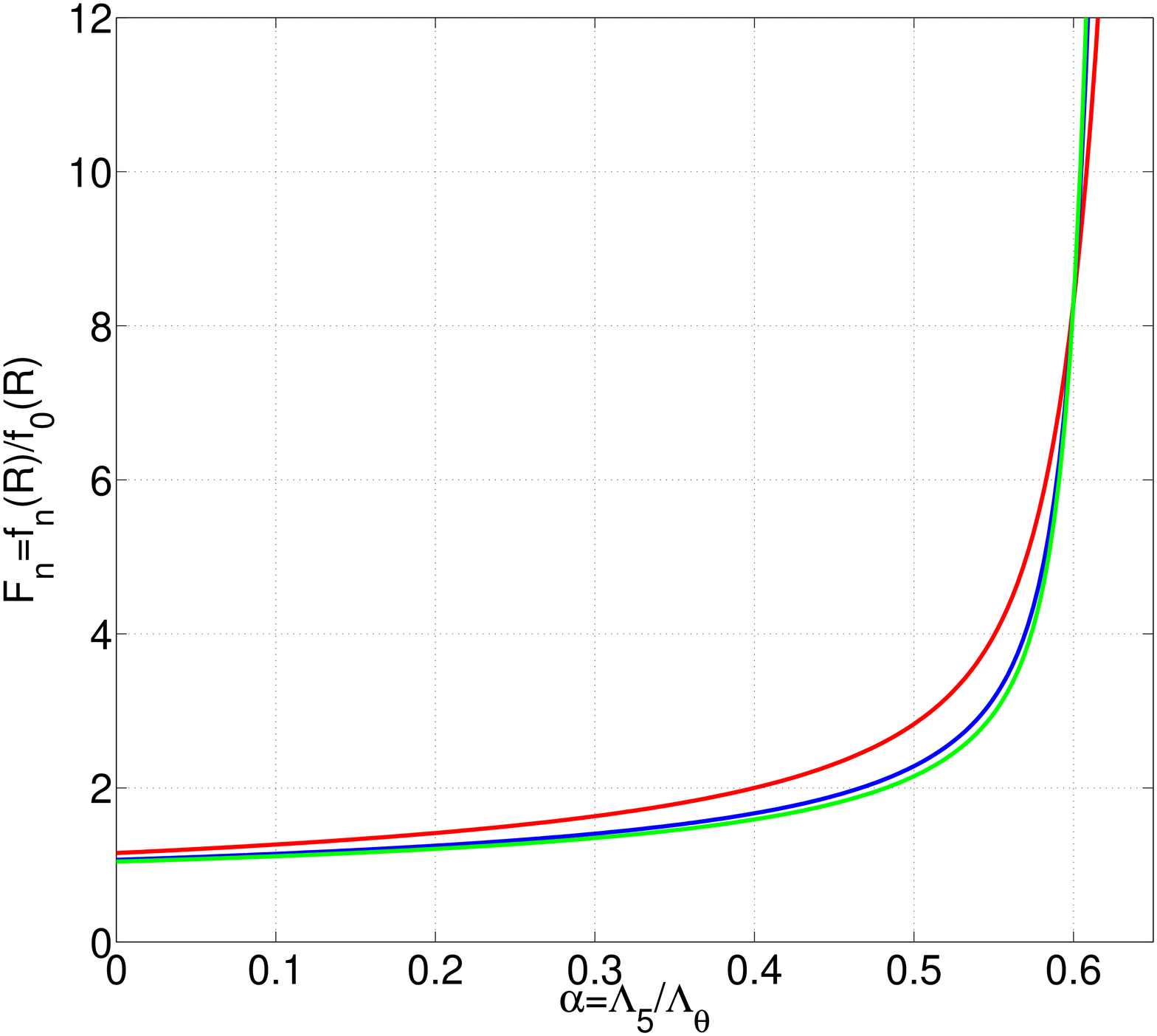}
        }    
\caption[Relative Gauge Couplings in Spaces with an Anisotropic Cosmological Constant]{The relative coupling of the $l_i=0$ KK gauge modes with a IR localised Higgs / fermion (on the left) and a UV localised field (on the right) in 6D (red), 8D (blue) and 10D (green). The lower graph is an enhancement of the couplings of IR localised fermions for small $\alpha$. Here $\Omega\equiv e^{kR}=10^{15}$ and $M_{\rm{KK}}\equiv\frac{k}{\Omega}=1$TeV. \cite{Archer:2010bm} }
\label{couplingfig }
\end{center}
\end{figure}

Likewise one can calculate the relative couplings for the $l_i=0$ KK modes, which are plotted in figure \ref {couplingfig }. The couplings for the $l_i\neq 0$ do not change significantly. Figure \ref{fig:UVcouplings} displays the relative coupling of the KK gauge fields to a particle localised on the UV brane. 

\subsection{Implications for Phenomenology}

So we can see that the phenomenological implications change significantly when the internal space is either growing or shrinking towards the IR. On one hand one can see that, as in section \ref{sect:powerGrow}, when the internal space is shrinking towards the IR brane ($\alpha>\frac{3}{5}$) the scaling in the volume factor in Eq. \ref{gaugeOrthog} causes the KK gauge modes to become significantly more strongly coupled to particles on the IR brane (see figure \ref{Fnbig}). However their coupling to a field localised towards the UV brane, such as a light fermion ($F_\psi^{(n)}$), would be suppressed by the same effect. In such spaces the gauge KK modes would be strongly coupled to the Higgs and the perturbative approach to computing EW constraints, taken in section \ref{EWChap}, would not be valid. None the less it is reasonable to assume that the EW constraints would be large and rule out such scenarios as resolutions to the hierarchy problem.

\begin{figure}[h]
\begin{center}
\subfigure[T Parameter]{%
           \label{fig:Tparam}
           \includegraphics[width=0.45\textwidth]{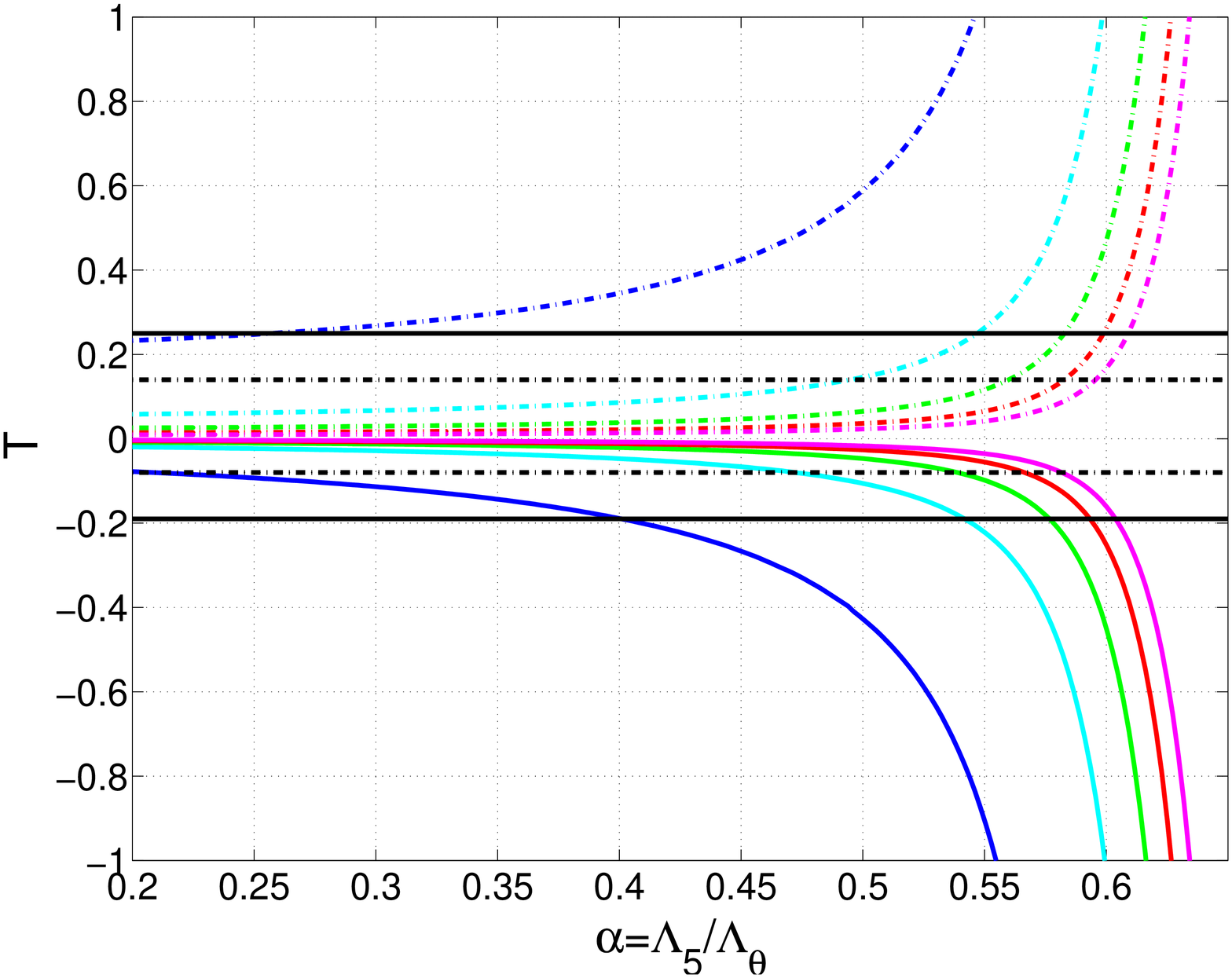}
        }
\subfigure[S Parameter]{%
           \label{fig:Sparam}
           \includegraphics[width=0.45\textwidth]{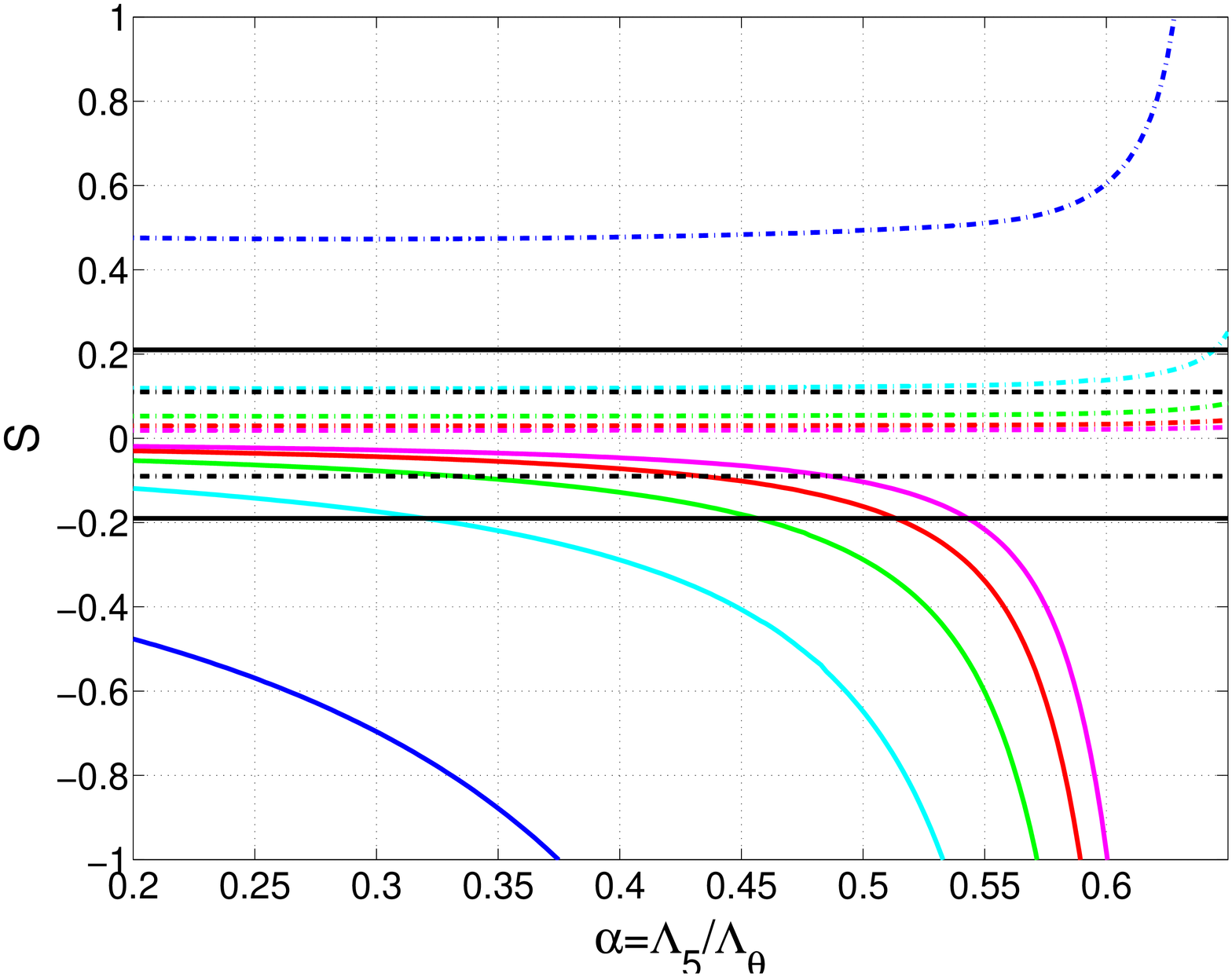}
        }
\caption[S and T parameters for 6D Spaces Without Custodial Symmetry]{The tree level contribution to the S and T parameters in six dimensional models, with out a custodial symmetry, where the KK scale ($M_{\rm{KK}}\equiv\frac{k}{\Omega}$) is 1 TeV (blue), 2 TeV (cyan), 3 TeV (green), 4 TeV (red) and 5 TeV (magenta). The fermions are localised on the IR, $r=R$ (solid lines) or on the UV brane $r=0$. (dot dash lines). In black are the $1\sigma$ and $2\sigma$ bounds ($M_H=117$ GeV) on S and T \cite{Nakamura:2010zzi}  $\Omega\equiv e^{kR}=10^{15}$. \cite{Archer:2010bm} }
\label{fig:STnoCustodial}
\end{center}
\end{figure}

\begin{figure}[h]
\begin{center}
\subfigure[T Parameter]{%
           \label{fig:custTparam}
           \includegraphics[width=0.45\textwidth]{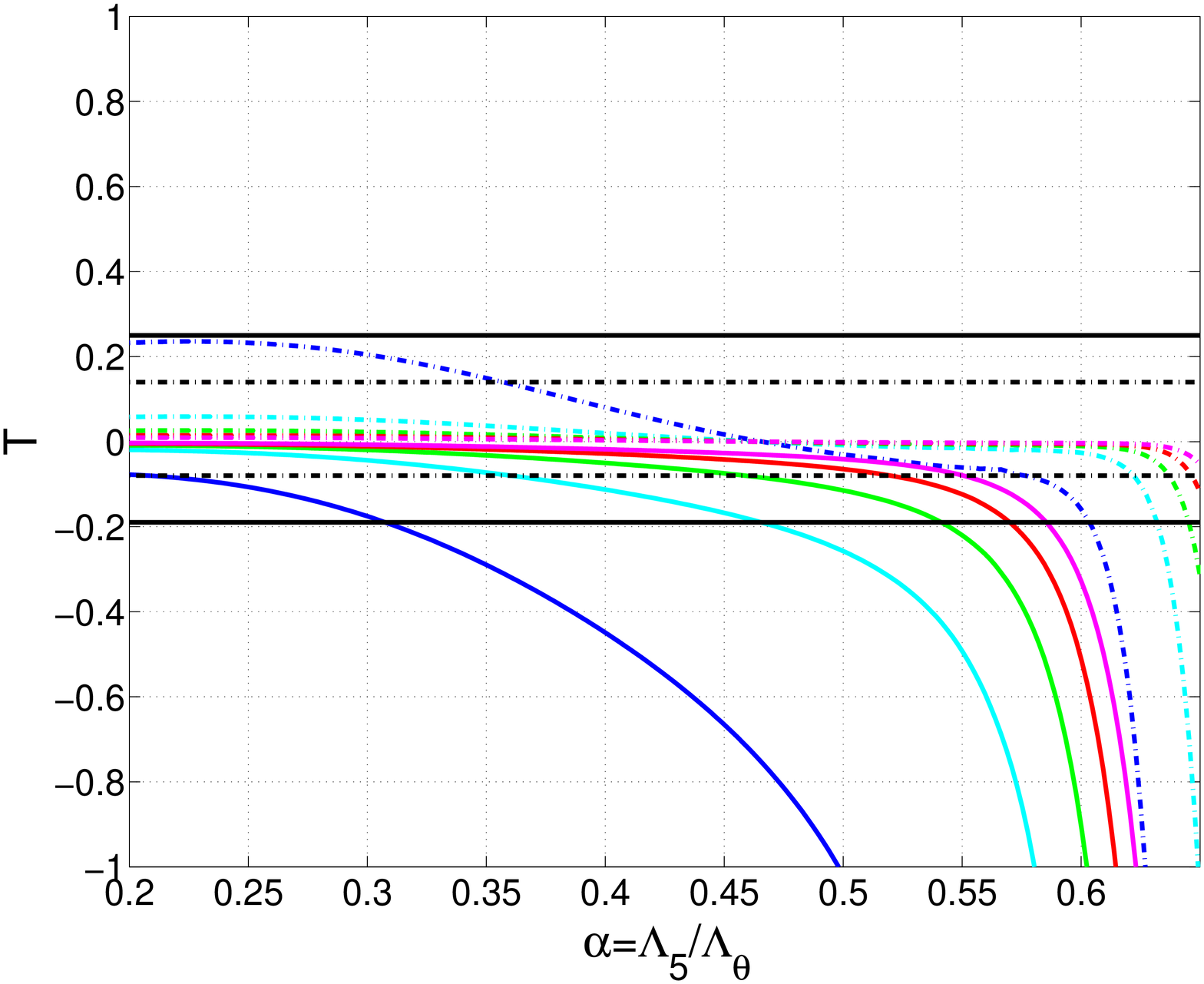}
        }
\subfigure[S Parameter]{%
           \label{fig:custSparam}
           \includegraphics[width=0.45\textwidth]{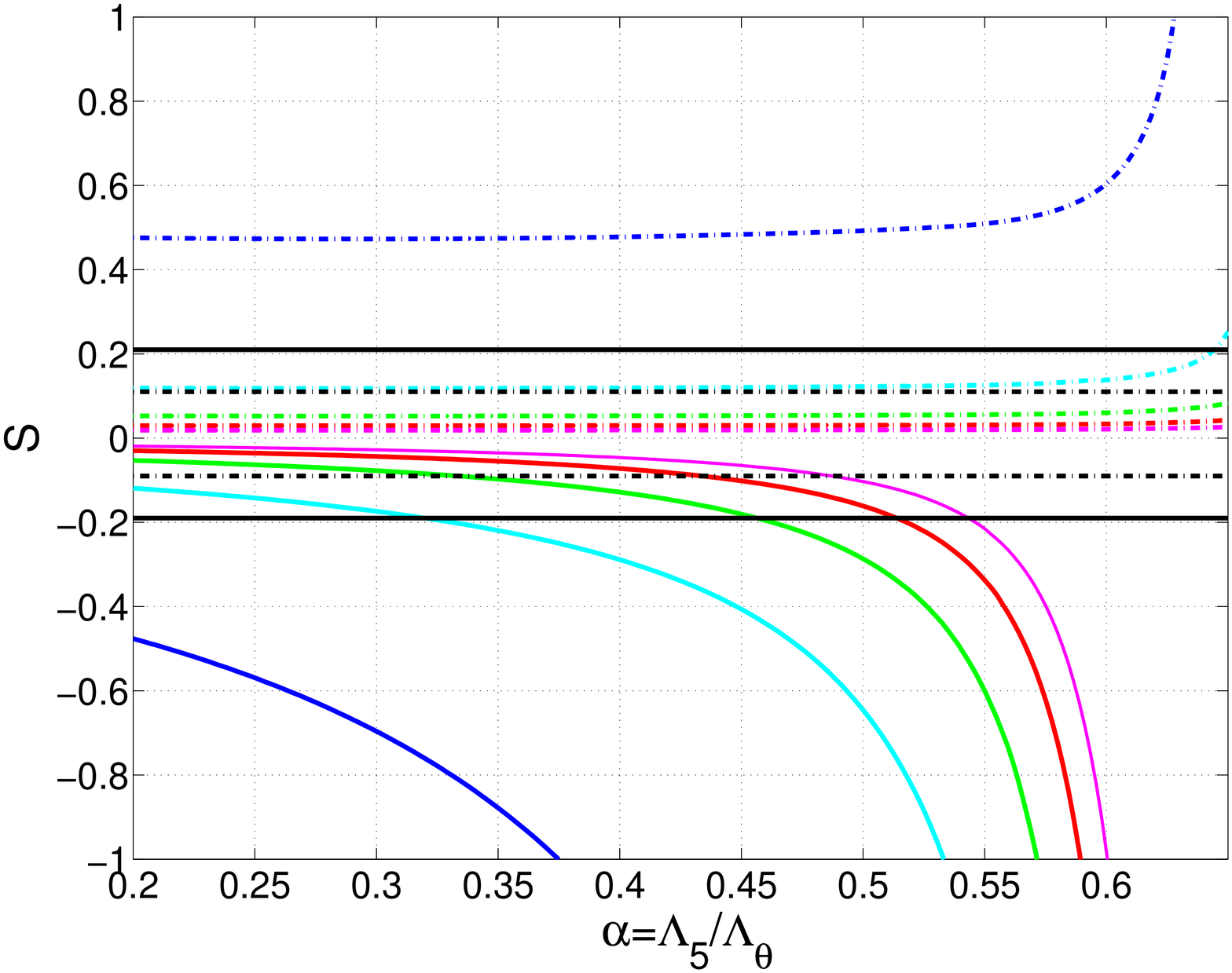}
        }
\caption[S and T Parameters for 6D spaces With Custodial Symmetry]{As in figure \ref{fig:STnoCustodial} the tree level S and T parameters but now with a bulk $\color{red}\mathrm{SU} (2)_R\times \color{red}\mathrm{SU} (2)_L\times \color{red}\mathrm{U}(1)$ custodial symmetry. \cite{Archer:2010bm}}
\label{fig:STparametersCustodial}
\end{center}
\end{figure}

\begin{figure}[!t]
\begin{center}
\begin{tabular}{c}
\includegraphics[height=2.8in]{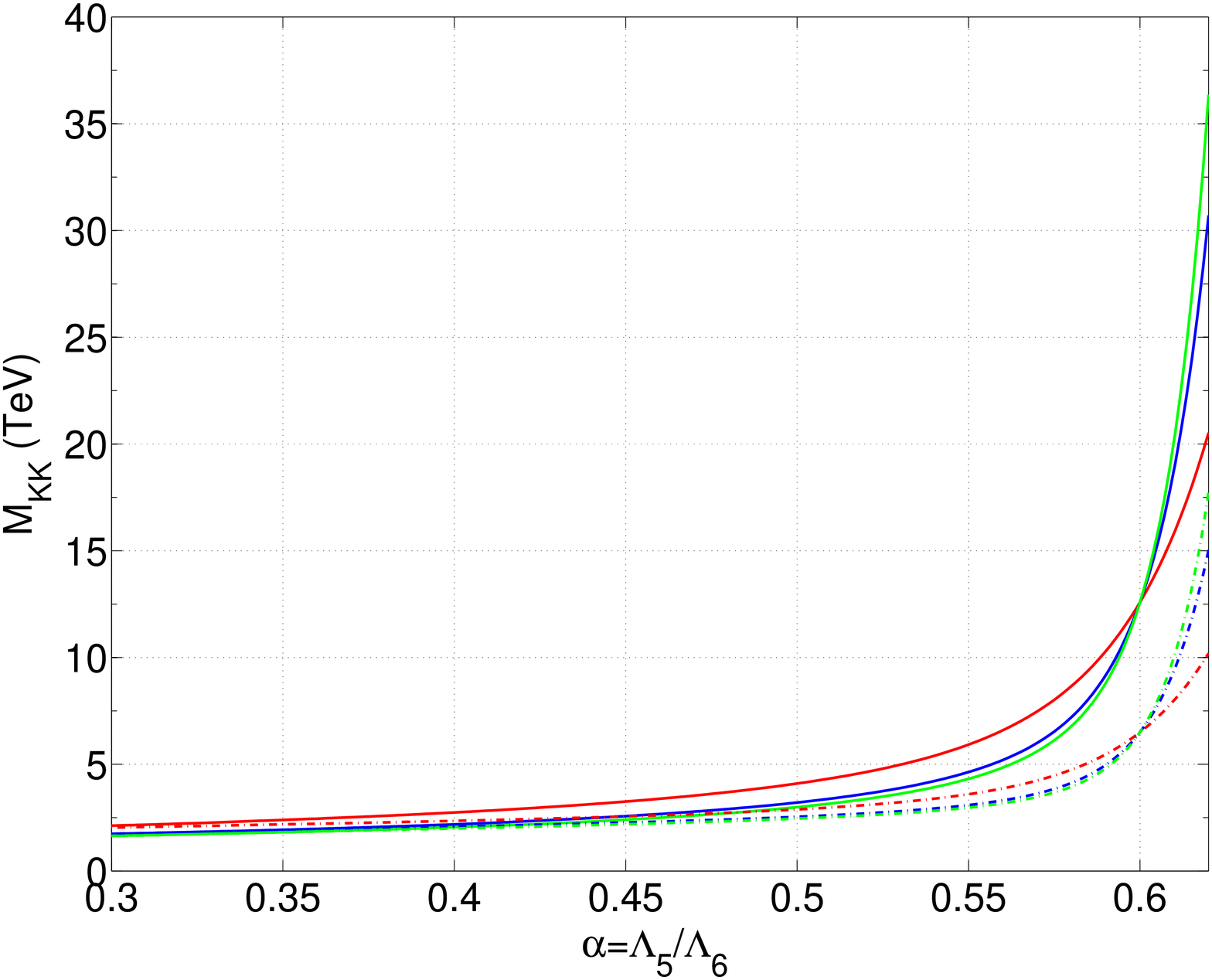}\\
\includegraphics[height=2.8in]{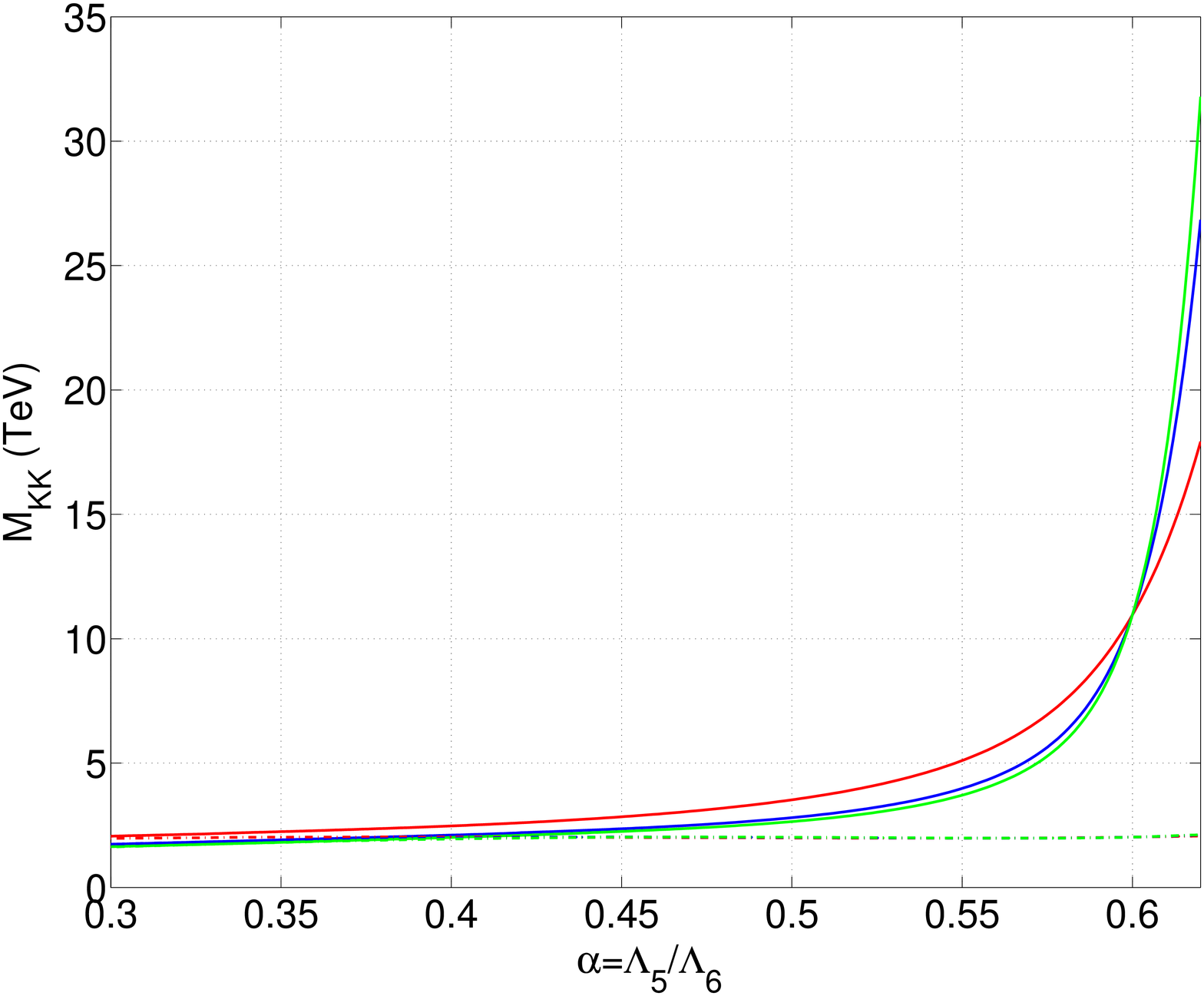}\\
\end{tabular}
\caption[The Constraint From $s_Z$ on Spaces With an Anisotropic Cosmological Constant]{The lower bound on $M_{KK}\equiv\frac{k}{\Omega}$ arising from the EW observable $s_Z^2$. With a bulk $\color{red}\mathrm{SU} (2)\times \color{red}\mathrm{U}(1)$ gauge symmetry (top) and a bulk $\color{red}\mathrm{SU} (2)_R\times \color{red}\mathrm{SU} (2)_L\times \color{red}\mathrm{U}(1)$ custodial symmetry (bottom). Here the fermions are localised to the IR brane (solid line) or the UV brane (dot-dash line) while the gauge fields propagate in 6D (red), 8D (blue) and 10D (green). The Higgs is localised w.r.t $r$ such that $\Omega=10^{15}$.  \cite{Archer:2010bm} }
\label{SZfig}
\end{center}
\end{figure}

On the other hand when the internal space is growing towards the IR brane ($\alpha<\frac{3}{5}$) the exact same volume effect suppresses the couplings between particles on the IR brane and hence $F_n$ tends to one, see figure \ref{Fnsmall}. The coupling to particles on the UV brane are also enhanced slightly but this effect is reduced by the KK gauge fields still being localised towards the IR brane. At first sight this looks very promising. In fact with the Higgs localised on to a codimension one brane and the fermions also localised on either the IR brane or the UV brane then one finds that the tree level contribution to the $S$ and $T$ parameters can be significantly reduced. These constraints have been computed using the method outlined in section \ref{EWChap}, both with and without a custodial symmetry and plotted in figures \ref{fig:STnoCustodial} and \ref{fig:STparametersCustodial}. Alternatively one can look at the constraints coming from individual EW observables. In section \ref{EWChap}, it was found that the tightest constraint on the RS model came from the weak mixing angle, $s_Z$. This constraint has been plotted in figure \ref{SZfig}. The constraints are again found to be slightly higher, than those arising from the $S$ and $T$ parameters, but generally the two approaches give similar results.

We have plotted these constraints for the six dimensional case but the results do not change significantly when one considers more than six dimensions. Readers may be surprised by the potentially large contribution to the T parameter. There are a number of effects contributing to this. Firstly, as discussed in section \ref{sect:Cust}, in any realistic scenario the custodial symmetry must be broken in order to reproduce the observed SM. In this case the custodial symmetry is broken by imposing DBC's on the $\color{red}\mathrm{SU} (2)_R$ fields and NBC's on the $\color{red}\mathrm{SU} (2)_L$ fields. The extent to which the custodial symmetry is broken can, to a certain extent, be parameterised by the difference between $F_n$ and $\tilde{F}_n$ and the difference between $m_n$ and $\tilde{m}_n$. This difference is enhanced when the root of the Bessel function is small and hence is enhanced when $J$ is negative ($\alpha<\frac{3}{5}$) as can be seen in figure \ref{gaugemassesfig}. Hence the contribution to the EW constraints, from this difference, is enhanced when $\alpha<\frac{3}{5}$. The second contribution arises from the fact that the custodial symmetry does not completely protect the gauge fermion vertex. If this correction is absorbed into the $S$ and $T$ parameters then one finds that terms proportional to $F_nF_\psi^{(n)}$ appear which become large when $\alpha>\frac{3}{5}$.         
    
\begin{figure}[ht!]
    \begin{center}
        \subfigure[Loops that would not contribute to Gauge propagator.]{
           \label{fig:NoContrib}        
    \begin{fmffile}{figures/feynNoContrib}
  \fmfframe(24,15)(24,15){ 	
   \begin{fmfgraph*}(190,110) 
    \fmfleft{i1}	
    \fmfright{o1}    
    \fmflabel{$A_\mu^{(0,0)}$}{i1} 
    \fmflabel{$A_\mu^{(0,0)}$}{o1} 
    \fmf{boson}{i1,v1} 
    \fmf{boson}{o1,v2} 
    \fmf{photon, label=$A_\mu^{(0\;l)}$, left, tension=.3}{v1,v2}
    \fmf{photon, label=$A_\mu^{(n\;0)}$, left, tension=.3}{v2,v1} 
   \end{fmfgraph*}
  }
\end{fmffile}}
        \subfigure[Loops that would contribute to Gauge propagator.]{
           \label{fig:ContribGauge}
      \begin{fmffile}{figures/ContribGauge} 	
  \fmfframe(24,15)(24,15){ 	
   \begin{fmfgraph*}(190,110) 
    \fmfleft{i1}	
    \fmfright{o1}    
    \fmflabel{$A_\mu^{(0,0)}$}{i1} 
    \fmflabel{$A_\mu^{(0,0)}$}{o1} 
    \fmf{boson}{i1,v1} 
    \fmf{boson}{o1,v2} 
    \fmf{photon, label=$A_\mu^{(0\;l)}$, left, tension=.3}{v1,v2}
    \fmf{photon, label=$A_\mu^{(0\;l)}$, left, tension=.3}{v2,v1} 
   \end{fmfgraph*}
  }
\end{fmffile}}
   \label{fig:LoopGauge}
    \end{center}
    \caption[Feynman Diagrams Contributing to Loop Corrections to Gauge Propagator]{\color{red} The top diagrams would not contribute to the gauge propagator due to the orthogonality of the gauge profiles. The bottom diagrams would contribute to the gauge propagator. Here the notation assumes six dimensions. \color{black}}
\end{figure}
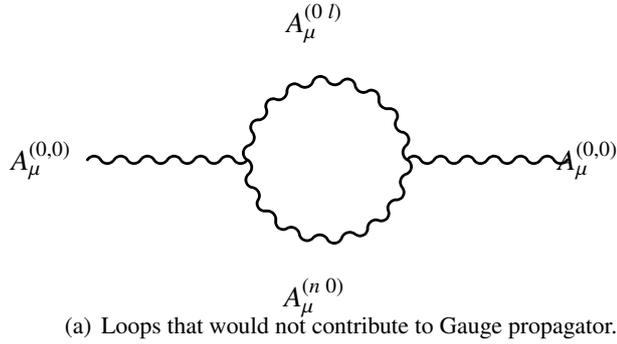
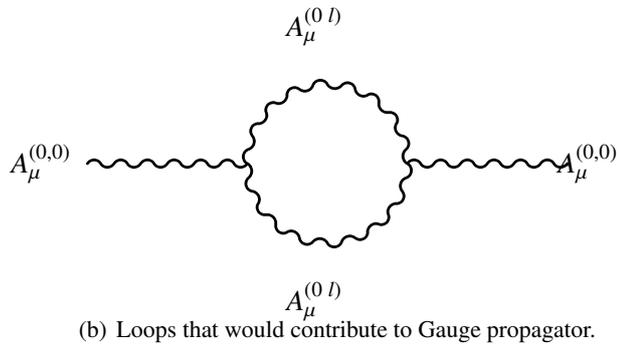

None the less based on these tree level results it would appear that, by considering spaces which are growing towards the IR brane $\alpha<\frac{3}{5}$, one can reduce the EW constraints to between $1.5-2$ TeV regardless of whether a custodial symmetry is included or not and regardless of the location of the fermions. However it is important to realise what has been computed. Here we have computed the tree level corrections to EW observables that arise from corrections to the W and Z masses as well as corrections to their couplings caused by the deformation of their profiles. What we have not included is, for example
\begin{itemize}
  \item The contribution from the gauge scalars. If the fermions and Higgs were localised to a codimension one brane and the gauge scalars had DBC's w.r.t $r$, such that the theory avoided any massless charged scalars, then one would forbid the coupling between the gauge scalars and the SM fermions and the Higgs. Hence the gauge scalars would not contribute at tree level to the EW constraints but would contribute at loop level via their coupling to the SM gauge fields. Although localising the fermions to a codimension one brane would not allow for the description of flavour discussed in section \ref{FlavourChap}.    
  \item The contribution from the KK modes of the Higgs. If both the Higgs and the fermions are localised to a codimension one brane then one would anticipate KK Higgs excitations, with repect to the internal space $\theta$, contributing to EW observables at tree level. One possibly could be to invoke the idea of KK parity (or conservation of higher dimensional momenta) within the internal space to forbid many of these interactions \cite{Cacciapaglia:2009pa, Appelquist:2000nn,  Agashe:2007jb}. Although typically the presence of a Higgs mass term would often deform the profile and cause, for example, the fermion and Higgs profiles to no longer be orthogonal. 
  \item The tree level exchange of gauge KK modes, see figure \ref{fig:KKgaugeexchange}. Such exchanges gives rise to four fermion operators. In the RS model such operators are suppressed by the RS GIM mechanism and are neglected when considering EW constraints. Although they prove to be problematic when considering constraints from flavour physics. Firstly if the fermions were localised towards the UV brane, in spaces with $\alpha<\frac{3}{5}$, then one can see from figure \ref{fig:UVcouplings} that such operators would no longer be as suppressed. Hence arguably such operators must now be included in EW analysis. Also if the fermions zero mode profiles are not flat w.r.t $\theta$ then one must also include the exchange of the $l_i\neq 0$ gauge modes, resulting in a divergent sum. If the profiles are flat then such couplings would be forbidden by orthogonality.   
  \item Loop corrections of any form. 
\end{itemize}
This last exclusion looks quite problematic to the above results. As we have already discussed if the Higgs is localised to a codimension one brane then only one tower of KK modes will contribute to EW observables at tree level. While many gauge interactions would be forbidden by the orthogonality of the profiles (see figure \ref{fig:NoContrib}) or equivalently conservation of KK number, one still finds that at loop level all KK towers will contribute. Hence one would expect, in six dimensions or more, the resulting summations to be divergent and it would be necessary to impose a cut off in the KK number. While a full study of loop contributions is beyond the scope of this thesis one can, following \cite{Appelquist:2000nn}, use naive dimensional analysis to estimate the scale at which this divergence would cause a break down in perturbative control of the theory.     
        
One can write down the loop expansion parameter for a gauge theory with a coupling $\sqrt{\alpha /4\pi}$ and $N_C$ colours
\begin{equation}
\label{LoopExpansion}
\epsilon=N_C\frac{\alpha(\Lambda_s)}{4\pi}N_{KK}(\Lambda_s)
\end{equation}   
where $N_{KK}(\Lambda_s)$ is the number of KK modes below the scale $\Lambda_s$. When $\epsilon>1$ it is taken as an indication that the perturbative effective theory has broken down. It is now not difficult to see the problem. As can be seen, from figure \ref{gaugedegfig}, where the internal space is growing towards the IR, the spacing between the KK modes with $l_i\neq 0$ becomes small and hence one would rapidly loose perturbative control of the theory. Table \ref{ tabBreakDown} lists the scales at which the effective theory of a bulk QCD theory would break down in six dimensions. Although it should be stressed that this would become far worse in more than six dimensions where the degeneracies of the KK spectrum increase significantly \cite{Cheng:1999fu}.   

\begin{table}
  \centering 
  \begin{tabular}{|c||c|c|c||c|c|c|}
\hline
   &  \multicolumn{3}{|c||}{$\alpha=0.6$}  & \multicolumn{3}{|c|}{$\alpha=0.55$}   \\
   \hline
  $M_{\rm{KK}}$ (TeV) & $R_\theta=R$ & $R_\theta=0.5R$ & $R_\theta=0.1R$&$R_\theta=R$ & $R_\theta=0.1R$ &$R_\theta=0.05R$   \\
 \hline
 \hline
 1  & 3.20 & 4.7 &10.4 & 2.31 & 2.33 & 2.40    \\
   2 & 6.61 & 9.7 &21.5 & 4.62 & 4.67 &4.83  \\
     3& 9.99 & 14.9 & 33.0 & 6.92 & 7.01& 7.27    \\
 4& 13.5 & 20.4 & 44.1 & 9.23 & 9.36 &9.71    \\
 5& 17.1 & 25.8 & 56.7 & 11.5 & 11.7 &12.2    \\
\hline
\hline
1st KK mode & 2.41$M_{\rm{KK}}$ & 2.41$M_{\rm{KK}}$ &2.45$M_{\rm{KK}}$ & 2.31$M_{\rm{KK}}$ & 2.31$M_{\rm{KK}}$ & 2.31$M_{\rm{KK}}$ \\
\hline
\end{tabular}
  \caption[Scale at Which Perturbation Theory Breaks Down]{Assuming one should include KK modes in loops, then the table give the scale (in TeV) at which one loses perturbative control of the six dimensional theory. The above numbers are based on the running of $\color{red}\mathrm{SU} (3)_C$. One would anticipate the situation being far worse for more than six dimensions and smaller values of $\alpha$.}\label{ tabBreakDown}
\end{table} 

Hence it looks naively like one would lose perturbative control of the theory before one reaches the scale of the first $Z^{\prime}$. One possible resolution to this problem is to simply make $R_\theta\ll R$. Although, with such an explanation, one must then explain how one can stabilise the different dimensions at different scales and still resolve the hierarchy problem. While this looks like a serious problem for these models it is also, arguably, incorrect.

To understand the mistake in the above argument it is important to remember in making a KK decomposition one is writing down a four dimensional effective theory based on an on shell description of a higher dimensional theory. When one goes to scales larger than the KK scale one would anticipate a break down in this effective theory. Hence in the above argument one was essentially computing the break down of an effective theory of an effective theory. Alternatively one can see the problem by considering the dual theory in which the KK modes are dual to bound states. In the same way that it is not correct to include bounds states in the loop computations of QCD, it is also not correct to include the KK modes in loop corrections. It is argued in \cite{Randall:2001gb}, that a correct approach is to impose a position dependent cut off on the higher dimensional propagator (section \ref{PosMomProp}). Here we have demonstrated that the higher dimensional propagator, for spaces with $\alpha<\frac{3}{5}$, would have a high density of poles. Although it is beyond the scope of this thesis to investigate how this would effect the propagator and consequently the running of the couplings.  

\subsection{Orthogonality, Conservation of Momentum and KK Parity}\label{sect:orthog}
When considering the other points on the above list we repeatedly alluded to the suppression of certain couplings using the orthogonality of the fields profiles or equivalently the conservation of momenta in the extra dimension. It is worth briefly spelling out when these coupling are zero and when they are not. In the four dimensional effective theory the couplings between KK modes are given by overlap integrals, for example
\begin{equation*}
\begin{array}{l}
\mbox{The gauge - fermion - fermion vertex:  }\int d^{1+\delta}x\;a^3bc^\delta\sqrt{\gamma} f_{L,R}^{(l)}\Theta_{L,R}^{(l)}f_m\Theta_m f_{L,R}^{(n)}\Theta_{L,R}^{(n)}\\
\mbox{The gauge - scalar - scalar - gauge vertex:  }\int d^{1+\delta}x\; a^2bc^\delta\sqrt{\gamma}f_l\Theta_l f_m^{(\Phi)}\Theta_m^{(\Phi)}f_n^{(\Phi)}\Theta_n^{(\Phi)}f_p\Theta_p\\
\mbox{The triple gauge vertex:  }\int d^{1+\delta}x\; bc^\delta\sqrt{\gamma}f_l\Theta_lf_m\Theta_mf_n\Theta_n\\
\mbox{The four gauge vertex:  }\int d^{1+\delta}x\; bc^\delta\sqrt{\gamma}f_l\Theta_lf_m\Theta_mf_n\Theta_nf_p\Theta_p.
\end{array}
\end{equation*} 
These taken along with the orthogonality relations Eqs. \ref{ScalarOrthog}, \ref{GaugeOrthog} and \ref{FermOrthog} result in a number of couplings being exactly zero. In the previous section it was demonstrated that when considering more than five dimensions the number of particles that are present at low energy increases significantly. Hence forbidding certain interactions could potentially be critical in preventing large corrections to existing observables. For example if a fermion zero mode is flat w.r.t $\phi$, (i.e. $\Theta=$ constant) then such a field will not interact with any of the $\alpha_n\neq 0$ modes. Although a description of flavour necessarily requires that the fermions are not flat in the extra dimension. In section \ref{sect:6dFerm} it was argued, in flat 6D space, that in fixing the bulk mass term to be real and neglecting the $M_5$ and $M_6$ mass terms then one could arrive at a scenario in which the fermions were only not flat with respect to $r$. Hence, assuming a 4D chiral theory could be obtained, then a warped extra dimensional description of flavour could be extended to six dimensions \color{red} without \color{black} too many additional contributions to the four fermion operators. However in making such a choice one is inherently choosing a preferred direction, $r$. 

\color{red}Likewise in the ansatz Eq. \ref{ansatz}, a preferred direction \color{black} was chosen such that the $a^2$ term was not dependent on $\phi$. If $a^2$ was warped w.r.t $\phi$ then clearly many of the orthogonality relations would not hold and the EW constraints would include many additional contributions. Once again a full study of the size of all these additional contributions to EW and flavour constraints is beyond the scope of this thesis and would of course be very model dependent. Here we simply wish to point out a potentially serious problem with considering warped spaces of more than five dimensions.      
 
 \section{Summary}
 In this chapter we have considered a wide range of possible warped geometries and hence we are now in a position to make a few generic statements. Firstly in five dimensions the KK modes are \color{red} likely to be \color{black} localised towards the IR tip of the space. Here we have demonstrated this using the equations of motion although one can equally consider the holographic description of warped spaces. In the 4D dual field theory, bulk fields are dual to elementary dynamical source fields coupled to a strongly coupled CFT.      
The KK modes are then mixtures of bound states of the CFT and the source field \cite{Gherghetta:2010cj,Contino:2004vy, Batell:2007jv}. A zero mode sitting towards the UV brane would be dominated by the source field while zero modes sitting towards the IR brane would largely be composed of bound states of the broken CFT. Like wise, one would speculate, that KK modes sitting towards the IR would be dual to confinement occurring in the IR of the field theory. A KK profile growing towards the UV could probably not arise in a conventional confining field theory. 

The extent to which the KK modes are localised towards the IR brane is dependent on the warp factor. Although a generic consequence of this, in five dimensions, is that the KK gauge modes will \color{red} typically \color{black} be more strongly coupled to fields localised in the IR, than the flat zero mode (i.e. a  large value of $F_n$). Hence one will always have sizeable EW corrections and always have the possibility for universal gauge fermion couplings.

In more than five dimensions, while the KK modes corresponding to $\alpha_n=0$ would still necessarily sit towards the IR, it is more difficult to demonstrate that all the KK modes would be IR localised. The problem being that there are unknown scales relating to the curvature and size of the internal space, although in all cases studied all the KK modes are IR localised as one would expect. However the size and the curvature of the internal space has two significant effects on the low energy phenomenology. 

Firstly the volume of the internal space can scale the couplings of the KK modes differently to that of the flat, or near flat, zero mode. This can cause a significant enhancement or suppression in the relative couplings. Here we demonstrated that this effect can be used to suppress corrections to the EW observables coming from corrections to the W and Z zero modes. It should be stressed that this is just one example of a potentially large deviation, in the phenomenological implications of extra dimensions, from that of the RS model.

The second effect is related to the spacing of the KK modes. Where there is more than one extra dimension there \color{red} is \color{black} essentially more than one KK scale. These additional KK scales are again related to the curvature and size of the additional dimensions and hence they need not all be at the same order of magnitude. This can result in the overall KK tower not being evenly spaced. In processes in which the full tower is contributing to a given observable, these additional modes could potentially be problematic. 

These two effects can be effectively removed by considering spaces in which the curvature and size of the internal space is significantly smaller than the rest of the space. For example, in section \ref{AnIsoLamb}, ensuring $|J|\ll |k|$ and $R_\theta\ll R$. However this then shifts the gauge hierarchy problem to explaining why a particular space has more than one scale. One must then demonstrate the stability of such spaces.\newline

Having considered deviation, from the RS model, in the dimensionality and geometry of the space we will now move on to look at deviation in a third characteristic of the model. \color{red} Namely \color{black} that the RS model considers a slice of AdS space with a sharp IR and UV cut off. These two cut offs leads to a discrete KK spectrum and determines the two scales present in the model. In the next section we shall consider the effect of smoothing out the IR cut off.         

%% file: SWchap.tex
\chapter{The Soft Wall Model}
\label{SWmodel} 
A generic feature of extra dimensional models which have `hard' cut offs is that the KK masses scale as $m_n^2\sim n^2$ for large $n$ (see for example \ref{RSscalarmass}). From the holographic perspective the KK modes are conjectured to be dual to bounds states of a CFT. The conventional way to describe such confinement, in QCD, is in terms of pairs of quarks being held together with `flux tubes' along which the energy density of the gluon is constant. This results in the total energy being linearly proportional to the distance between the two quarks, $L$, i.e.
\begin{displaymath}
E\sim p\sim M_n\sim KL
\end{displaymath}     
where $K$ is the tension in the flux tube and $M_n$ is the mass of the $n$th meson with momentum $p$. If one then assumes that the meson can be described using the Bohr-Sommerfield quantization \cite{LandauLifshitz} and uses the above relation then 
\begin{displaymath}
\int pdx\sim pL \sim \frac{M_n^2}{K}\sim n
\end{displaymath}  
Likewise the angular momentum of the meson should scale as $J\sim KL^2\sim KM_n^2$. These quite crude arguments fit the observed mesons remarkably well and are referred to as the Regge trajectories.

As pointed out in \cite{Schreiber:2004ie, Shifman:2005zn}, bound states with masses scaling as $m_n^2\sim n^2$ would require a different type of confinement to that which is observed in QCD. It was this unusual scaling in the bound states masses that led to the proposal of the soft wall model in which the hard IR brane is replaced by a smooth space-time cut off \cite{Karch:2006pv}. This cut off is provided by the background value of a `dilaton' field, $\Phi$,
\begin{equation}
\label{ }
S=\int d^5x \sqrt{g}e^{-\Phi}\mathcal{L}.
\end{equation}    
Here we will restrict our discussion to only considering AdS${}_5$,
\begin{equation}
\label{SWmetric}
ds^2=\left (\frac{R}{z}\right )^2\left (\eta^{\mu\nu}dx_\mu dx_\nu-dz^2\right ),
\end{equation}
where now $z$ runs from a UV brane at $z=R$ to infinity. If one assumes that the dilaton field scales as a power law 
\begin{equation}
\label{SWdilaton}
\Phi=\left (\frac{z}{R^{\prime}}\right )^\nu
\end{equation} 
then a discrete KK spectrum is obtained when $\nu> 1$. The KK masses then scale as \cite{Falkowski:2008fz, Batell:2008me} 
\begin{equation}
\label{ }
m_n^2\sim \frac{n^{2-\frac{2}{\nu}}}{R^{\prime\;2}}.
\end{equation}
So the Regge scaling of the KK masses can be obtained with a dilaton scaling quadratically. Note that now the KK scale is determined by the form of the dilaton $M_{\rm{KK}}=\frac{1}{R^{\prime}}$. Also as $\nu \rightarrow \infty$ then one regains the RS model. Also when $\nu\leqslant 1$ then the KK profiles are not normalisable, w.r.t the orthogonality relations Eqs. \ref{ScalarOrthog}, \ref{GaugeOrthog} and \ref{FermOrthog}, and one obtains a continuous KK spectrum. When $\nu=1$ this continuous spectrum begins after a mass gap and is considered to be dual to a hidden sector of unparticles. While if $\nu<1$ then the spectrum is continuous above zero mass. Such a scenario would be dual to unparticles existing without a mass gap \cite{Falkowski:2008fz, Batell:2008me,Cabrer:2009we}.

\section{Resolving the Gauge Hierarchy Problem}
However with $\nu=2$ one obtains Regge scaling in the bound states. A natural question, one can then ask, is can many of the phenomenological features of the RS model be extended to spaces with a soft wall (SW). In other words can such spaces offer explanations of the gauge hierarchy and the fermion mass hierarchy. Here we shall briefly review some of the existing work which has been carried out on this question before looking in more detail at extending the RS description of flavour (outlined in section \ref{FlavourChap}) to the SW model.  

\subsection{Can the Space be Stabilised in Order to Generate a Large Warp Factor}
Before a space can claim to be able to resolve the gauge hierarchy problem it must be demonstrated that it can be stabilised, such that $\Omega\approx 10^{15}$, without fine tuning. In the RS model this is done using the Goldberger Wise mechanism, in which a gravity-scalar system is considered such that the separation between the two branes is then determined by the VEV of a scalar \cite{Goldberger:1999uk}. It can then be shown that a large hierarchy can be generated with out unnaturally large parameters in the bulk and brane potentials. 

This question has been studied, for the SW model, in both \cite{Cabrer:2009we} and \cite{Gherghetta:2010he}. Here we shall not reproduce the calculation but rather summarise the results. Both papers again studied a gravity-scalar system which was solved using the `superpotential' method of \cite{DeWolfe:1999cp}. With the absence of an IR cut off, the SW model suffers from a naked curvature singularity in the IR \cite{Aybat:2010sn, George:2011gs}. Hence, in any valid solution, the boundary terms of the equations of motion must vanish at this singularity. In \cite{Gherghetta:2010he} this is done directly from the equations of motion where as in \cite{Cabrer:2009we} this is achieved by requiring that the superpotential does not grow too quickly. This results in the two papers studying a subtly different class of solutions. Both papers go on to conclude that a large hierarchy can be obtained however \cite{Gherghetta:2010he} conclude that, if one wishes to avoid tuning the parameters in the potential on the UV brane, then $\nu<1$ and hence one cannot obtain a large hierarchy along with a discrete KK spectrum and Regge scalings of the KK masses. Alternatively \cite{Cabrer:2009we} conclude that the hierarchy problem can be resolved with $\nu<2$ and hence a discrete KK spectrum and Regge scalings are achievable. When $\nu\geqslant 2$ the equations of motion do not vanish at the singularity.

For the remainder of this section we will focus on the solutions, with $\nu=2$, that give a large hierarchy and Regge scaling. Hence we will not consider stabilisation further but instead impose a large hierarchy by hand such that 
\begin{equation}
\label{ }
\Omega\equiv \frac{R^{\prime}}{R}\sim 10^{15}.
\end{equation} 
By considering $\nu=2$ we are considering the extremal case of solutions that avoid contributions from the singularity. From a phenomenological perspective, solutions with $\nu=2$ will be roughly equivalent to $\nu=1.999\dots$, but one finds that the KK profiles have analytical solutions when $\nu=2$. It is interesting to note that it is this extremal case that gives the Regge scalings in the KK spectrum.
 
\subsection{Electroweak Constraints}
It was argued in section \ref{sect:5dFn}, that five dimensional warped spaces that resolve the gauge hierarchy problem will always have sizeable EW constraints, giving rise to a little hierarchy problem. However this argument assumes the Higgs is localised on the IR brane. In the SW model, with no IR brane, the Higgs must be placed in the bulk hence $F_n$ is now replaced by an overlap integral between the Higgs VEV and the gauge profiles. It is this that leads to a significant reduction in the constraints from EW observables. In particular in \cite{Falkowski:2008fz} an extension of the Gauge-Higgs unification scenario was studied in the SW model. With the absence of an IR brane, it is necessary to include an additional charged bulk scalar in order to break the gauge symmetry. The `Higgs' is then a mixture of the fifth component of the gauge field and this bulk scalar. In this model it is found that with $\nu=2$ the constraints, from EW observables, on the mass of the first KK gauge field are reduced to about 2 TeV. While with $\nu=1$ they are reduced to about $1$ TeV.

An alternative model was considered in \cite{Cabrer:2010si, Cabrer:2011fb} in which EW symmetry is broken by a bulk Higgs. There an IR brane was included but the space was continued to the singularity. The resulting constraints are then dependent on $\nu$ as well as the spacing between the IR brane, the singularity and the exponent of the Higgs VEV, $\alpha$, where the Higgs VEV is found to take the form $h(z)\sim h_0\left(\frac{z}{R}\right )^\alpha$. It is found that the lower $\alpha$ the lower the EW constraints. However if $\alpha\lesssim 2$ then additional terms contribute to the Higgs VEV and a fine tuning is reintroduced. Hence the gauge hierarchy problem has not been resolved. Although this model differs from the one we will study by the presence of an IR brane, we shall none the less use this result and principally consider a quadratic Higgs VEV. Once again the constraints on the mass of the first KK gauge field can be reduced to about $1$ TeV \cite{Cabrer:2011fb}. 

When the IR brane is removed the EW constraints again remain relatively small and the Higgs VEV is again found to have the form $h(z)\sim h_0\left(\frac{z}{R}\right )^\alpha$ \cite{Batell:2008me}. 
\newline

The point we wish to make in this section is that, in removing the IR brane and changing the way the RS model is cut off, one can still resolve the gauge hierarchy problem. Further still with a bulk Higgs the EW constraints are typically significantly lower. However, as with extending the RS model into more than five dimensions, in the SW model there are a number of new unknown parameters. Notably the form of the dilaton and the form of the Higgs VEV. None the less the SW model does look like a potentially very attractive framework for discussing EW symmetry breaking. As we saw in section \ref{FlavourChap} some of the tightest constraints on the RS model comes from the flavour sector. Hence the central question of this chapter is can the RS description of flavour be extended into models with a SW? In other words
\begin{itemize}
  \item Can the hierarchy of fermion masses be generated in models with a soft wall?
  \item Are FCNC's still suppressed?
  \item In Eq. \ref{ILLLL} one makes the sum $\sum_n\frac{1}{m_n^2}$ which naively is divergent (in five dimensions) when $m_n^2\sim n$. If this sum is divergent at what scale does one lose perturbative control of the theory?
\end{itemize}         
This work very much continues on from work done in \cite{MertAybat:2009mk, Gherghetta:2009qs, Delgado:2009xb, Atkins:2010cc}.

\section{Fermions in a Soft Wall}
One of the central problems in extending the RS description of flavour to the SW model is that of dealing with the fermions. As explained in \cite{Batell:2008me,MertAybat:2009mk, Delgado:2009xb}, the fermion's bulk profile is determined from the extra dimensional geometry via the presence of a mass term. In the RS model the Higgs is localised on the brane and the Yukawa couplings contribute to the fermion profiles only through the boundary conditions. Whereas, in the SW model, the Higgs is in the bulk and hence one must account for the back reaction of the Yukawa couplings in the fermion equations of motion. This makes the full study of flavour, with anarchic couplings, computationally challenging. Another related feature of the SW model is that the extra dimension extends to infinity. Hence in order for the KK modes of the fermion to be discrete and normalisable, one must couple them to something that bounds them to the space. In this case it is the coupling to the Higgs that gives rise to a discrete fermion KK spectrum. All this makes a full study of flavour potentially quite involved. However we will simplify the situation considerably by considering a $z$ dependent mass term as explained in the next section.

\subsection{The Fermion KK Decomposition}
Many of the results of section \ref{chap:KKRed.} assumed an IR cut off and did not include the dilaton and hence must be rederived for the soft wall model. If we begin by considering two bulk fermions, $\Psi$ a doublet under $\color{red}\mathrm{SU} (2)$ and $\Upsilon$ a singlet, propagating in Eq. \ref{SWmetric}
\begin{eqnarray}
S=\int d^5x\; \sqrt{g}e^{-\Phi}\Bigg [ \frac{1}{2}(i\bar{\Psi}\Gamma^M\nabla_M\Psi-i\nabla_M\bar{\Psi}\Gamma^M\Psi)-M_\Psi\bar{\Psi}\Psi\hspace{2cm}\nonumber\\
+\frac{1}{2}(i\bar{\Upsilon}\Gamma^M\nabla_M\Upsilon-i\nabla_M\bar{\Upsilon}\Gamma^M\Upsilon)
-M_\Upsilon\bar{\Upsilon}\Upsilon+Y( h\bar{\Psi}\Upsilon+\rm{h.c.})\Bigg ]
\end{eqnarray}
where $Y$ are the Yukawa couplings to a bulk Higgs (assumed to be a doublet under $\color{red}\mathrm{SU} (2)$) with a VEV $h(z)$. The remaining notation is as used in section \ref{sect:5dFerm}. We can now see that a bulk Higgs mixes the two representations together. Hence if we make the KK decompositions
\begin{equation}
\label{  FermKKdeomp}
\Psi_{L,R}=\sum_n \frac{z^2}{R^2}e^{\Phi/2}f^{(n)}_{\Psi\;L,R}(z)\Psi_{L,R}^{(n)}\mbox{   and   }\Upsilon_{L,R}=\sum_n \frac{z^2}{R^2}e^{\Phi/2}f^{(n)}_{\Upsilon\;L,R}(z)\Upsilon_{L,R}^{(n)}
\end{equation}      
then the equations of motion are given by \cite{Gherghetta:2009qs, Atkins:2010cc}
 \begin{equation}
\label{offDiagSWferms}
\pm\partial_z\left(\begin{array}{c}f_{\Psi\;R,L}^{(n)} \\f_{\Upsilon\;R,L}^{(n)}\end{array}\right)+\frac{R}{z}\left(\begin{array}{cc}M_{\Psi} & Y_\Upsilon h(z) \\ Y_\Psi h(z)& M_{\Upsilon}\end{array}\right)\left(\begin{array}{c}f_{\Psi\;R,L}^{(n)} \\f_{\Upsilon\;R,L}^{(n)}\end{array}\right)=m_n\left(\begin{array}{c}f_{\Psi\;L,R}^{(n)} \\f_{\Upsilon\;L,R}^{(n)}\end{array}\right)
\end{equation}
while the orthogonality relations are given by
\begin{equation}
\label{SWorthogOfDiag}
\int^{\infty}_R \left (f_{\Psi\;L,R}^{(n)}f_{\Psi\;L,R}^{(m)}+f_{\Upsilon\;L,R}^{(n)}f_{\Upsilon\;L,R}^{(m)}\right )=\delta^{nm}.
\end{equation}
It is now possible to see the double role of the Higgs. Firstly it is giving mass to the fermion zero modes which are associated with the SM particles. Secondly, in order to arrive at a discrete fermion KK spectrum, it is necessary for Eq. \ref{SWorthogOfDiag} to be convergent and hence the fermion profiles must go to zero as $z\rightarrow\infty$. Here this is achieved by coupling the fermions to the Higgs such that $\lim_{z\rightarrow\infty}\frac{h(z)}{z}\rightarrow \infty$ \cite{Gherghetta:2009qs}. One can also see the difficulties that would arise in scanning over many anarchic Yukawa couplings, since for each $3\times 3$ Yukawa matrix one would have to solve the six coupled differential equations and normalise all the solutions.

Hence here we shall use the approximation introduced in \cite{MertAybat:2009mk} and assume that Eq. \ref{offDiagSWferms} can be diagonalised such that the fermion profiles can be obtained from   
\begin{equation}
\label{ }
S=\int d^5x \sqrt{g}e^{-\Phi}\left (\bar{\Psi} (i\Gamma^M\nabla_M-M(z)) \Psi \right ),
\end{equation} 
where we have now included the back reaction of the Higgs VEV in a $z$ dependent mass term
\begin{equation}
\label{bulkmass}
M(z)=\frac{c_0}{R}+\frac{c_1}{R}\frac{z^\alpha}{R^{\prime\;\alpha}}.
\end{equation}
Through a slight abuse of notation, we shall refer to the $c_0$ term as a bulk mass parameter and the $c_1$ term as the term arising from the Yukawa coupling. After making the same KK decomposition, Eq. \ref{ FermKKdeomp}, such that  
\begin{equation}
\label{orthogferm}
\int_R^\infty f_{L,R}^{(n)}f_{L,R}^{(m)}=\delta_{nm},
\end{equation} 
and the equations of motion are now
\begin{equation}
\label{ }
\left (\partial_z \pm \frac{R}{z}M(z)\right )f_{L,R}^{(n)}=\pm m_nf_{R,L}^{(n)},
\end{equation}
where the $\pm$ act on $f_L$ and $f_R$ respectively. The zero mode profile ($m_n=0$) can now be solved for
\begin{equation}
\label{fermionzeromode}
f_{L,R}^{(0)}=\sqrt{\frac{\frac{\pm 2c_1}{(R^{\prime})^{1\mp 2c_0}}\left (\frac{\alpha}{\pm 2c_1}\right )^{1-\frac{1\mp 2c_0}{\alpha}}}{\Gamma\left (\frac{1\mp 2c_0}{\alpha},\pm \frac{2c_1}{\alpha}\Omega^{-\alpha}\right )}}\;z^{\mp c_0}\exp \left (\mp\frac{c_1}{\alpha}\frac{z^\alpha}{R^{\prime\;\alpha}}\right ).
\end{equation} 
where $\Gamma(a,x)=\int_x^\infty e^{-t}t^{a-1}dt$ is the incomplete gamma function. Note the zero mode only exists if $\frac{c_1}{\alpha}>0$ for $\psi_L$ or $\frac{c_1}{\alpha}<0$ for $\psi_R$. Also, as in the RS case, the $\psi_L(\psi_R)$ profile will sit towards the UV when $c_0>\frac{1}{2}$ $(c_0<-\frac{1}{2})$.  

\subsection{The Fermion Masses}

Having obtained the fermion zero mode profile we will treat the fermion mass, arising from the coupling to the Higgs, as a perturbation. I.e. we let the fermion zero mode masses be approximated by 
\begin{equation}
\label{MassIntegral}
M_{ij}=\int_R^{\infty}dz\frac{R}{z}Y_{ij}h(z)f_{L}^i(z)f_R^j(z),
\end{equation}     
where $i,j$ are flavour indices and $f_{L,R}^i$ is the zero mode profile, Eq. \ref{fermionzeromode} with $c_0=c_0^{L,R\;i}$ and $c_1=c_1^{L,R\;i}$. Here we parameterise the Yukawa couplings by $Y_{ij}=\lambda_{ij}\sqrt{R}$, where $\lambda_{ij}$ are taken to be order one. The Higgs VEV is taken to be of the form \cite{Batell:2008me}\color{red}
\begin{equation}
\label{higgsvev}
h(z)=h_0R^{-\frac{3}{2}}\left (\frac{z}{R^{\prime}}\right )^\alpha,
\end{equation}
where $h_0$ is a dimensionless constant.  If one assumes that the $c_1$ term in Eq. \ref{bulkmass} has arisen from the Higgs VEV then there will be a non trivial relationship between the $c_1^{L,R\;i}$ values and $h_0\lambda_{ij}$. Since the only way to know such a relationship would be to solve Eq. \ref{offDiagSWferms}, one would wonder whether the situation has been improved. However fortunately it is found that, when the fermions are sitting towards the UV brane, the four dimensional effective couplings, that govern the low energy phenomenology, are far more sensitive to changes in the $c_0$ bulk mass parameters than changes in the $c_1$ parameters. This allows us to make the approximation of assuming a universal $c_1$ value, i.e. assuming that $c_1^{Li}=-c_1^{Ri}=c_1$. 

Turning back to the fermion masses Eq. \ref{MassIntegral} and substituting in Eq. \ref{fermionzeromode} gives
\begin{equation}
\label{ } 
M_{ij}=h_0\frac{\lambda_{ij}}{R^{\prime}}\frac{2^{(\frac{1-c_0^{Li}+c_0^{Rj}}{\alpha})}(c_1^{Li})^{\frac{1-2c_0^{Li}}{2\alpha}}(-c_1^{Rj})^{\frac{1+2c_0^{Rj}}{2\alpha}}}{\alpha^{\frac{1-\alpha}{\alpha}}(c_1^{Li}-c_1^{Rj})^{\frac{\alpha-c_0^{Li}+c_0^{Rj}}{\alpha}}}\frac{\Gamma\left (\frac{\alpha-c_0^{Li}+c_0^{Rj}}{\alpha},\frac{c_1^{Li}-c_1^{Rj}}{\alpha}\Omega^{-\alpha}\right )}{\sqrt{\Gamma\left (\frac{1-2c_0^{Li}}{\alpha},\frac{2c_1^{Li}}{\alpha}\Omega^{-\alpha}\right  )\Gamma\left (\frac{1+2c_0^{Rj}}{\alpha},\frac{-2c_1^{Rj}}{\alpha}\Omega^{-\alpha}\right  )}}.
\end{equation} 
Note that the integral in Eq. \ref{MassIntegral} is convergent only when $\alpha>0$, $(c_1^{Li}-c_1^{Rj})>0$ and $\frac{\alpha-c_0^{Li}+c_0^{Rj}}{\alpha}>0$. One can also see that the $c_0$ terms always appear in the exponent while the $c_1$ terms appear in the base. If one now assumes universal $c_1$ values then this expression simplifies to
\begin{equation}
\label{SWMasses}
M_{ij}=h_0\frac{\lambda_{ij}}{R^{\prime}}\left (\frac{\alpha}{2c_1}\right )^{\frac{\alpha-1}{\alpha}}\frac{\Gamma\left (\frac{\alpha-c_0^{Li}+c_0^Rj}{\alpha},\frac{2c_1}{\alpha}\Omega^{-\alpha}\right )}{\sqrt{\Gamma\left (\frac{1-c_0^{Li}}{\alpha},\frac{2c_1}{\alpha}\Omega^{-\alpha}\right )\Gamma\left (\frac{1+c_0^{Ri}}{\alpha},\frac{2c_1}{\alpha}\Omega^{-\alpha}\right )}}.
\end{equation}
\color{black} This should be compared with the analogous expression for the RS model, Eq. \ref{RSMassMat}
\begin{equation}
\label{RSMasses}
M_{ij}=\frac{\lambda_{ij}v\Omega}{k}f_L^i(R)f_R^j(R)=\lambda_{ij}v\sqrt{(1-2c_L^i)(1+2c_R^j)}\;\frac{\Omega^{1-c_L^i+c_R^j}}{\sqrt{(\Omega^{1-2c_L^i}-1)\;(\Omega^{1+2c_R^j}-1)}},
\end{equation}
where $v\approx 174$ GeV is the Higgs VEV.

\color{red}
\begin{figure}[ht!]
    \begin{center}
        \subfigure[The RS Model]{%
            \label{fig:FermMassRS}
            \includegraphics[width=0.52\textwidth]{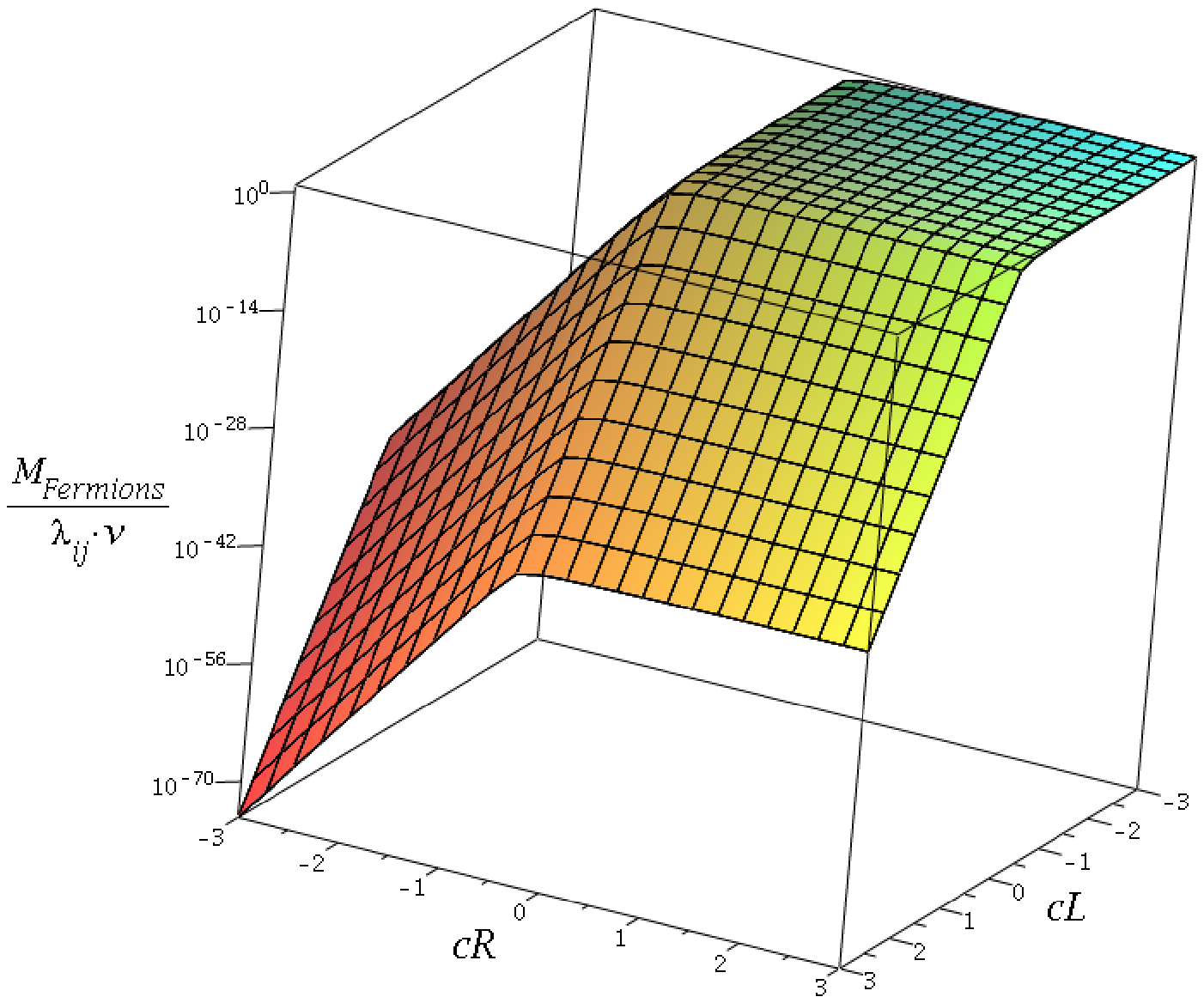}
        }
        \subfigure[$\alpha=1$]{%
           \label{fig:second}
           \hspace{-1.5cm}\includegraphics[width=0.52\textwidth]{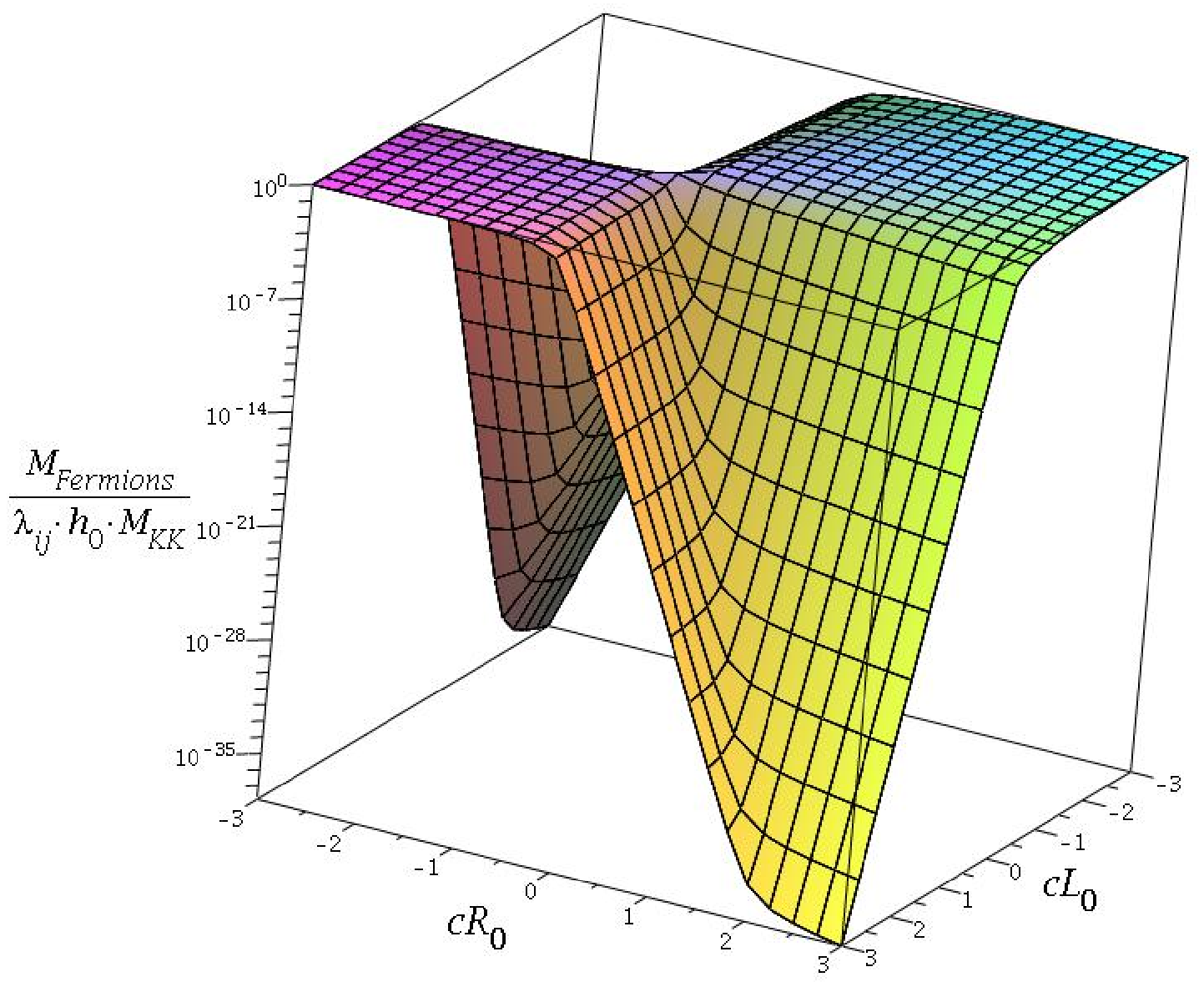}
        }\\ 
        \subfigure[$\alpha=2$]{%
            \label{fig:FermMass2}
            \includegraphics[width=0.52\textwidth]{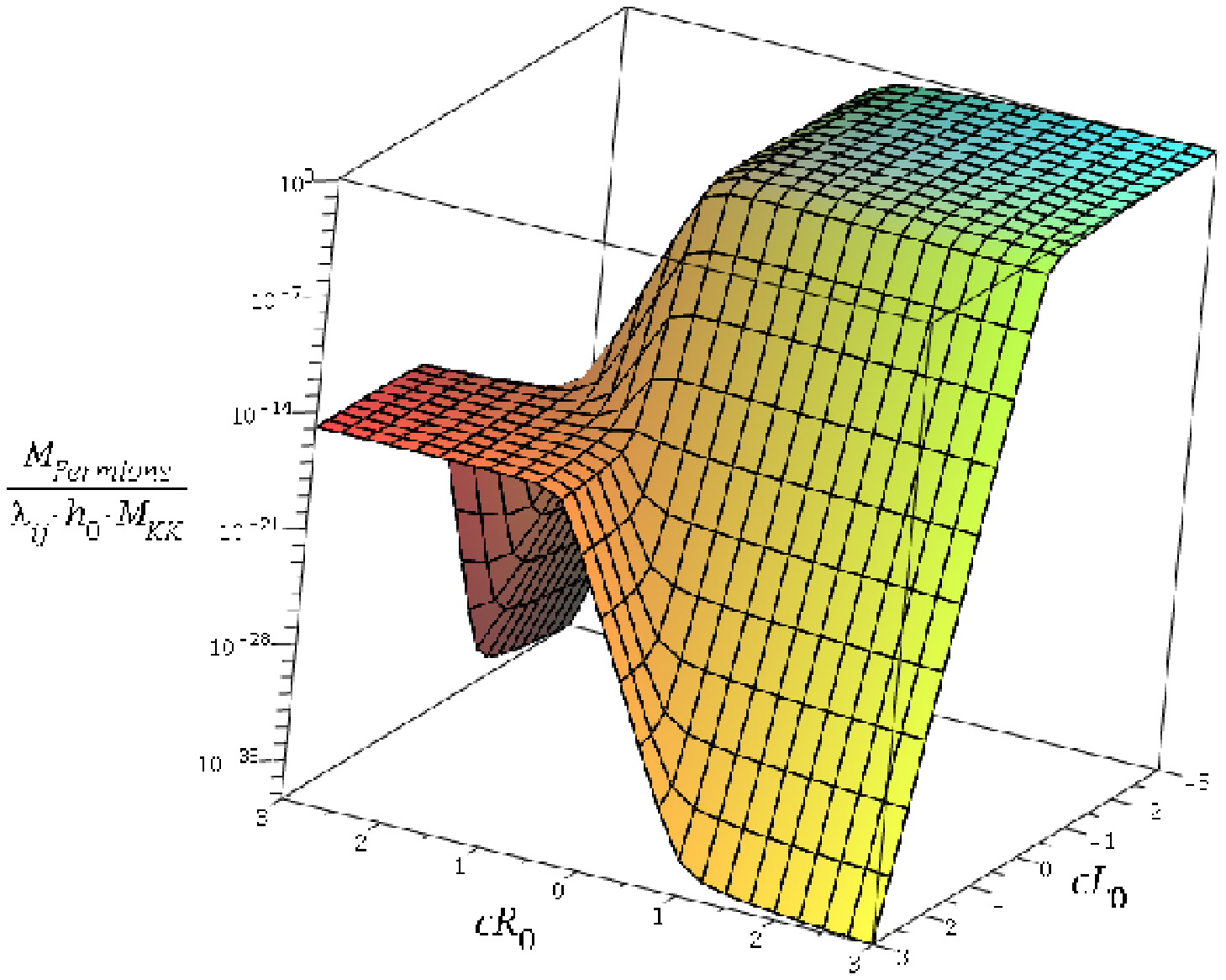}
        }%
        \subfigure[$\alpha=4$]{%
            \label{fig:FermMass4}
             \hspace{-1.5cm}\includegraphics[width=0.52\textwidth]{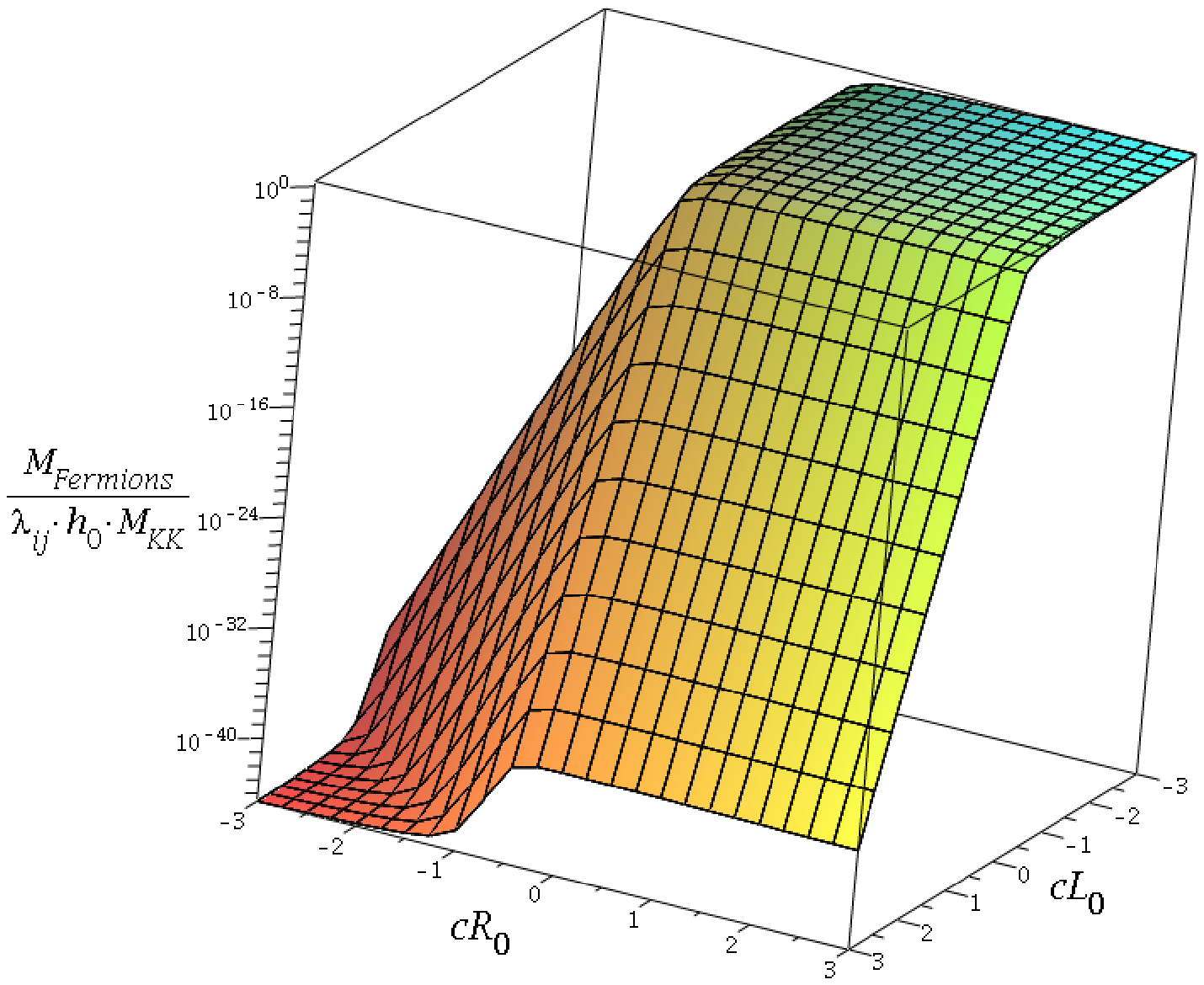}
        }%
   \label{fermmass}
    \end{center}
    \caption[The hierarchy of fermion masses in the SW model.]{The hierarchy of fermion masses for different Higgs VEV exponents. Here $c_1=1$ but it is found that the results are not sensitive to the value of $c_1$, while $c_{L,R}$ refer to the $c_0$ constant bulk mass terms. We take $\Omega=10^{15}$ and $M_{\rm{KK}}\equiv \frac{1}{R^{\prime}}=1\;\rm{TeV}$.  } \label{fermmass}
\end{figure}

The ranges of possible masses have been plotted in figure \ref{fermmass}. One can see two significant differences from the RS case. The first and most obvious, is that as one localises the fermions closer and closer to the UV one hits a minimum fermion mass. The location of this minimum is very sensitive to both the warp factor and the form of the Higgs VEV. In particular one would struggle to generate the hierarchy of fermion masses with a linear Higgs VEV or, for that a matter, a small warp factor. What is particularly interesting is that the optimum solutions to the gauge hierarchy problem, found in \cite{Cabrer:2011fb}, appear to correspond to $\Omega\approx 10^{15}$ and $\alpha\approx 2$. This would put the minimum fermion mass about fifteen orders of magnitude lower than the mass of an IR localised fermion. If one took $h_0M_{\rm{KK}}\sim 100$ GeV then this would correspond to a minimum fermion mass of the order $0.1$ meV which is roughly the scale of the observed neutrino masses. It is curious to note that this approximately corresponds to models which are found to have minimal electroweak constraints \cite{Cabrer:2011fb}. One could obtain a lower mass by considering a split fermion scenario, in which the left-handed fermions are localised towards the opposite side of the space to the right-handed fermions. However one would suspect that this scenario would give rise to large FCNC's and hence would not be phenomenologically viable. 

The second difference, from the RS model, is the gradient of the
slope. In both the RS model and the SW model the possible masses drop
away exponentially. However, in the SW model, the exponent is no
longer constant. This leads us to anticipate small mass hierarchies for light fermions, such as neutrinos, and larger mass hierarchies between heavier fermions, such as the the top and bottom quarks. This reduced $c_0$ dependence also has implications in the suppression of FCNC's which we shall consider in later sections.

It should be noted that, although here we have treated the Yukawa couplings as a perturbation, these results are in good agreement with the corresponding plot in \cite{Gherghetta:2009qs} (with $\alpha=2$ and $\Omega=10^3$). One would also expect that these differences would be present in any model with a bulk Higgs (see for example \cite{Medina:2010mu}). Since this notion of a minimum in the 4D effective Yukawa coupling is generic to all couplings between fermion zero modes and profiles sitting towards the IR. It is, for example, well known that the same effect is present in the fermion couplings to KK gauge fields (see figure \ref{RelFermCoupl}). \color{black}

\section{Gauge Fields in a Soft Wall} 
Having demonstrated that, with $\alpha\gtrsim 2$ and $\Omega=10^{15}$, one can generate the hierarchy of fermion masses we shall now move on to look at FCNC's. As demonstrated in section \ref{FlavourChap}, some of the most stringent constraints on the RS model come from FCNC's in the kaon sector. However, as already mentioned, one may be concerned that, with $m_n^2\sim n$, Eq. \ref{ILLLL} may be logarithmically divergent. If such processes are divergent then this would mean the four fermion operators are dominated by the higher KK modes and one must include a cut off in the KK number. Since it is not really known how to renormalise such a scenario and results are typically sensitive to such a cut off \cite{Appelquist:2000nn}, one should be concerned about the validity of any results with such a divergence. If such processes are not divergent then clearly it is necessary to include more KK modes than in the RS model in order to arrive at an accurate result. Although it is not clear how many need to be included. With this in mind we shall now compute the 5D gauge propagator that inherently includes the full KK tower. For the remainder of this section we shall restrict ourselves to studying models with a quadratic Higgs VEV ($\alpha=2$).

\subsection{The Photon Propagator}
We start by considering a $U(1)$ gauge field in the space Eq. \ref{SWmetric},
\begin{equation}
\label{guageaction}
S=\int d^5x \sqrt{g}e^{-\Phi}\left (-\frac{1}{4}F_{MN}F^{MN}\right ).
\end{equation}
Following the analysis of section \ref{PosMomProp}, i.e. integrating by parts and introducing a gauge fixing term of the form $\mathcal{L}_{GF}=-\frac{1}{2\xi}\frac{R}{z}e^{-\Phi} \left(\partial_\mu A^\mu -\xi e^{\Phi}\frac{z}{R}\partial_5 ( \frac{R}{z}e^{-\Phi} A^5) \right )^2$, one gets
\begin{eqnarray}
\label{ }
S=\int d^5x \bigg [ \frac{e^{-\Phi}}{2}\frac{R}{z}A_\mu\left (\eta^{\mu\nu}\partial^2-\left (1-\frac{1}{\xi}\right )\partial^\mu\partial^\nu-\frac{e^{\Phi}z}{R}\eta^{\mu\nu}\partial_5\left (e^{-\Phi}\frac{R}{z}\partial_5 \right )\right )A_\nu\nonumber\\
-\frac{e^{-\Phi}}{2}\frac{R}{z}A_5\partial^2A_5 +\frac{\xi e^{-\Phi}}{2}\frac{R}{z}A_5\;\partial_5\left (e^{\Phi}\frac{z}{R}\partial_5\left (e^{-\Phi}\frac{R}{z}A_5 \right )\right )\bigg ].
\end{eqnarray}
Again we shall work in the unitary gauge ($\xi\rightarrow\infty$) and so neglect contributions from ghosts and the unphysical $A_5$ goldstone bosons. After Fourier transforming with respect to the four large dimensions, such that $p_\mu=i\partial_\mu$, the gauge propagator is given by
\begin{equation}
\label{ }
\langle A^\mu A^\nu\rangle=-iG_p(z,z^{\prime})\left (\eta^{\mu\nu}-\frac{p^\mu p^\nu}{p^2}\right )
\end{equation}
where
\begin{equation}
\label{Propag}
\left (e^{\Phi}\frac{z}{R}\partial_5\left (e^{-\Phi}\frac{R}{z}\partial_5 \right )+p^2 \right )G_p(z,z^{\prime})=e^{\Phi}\frac{z}{R}\delta(z-z^{\prime}).
\end{equation}
The point of all this is that in order to obtain a 4D effective theory one just has to integrate out the Green's function $G_p(z,z^{\prime})$. With a dilaton of the form $\Phi=\frac{z^2}{R^{\prime\; 2}}$, the most general solution of Eq. \ref{Propag} is
\begin{equation}
\label{ }
G_p(z,z^{\prime}) =
\begin{cases}
Az^2M(1-a,2,\frac{z^2}{R^{\prime\;2}})+Bz^2U(1-a,2,\frac{z^2}{R^{\prime\;2}}) & \text{if }z<z^{\prime} \\
 Cz^2M(1-a,2,\frac{z^2}{R^{\prime\;2}})+Dz^2U(1-a,2,\frac{z^2}{R^{\prime\;2}}) & \text{if }z>z^{\prime}
\end{cases}
\end{equation}
where $A,B,C,D$ are constants of integration, we have also introduced the quantity
\begin{displaymath}
a\equiv\frac{R^{\prime\; 2}p^2}{4}.
\end{displaymath}
 While $M(\alpha,\beta,x)$ and $U(\alpha,\beta,x)$ are confluent hypergeometric or Kummer functions\footnote{Note here we use the notation of \cite{abramowitz+stegun} although $M(\alpha,\beta,x)$ can be alternatively denoted by ${}_1F_1(\alpha,\beta,x)$ or $\Phi(\alpha,\beta,x)$ and likewise $U(\alpha,\beta,x)$ can be denoted $x^{-\alpha}{}_2F_0(\alpha,1+\alpha-\beta);\; ;-\frac{1}{x})$ or $\Psi(\alpha,\beta,x)$.}. Throughout the remainder of this section we will repeatedly use relations taken from \cite{ abramowitz+stegun,Slater, Buchholz}. To proceed further, following \cite{Randall:2001gb}, we introduce $u=\min(z,z^{\prime})$ and $v=\max(z,z^{\prime})$ and impose the `continuity' condition that matches the two solutions at $z=z^{\prime}$, i.e. $G_p(u,u)=G_p(v,v)$
 \begin{eqnarray*}
 \normalsize
G_p(u,v)=Nu^2v^2\left (AM\left (1-a,2,\frac{u^2}{R^{\prime\;2}}\right )+BU\left (1-a,2,\frac{u^2}{R^{\prime\;2}}\right ) \right )\hspace{2cm}\\
\times\left (CM\left (1-a,2,\frac{v^2}{R^{\prime\;2}}\right )+DU\left (1-a,2,\frac{v^2}{R^{\prime\;2}}\right ) \right ).
\normalsize
\end{eqnarray*}
Here $N$ is a normalisation constant found by integrating over Eq. \ref{Propag} to get the `jump' condition
\begin{displaymath}
\lim_{\epsilon\rightarrow 0}\partial_zG_p(u,v)\lvert_{z^{\prime}-\epsilon}^{z^{\prime}+\epsilon}=\exp \left(\frac{z^{\prime\;2}}{R^{\prime\;2}}\right )\frac{z^{\prime}}{R} 
\end{displaymath}
giving
\begin{displaymath}
N=\frac{\Gamma(1-a)}{2(BC-AD)RR^{\prime\;2}}
\end{displaymath}
where we have used the relation
\begin{displaymath}
\frac{e^x}{x\left (M(1-a,2,x)U(-a,2,x)+(1+a)U(1-a,2,x)M(-a,2,x)\right )}=\Gamma(1-a).
\end{displaymath}
To fix the constants of integration we impose Neumann boundary conditions on the UV brane, i.e. $\partial_uG_p(u,v)\vert_{u=R}=0$, to get
\begin{eqnarray}
A(a)=(\Omega^{-2}-a)U(1-a,2,\Omega^{-2})-U(-a,2,\Omega^{-2})\label{Aconst},\\
B(a)=(a-\Omega^{-2})M(1-a,2,\Omega^{-2})-(1+a)M(-a,2,\Omega^{-2})\label{Bconst}.
\end{eqnarray}
Since there is no IR brane the determination of $C$ and $D$ is a little more subtle. Here we replace one of the boundary conditions with a `normalisability' condition that dictates that the propagator be comprised of KK modes which are normalisable with respect to $\int_R^{\infty}dz\;e^{-\Phi}\frac{R}{z}f_n^2=1$. This condition implies that
\begin{equation}
\label{IRNorm}
\int_R^{\infty}dz\;e^{-\Phi}\frac{R}{z}G_p(z,z)=\sum_n\int_R^{\infty}dz\;e^{-\Phi}\frac{R}{z}\frac{f_n^2(z)}{p^2-m_n^2}\sim\sum_n\frac{1}{n}.
\end{equation}
Hence we require that the integral, Eq. \ref{IRNorm}, be logarithmically divergent. For large $x$ Kummer functions scale as $M(\alpha,\beta,x)\sim\frac{\Gamma(\beta)}{\Gamma(\alpha)}e^{x}x^{\alpha-\beta}$ and $U(\alpha,\beta,x)\sim x^{-\alpha}$. The integrand will then scale as
\begin{displaymath}
\int_R^{\infty}dz\;e^{-\Phi}\frac{R}{z}G_p(z,z)\sim \int_R^{\infty}dz \left (AC e^{\frac{z^2}{R^{\prime\;2}}}z^{-4a-1}+(AD+BC)z^{-1}+BDe^{-\frac{z^2}{R^{\prime\;2}}}z^{4a-1}\right ).
\end{displaymath}
The last term is clearly convergent and so with $A$ and $B$ already fixed, only by setting $C=0$ can one obtain a logarithmic divergence. This then results in D being arbitrary and the full propagator being given by
\begin{equation}
\label{ }
G_p(u,v)=-\frac{\Gamma(1-a)u^2v^2}{2ARR^{\prime\;2}}\left (AM\left(1-a,2,\frac{u^2}{R^{\prime\;2}}\right )+BU\left (1-a,2,\frac{u^2}{R^{\prime\;2}}\right )\right )U\left (1-a,2,\frac{v^2}{R^{\prime\;2}}\right )
\end{equation}
Note the KK masses will be given by the poles of the propagator, i.e. when $A(a)=0$. Clearly $G_p(u,v)$ will have the same form for the gluons, although the W/Z propagator will be slightly deformed.

\subsection{ The W/Z Gauge Fields}\label{sect:Zmass}
\color{red}
Before considering the W/Z propagator it is useful to first consider the individual KK modes. Working post spontaneous symmetry breaking, we add to Eq. \ref{guageaction} a mass term of the form
\begin{displaymath}
\Delta\mathcal{L}=\frac{1}{4}h(z)^2g^2A_\mu A^\mu,
\end{displaymath} 
where $g$ is the 5D coupling to the Higgs (or for the Z boson $\frac{1}{4}h(z)^2(g^2+g^{\prime\;2})A_\mu A^\mu$) and $h$ is the Higgs VEV (\ref{higgsvev}). Carrying out the usual KK decomposition, $A_\mu=\sum_n f_n(z)A_\mu^{(n)}(x^\mu)$, such that 
\begin{equation}
\label{SWGaugeOrthog}
\int_R^{\infty}dz\;e^{-\Phi}\frac{R}{z}f_nf_m=\delta_{nm},
\end{equation}
the gauge profile will then be given by
\begin{equation}
\label{Zeqn}
\left (\partial_5^2-\left(\frac{2z}{R^{\prime\;2}}+\frac{1}{z}\right )\partial_5-\frac{g^2h_0^2}{2Rz^2}\left(\frac{z}{R^{\prime}}\right )^{2\alpha}+m_n^2\right )f_n=0.
\end{equation}
When $\alpha=2$ this can be solved to give
\begin{equation}
\label{Zprofile}
f_n(z)=Nz^2\exp \left(\frac{1}{2}\frac{z^2}{R^{\prime\;2}}(1-\zeta)\right )\;U\left (1-\tilde{a}_n,2,\zeta\frac{z^2}{R^{\prime\;2}}\right ),
\end{equation}
where we have introduced the quantities
\begin{displaymath}
\zeta\equiv\sqrt{\frac{g^2h_0^2}{2R}+1} \quad\mbox{and}\quad\tilde{a}_n\equiv\frac{m_n^2R^{\prime\;2}}{4\zeta}. 
\end{displaymath}
Note on sending $g^2h_0^2\rightarrow 0$, $\zeta\rightarrow 1$ and bearing in mind that $U(\alpha,\beta,x)=x^{1-\beta}U(1+\alpha-\beta,2-\beta,x)$, one regains the gauge profiles for massless gauge fields found in \cite{Batell:2008me}. We have also imposed the normalisability condition Eq. \ref{SWGaugeOrthog} resulting in the coefficient in front of the $M(\alpha,\beta,x)$ part of the solution being set to zero. Imposing Neumann boundary conditions on the UV brane ($\partial_5f_n\lvert_{z=R}=0$) then gives 
\begin{displaymath}
\left ((1+\zeta)\Omega^{-2}-2\tilde{a}_n\right )U\left (1-\tilde{a}_n,2,\zeta\Omega^{-2}\right )-2U\left (-\tilde{a}_n,2,\zeta\Omega^{-2}\right )=0.
\end{displaymath}  
Once again as $\zeta\rightarrow 1$, the KK masses are the same as those given by the poles of the propagator Eq. \ref{Aconst}. $\zeta$ can then be found by solving
\begin{displaymath}
\left ((1+\zeta)\Omega^{-2}-\frac{M_{W/Z}^2R^{\prime\;2}}{2\zeta}\right )U\left (1-\frac{M_{W/Z}^2R^{\prime\;2}}{4\zeta},2,\zeta\Omega^{-2}\right )=2U\left (-\frac{M_{W/Z}^2R^{\prime\;2}}{4\zeta},2,\zeta\Omega^{-2}\right ).
\end{displaymath}
With $M_{\rm{KK}}\approx 1-10$ TeV, then $\zeta_Z\approx 1.28-1.0028$ and hence when $M_{\rm{KK}}\ngg M_{W/Z}$ then $g^2\sim\mathcal{O}(R)$. In other words, provided the KK scale is not too large, one does not have to introduce any couplings that are very large or very small in order to generate the correct W and Z masses. 

We can now estimate $h_0$ by comparison with electroweak observables. Ideally one should compare to all observables in particular the Fermi constant. However, since one of the motivations for studying this model was its relatively small electroweak corrections, it is reasonable to just fit the gauge couplings and $h_0$ to three observables. In particular if one fits to the Z mass, fine structure constant, $\alpha$, and weak mixing angle, $\theta_w$, then
\begin{displaymath}
4\pi \alpha=g^2s_w^2f_0^2,
\end{displaymath}
where $s_w^2=\sin^2 \theta_w$ and $c_w^2=1-s_w^2$. We have also introduced the flat normalised photon gauge profile
\begin{equation}
\label{Fzero}
f_0=\sqrt{\frac{2}{R\;\mbox{E}_1(\Omega^{-2})}}\quad \mbox{and} \quad \mbox{E}_1(x)=\int_x^\infty dt\frac{e^{-t}}{t}.
\end{equation} 
Hence for a quadratic Higgs VEV we find
\begin{displaymath}
h_0^2\approx\frac{(\zeta_Z^2-1)c_w^2s_w^2}{\pi\alpha\; \mbox{E}_1(\Omega^{-2})}.
\end{displaymath}
By fitting the couplings of the five dimensional gauge field, $g$ and $g^{\prime}$, directly to observables we are assuming that the ratio of the 5D couplings is the same as the ratio of the 4D effective couplings. Hence we are neglecting $\mathcal{O}(M_{W/Z}^2/M_{\rm{KK}}^2)$ corrections to both the W and Z couplings and masses. One can check the validity of this approximation by numerically verifying that
\begin{displaymath}
\frac{\zeta_W^2-1}{\zeta_Z^2-1}=\hat{c}_w^2\approx c_w^2,
\end{displaymath}    
which gives $\hat{c}_w^2=0.7564,\;0.7716,\;0.7759$ and $0.7771$ for $M_{\rm{KK}}=1,\;2,\;4$ and $10$ TeV. Using these relations $h_0M_{\rm{KK}}\approx 262-245$ GeV for $\Omega=10^{15}$ and $M_{\rm{KK}}=1-10$ TeV. Having obtained $h_0$ one can now repeat the analysis of the previous section to obtain the W/Z propagator using
\begin{displaymath}
\left (\partial_5^2-\left(\frac{2z}{R^{\prime\;2}}+\frac{1}{z}\right )\partial_5-\frac{g^2}{Rz^2}\left(\frac{z}{R^{\prime}}\right )^{4}+p^2\right )G_p^{(W,Z)}(z,z^{\prime})=e^\Phi\frac{z}{R}\delta(z-z^{\prime}),
\end{displaymath}
to give
\begin{eqnarray}
G_p^{(W/Z)}(u,v)=-\frac{\zeta\Gamma(1-\zeta)u^2v^2\exp\left(\frac{1}{2}\frac{u^2}{R^{\prime\;2}}(1-\zeta)+\frac{1}{2}\frac{v^2}{R^{\prime\;2}}(1-\zeta)\right )}{2ARR^{\prime\;2}}\Bigg (AM\left (1-\tilde{a},2,\frac{\zeta u^2}{R^{\prime\;2}}\right )\nonumber\\
+BU\left (1-\tilde{a},2,\frac{\zeta u^2}{R^{\prime\;2}}\right )\Bigg)\;U\left (1-\tilde{a},2,\frac{\zeta v^2}{R^{\prime\;2}}\right ),
\end{eqnarray}
where $\tilde{a}\equiv\frac{p^2R^{\prime\;2}}{4\zeta}$ and 
\begin{eqnarray}
A(\tilde{a})=\left (\Omega^{-2}(1+\zeta)-2\tilde{a}\right )U\left (1-\tilde{a},2,\zeta \Omega^{-2}\right )-2 U\left (-\tilde{a},2,\zeta \Omega^{-2}\right ),\\
B(\tilde{a})=\left (2\tilde{a}-\Omega^{-2}(1+\zeta)\right )M\left (1-\tilde{a},2,\zeta\Omega^{-2}\right )-2(1+\tilde{a})M\left (-\tilde{a},2,\zeta\Omega^{-2}\right ).
\end{eqnarray}
As already mentioned, for the majority of this paper we shall focus on the quadratic Higgs VEV ($\alpha=2$). None the less it is worth pointing out the special case of a linear VEV. With $\alpha=1$, Eq. \ref{Zeqn} can be solved to give 
\begin{equation}
\label{ }
f_n=Nz^2U\left (\hat{\zeta}-a_n,2,\frac{z^2}{R^{\prime\;2}}\right ),\quad\mbox{where}\quad\hat{\zeta}=1+\frac{g^2h_0^2}{8R}\quad\mbox{and}\quad a_n=\frac{m_n^2R^{\prime\;2}}{4}.
\end{equation}
Once again imposing Neumann boundary conditions on the UV brane gives the condition
\begin{displaymath}
2(1-\hat{\zeta}+a_n)zN\left ((a_n-\hat{\zeta})U\left (1+\hat{\zeta}-a_n,2,\Omega^{-2}\right )+U\left (\hat{\zeta}-a_n,2,\Omega^{-2}\right )\right )=0,
\end{displaymath} 
which in turn implies $\hat{\zeta}=1+\frac{M_{W/Z}^2R^{\prime\;2}}{4}$ and
\begin{displaymath}
f_0=Nz^2U\left (1,2,\frac{z^2}{R^{\prime\;2}}\right )=N.
\end{displaymath}
Hence in the case of a linear VEV ($\alpha=1$), the zero mode profile of the W and Z gauge fields are flat. This can alternatively be seen by noting that $\partial_5f_0=0$ satisfies Eq. \ref{Zeqn} when $g^2h_0^2=2m_0^2RR^{\prime\;2}$. In models with warped extra dimensions, the deformation of the W and Z zero mode gives rise to a number of significant constraints, for example corrections to the $Z\bar{b}b$ vertex, rare lepton decays and corrections to electroweak observables, in particular the $S$ parameter. One would anticipate that a flat $W/Z$ profile would lead to a significant suppression of such constraints, although one would expect that it would also be difficult to generate the fermion mass hierarchy with a linear VEV. 
\color{black}

\subsection{Convergence of Four Fermion Operators}

We are now in a position to test whether or not the coefficients in front of the four fermion operators, of interest to flavour physics, are divergent (at tree level) with respect to KK number. As demonstrated in section \ref{FlavourChap}, these coefficients are determined by the integral
\begin{equation}
\label{PropILLLL}
\mathcal{I}=\int_R^{\infty}dz\int_R^{\infty}dz^{\prime}\; f_{L/R}^i(z)f_{L/R}^i(z)G_p(u,v)f_{L/R}^j(z^{\prime})f_{L/R}^j(z^{\prime}).
\end{equation} 
Typically this integral cannot be done analytically although one can make a small momentum (i.e. small $a$) approximation\footnote{Here we have used that $M(1,2,x)=\frac{e^x-1}{x}$ and Taylor expanded
\begin{eqnarray*}
x^2U\left (1-a,2,\frac{x^2}{R^{\prime\;2}}\right )\approx \frac{R^{\prime\;2}}{\Gamma(1-a)}\left (1+\frac{ax^2}{R^{\prime\;2}}\left [1-2\gamma-\Psi(1-a)-\ln (\frac{x^2}{R^{\prime\;2}})+\frac{x^2}{R^{\prime\;2}}\left (\frac{5-2\Psi(2-a)-4\gamma-2\ln(\frac{x^2}{R^{\prime\;2}}}{4}\right )\right ]+\mathcal{O}\left (\frac{a^2x^4}{R^{\prime\;4}}\right )\right )\\\approx\frac{R^{\prime\;2}}{\Gamma(1-a)}\left (1+\frac{ax^2}{R^{\prime\;2}}\right )+\mathcal{O}(a^2),\hspace{9.35cm}\end{eqnarray*}
where $\Psi$ is a digamma function and $\gamma$ is the Euler constant.} 
\begin{equation}
\label{ }
G_p(u,v)\approx\frac{R^{\prime\;2}}{2R}\left (1-\exp\left (\frac{u^2}{R^{\prime\; 2}}\right )-\frac{B(a)}{A(a)}\right )\left (1+\frac{av^2}{R^{\prime\;2}}+\mathcal{O}(a^2)\right ).
\end{equation}
This allows Eq. \ref{PropILLLL} to be approximated as
\begin{eqnarray}
\mathcal{I}\approx N^{2}_iN^{2}_j\frac{R^{\prime\;2}}{4R}\Bigg [\int_R^{\infty}dz z^{-2c_0^i}\exp\left (\frac{-c_1^iz^2}{R^{\prime\;2}}\right )\left (1-\exp\left (\frac{z^2}{R^{\prime\; 2}}\right )-\frac{B}{A}\right )\Bigg (\left (\frac{c_1^j}{R^{\prime\;2}}\right )^{c_0^j-\frac{1}{2}}\Gamma\left (\frac{1}{2}-c_0^j,\frac{z^2c_1^j}{R^{\prime\;2}}\right ) \nonumber\\
+\frac{a}{R^{\prime\;2}}\left (\frac{c_1^j}{R^{\prime\;2}}\right )^{c_0^j-\frac{3}{2}}\Gamma\left (\frac{3}{2}-c_0^j,\frac{z^2c_1^j}{R^{\prime\;2}}\right )\Bigg )+\int_R^{\infty}dz^{\prime}\left ((z\rightarrow z^{\prime}),\;(i\leftrightarrow j) \right )\Bigg ].
\end{eqnarray} 
The point is that this can be approximately evaluated using
\begin{displaymath}
\int_0^{\infty}dx\;x^{\mu-1}e^{-\beta x}\Gamma(\nu,\alpha x)=\frac{\alpha^\nu\Gamma(\mu+\nu)}{\mu(\alpha+\beta)^{\mu+\nu}}{}_2F_1\left (1,\mu+\nu;\mu+1;\frac{\beta}{\alpha+\beta}\right )
\end{displaymath}
 but only when $c_1^i+c_1^j-1>0$ and $c_0^i,c_0^j<\frac{1}{2}$ (i.e. only when the fermions are localised towards the IR) \cite{Gradshteyn+Ryzhik}. Alternatively one can look to see when the integrand blows up. Making a large $x$ expansion of the Kummer functions gives 
\begin{eqnarray}
\mathcal{I}\approx\frac{N_i^2\Gamma(1-a)N_j^2}{2ARR^{\prime\;2}}\Bigg [\int_R^\infty dz \int_{z^{\prime}=z}^\infty dz^{\prime} z^{2-2c_0^i}\exp\left (\frac{-c_1^iz^2}{R^{\prime\;2}}\right )\Bigg [\frac{A}{\Gamma(1-a)}\exp\left (\frac{z^2}{R^{\prime\;2}}\right )\left (\frac{z^2}{R^{\prime\;2}}\right )^{-1-a}\nonumber \\
+B\left (\frac{z^2}{R^{\prime\;2}}\right )^{a-1}\Bigg ]\left (\frac{z^{\prime\;2}}{R^{\prime\;2}}\right )^{a-1}z^{\prime\;2-2c_0^j}\exp\left (\frac{-c_1^jz^{\prime\; 2}}{R^{\prime\;2}}\right )+\int_R^\infty dz^{\prime} \int_{z=z^{\prime}}^\infty dz \left ((z\leftrightarrow z^{\prime}),\;(i\leftrightarrow j) \right )\Bigg ].
\end{eqnarray}
Clearly for large $z$ the integrand will be dominated by the exponential and so will go to zero only when, once again,
\begin{equation}
\label{divergenceCondition}
c_1^i+c_1^j-1>0.
\end{equation} 
It is straight forward to check this empirically, for the exact integrand, and it is found to hold for every case checked. So, although we have not been able to prove it explicitly, here we strongly suspect that the coefficients of four fermion operators will only be convergent when Eq. \ref{divergenceCondition} is satisfied. That is to say the four fermion operator coefficients will be divergent if the fermions are too weakly coupled to what ever is bounding them to the space.

In practice, for large warp factors, the integral Eq. \ref{PropILLLL} is difficult to do even numerically. In order to carry out the scans over parameter space required to study flavour one finds it is easier to work with the individual KK modes. In other words Eq. \ref{PropILLLL} can be equated to
\begin{equation}
\label{ }
\mathcal{I}=\sum_{n=0}^\infty\frac{g_n^ig_n^j}{p^2-m_n^2}\quad\mbox{where}\quad g_n^i=\int_R^{\infty}dz f_{L/R}^i(z)f_n(z)f_{L/R}^i(z)
\end{equation} 
 where $f_n$ is the gauge profile (Eq. \ref{Zprofile} with $\zeta=1$). The question then remains, how many KK modes should be summed over? In table \ref{PropCoverge} we compare this convergence for a low warp factor. Note, at momenta much lower than the KK scale, the propagator is dominated by the zero mode which we have subtracted off. 
 
 As one would have expected the convergence appears to be logarithmic and hence one would always anticipate a slight error in using an expansion in KK modes. In practice we sum over the first 15 KK modes and hence would expect an error of the order of a few percent which we consider acceptable. 
  
  \begin{table}[h!]
  \centering 
  \begin{tabular}{|c|c||c|c|c|c|}
  \hline
$c_0^i$ & $c_0^j$ & Propagator & 5 KK modes & 10 KK modes & 20 KK modes \\
\hline
\hline
   $ 0.3 $& $0.4$ & $1.7550\times 10^{-8}$ & $1.6729\times 10^{-8}$ & $1.7158\times 10^{-8}$ & $1.7325\times 10^{-8} $\\
     $ 0.3 $& $0.65$ & $-1.8710\times 10^{-8}$ & $-1.7570\times 10^{-8}$ & $-1.8202\times 10^{-8}$ & $-1.8471\times 10^{-8} $\\
  $ 0.6 $& $0.65$ & $6.8463\times 10^{-9}$ & $6.1236\times 10^{-9}$ & $6.4615\times 10^{-9}$ & $6.6265\times 10^{-9} $ \\
\hline
\end{tabular}
\caption{In the propagator column is the quantity $\frac{1}{p^2}-\frac{\mathcal{I}}{f_0^2}$. While in the remaining columns is the quantity $-\frac{1}{f_0^2}\sum_{n=1}^{5, 10, 20}\frac{g_n^ig_n^j}{p^2-m_n^2}$. Here $\Omega=10^2$, $c_1=1$, $p=10$ GeV and $M_{\rm{KK}}\equiv\frac{1}{R^{\prime}}=1$ TeV.} \label{PropCoverge} 
\end{table} 

  \begin{figure}[ht!]
    \begin{center}
        \subfigure[]{%
           \label{fig:FermCoupl}
           \includegraphics[width=0.73\textwidth]{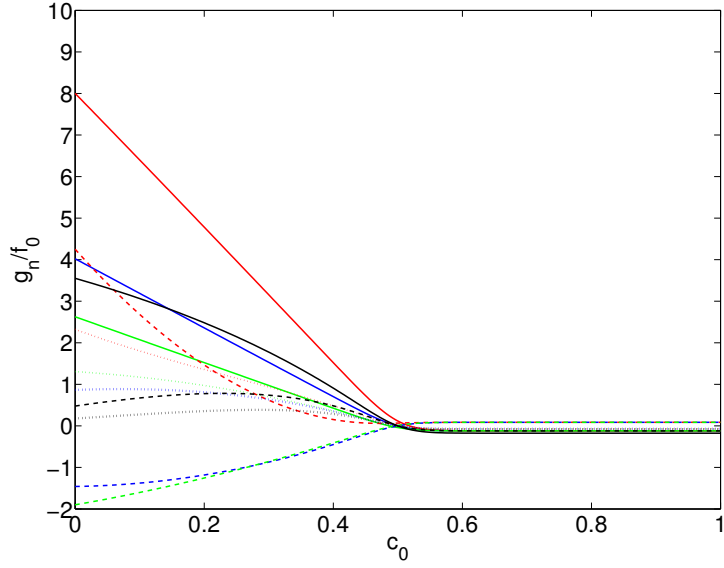}
        }
        \subfigure[]{%
           \label{fig:UVFermCoupl}
           \includegraphics[width=0.73\textwidth]{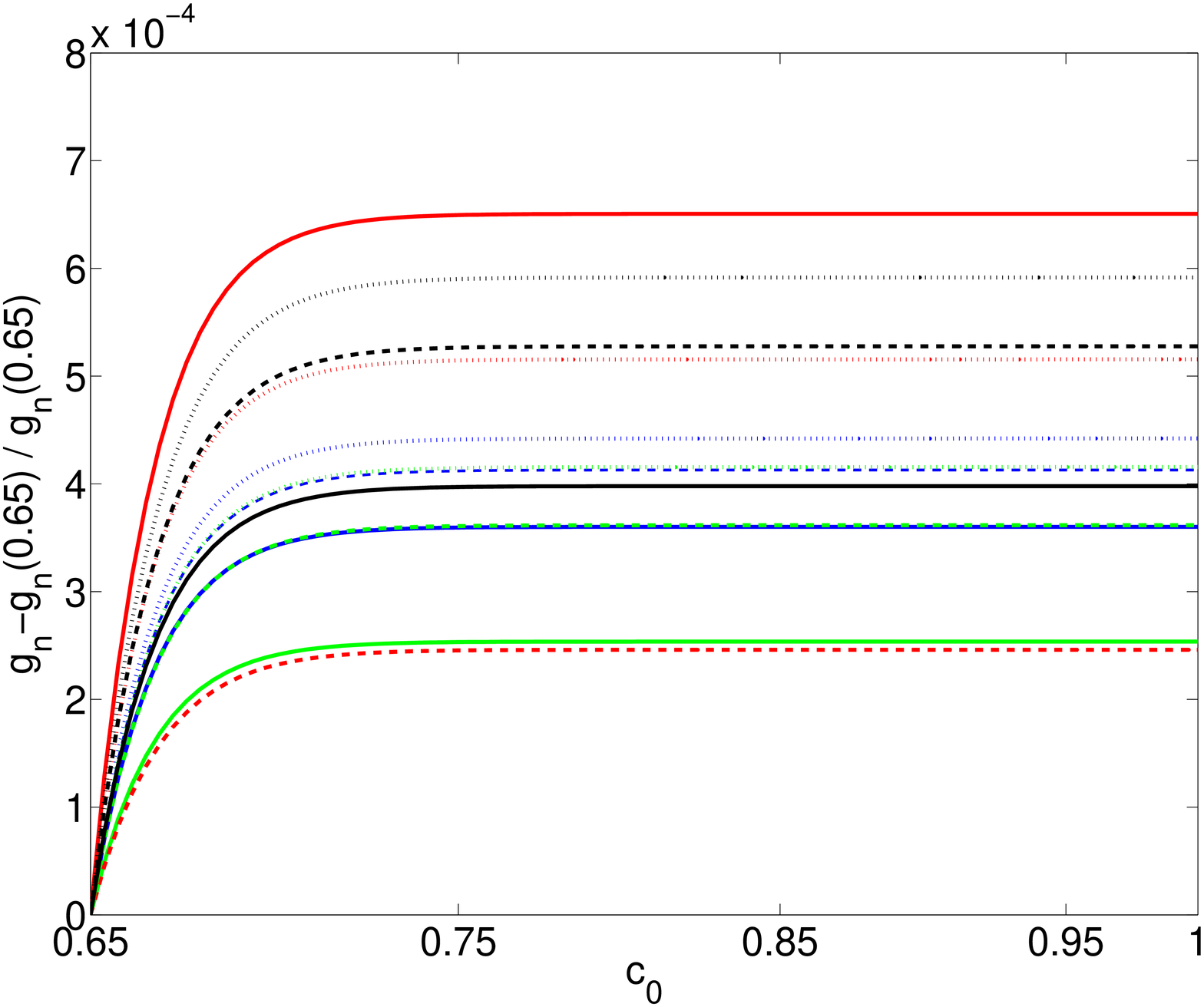}
        }
    \end{center}
    \caption[Relative Gauge Fermion Coupling in the SW Model]{The relative coupling between the fermions and the first (solid lines), second (dashed lines) and third (dots) gauge KK modes. For the RS model (black) and the SW model with $c_1=0.5$ (red), $c_1=1$ (blue) and $c_1=1.5$ (green). Here $\Omega=10^{15}$ and $M_{KK}\equiv \frac{1}{R^{\prime}}=1\;\rm{TeV}$. The lower graph focuses on the universality of the couplings of fermions localised towards the UV brane.} \label{fermCoupl}
\end{figure}

Before moving on to look at the results it is worth briefly looking at the relative gauge fermion couplings plotted in figure \ref{fig:FermCoupl}. The only reason why a convergence of Eq. \ref{PropILLLL} is possible is because the higher KK modes are increasingly weakly coupled to the fermion zero modes. One can also see that the couplings of the RS model do appear to fall away more rapidly than those of the SW model, although it is not really possible to say anything concrete when considering just 3 modes. 

As discussed in section \ref{FlavourChap}, the scale of constraints from flavour physics is partly determined by the difference or non-universality of the coupling of different flavours, particularly when they are localised towards the UV brane ($c_0>\frac{1}{2}$). This has been plotted in figure \ref{fig:UVFermCoupl} for an arbitrary coupling of $c_0=0.65$. Firstly one can see that, as in the RS model, when the the fermions are sitting towards the UV brane the gauge fermion coupling is approximately universal. One can also see, from figure \ref{fig:UVFermCoupl}, that the SW model has an equivalent level of universality to that of the RS model. Critically one can also see, from figure \ref{fig:FermCoupl}, that when the fermions are localised towards the UV, differences in the couplings are dominated by differences in the $c_0$ bulk mass term and not the $c_1$ term. Hence here assuming universal $c_1$ parameters would not significantly change the results. However as one localises the fermions further and further towards the IR then one can see that the coupling becomes increasingly sensitive to the $c_1$ parameter.

\section{The Lepton Sector}\label{sect:Lept}
With respect to studies of flavour physics, the SW model suffers from a significant complication which can be seen by comparing Eq. \ref{SWMasses} with Eq. \ref{RSMasses}. In the RS Model the fermion zero mode masses are independent of the KK mass scale, hence one can find a configuration of $c$ parameters that gives the correct masses and mixings and use that throughout a study. This is no longer the case in the SW model and one is forced to solve for the $c$ parameters at each KK scale. This can make studies including scans over anarchic Yukawas quite involved.

Before considering the quark sector we shall first look at the tree level decays involving just the charged leptons. The advantage to this is that one need only fit to the lepton masses and this will allow us to demonstrate the central physics a little more clearly. In light of the current experimental status of the PMNS matrix one is inclined to favour configurations with large charged lepton mixings (i.e. closely spaced $c$ parameters). However here we will consider scenarios with both large and small mixings.  

We shall also consider only the zero modes of the fermions. One may be concerned that, when the full mass matrix $MM^\dag$ is diagonalised, the mixing of the zero modes would receive contributions from the terms that are off diagonal with respect to KK number. In the RS model these off diagonal terms are partially suppressed by the orthogonality of the fermion profiles, Eq. \ref{FermOrthog} \cite{Huber:2003tu}. One would expect a similar effect  being present in the SW model although it would be partially reduced due to the presence of the Higgs profile. As in the case of the RS model, one should be particularly cautious, about neglecting the contribution from the fermion KK modes, when one has matching $c_0$ values. Since this gives rise to particularly universal couplings which can, if the KK modes are not considered, exaggerate the suppression of FCNC's.          

\begin{figure}[ht!]
\begin{center}
\includegraphics[width=0.6\textwidth]{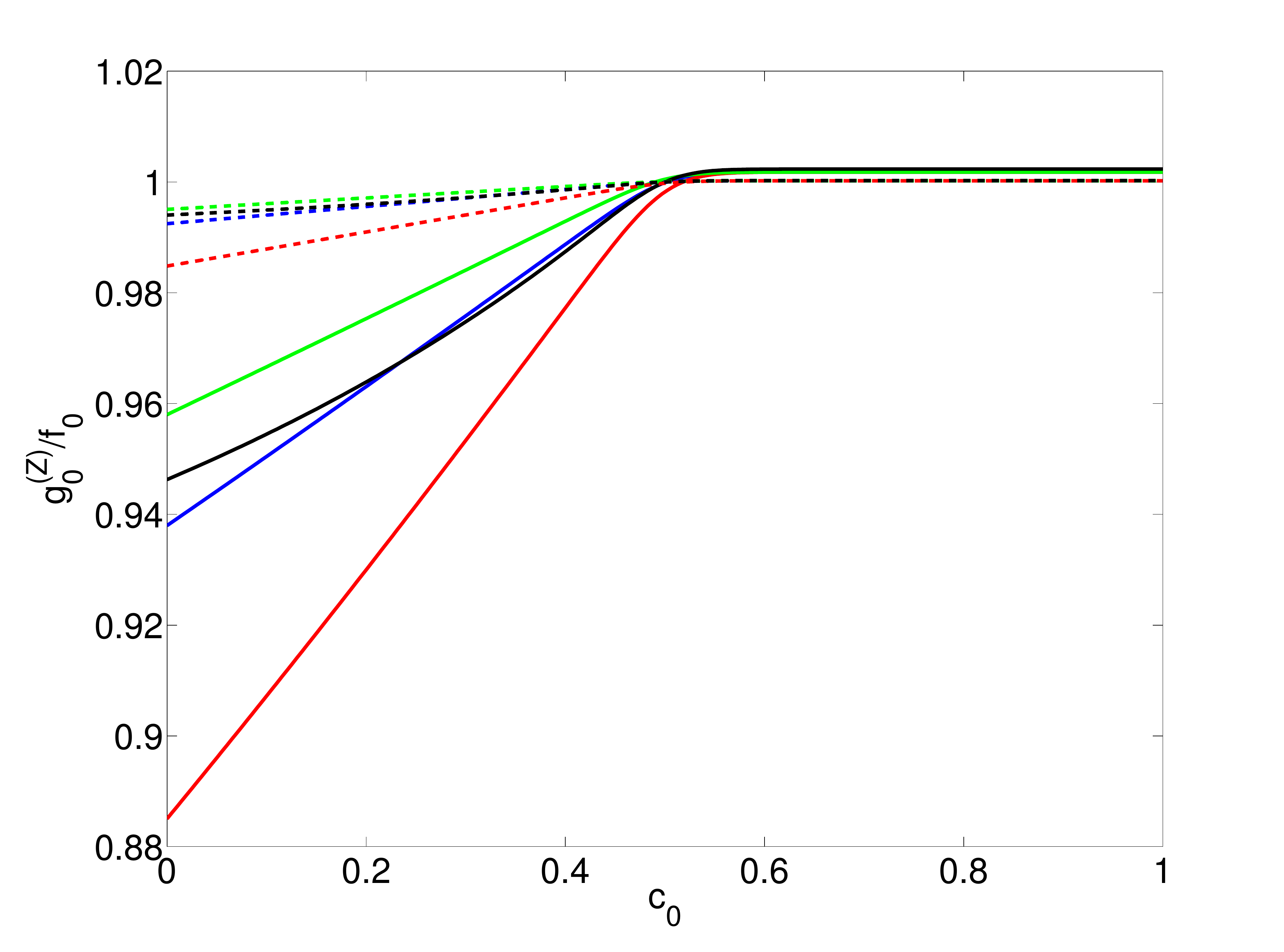}
\caption[Coupling of the Z Zero Mode in the SW Model]{The coupling of the Z zero mode for the RS model (black) and the SW model with $c_1=0.5$ (red), $c_1=1$ (blue) and $c_1=1.5$ (green). Here $\Omega=10^{15}$ while $M_{KK}= 1$ TeV (solid lines) and $M_{KK}= 3$ TeV (dashed lines). }
\label{fig:Zcoupling}
\end{center}
\end{figure}

The processes we will consider here is the tree level decay $l_j\rightarrow l_il_i\bar{l}_i$ which is given by \cite{Langacker:2000ju}
\begin{displaymath}
\Gamma(l_j\rightarrow l_il_i\bar{l}_i)=\frac{G_F^2m_{l_j}^5}{48\pi^3}\left (2|\mathcal{C}^L_{ij}|^2+2|\mathcal{C}^R_{ij}|^2+ |\mathcal{D}^L_{ij}|^2+|\mathcal{D}^R_{ij}|^2\right )
\end{displaymath}
where 
\begin{displaymath}
\mathcal{C}_{ij}^{L/R}=\sum_n\frac{M_Z^2}{m_n^2}\left (\mathcal{B}_{(n)}^{L/R}\right )_{ij}\left (\mathcal{B}_{(n)}^{L/R}\right )_{ii}\quad\mbox{and}\quad \mathcal{D}_{ij}^{L/R}=\sum_n\frac{M_Z^2}{m_n^2}\left (\mathcal{B}_{(n)}^{L/R}\right )_{ij}\left (\mathcal{B}_{(n)}^{R/L}\right )_{ii} 
\end{displaymath}
and
\begin{displaymath}
\mathcal{B}_{(n)}^{L/R}=\frac{1}{f_0}U_{L/R}\;g_n\;U_{L/R}^\dag.
\end{displaymath}
$U_{L,R}$ are the unitary matrices that diagonalise the fermion mass matrix. This process receives contributions from both the exchange of KK photons and KK Z bosons but in practice it is found to be completely dominated by the deformation of the Z zero mode plotted in figure \ref{fig:Zcoupling}. We do not include the contribution from the Higgs or KK modes of the Higgs. The current experimental bounds on these processes are \cite{Nakamura:2010zzi} 
\begin{eqnarray}
\mbox{Br}(\mu^{-}\rightarrow e^-e^+e^-)<1.0\times10^{-12} \nonumber\\ 
  \mbox{Br}(\tau\rightarrow \mu^-\mu^+\mu^-)<3.2\times10^{-8} \nonumber\\    
  \mbox{Br}(\tau^{-}\rightarrow e^-e^+e^-)<3.6\times10^{-8}.\nonumber
\end{eqnarray}

\subsection{Numerical Analysis and Results}
This slightly simplified study essentially serves two purposes. Firstly it allows us to look at the central physics for a relatively simple example. Secondly it allows us to test the two significant approximations made in this section, notably the introduction of the $z$ dependent mass, Eq. \ref{bulkmass}, and the assumption of a universal $c_1$ value. The first approximation is tested by comparing this study with some earlier work \cite{Atkins:2010cc}. The second approximation is rather crudely tested by assuming a universal $c_1$ value and then examining the results for any $c_1$ dependence.  
By considering just these three decays, it is reasonable to just fit to the three charged lepton masses. However one still has a sizeable number of input parameters. Although we assume a universal $c_1$ value we still allow for anarchic Yukawa couplings $\lambda$ in Eq. \ref{SWMasses}. We shall also assume real flavour diagonal $c_0$ values.

\begin{figure}[ht!]
    \begin{center}
        \subfigure[]{%
           \label{fig:ueee}
           \includegraphics[width=0.85\textwidth]{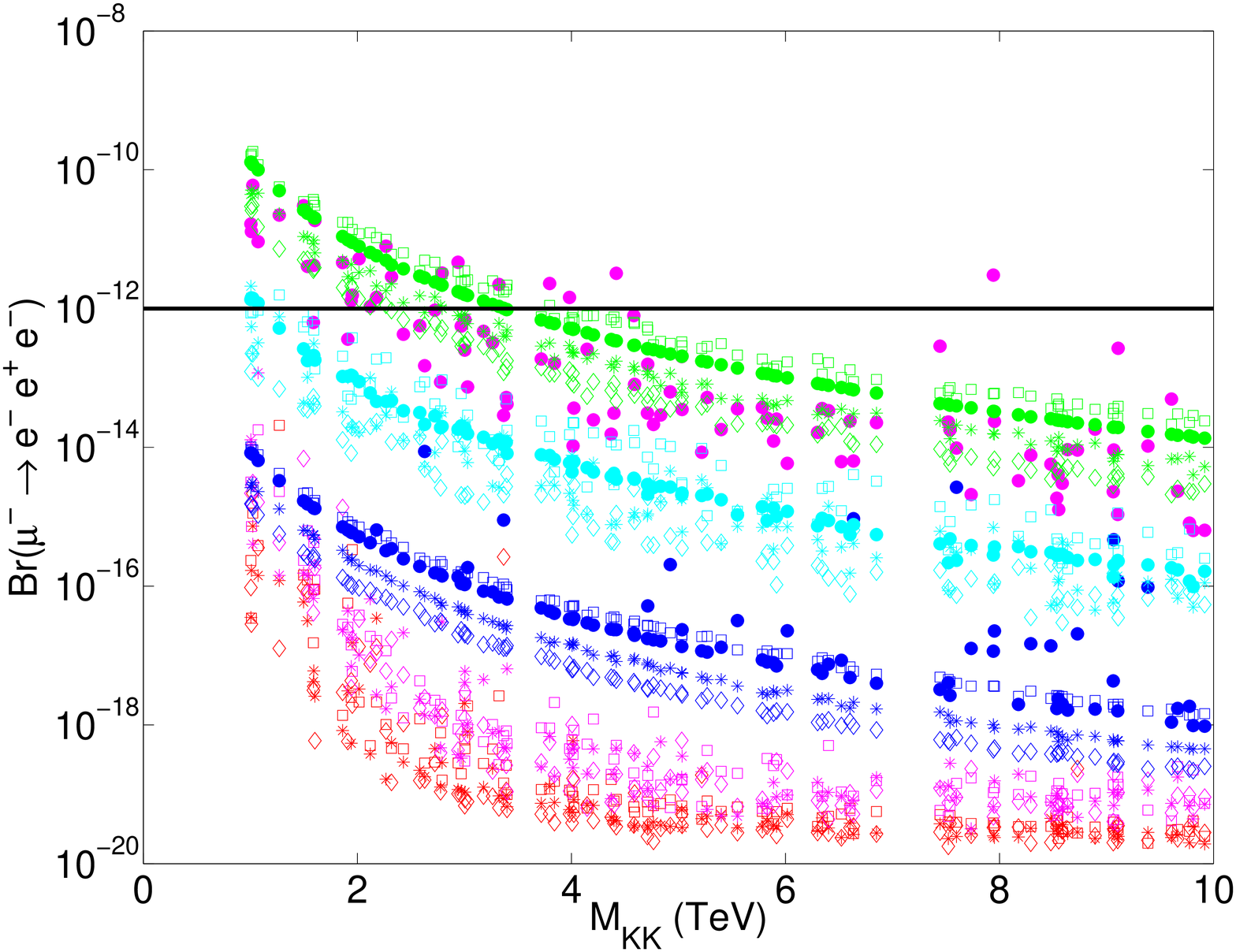}
        }\\
        \subfigure[]{%
           \label{fig:fig:teee}
           \includegraphics[width=0.47\textwidth]{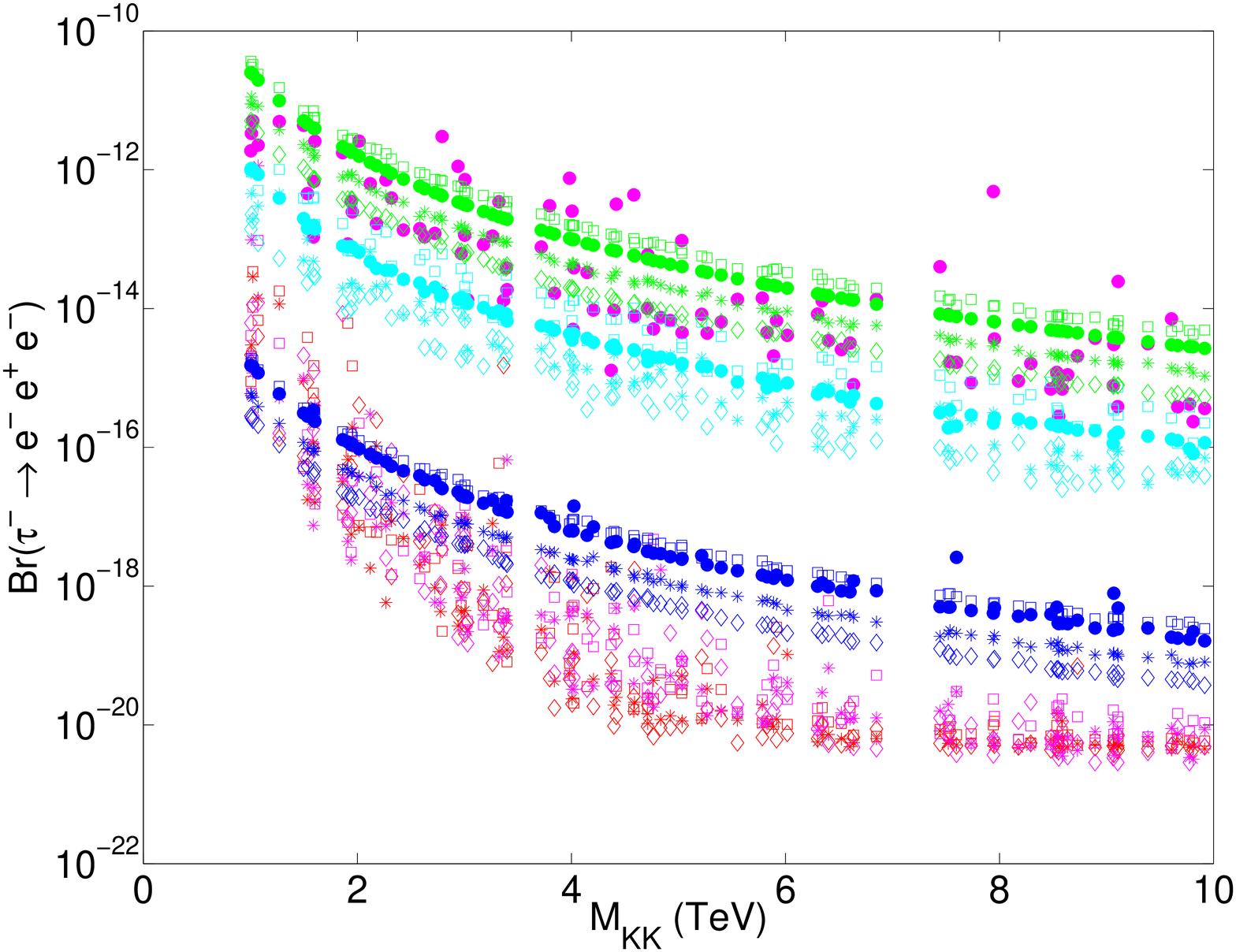}
        }
        \subfigure[]{%
           \label{fig:teee}
           \includegraphics[width=0.47\textwidth]{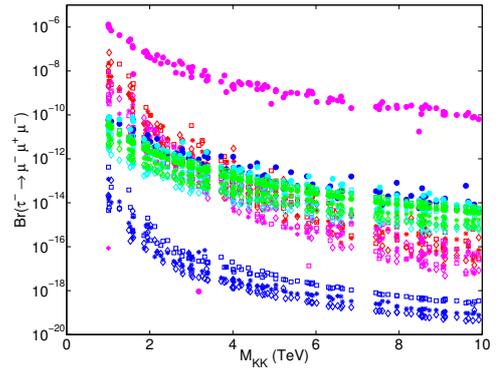}
        }
    \end{center}
    \caption[Rare Lepton Decays in the SW Model]{Branching ratios for rare lepton decays in the RS model (solid dots) and the SW model with $c_1=0.5$ (square), $c_1=1$ (star) and $c_1=1.5$ (diamond). The five configurations considered are given in Eq. \ref{leptConfig}, i.e. (A) in red, (B) in magenta, (C) in blue, (D) in cyan and (E) in green. Note the RS configuration (A) points have not been plotted due to the difficulty in obtaining a reasonable fit to the masses. $\Omega=10^{15}$.} \label{fig:leptonDecay}
\end{figure}

As already mentioned, in the SW model we must solve for the relevant $c_0$ values at each KK scale. Hence here we take the $c_L\;(c_0^{(L)})$ as an input parameter and randomly generate one $3\times 3$ complex Yukawa coupling, $|\lambda_{ij}|\in[1,3]$, allowing the $c_R$ values to be solved for by fitting to the lepton masses \cite{Nakamura:2010zzi}.
\begin{displaymath}
m_e=0.511\;\mbox{MeV}\quad m_\mu=106\;\mbox{MeV}\quad m_\tau=1780\;\mbox{MeV}. 
\end{displaymath}     
We then proceed to generate a further 10,000 Yukawa couplings and for each one compute the branching ratios and lepton masses. Inevitably most of these configurations will not give the correct masses and so we take the 100 configurations which give the most accurate masses. From these 100 configurations we plot the average of the branching ratios in figure \ref{fig:leptonDecay}. We then repeat this for a hundred random KK scales $M_{\rm{KK}}\equiv \frac{1}{R^{\prime}}\in [1,10]\;$TeV, three $c_1$ values and five $c_L$ values.
\begin{eqnarray}
(A)\quad c_L=[0.710,\;0.700,\;0.690]\quad\quad\quad(B)\quad c_L=[0.750,\;0.700,\;0.650] \nonumber\\
(C)\quad c_L=[0.601,\;0.600,\;0.599]\quad\quad\quad(D)\quad c_L=[0.650,\;0.600,\;0.550] \nonumber\\
(E)\quad c_L=[0.560,\; 0.550,\; 0.540]\label{leptConfig}
\end{eqnarray} 
We then proceed to compute the branching ratios for the RS model using exactly the the same method. In the literature there always appears to be a slight debate over what should be referred to as the KK scale. Here we define the KK scale to be $\frac{1}{R^{\prime}}$ for the SW model and $\frac{k}{\Omega}$ for the RS model. However in the interests of comparison, when plotting the RS points, we scale $M_{\rm{KK}}$ by a factor of $\frac{2.0}{2.45}$ such that the mass of the first KK gauge mode will, for both models be about two times $M_{\rm{KK}}$. The results are plotted in figure \ref{fig:leptonDecay}.     

Firstly it should be noted that the (C) configuration, with $c_1\approx 1$, is in good agreement with the equivalent calculation done without the $z$ dependent mass term \cite{Atkins:2010cc}. Secondly, as one would expect, there is an increasing $c_1$ dependence as one moves the $f_L$ profile towards the IR. Although this dependence is still small compared to changes in the $c_0$ parameters, e.g. going from configuration (C) to (D). None the less it is found that this dependence is negligible for configurations (A) and (B). Bearing in mind that these configurations typically have $c_0^R$ values sitting further towards the UV then here it is suspected that the assumption of universal $c_1$ values is good when $c_0^L\gtrsim 0.6\;(\mbox{ or }\;c_0^R\lesssim-0.6)$.    

These figures also highlight the implications of the reduced gradient in the range of fermion masses (figure \ref{fermmass}). One can clearly see that as one localises the left handed fermions closer and closer towards the UV brane (i.e. larger $c_L$ values) the scale of the branching ratios is reduced. In order to maintain the correct masses the corresponding right handed fermions must sitter closer towards the IR. Sooner or later this leads to a problem via either a large value of $\mbox{Br}(\tau^-\rightarrow \mu^-\mu^+\mu^-)$ or even a difficulty obtaining the correct masses. However, due to this reduced gradient, the corresponding $c_0^R$ are not as extreme in the SW model as in the RS model. Hence the RS model has problematic points in parameter space before the SW model. In other words the reduced gradient, in the SW model, gives rise to a larger phenomenologically viable parameter space than the RS model.  

Here we have considered a deliberately simplified scenario. In a more complete study, that included processes such as $\mu\rightarrow e\gamma$, it would be necessary to include both neutrino masses as well as mixings. This would restrict the range of possible configurations (\ref{leptConfig}) but one would anticipate that the basic results still hold. However we will now move on to consider the quark sector, or in particular the kaon sector, where clearly it is necessary to fit to mixings. 

\section{The Quark Sector}\label{sect:SWQuarksect}
\color{red}
Here we wish to investigate two points. Firstly, can the SW model reproduce the correct quark masses and mixing angles and secondly, to what extent are FCNC's suppressed. However, even if one assumes real, flavour-diagonal, bulk mass parameters and universal, order unity, $c_1$ values then one still has 18 complex Yukawas and 9 real $c_0$ parameters to fit. Such a large parameter space gives rise to an under constrained problem or in other words the existing constraints from flavour physics can always be satisfied with sufficient tuning of the free parameters. Hence the relevant question, we wish to address here, is which of the two models requires the least tuning in order to reproduce all existing observables. Ideally one should carry out a full Monte Carlo analysis, although accurately carrying out the integrals is numerically too slow for this to be a computationally viable option. Another possible approach would be to compute the fine tuning parameter (introduced in \cite{Barbieri88}) including all know observables. However this approach would offer no indication as to the `typical' size of a given observable in a given model. 

The approach taken here is to find points in parameter space that give the correct masses, mixing angles and Jarlskog invariant and then proceed to calculate the size of the additional contributions, to $\epsilon_K$ and $\Delta m_K$, from the KK gauge fields. In selecting such points one should be aware of two factors. Firstly, is the point fine tuned, i.e. are the output observables sensitive to small changes in the input parameters. Secondly, is the point a particularly rare point in parameter space. In order to address the second issue, here we endeavour to scan over as wide a range of the parameter space as is computationally viable. For the first point we will compute the fine tuning parameter at each point considered. To be a little more explicit our method is as follows.

\begin{table}[t!]
\small
  \flushleft
  \begin{tabular}{|c|c|c|c|}
\hline
   \multicolumn{2}{|c|}{Configuration}&$c_R^u$&$c_R^d$  \\
   \hline\hline
\multirow{4}{*}{(A)}&$c_1=0.5$&$[-0.66\pm0.04,\;-0.47\pm0.12,\;0.46\pm0.10]$& $[-0.63\pm0.01,\;-0.61\pm0.01,\;-0.57\pm0.01]$ \\
&$c_1=1$&$[-0.65\pm0.04,\;-0.45\pm0.11,\;0.47\pm0.07]$& $[-0.62\pm0.01,\;-0.60\pm0.01,\;-0.56\pm0.01]$ \\
&$c_1=1.5$&$[-0.64\pm0.03,\;-0.43\pm0.13,\;0.45\pm0.13]$& $[-0.62\pm0.01,\;-0.59\pm0.01,\;-0.56\pm0.01]$  \\ \cline{2-4}
&RS&$[-0.62\pm0.01,\;-0.44\pm0.05,\;4.63\pm1.98]$& $[-0.60\pm0.01,\;-0.58\pm0.01,\;-0.55\pm0.01]$  \\
\hline\hline
\multirow{4}{*}{(B)}&$c_1=0.5$&$[-0.69\pm0.03,\;-0.52\pm0.03,\;0.38\pm0.18]$& $[-0.66\pm0.01,\;-0.61\pm0.01,\;-0.60\pm0.01]$ \\
&$c_1=1$&$[-0.69\pm0.07,\;-0.46\pm0.14,\;0.46\pm0.13]$& $[-0.65\pm0.01,\;-0.61\pm0.01,\;-0.58\pm0.01]$ \\
&$c_1=1.5$&$[-0.67\pm0.01,\;-0.46\pm0.12,\;0.45\pm0.15]$& $[-0.65\pm0.01,\;-0.60\pm0.01,\;-0.58\pm0.01]$  \\ \cline{2-4}
&RS&$[-0.65\pm0.01,\;-0.48\pm0.04,\;1.09\pm0.67]$& $[-0.63\pm0.01,\;-0.59\pm0.01,\;-0.57\pm0.01]$  \\
\hline\hline
\multirow{4}{*}{(C)}&$c_1=0.5$&$[-0.72\pm0.04,\;-0.56\pm0.01,\;-0.25\pm0.22]$& $[-0.69\pm0.01,\;-0.65\pm0.01,\;-0.62\pm0.01]$ \\
&$c_1=1$&$[-0.71\pm0.01,\;-0.53\pm0.08,\;0.06\pm0.33]$& $[-0.68\pm0.01,\;-0.64\pm0.01,\;-0.61\pm0.01]$ \\
&$c_1=1.5$&$[-0.70\pm0.03,\;-0.52\pm0.06,\;0.30\pm0.32]$& $[-0.68\pm0.01,\;-0.63\pm0.01,\;-0.60\pm0.01]$  \\ \cline{2-4}
&RS&$[-0.68\pm0.01,\;-0.53\pm0.01,\;-0.06\pm0.14]$& $[-0.66\pm0.01,\;-0.62\pm0.01,\;-0.60\pm0.01]$  \\
\hline\hline
\multirow{4}{*}{(D)}&$c_1=0.5$&$[-0.74\pm0.01,\;-0.60\pm0.01,\;-0.37\pm0.08]$& $[-0.72\pm0.01,\;-0.67\pm0.01,\;-0.62\pm0.01]$ \\
&$c_1=1$&$[-0.73\pm0.01,\;-0.58\pm0.01,\;-0.23\pm0.21]$& $[-0.71\pm0.01,\;-0.67\pm0.01,\;-0.61\pm0.01]$ \\
&$c_1=1.5$&$[-0.73\pm0.01,\;-0.56\pm0.06,\;-0.15\pm0.24]$& $[-0.71\pm0.01,\;-0.66\pm0.01,\;-0.61\pm0.01]$  \\ \cline{2-4}
&RS&$[-0.72\pm0.01,\;-0.57\pm0.01,\;-0.29\pm0.06]$& $[-0.69\pm0.01,\;-0.65\pm0.01,\;-0.61\pm0.01]$  \\
\hline\hline
\multirow{4}{*}{(E)}&$c_1=0.5$&$[-0.77\pm0.01,\;-0.63\pm0.01,\;-0.43\pm0.03]$& $[-0.75\pm0.01,\;-0.71\pm0.01,\;-0.63\pm0.01]$ \\
&$c_1=1$&$[-0.76\pm0.01,\;-0.62\pm0.01,\;-0.36\pm0.07]$& $[-0.74\pm0.01,\;-0.71\pm0.01,\;-0.62\pm0.01]$ \\
&$c_1=1.5$&$[-0.76\pm0.01,\;-0.62\pm0.01,\;-0.21\pm0.24]$& $[-0.73\pm0.01,\;-0.70\pm0.01,\;-0.61\pm0.01]$  \\ \cline{2-4}
&RS&$[-0.74\pm0.01,\;-0.61\pm0.01,\;-0.35\pm0.05]$& $[-0.72\pm0.01,\;-0.69\pm0.01,\;-0.61\pm0.01]$  \\
\hline
\end{tabular}
\normalsize
  \caption[The bulk mass parameters for the right handed fermions]{Bulk mass parameters ($c_R=c_0^{Ri}$) of the right-handed fermions obtained by fitting to quark masses and mixing angles with the left-handed fermions having the bulk configurations; $(A)\; c_L=[0.72,\;0.64,\;0.52]$, $(B)\; c_L=[0.69,\;0.63,\;0.49] $, $(C)\; c_L=[0.66,\;0.60,\;0.42]$, $(D)\; c_L=[0.63,\;0.57,\;0.34] $, $(E)\; c_L=[0.60,\;0.52,\;0.25]$. Quoted is the mean and standard deviation taken over 50 random points, see text.}\label{Tab:BulkmassParams}
\end{table}

\begin{figure}[t!]
\begin{center}
\vspace{-1cm}
 \subfigure[]{%
           \label{Fig:Epk}
           \vspace{-4cm}
\includegraphics[width=0.65\textwidth]{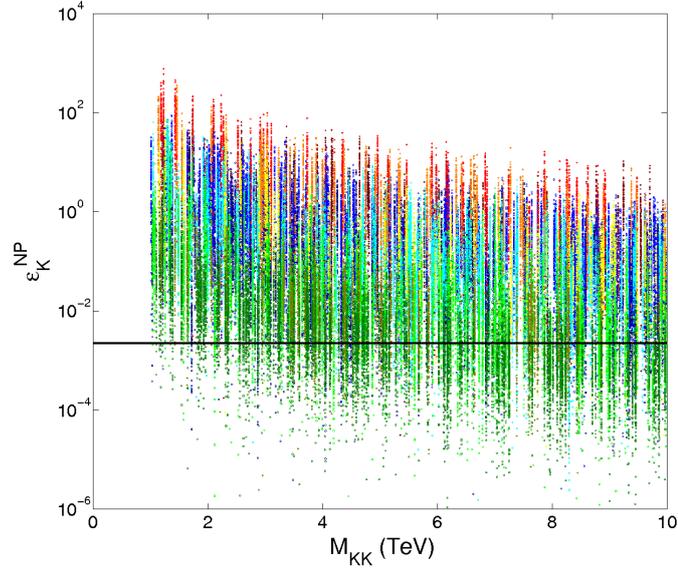}
        }
        \subfigure[]{%
           \label{Fig:MK}
           \vspace{-1cm}
           \includegraphics[width=0.65\textwidth]{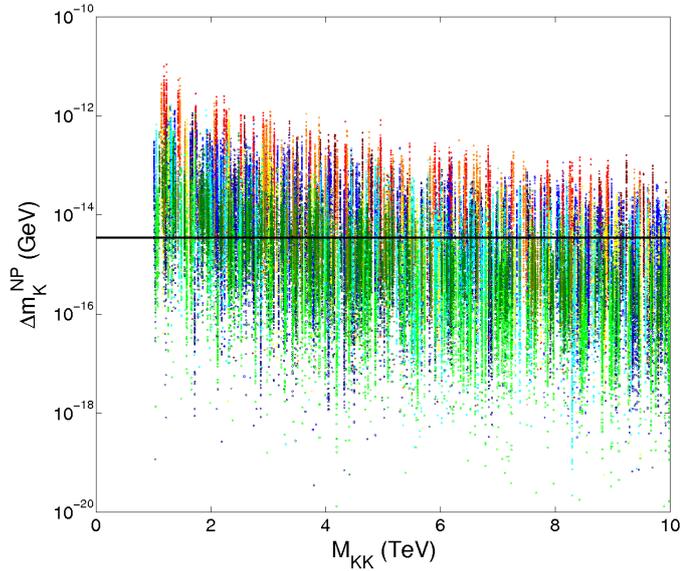}
        }
\caption[$\epsilon_K$ and $\Delta m_K^{\rm{NP}}$ in the RS and the SW Models]{$\epsilon_K^{\rm{NP}}$ and $\Delta m_K^{\rm{NP}}$ for the RS model (stars) and the SW model with $c_1=1.5$ (circles), $c_1=1$ (squares) and $c_1=0.5$ (diamonds). The $c_L$ values are given in Eq. \ref{SWquarkConfig}. For the SW model configuration (A) is plotted in dark blue, (B) is plotted in light blue, (C) is plotted in cyan, (D) is plotted in light green and (E) is plotted in dark green. While for the RS model (A) is plotted in dark red, (B) is plotted in light red, (C) is plotted in orange, (D) in yellow and (E) in dark yellow. For both the RS model and the SW model the mass of the first gauge KK mode will be about two times $M_{\rm{KK}}$. $\Omega=10^{15}$.      }
\label{ fig:epK}
\end{center}
\end{figure}

\begin{figure}[ht!]
    \begin{center}
        \subfigure[]{%
           \label{Fig:NoEpkPoi}
           \includegraphics[width=0.68\textwidth]{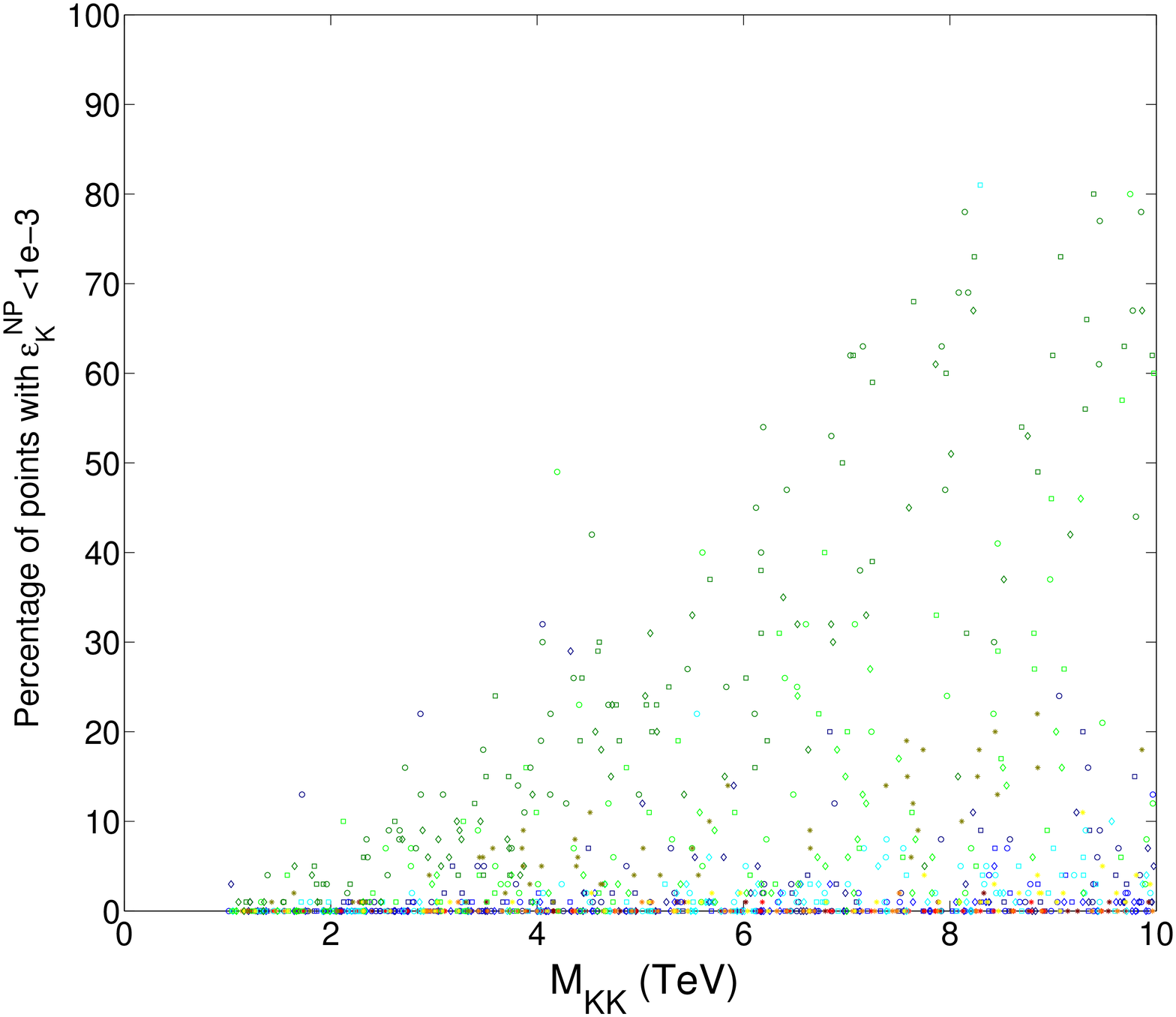}
        }
        \subfigure[]{%
           \label{Fig:NoMkPoi}
           \includegraphics[width=0.68\textwidth]{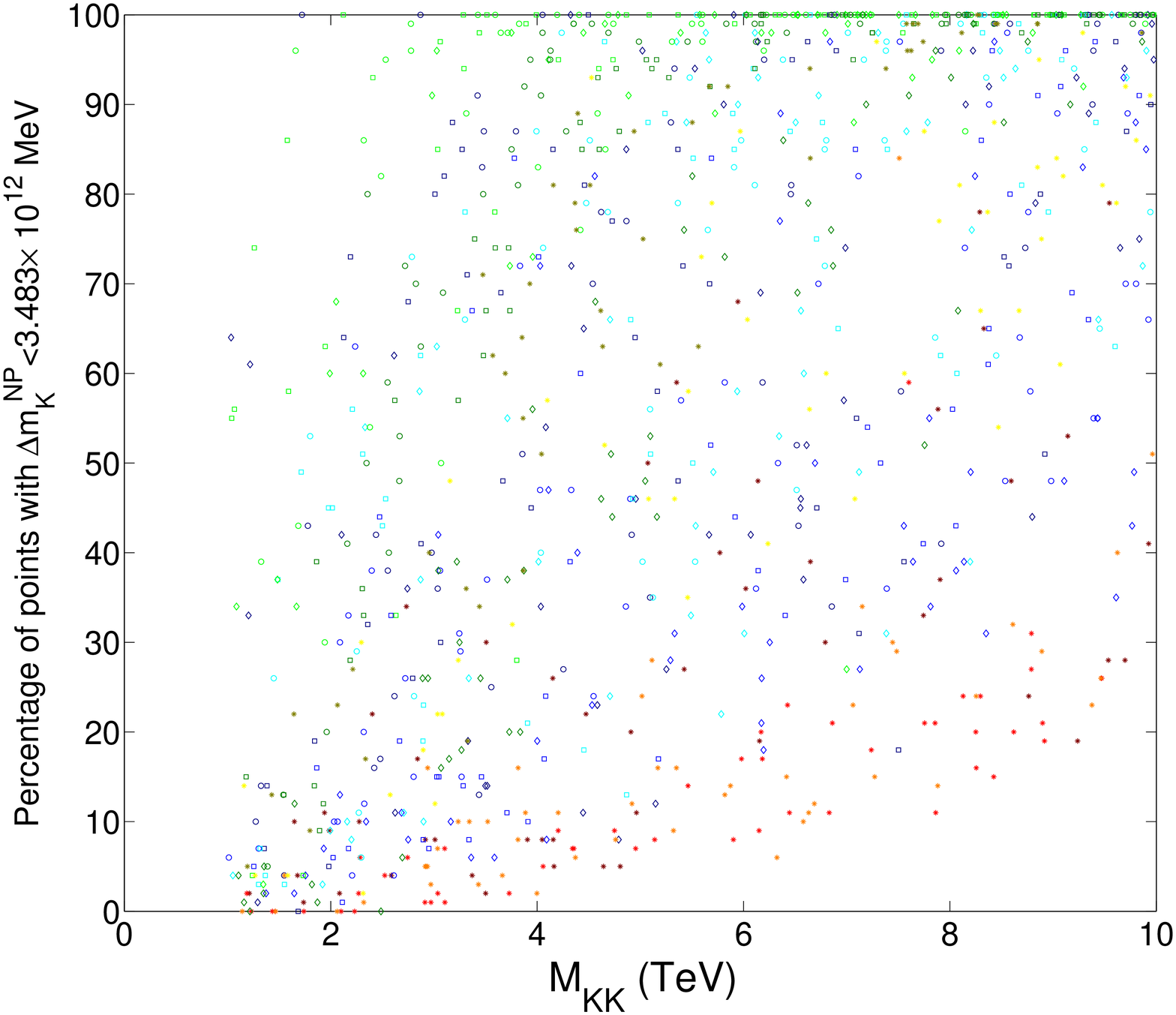}
        }
         \end{center}
    \caption[Percentage of points consistent with experimental values in the RS and the SW model]{The percentage of points that are consistent with experimental values. The colours and points correspond to those used in figure \ref{ fig:epK}. } \label{fig:NumPoints}
\end{figure}

\begin{figure}[t!]
\begin{center}
\includegraphics[width=0.8\textwidth]{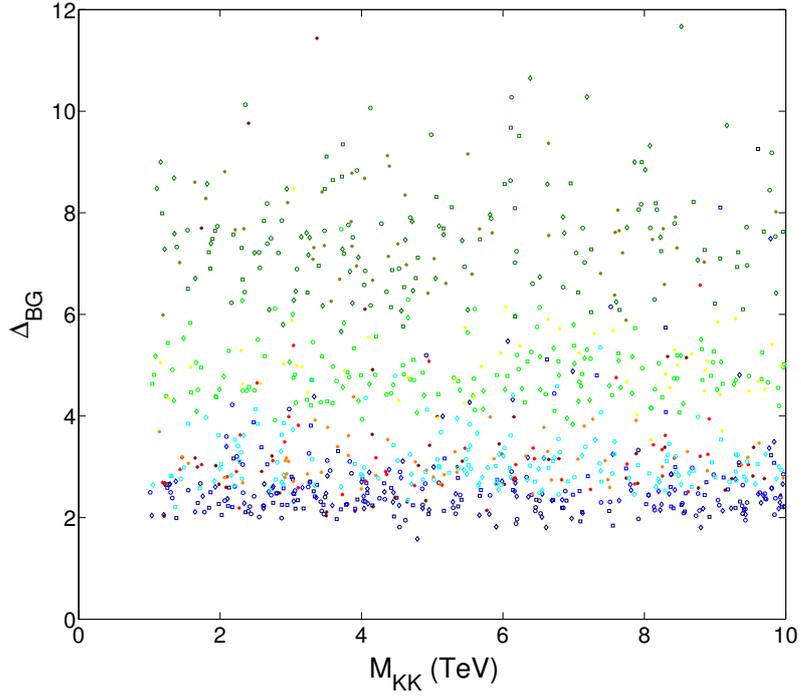}
\caption[Fine tuning in the SW and the RS model]{The fine tuning parameter giving an indication of the sensitivity of the quark masses, mixing angles and Jarlskog invariant to variations in the Yukawa couplings. The colours and points correspond to those used in figures \ref{ fig:epK}. }
\label{fig:finetune}
\end{center}
\end{figure}

 \begin{itemize}
 \item With the exception of the overall warp factor ($\Omega=10^{15}$) and the KK scale, the only input parameters we fix by hand are the bulk mass parameters $c_L$ and the universal $c_1$ parameter. Here we consider five configurations of $c_L$
  \begin{eqnarray}
(A)\quad c_L=[0.72,\;0.64,\;0.52]\quad\quad\quad(B)\quad c_L=[0.69,\;0.63,\;0.49] \nonumber\\
(C)\quad c_L=[0.66,\;0.60,\;0.42]\quad\quad\quad(D)\quad c_L=[0.63,\;0.57,\;0.34] \nonumber\\
(E)\quad c_L=[0.60,\;0.52,\;0.25]\hspace{3cm}\label{SWquarkConfig}
\end{eqnarray}   
 and three $c_1$ values, $c_1=0.5,\;1,\;1.5$. The five $c_L$ configurations have been chosen such that they give roughly the correct mixing angles. Note that for configurations with $0.74<c_0^{L1}(c_L^1)<0.60$ it becomes increasingly difficult to get a good fit to the quark masses without including quite large bulk mass parameters. 
 \item Next we find the `natural' $c_R^u$ and $c_R^d$ values. By natural we mean the bulk mass parameters that give the correct masses and mixing angles assuming that there is no hierarchy in the Yukawa couplings. For this we generate ten sets of two Yukawa matrices and for each one solve for $c_R^{u/d}$ by fitting to the quark masses. To avoid accidentally using a fairly extreme Yukawa, the $c_R$ values used is the median of these ten values. The Yukawa matrices are generated such that $|\lambda_{ij}|\in[1,\;3]$. As discussed in chapter \ref{FlavourChap} we avoid using large Yukawa couplings. These average $c_R$ values are given in table \ref{Tab:BulkmassParams}.
\item With the nine bulk mass parameters fixed, we proceed to find 100 points in parameter space, that give the correct quark masses, mixing angles and Jarlskog invariant by solving for the Yukawa couplings. The quark masses are run up from $2$ GeV to the mass of the first KK gauge mode ($2M_{\rm{KK}}$) using the $2$ GeV values
 \begin{eqnarray}
m_u=2.5\;\rm{MeV}\quad m_c(3\;\rm{GeV})=0.986\;\rm{GeV}\quad m_t=164\;\rm{GeV}\hspace{0.35cm}\nonumber\\
m_d=4.95\;\rm{MeV}\quad m_s=96.2\;\rm{MeV}\hspace{1.8cm} m_b=4.163\;\rm{GeV}\label{Quarkmasses}
\end{eqnarray}
and the mixing angles are \cite{Nakamura:2010zzi,  Bona:2007vi}
 \begin{eqnarray}
\label{expMixangle}
V_{us}=0.2254\pm0.00065\quad\quad\quad V_{cb}=0.0408\pm0.00045\nonumber\\
 V_{ub}=0.00376\pm0.0002\quad\quad\quad J=2.91^{+0.19}_{-0.11}\;\times 10^{-5}.
\end{eqnarray} 

\item By randomly generating the initial guess of the solver one would anticipate that these points would be spread evenly over the parameter space. However one still needs to check the level of tuning required to obtain masses and mixing angles. Hence here we compute the fine tuning parameter \cite{Barbieri88}  
\begin{equation}
\label{ }
\Delta_{BG}(O_i,p_j)=\left |\frac{p_j}{O_i}\frac{\Delta O_i}{\Delta p_j}\right |,
\end{equation}
where the observables, $O_i$, run over all the masses, mixing angle and Jarlskog invariant and the input parameters, $p_j$, run over the all the Yukawas. In practice we vary the Yukawas over a range of $0.1+0.1i$ (i.e. $\Delta p_j$). The fine tuning parameter is then taken as the maximum value with respect to both input parameters and output parameters. Plotted in figure \ref{fig:finetune} is the mean value taken over the 100 points.  
  
\item Having found these 100 viable points in parameter space we proceed to compute the size of the contributions to $\epsilon_K$ and $\Delta m_K$. These results have been plotted in figure \ref{ fig:epK}. For clarity we have also plotted, in figure \ref{fig:NumPoints}, the percentage of the points that are consistent with experimental results. Again in the case of the RS model the KK mass has been scaled by a factor of $\frac{2}{2.45}$ such that the mass of the first gauge KK  mode would be about the same in the two models.

\item This process is then repeated for 50 KK scales, randomly chosen such that $M_{\rm{KK}}\in[1,\;10]$ TeV, for both the RS model and the SW model with $c_1=0.5$, $c_1=1$ and $c_1=1.5$ and also at each configuration in Eq. \ref{SWquarkConfig}.       
 \end{itemize}
 
 As mentioned before, it is hoped that this approach will offer an unbiased spread of points over the region of parameter space of relevance to generating SM fermion masses and mixing angles. Let us turn now to the first question, posed at the beginning of this section, can the SW model generate the correct fermion masses and mixings? As one would expect the answer is yes. For both the RS model and the SW model, not one of the points plotted in figure \ref{fig:finetune} show any sign of significant fine tuning (i.e $\Delta_{BG}>10$). This is of course because of bulk configurations Eq. \ref{SWquarkConfig} taken as input parameters. Had one chosen more UV or IR localised configurations then clearly more tuning would have been required in order to get the correct masses and mixing angles.   
 
This then brings us to the second question, related to the the suppression of FCNC's. By grouping points in parameter space according to their configurations of bulk mass parameters Eq. \ref{SWquarkConfig}, one is essentially comparing points which require equivalent levels of tuning in order to obtain the correct masses and mixing angles (see figure \ref{fig:finetune}). Even with this comparison it is still difficult to meaningfully quantify the extent to which FCNC's are suppressed. None the less one can see that, for all points considered, FCNC's are considerably more suppressed in the SW model than in the RS model with an equivalent level of tuning. For example, one can see from figure \ref{Fig:NoEpkPoi} that, in the SW model with configuration (E), in order to have about 20\% of the points consistent with $\epsilon_K$ one would require $M_{\rm{KK}}\gtrsim 4$ TeV (corresponding to a KK gluon mass of $\sim 8$ TeV). Where as in the RS model one would require $M_{\rm{KK}}\gtrsim 10$ TeV (corresponding to a KK gluon mass of $\sim 25$ TeV). Alternatively one can look at the total number of points that satisfy the $\epsilon_K$ constraint regardless of KK scale. In the SW for configuration (E) with $c_1=1.5,\;1,\;0.5$ this is about $31\% ,\; 31\% ,\; 19\%$ respectively. While for configuration (A) this is about  $4.2\% ,\; 2.5\% ,\; 2.6\%$. This can be compared to the RS model which is about $8\%$ for configuration (E) and $0.22\%$ for configuration (A).           

There are a number of factors contributing to this increased suppression of FCNC's. The extent to which FCNC's are suppressed is largely determined, at tree level, by how universal the gauge-fermion couplings, that appear in Eq. \ref{fourfermInt}, are. As can be seen in figure \ref{fig:FermCoupl}, as one localises the fermions further and further towards the UV, the gauge fermion coupling becomes increasingly universal. While this is true for both the SW model and the RS model one can also see, from figure \ref{fig:UVFermCoupl}, that in the SW model, with $c_1\gtrsim 1$ that the gauge fermion couplings are slightly more universal than the RS model. In addition to this in the SW model, due to the presence of the bulk Higgs, the fermions will typically sit slightly further towards the UV than in the RS model (see table \ref{Tab:BulkmassParams}). This effect should be combined with the reduced $c_0$ dependence in the range of possible fermion masses (the gradient in figure \ref{fermmass}). This results in avoiding extreme bulk mass parameters in configurations in which either the left-handed or right-handed fermions are localised quite far towards the UV. For example, see $c_R^{u3}$ for configurations (A) and (B) in table \ref{Tab:BulkmassParams} for an extreme example. When all these effects are combined one finds that, for all points in parameter space considered, FCNC's are more suppressed in the SW model than in the RS model. 

One of the underlying assumptions in this analysis is that of a universal $c_1$ value. However, upon examining the results one can see that, despite most of the bulk mass parameters of relevance to kaon physics having $c_0^R\lesssim-0.6$, there is still a very small $c_1$ dependence. In particular smaller $c_1$ values tend to give slightly more UV localised fermions and slightly less universal gauge fermion couplings. These effects are quite small and hence negligible when compared to varying the bulk mass parameters, $c_0$. So here we would argue that these results would not change significantly if one was to relax this assumption. None the less, since the exact origin of the $c_1$ term has not been clearly defined it is not clear if this number can be quite large or small. Here we shall leave investigation of this to future work.     

It should also be stressed that this cannot be considered a complete study for a number of reasons. Firstly we only consider tree level gauge mediated FCNC's. One would also anticipate additional contributions to FCNC's arising from, for example, the dilaton or the Higgs \cite{Casagrande:2008hr, Azatov:2009na, Agashe:2009di, Buras:2010mh}. The inclusion of such additional contributions would inevitably enlarge the parameter space and hence make a fair comparison with the RS model more difficult. However one would anticipate that such fields would be IR localised and hence such FCNC's should also be suppressed. Secondly we also only focus on kaon physics. One can see from table \ref{Tab:BulkmassParams} that a study involving top physics or B physics would involve fermions sitting further towards the IR where the assumption of universal $c_1$ values is arguably less valid. So here we will leave a more comprehensive study to future work.   

It is important to realise that the central physics involved in this result is not necessarily related to the soft wall but rather the change in the relationship between fermion masses and their positions. In particular the reduction of the gradient in figure \ref{fermmass}. One would anticipate such an effect showing up, to some extent, in any bulk Higgs scenario. However soft wall models offer a framework in which the Higgs can propagate in the bulk and the gauge hierarchy problem can still potentially be resolved.  
\color{black}

\section{Summary}
We began this chapter by wanting to investigate the phenomenological implications of changing the nature of the IR cut off in the RS model. As with changing the geometry, when one removes the IR brane one introduces many new free parameters to the model. In particular the exponent of the dilaton and the form of the Higgs potential (or the form of whatever is responsible for EW symmetry breaking). A consequence of this second point, of relevance to flavour physics, is that one can obtain a change in the relationship between bulk mass terms and zero mode masses. To be more specific, one can quite generically anticipate that a $z$ dependent bulk Higgs VEV would give rise to a minimum in the zero mode masses of UV localised fermions. Further still the possible masses no longer need to scale down, from the SM Higgs VEV, in order to be compatible with EW observables. Here, based on earlier work, we have studied the implications on flavour physics of the particular case of a quadratic Higgs VEV and we have found a significant deviation, in the implication for flavour physics, from that of the RS model. Clearly a more complete study should really be carried out in conjunction with a study of the EW sector. Nonetheless this rather limited study has hinted towards some potentially interesting physics. For example, potential explanations of the small mass hierarchy for light fermions compared with the large mass hierarchy for heavy fermions, as well as an increased suppression in gauge mediated FCNC's, relative to that of the RS model. 

In terms of addressing the question of what spaces offer resolutions to the gauge hierarchy problem, then one can see that there are a number of related factors. In the SW model, how sharply one cuts off the space will in turn be related to the form of the Higgs VEV. In models with a bulk Higgs, the form of the Higgs VEV will determine whether or not a small EW scale can be generated, without fine tuning, as well as determining the size of corrections to EW observables and the range of possible fermion masses. These last two points combined will dictate whether it is possible to locate all the fermions in a region of parameter space with universal gauge fermion couplings and hence suppress constraints from FCNC's. At this stage it is not clear how all these factors are related and so we are unable to offer a full answer to the initial question.

%% file: Conclusion.tex
\chapter{Conclusion}
\label{chap:conc}

Over the past decade there has been a surge of interest in the phenomenological implications of models with warped extra dimensions for a number of reasons. Although, arguably, the primary reason is the ability of such models to offer geometrical explanations of the two apparent hierarchies in the SM: Firstly, by localising the Higgs towards the IR tip of the extra dimension, one can explain why the Planck scale is so much larger than the EW scale and hence offer a potential resolution to the gauge hierarchy problem. Secondly, by allowing the fermions to propagate in the bulk and including a bulk mass term, one can localise the lighter fermions away from the, IR localised, Higgs and hence offer an explanation for the apparent hierarchy in the effective Yukawa couplings.  

Up to now the bulk of this work has focused on the original RS model based on AdS${}_5$. However there are a number of reasons for supposing a more complex geometry. In particular, superstring theory requires ten and not five dimensions. If those five extra dimensions are included as a fibered $S_5$ manifold then the corresponding solution of type IIb string theory is conjectured to be dual to a conformal field theory. In any realistic scenario one would anticipate any conformal symmetry would be broken in the low energy regime and it is reasonable to speculate that this would correspond to some deformation of the AdS${}_5\times S_5$ geometry. Alternatively from a `bottom up' perspective, one can simply ask what spaces offer viable resolutions to the hierarchy problem.  This thesis has attempted to investigate what would be the phenomenological implications of such deformations, assuming the spaces still offer explanations of the two hierarchies mentioned above.  

Before a space can hope to explain these two hierarchies it must have a large warp factor while at the same time, we require, that all the dimensionful parameters (e.g. curvature and radius) exist at about the same order of magnitude. In all cases considered, these two requirements implies that the KK scale should be close to that of the EW scale. Hence for a space to offer a viable resolution to the gauge hierarchy problem two criteria must be met: Firstly, one must be able to demonstrate that the space can be stabilised, without fine tuning, so as to give a large warp factor. Secondly, one must be able to satisfy all existing experimental constraints without resorting to a KK scale many orders of magnitude larger than the EW scale. This thesis has primarily focused on the second criteria and in doing so one is essentially excluding spaces that do not resolve the gauge hierarchy problem rather than proposing ones that do. Hence we have often considered a `best case scenario' in which corrections to the SM observables are dominated by the gauge fields and additional contributions, from for example the gauge scalars, are assumed to be small. None the less the stability of the spaces and size of additional contributions can be checked in future work. 

Our strategy has been to compute the size of the corrections to EW observables and flavour observables arising from the exchange of KK gauge fields propagating in a reasonably generic space. This of course means our conclusions are only valid for spaces in which the SM gauge fields propagate in the extra dimensions. It is found, rather unsurprisingly, that the size of corrections to EW observables are essentially determined by the mass of the KK gauge fields and their relative coupling to the Higgs and to the SM fermions. Likewise it is found that the coefficients of four fermion operators, that give rise to FCNC's, are largely determined by the level of universality of the couplings between KK gauge fields and SM fermions. 

Next we proceeded to demonstrate, for generic five dimensional space with a large warp factor, that the KK modes are heavily peaked towards the IR tip of the space. This is possibly equivalent to confinement, in the dual theory, always occurring in the IR of the theory. This has a number of important consequences. Firstly the couplings between KK gauge fields and fermions sitting towards the UV tip of the space will not be sensitive to how UV localised the fermions are. I.e. the couplings will be close to universal and  FCNC's will be partially suppressed. Secondly the same gauge fermion coupling will be suppressed relative to the gauge zero mode coupling. This can be used to partially suppress EW constraints, in particular the contributions to the $S$ parameter. Thirdly if the Higgs is localised on the IR brane then its coupling to the KK gauge fields will be enhanced. This gives rise to sizeable corrections to EW observables, in particular the $T$ parameter. 

In more than five dimensions, while the KK modes are still peaked towards the UV, there \color{red} are \color{black} now two new effects: Firstly the spacing between the KK modes, which is determined by the radius and curvature of the extra dimension, is no longer necessarily of a similar order of magnitude for all of the extra dimensions. Hence the overall KK spectrum need not be evenly spaced. Secondly the volume of the internal space will now scale the 4D effective couplings. Critically when this radius is dependent on the additional dimension, i.e. the internal space is also warped, then this effect will scale the KK modes couplings differently to that of the flat zero mode. If the Higgs is localised on the IR brane then this has a significant effect on the gauge Higgs couplings. In particular when the internal space is shrinking towards the IR brane the couplings between KK gauge modes and the Higgs is enhanced relative to the zero mode. This leads to strongly coupled gauge modes and potentially large EW corrections and hence such spaces would probably not offer resolutions to the gauge hierarchy problem. This includes a large class of possible spaces including spaces such as the Klebanov-Strassler solution and spaces that arise from an isotropic cosmological constant. Although here we have not looked at the dual theories one can speculate that this class of spaces may be dual to any field theory in which the number of colours is being reduced as one moves into the IR. 

Conversely if the internal space is growing towards the IR then the couplings of the gauge KK modes will be suppressed by the same effect and hence the contribution to EW observables from KK gauge fields can be significantly reduced. However in such spaces the new spacings in KK masses are typically found to be much smaller than the KK scale. Hence such spaces would have a high density of KK modes or in a potential dual theory a high density of bound states. It is not clear what the phenomenological implications of this high density of states would be. It is also not clear what the dual field theory would be, if one existed. 

It is important to note that the above conclusions are sensitive to the location of the Higgs. One would anticipate that if the Higgs was localised w.r.t more than one direction then the EW constraints would increase, due to an increase in the number of KK modes contributing to EW observables. Conversely if the Higgs was not localised w.r.t the `warped direction' $r$, then the 4D effective gauge Higgs coupling would arise via an overlap integral and one would anticipate that the above effect could be slightly reduced. The extent to which the effect is reduced would depend on the extent to which the Higgs sits towards the IR. 

Although models with a bulk Higgs must then demonstrate that a small EW scale can still be generated. A model in which this has been shown is the soft wall model and as one would anticipate such models are found to have reduced EW constraints. However in considering a bulk Higgs one finds that the 4D effective Yukawa couplings, of the fermion zero modes, scale relative to the KK scale. Also one finds that, exactly analogous to the minimum gauge fermion coupling, that gives rise to the partial suppression of FCNC's, there now exists a minimum 4D effective Yukawa coupling. Hence models with a bulk Higgs will inherently give rise to a minimum fermion mass. In the SW model studied here, with a quadratic Higgs VEV, this minimum mass sits about fifteen orders of magnitude lower than the KK scale which appears to be in rough agreement with the observed fermion masses. Further still in such a scenario it is possible for all the fermion zero modes to sit towards the UV tip of the space where their couplings, to KK gauge fields, are approximately universal. Hence the constraints from FCNC's are significantly reduced compared to those of the RS model. This is again all very sensitive to the form of the bulk Higgs VEV and hence further work is really needed to understand what Higgs VEVs are possible in models that resolve the hierarchy problem.

So in conclusion it has been found that, if one assumes that a given space resolves the gauge hierarchy problem, then any deviation from AdS${}_5$ that includes additional warped dimensions results in a significant change in the phenomenological implications, from that of the RS model. In particular either an increased density of KK modes or KK modes that are strongly coupled to fields peaked in the IR. Likewise small changes in the location of the Higgs (or whatever is responsible for breaking EW symmetry) can \color{red} result \color{black} in significant changes in the relationship between fermion masses and fermion locations. This would, in turn, lead to a very different descriptions of flavour than that of the RS model.   Hence we would argue that in an `LHC era', in which one should keep as open a mind as possible, it is not sufficient to restrict phenomenological studies, of warped extra dimensions, to that of just studying the RS model.